\documentclass[11pt]{article}
\RequirePackage{amsthm,amsmath,amsfonts,amssymb,mathbbol,float,enumitem,verbatim,latexsym,xcolor,comment,babel,lmodern}
\RequirePackage[numbers]{natbib}
\usepackage{etoolbox}
\usepackage{xstring}
\usepackage{xr}
\usepackage[affil-it]{authblk}

\RequirePackage{graphicx}

\theoremstyle{plain}
\newtheorem{theorem}{Theorem}
\newtheorem{lemma}{Lemma}
\newtheorem{proposition}{Proposition}
\newtheorem{assumption}{Assumption}
\newtheorem{remark}{Remark}
\newtheorem{corollary}{Corollary}

\graphicspath{{./fig/}}

\newcommand{\cdummy}{\cdot}
\newcommand{\infixand}{\text{ and }}
\newcommand{\nosymbol}{}
\newcommand{\tmmathbf}[1]{{\boldsymbol{#1}}}
\newcommand{\tmop}[1]{{\operatorname{#1}}}
\newcommand{\tmtextbf}[1]{\text{{\bfseries{#1}}}}

\usepackage[paper=a4paper,dvips,top=2cm,left=2cm,right=2cm,
foot=1cm,bottom=2cm]{geometry}
\allowdisplaybreaks

\title{Consistent Bayesian Spatial Domain Partitioning Using Predictive Spanning Tree Methods}

\author[1]{Kun Huang}
\author[2]{Huiyan Sang}

\affil[1]{Department of Statistics, Texas A\&M University, College Station, US. Email: k-huang@tamu.edu}
\affil[2]{Department of Statistics, Texas A\&M University, College Station, US. Email: huiyan@stat.tamu.edu}

\date{} % Or use \date{} for no date

\begin{document}

\maketitle

% Abstract section
\begin{abstract}

Bayesian model-based spatial clustering methods are widely used for their flexibility in estimating latent clusters with an unknown number of clusters while accounting for spatial proximity. Many existing methods are designed for clustering finite spatial units, limiting their ability to make predictions, or may impose restrictive geometric constraints on the shapes of subregions. Furthermore, the posterior clustering consistency theory of spatial clustering models remains largely unexplored in the literature. In this study, we propose a Spatial Domain Random Partition Model (Spat-RPM) and demonstrate its application for spatially clustered regression, which extends spanning tree-based Bayesian spatial clustering by partitioning the spatial domain into disjoint blocks and using spanning tree cuts to induce contiguous domain partitions. Under an infill-domain asymptotic framework, we introduce a new distance metric to study the posterior concentration of domain partitions. We show that Spat-RPM achieves a consistent estimation of domain partitions, including the number of clusters (which may go to infinity), and derive posterior concentration rates for partition, parameter, and prediction. We also establish conditions on the hyperparameters to achieve consistency, offering important practical guidance for hyperparameter selection. 
Finally, we examine the asymptotic properties of our model through simulation studies and apply it to Atlantic Ocean data.

% We propose a Bayesian framework to model spatially heterogeneous data with an unknown piecewise constant parameter, where a latent partition defines regions of constant values. We segment the spatial domain into disjoint blocks and construct spanning trees on these blocks. A contiguous domain partition is then induced by removing some edges and assigning locations within the same block to the same cluster.
% We assign priors to the spanning trees, the number of clusters, and the induced partitions, after which we derive the posterior distribution of the unknown parameter and domain partition. We establish conditions on the priors and the number of blocks, under which the spatial domain partition and parameter estimation achieve posterior consistency.
% Based on the partition results, we propose a prediction method for newly observed locations and analyze the contraction rate of the prediction error. We conduct simulation studies to examine the asymptotic properties of our model and compare its performance with the BSCC model. Finally, we apply our model to Atlantic Ocean data to study its temperature-salinity relationship.

\end{abstract}

%%%%%%%%%%%%%%%%%%%%%%%%%%%%%%%%%%%%%%%%%%%%%%%%%%%%%%%%%%%%%%%%%%%%%%%%%%%%%%%%%%%%%%%%%%%%%%%%%%%%%%%%%%%%%%%%%%%%%%%%%%%%
\newpage
\section{Introduction}

% Spatial heterogeneity is increasingly recognized as a pervasive characteristic
% in data collected from space, particularly in fields such as urban studies and epidemiology (e.g., {\cite{pickett2017dynamic}},
% {\cite{yin2022analysis}}). 
%This variability across space is often driven by underlying differences in physical, social, or biological factors, resulting
%in distinct spatial patterns.
Spatial clustering models ~\cite{dasgupta1998detecting,fraley2002model,cadez2000general,paci2018dynamic} are essential for identifying and characterizing spatial
 heterogeneity, enabling a deep understanding of spatial patterns and the underlying differences in physical, social, or biological driving factors. 
%for improving the accuracy of models and ensuring
%robust inferences (see {\citealp{wang2024statistical}}).
% Model-based spatial clustering methods (MSCM) assume a probabilistic framework to partition spatial units or a spatial domain into clusters, taking into account their spatial proximity and similarities. %These methods have attracted significant attraction and widespread applications for modeling spatial heterogeneity, leading to a rich literature, see, e.g.,~\cite{dasgupta1998detecting,fraley2002model,cadez2000general,paci2018dynamic}. 
%In particular, 
Bayesian spatial clustering models have gained great popularity due to their flexibility in modeling latent clustered variables within Bayesian hierarchical frameworks while accounting for spatial information. % and their ability to estimate an unknown number of clusters. 
These models specify a prior distribution over the partition space and fit probabilistic models to the data to estimate underlying cluster memberships and other model parameters. 
One concrete example is the spatially clustered regression model (see \cite{lee2017cluster,shenweining,li2019spatial,zhong2023sparse}), where the goal is to study the spatial heterogeneity in the latent relationship between covariates and spatial response. Let $\{\tmmathbf{s}_i,
\tmmathbf{x}(\tmmathbf{s}_i), y (\tmmathbf{s}_i)\}_{i = 1}^n$ be the spatial
data observed at locations $\tmmathbf{s}_1, \ldots, \tmmathbf{s}_n \in
\mathcal{D} \subset \mathbb{R}^2$, where $\tmmathbf{x}(\tmmathbf{s}_i)$ is d-dimensional covariate and $y(\tmmathbf{s}_i)$ is the response variable at $\tmmathbf{s}_i$. Conditional on $\{
\tmmathbf{x}(\tmmathbf{s}_i)\}_{i = 1}^n$, we write the likelihood of  $\{{y}(\tmmathbf{s}_i)\}_{i=1}^{n}$ as $\prod_{i =
1}^n \mathbb{P}_{\tmmathbf{\theta} (\tmmathbf{s}_i)} \{y (\tmmathbf{s}_i) \mid
\tmmathbf{x}(\tmmathbf{s}_i)\}$, where $\mathbb{P}_{\tmmathbf{\theta}
(\tmmathbf{s}_i)} \{\cdummy \mid \tmmathbf{x}(\tmmathbf{s}_i)\}$ is
the conditional probability density function of $y (\tmmathbf{s}_i)$ with unknown
parameter $\tmmathbf{\theta} (\tmmathbf{s}_i)$. To account for spatial heterogeneity, we assume $\tmmathbf{\theta} (\tmmathbf{s}_i) =\tmmathbf{\theta} (\tmmathbf{s}_{i'})$ if
$\tmmathbf{s}_i$ and $\tmmathbf{s}_{i'}$ belong to the same cluster, and
$\tmmathbf{\theta} (\tmmathbf{s}_i) \neq \tmmathbf{\theta}
(\tmmathbf{s}_{i'})$ otherwise. 

Most existing Bayesian spatial clustering methods consider a finite random partition prior model to cluster the $n$ observed locations into $k_0$ disjoint sub-clusters~\citep{page_spatial_2016, sugasawa2021spatially,  quintana2022dependent,hu2023bayesian}, but they cannot be used to predict cluster memberships, regression parameters, or responses at new locations. Alternatively, a domain partition model considers a random partition of the entire domain $\mathcal{D}$ into $k_0$ disjoint sub-domains, say $\{\mathcal{D}_{l, 0} \}_{l= 1}^{k_0}$, such that $\cup_{l = 1}^{k_0} \mathcal{D}_{l, 0} =\mathcal{D}$, making it suitable for prediction tasks. Popular domain partition models include binary decision trees \citep{denison1998bayesian}, which recursively split the domain into non-overlapping hyper-rectangular regions, and Voronoi tessellation models \citep{knorr2000bayesian, kim2005analyzing,feng2016spatial}, which partition the domain into convex polygons. However, these shape constraints might be too restrictive for some applications. 
%However, both methods impose restrictive shape constraints on subregions. % and are not suitable for domains with irregular boundaries.  

% Binary decision-tree based methods~\citep{denison1998bayesian,gramacy2008bayesian} partition the domain into non-overlapping regions by making binary splits recursively. Voronoi tessellation based models \citep{knorr2000bayesian, kim2005analyzing,feng2016spatial} employ a set of central locations to define regions such that every point within a given region is closer to its assigned center than to any other central locations. However, Binary decision trees produce rectangular-shaped partitions while Voronoi tessellations imply a convexity assumption on the region shapes, limiting their applicability. 

While spatial clustering methods have been widely studied and applied, their theoretical development remains relatively limited. Existing theoretical work of spatially clustered regression model mostly focuses on showing that the posterior is a consistent estimate of the true regression parameter or data-generating density%governing the data-generating mechanism, 
~\cite{luo2023nonstationary,nguyen2013convergence}, while the more relevant problem of clustering consistency has not been well studied due to its theoretical challenges. Many existing work on Bayesian posterior clustering consistency (see, e.g., \cite{guha2021posterior,miller2018mixture,zeng2023consistent,ascolani2023clustering})  focuses on exchangeable random set partition models~\cite{gnedin2006exchangeable}. % such as the Chinese restaurant process model~\cite{ascolani2023clustering} or the mixture of finite mixture model~\cite{miller2018mixture}. 
%\cite{}
%Furthermore, much of the literature studies the posterior consistency of the number of clusters (), rather than the consistency of cluster memberships. 
Nevertheless, these exchangeable random partition priors differ fundamentally from those used in spatial clustering models, making some of the existing theoretical tools unsuitable for directly analyzing spatial clustering. Most recently, \cite{zheng2024consistency} establishes clustering consistency under a mixture model framework, assuming that the data are generated dependently from a disjoint union of component graphs. \cite{shenweining} establishes clustering consistency for spatial panel data assuming the number of repeated measurements goes to infinity. However, in spatial statistics, it is more common and reasonable to assume either an infill-domain asymptotic or an increasing-domain asymptotic framework, where the number of spatial locations goes to infinity. 
% For the data with form $[\{\tmmathbf{s}_i,
% \tmmathbf{x}(\tmmathbf{s}_i), y (\tmmathbf{s}_i)\}_{i = 1}^n$, 

%Spanning trees have been extensively utilized for clustering in the existing literature, see .

% \citep{murtagh1985survey, legendre1987constrained, shirabe2005model, page2016spatial, validi2022imposing,hu2023bayesian}, and those with hard contiguity constraint where each cluster is a spatially connected component
%  \citep{ambroise1998convergence,assunccao2006efficient, blei2011distance,li2019spatial, wang2023spatial}

% Partition consistency refers to that the latent domain partition $\{\mathcal{D}_{l, 0} \}_{l
% = 1}^{k_0}$ can be consistently estimated as sample size $n$ goes to infinity.
 % a widely used tool in model-based partitioning analysis (see \cite{zahn1971graph,grygorash2006minimum,luo2021bayesian,luo2023nonstationary,lee2021t,li2019spatial}),
%our model
We propose a spatial domain random partition model (Spat-RPM) in the context of spatially clustered regression for data $[\{\tmmathbf{s}_i,
\tmmathbf{x}(\tmmathbf{s}_i), y (\tmmathbf{s}_i)\}_{i = 1}^n$. The model extends the finite spanning tree prior
\citep{teixeira2019bayesian,luo2021bayesian,lee2021t}
%used in a Bayesian spatially clustered varying coefficient model regression (BSCC, \cite{luo2021bayesian}) 
by modeling latent domain partition  $\{\mathcal{D}_{l, 0} \}_{l
= 1}^{k_0}$. In Spat-RPM, we first discretize the domain into small
disjoint blocks. We construct spanning trees on blocks, based on which a contiguous domain partition is induced after removing some edges and assigning locations within the same block to the same cluster. 
We assign priors on spanning trees, the number of clusters, and the induced partitions. We design an efficient Bayesian inference algorithm to draw posterior samples of domain partitions and $\tmmathbf{\theta}(\cdot)$. We show in our numerical examples that Spat-RPM %inherits the advantageous properties of spanning tree-based finite spatial clustering models \citep{teixeira2015generative,li2019spatial,duan2023bayesian}, including 
%flexibility in accommodating irregular domains, 
produces spatially contiguous clusters with more flexible shapes while enabling spatial predictions. % can be derived afterward. 
% Spat-RPM inherits the advantageous properties of spanning tree-based finite spatial clustering models \citep{teixeira2015generative,li2019spatial,duan2023bayesian}, including flexibility in accommodating irregular domains, producing clusters with flexible shapes, and guaranteeing spatial contiguity, while substantially broadening the applicability
% by enabling spatial predictions. Furthermore, the blocking strategy enables Spat-RPM to handle larger-scale data by reducing the computational burden with a smaller partition space. 

The blocking technique in Spat-RPM reduces the infinite domain partition space to the finite blocking partition space, making the estimation practical. The blocking also enables Spat-RPM to handle larger-scale data by reducing the computational burden. Under a mild assumption on the Minkowski dimension of the true partition boundary set and the shape of each subregion, we study the approximation error between the blocking partition space and the true domain partition. In our theoretical analysis, we establish conditions on the asymptotic rate of the number of blocks, balancing the trade-off between approximation error and partition space dimensionality to achieve partition consistency.

% Among these, spanning tree-based spatial clustering models ~\citep{teixeira2015generative, teixeira2019bayesian,luo2023nonstationary,li2019spatial,lee2021t} have gained considerable attention due to their flexibility in accommodating irregular domains, ability to produce clusters with flexible shapes, and guarantee of spatial contiguity such that data points within a cluster are geographically adjacent or connected. %~\citep{li2019spatial,validi2022imposing,hu2023bayesian,wang2023spatial}. A popular approach to ensuring contiguity constraints involves the use of spanning tree cuts , %which relies on partitions of connected graphs to guarantee contiguity and can adapt to irregular domains. 
% Given a connected spatial graph where each vertex represents an observed spatial unit, a spanning tree is defined as a subgraph that contains no cycles and connects all vertices. Contiguous clusters of spatial units are obtained by removing some edges from the spanning tree to form connected components of graph vertices. This nice property has motivated several generative random graph partition models prior for spatial clustering~\citep{teixeira2019bayesian,luo2021bayesian,luo2021bast}. 

We conduct Bayesian posterior theoretical analysis for Spat-RPM, assuming an infill-domain asymptotic framework. We establish the domain partition consistency theory from the ground up. To study partition consistency, we formally define a valid distance metric for comparing two domain partitions. We show that under the defined metric, the posterior domain partition converges to the true partition, given certain conditions including the growing rate of the cluster number, minimum separation between cluster-specific parameters, regularity constraints on the domain boundary, and orders of hyperparameters. The clustering consistency of the observed locations follows directly from the domain partition consistency result. We also show that the number of clusters can be consistently estimated. 
Since the blocking partition prior model involves spanning trees with a diverging number of nodes as $n$ increases, we derive several original graph-theoretical results related to spanning trees to establish partition consistency, which may be of independent interest for future research. Furthermore, based on the partition consistency, we show the Bayesian posterior contraction rate of $\tmmathbf{\theta}(\cdot)$ and prediction error. To the best of our knowledge,  our work is among the first to develop Bayesian spatial domain partitioning consistency under the spatial infill domain asymptotic framework. %aeither of these two asymptotic frameworks.

The rest of the paper is organized as follows. We introduce Spat-RPM in Section \ref{SEC:MODEL} and present theoretical results in
Section \ref{SEC:mainresult}. In Section \ref{SEC:simu}, we conduct numerical
simulations to examine the properties of Spat-RPM.
%illustrate the effectiveness of our model.% and make a comparison
%with the BSCC model proposed by {\cite{luo2021bayesian}}. 
In Section
\ref{SEC:realdata}, we apply our model to real data to demonstrate results. Supplementary Material contains all technical proofs and additional model details.

\section{Methodology}\label{SEC:MODEL}

% Recall that the data is $[\{\tmmathbf{s}_i, \tmmathbf{x}(\tmmathbf{s}_i), y
% (\tmmathbf{s}_i)\}_{i = 1}^n$ with probability model $\{y (\tmmathbf{s}_i)
% \mid \tmmathbf{x}(\tmmathbf{s}_i)\}_{i = 1}^n \sim \prod_{i = 1}^n
% \mathbb{P}_{\tmmathbf{\theta} (\tmmathbf{s}_i)} \{y (\tmmathbf{s}_i) \mid
% \tmmathbf{x}(\tmmathbf{s}_i)\}$, where
% $\{\tmmathbf{\theta}(\tmmathbf{s}_i)\}_{i = 1}^n$ are spatially piecewise
% constants. 
%In this section, we introduce the proposed Spat-RPM and the corresponding Bayesian inference algorithm.  

\subsection{Background of graphs and spanning trees}
We start by introducing some concepts and notations of graphs. Let
$\mathcal{V} = \{\tmmathbf{v}_1, \ldots,
\tmmathbf{v}_{n^{\ast}} \}$ be $n^{\ast}$ vertices and $\mathcal{G}=
(\mathcal{V}, \mathcal{E})$ be an undirected graph, where the edge set
$\mathcal{E}$ is a subset of $\{(\tmmathbf{v}_i,
\tmmathbf{v}_{i'}) : \tmmathbf{v}_i, \tmmathbf{v}_{i'}
\in \mathcal{V}, \tmmathbf{v}_i \neq \tmmathbf{v}_{i'}
\}$. We call a sequence of edges $\{(\tmmathbf{v}_{i_0},
\tmmathbf{v}_{i_1}),(\tmmathbf{v}_{i_1},
\tmmathbf{v}_{i_2}), \ldots, (\tmmathbf{v}_{i_{t - 1}},
\tmmathbf{v}_{i_t})\} \subseteq \mathcal{E}$ as a path of length $t$
between $\tmmathbf{v}_{i_0}$ and $\tmmathbf{v}_{i_t}$, if all
$\{\tmmathbf{v}_{i_j}\}_{j=0}^{t}$ are distinct. A path is called a cycle if
$\tmmathbf{v}_{i_0} =\tmmathbf{v}_{i_t}$ and all other vertices
are distinct. A subgraph $(\mathcal{V}_0, \mathcal{E}_0)$, where $\mathcal{V}_{0}
\subseteq \mathcal{V}$ and $\mathcal{E}_0 \subseteq \mathcal{E}$, is called a
connected component of $\mathcal{G}$ if there is a path between any two vertices and there is no path
between any vertex in $\mathcal{V}_0$ and any vertex in
$\mathcal{V} \setminus \mathcal{V}_0$, the difference between $\mathcal{V}$ and $\mathcal{V}_0$. Given an undirected
graph $\mathcal{G}= (\mathcal{V}, \mathcal{E})$, a subset
$\mathcal{V}_0 \subseteq \mathcal{V}$ is a contiguous
cluster if there exists a connected subgraph $\mathcal{G}_0 =
(\mathcal{V}_0, \mathcal{E}_0)$, where $\mathcal{E}_0 \subseteq
\mathcal{E}$. We say $\pi (\mathcal{V}) = \{\mathcal{V}_1,
\ldots, \mathcal{V}_k \}$ is a  contiguous
partition of $\mathcal{V}$ with respect to $\mathcal{G}$, if $\mathcal{V}_j \subseteq
\mathcal{V}$ is a contiguous cluster for $j = 1, \ldots, k$, $\cup_{j =
1}^k \mathcal{V}_j =\mathcal{V}$, $\mathcal{V}_j \cap
\mathcal{V}_{j'} = \emptyset$ for $j \neq j'$. For simplicity, we refer to contiguous partitions (clusters) simply as
partitions (clusters) in the following context. 

% We say a vertex set
% $\mathcal{S}_0^{\ast}$ is connected under $\mathcal{G}$, if there exists a
% path from $\tmmathbf{s}_m^{\ast}$ to $\tmmathbf{s}_{m'}^{\ast}$ with all the
% vertices in path contained in $\mathcal{S}_0^{\ast}$, for any two vertices
% $\tmmathbf{s}_m^{\ast}, \tmmathbf{s}_{m'}^{\ast} \in \mathcal{S}_0^{\ast}$.

\begin{figure}[htbp]
		\centering
		\begin{tabular}{ccc}
			{\includegraphics[width=0.28\linewidth,height=0.15\textheight]{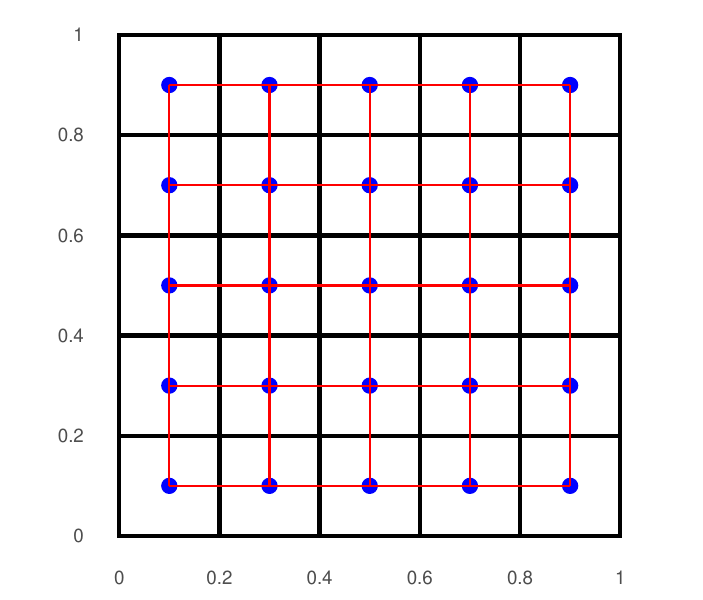}}&
			{\includegraphics[width=0.28\linewidth,height=0.15\textheight]{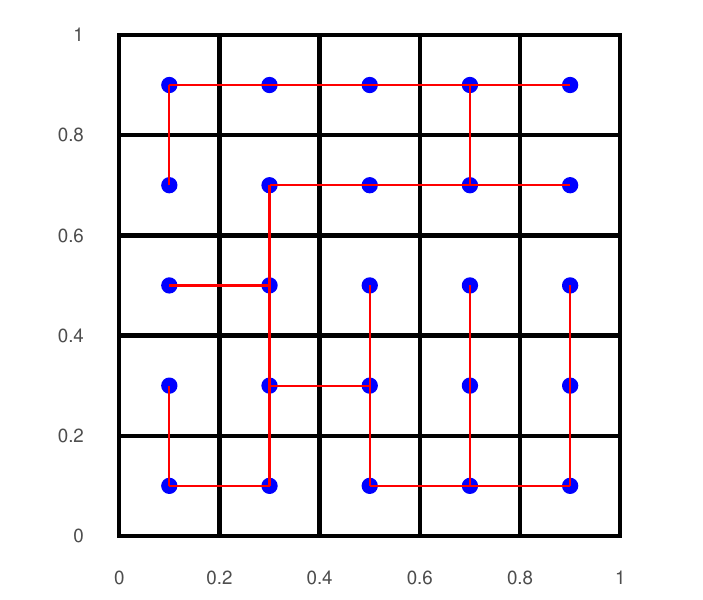}}&
			{\includegraphics[width=0.28\linewidth,height=0.154\textheight]{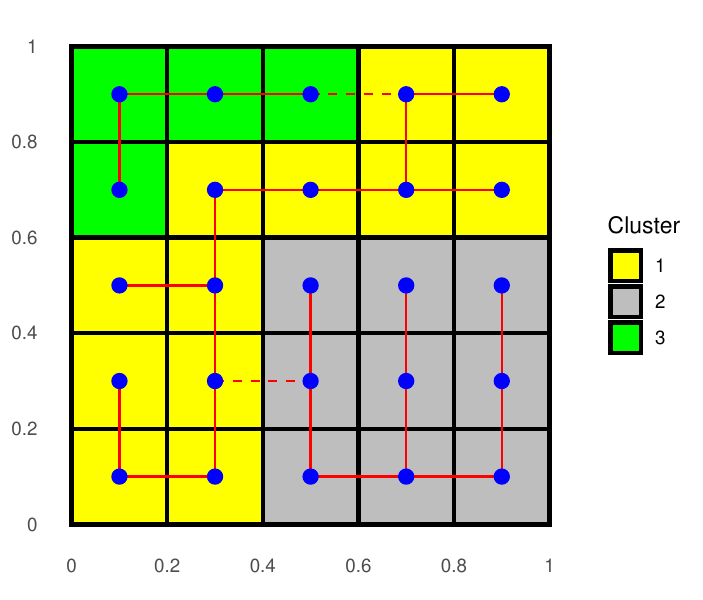}}
			\\
			{\small (a)} & {\small (b)} &{\small (c) } 
		\end{tabular}
		\caption{Illustration of our partition model with $K=5$. (a) Graph $\mathcal{G}=
(\mathcal{V}, \mathcal{E})$, where $\mathcal{V}$ is the set of blocks, and edges in $\mathcal{E}$ are denoted by the red lines between adjacent blocks. (b) One spanning tree obtained by cutting some edges of graph $\mathcal{G}$ in (a). (c) Domain partition $\pi^\ast(\mathcal{D})$ induced by cutting two edges (dashed lines) of the spanning tree in (b). Each cluster (denoted by different colors) is a connected component. Locations within the same block have the same cluster membership.}
		\label{FIG:illustration}
        \vspace{-10pt}
	\end{figure}

A spanning tree of $\mathcal{G}$ is defined as a subgraph $\mathcal{T}=
(\mathcal{V}, \mathcal{E}_{\mathcal{T}})$, where the edge set
$\mathcal{E}_{\mathcal{T}} \subseteq \mathcal{E}$ has no cycle and connects
all vertices.  Hence, a spanning tree has $n^{\ast}$ vertices and $n^{\ast} -
1$ edges. See Figure~\ref{FIG:illustration} for an example of a spanning tree of a lattice graph. A well-known property of the spanning tree is that we obtain $k$
connected components of $\mathcal{T}$, if $k - 1$ edges are deleted from $\mathcal{T}$.
This property has motivated the development of hierarchical generative prior models of spatial clusters.
These models begin with the construction of a spatial graph $\mathcal{G}$, based on which a prior is defined over the spanning tree space of $\mathcal{G}$. Conditional on the spanning tree,  prior models are assumed for the number of clusters and a partition of the spanning tree. The likelihood of $\{y (\tmmathbf{s}_i)\}_{i = 1}^n$ can be derived afterward given the partition. Following this path, we describe below a domain partition prior model based on spanning trees. 

    \subsection{A prior model for partitions}\label{SEC:Model1}

Without loss of generality, we assume $\mathcal{D}= [0, 1]^2$. Note that our method and
theoretical results can be easily extended to a more general domain that is
homeomorphic to $[0, 1]^2$ with the Euclidean metric and a bi-Lipschitz
homeomorphism. We first select an integer $K$, and segment $\mathcal{D}$ into
$K^2$ disjoint blocks, say $\{B_m \}_{m = 1}^{K^2}$, where each
block $B_m$ is a $K^{- 1} \times K^{- 1}$ rectangle. We construct $\mathcal{G}= (\mathcal{V}, \mathcal{E})$ as a mesh grid graph of blocks, where $\mathcal{V}=\{B_m \}_{m = 1}^{K^2}$ and $\mathcal{E}$ is the set of edges connecting only adjacent blocks (see Figure \ref{FIG:illustration}(a) for the constructed $\mathcal{G}$ with $K=5$). Given graph $\mathcal{G}$, write $\Delta$ as the space of
spanning trees induced from $\mathcal{G}$. The following describes two popular approaches in
the existing literature to assign priors on spanning trees.

The first approach is the uniform spanning tree prior (UST,
\cite{aldous1990random, schramm2000scaling, teixeira2019bayesian}), which
assumes 
\begin{equation}
  \mathbb{P}(\mathcal{T}) \propto \mathbb{I} (\mathcal{T} \in \Delta),
  \label{DEF:spanningtree1}
\end{equation}
where $\mathbb{I}(\cdot)$ is the indicator function. {\cite{burton1993local}} establishes foundational principles for understanding
UST.
%and Wilson's algorithm {\citep{wilson1996generating}} can be used to generate samples from it. 
The second approach is the random minimum spanning
tree (MST) prior, where a spanning tree is generated based on some random weights added to the edge set $\mathcal{E}$ \citep{frieze1985value, dobrin2001minimum, luo2021bayesian}. For simplicity in the theoretical analysis, we consider only UST in this paper. See Figure \ref{FIG:illustration}(b) for an example of obtaining $\mathcal{T}$ from $\mathcal{G}$.

% Given graph $\mathcal{G}$, RST assigns a weight $w_e$ for  $\forall e \in
% \mathcal{E}$ with a uniform prior, and obtains the minimum spanning tree (MST),
% defined as the spanning tree with the minimal $\Sigma_{e \in
% \mathcal{E}_{\mathcal{T}}} w_e$ over all spanning trees induced from
% $\mathcal{G}$, i.e.,
% \begin{equation}
%   \mathcal{T}= \tmop{MST} (\{w_e \}_{e \in
% \mathcal{E}}),\text{ } w_e \overset{i.i.d.}{\sim} \tmop{Unif}
%   (0, 1), \label{DEF:spanningtree2}
% \end{equation}
% where $\tmop{Unif} (\cdummy, \cdummy)$ is the uniform distribution.
% %Prim's algorithm {\citep{prim1957shortest}} can be used to construct MST with a computational complexity of $O(K^2\log K)$.

% Either of priors (\ref{DEF:spanningtree1}) and (\ref{DEF:spanningtree2}) can
% be used in our model. It has been shown in Proposition 7 of
% {\cite{luo2021bayesian}} that the MST
% algorithm (\ref{DEF:spanningtree2}) generates $\mathcal{T}$ with a strictly
% positive probability, for $\forall \mathcal{T} \in \Delta$. Thus, priors (\ref{DEF:spanningtree1}) and
% (\ref{DEF:spanningtree2}) have the same spanning tree support (see Figure \ref{FIG:illustration}(b) for an example of obtaining $\mathcal{T}$ from $\mathcal{G}$). 
Next, we assume
the number of clusters $k$ follows a Poisson distribution with mean
parameter $\lambda$:
\begin{equation}
  k \sim \tmop{Poisson} (\lambda) \cdot \mathbb{I}(1\leq k \leq K^2). \label{DEF:k}
\end{equation}
Conditional on
$\mathcal{T}$ and $k$, we assume a uniform distribution on
all possible partitions induced by $\mathcal{T}$:
\begin{equation}
  \mathbb{P}\{\pi (\mathcal{V}) |k, \mathcal{T}\} \propto \mathbb{I}
  \left\{ \pi (\mathcal{V}) \text{ is induced from } \mathcal{T} \text{ and
  has } k \text{ clusters} \right\} . \label{DEF:pi}
\end{equation}
Note that $\pi(\mathcal{V})$ is a partition of blocks. Conditional on $\pi(\mathcal{V})$, the partition of $\mathcal{D}$, say $\pi^\ast(\mathcal{D})$, is obtained immediately by assigning locations within the same block to the same cluster. Note that we use the notation $\cdot^{\ast}$ to emphasize that $\pi^\ast(\mathcal{D})$ is induced from $\pi(\mathcal{V})$.  See Figure \ref{FIG:illustration}(c) for an example of obtaining $\pi^\ast(\mathcal{D})$ from $\mathcal{T}$. 
% For the simplicity of notations, we do not distinguish between $\pi^\ast(\mathcal{D})$ and $\pi(\mathcal{V})$ in the later context. 

Let $\mathcal{S}=
\{\tmmathbf{s}_i \}_{i = 1}^n$ be the set of observed locations. Conditional on $\pi^\ast(\mathcal{D})$, the partition of $\mathcal{S}$, say $\pi(\mathcal{S})$, is obtained immediately by assigning locations in $\mathcal{S}$ the same cluster memberships as in $\pi^\ast(\mathcal{D})$. 

\begin{remark}
    \label{RM:K}
    Note that in this paper, we consider $K$ as a pre-specified hyperparameter. An alternative is to assume a prior on $K$ and draw posterior samples of it. From a theoretical perspective, both approaches can achieve partition consistency, provided that the specified order of $K$ (Assumption \ref{AS:lorder} in Section \ref{SEC:mainresult}) or the prior on $K$ is chosen appropriately. However, assuming a prior on $K$ increases the computational burden of posterior sampling, since every draw of $K$ corresponds to a new graph $\mathcal{G}$, and therefore requires additional sampling of the corresponding $\mathcal{T}$ and $\pi^\ast(\mathcal{D})$.
\end{remark}

\subsection{The proposed Spat-RPM}\label{SEC:Model2}

Let $\pi^\ast(\mathcal{D})=\{\mathcal{D}_{1}^{\ast}, \ldots,
\mathcal{D}_k^{\ast} \}$ be the domain partition with $k$ clusters. Given a cluster $\mathcal{D}_{j}^\ast\in \pi^\ast (\mathcal{D})$, let $\tmmathbf{\theta}_j$ be the regression coefficient within it and write $\tmmathbf{\theta}= \{\tmmathbf{\theta}_j \}_{j = 1}^k$. Conditional on $\pi^\ast(\mathcal{D})$, we write our hierarchical
model as
\begin{gather}
  \tmmathbf{\theta} \mid \pi^\ast (\mathcal{D}), k, \mathcal{T}, \tmmathbf{s},
  \tmmathbf{x} \sim \prod_{j = 1}^k \mathbb{P} (\tmmathbf{\theta}_j), \text{ and }
  \label{EQ:priortheta0}\\
  \tmmathbf{y} \mid \pi^\ast (\mathcal{D}), k, \mathcal{T}, \tmmathbf{s},
  \tmmathbf{x}, \tmmathbf{\theta} \sim \prod_{j = 1}^k \prod_{\tmmathbf{s}_i
  \in \mathcal{D}_j^\ast} \mathbb{P}_{\tmmathbf{\theta}_j} \{y (\tmmathbf{s}_i)
  \mid \tmmathbf{x}(\tmmathbf{s}_i)\}, \label{EQ:priory0}
\end{gather}
where $\tmmathbf{s}= \{\tmmathbf{s}_i \}_{i =
1}^n$, $\tmmathbf{x}= \{\tmmathbf{x}(\tmmathbf{s}_i)\}_{i = 1}^n$,
$\tmmathbf{y}= \{y (\tmmathbf{s}_i)\}_{i = 1}^n$ and $\mathbb{P} (\tmmathbf{\theta}_j)$ is the prior model we assume for the
unknown parameter $\tmmathbf{\theta}_j$.

\subsection{Prediction at new locations}\label{SEC:prediction}

Following priors in Section \ref{SEC:Model1} and the hierarchical model in Section \ref{SEC:Model2}, we obtain the posterior distribution of  $\{ \tmmathbf{\theta}, \pi^\ast (\mathcal{D})
\}$, based on which we predict the distribution of the response variable $y(\mathbb{s}) \mid \mathbb{s}, \tmmathbf{x}(\mathbb{s})$ for a new location $\mathbb{s}$ and covariate $\tmmathbf{x} (\mathbb{s})$. Given $\{ \tmmathbf{\theta}, \pi ^\ast(\mathcal{D})
\}$, write $\tmmathbf{\theta}
(\mathbb{s}) =
\sum_{j = 1}^k \tmmathbf{\theta}_j \mathbb{I} (\mathbb{s} \in
\mathcal{D}^{\ast}_j)$ as the predicted regression coefficient at $\mathbb{s}$. The conditional predictive distribution of $y
(\mathbb{s}) \mid \mathbb{s}, \tmmathbf{x}(\mathbb{s}),\tmmathbf{\theta}
(\mathbb{s}) $ is made by $\mathbb{P}_{\tmmathbf{\theta}
(\mathbb{s} )} \{\cdot \mid \tmmathbf{x}(\mathbb{s})\}$, based on which we predict the distribution of $y(\mathbb{s}) \mid \mathbb{s}, \tmmathbf{x}(\mathbb{s})$ by
\[
\int \mathbb{P}_{\tmmathbf{\theta}
(\mathbb{s} )} \{\cdot\mid \tmmathbf{x}(\mathbb{s})\}\mathbb{P}\{\tmmathbf{\theta}
(\mathbb{s})\mid \mathfrak{D}\}d\tmmathbf{\theta}
(\mathbb{s}),
\]
where $\mathfrak{D}= [\{\tmmathbf{s}_i, \tmmathbf{x}(\tmmathbf{s}_i), y
(\tmmathbf{s}_i)\}_{i = 1}^n$ is the observed data. In practice, the integral above can be approximated with posterior samples of $\{ \tmmathbf{\theta}, \pi^\ast (\mathcal{D})\}$. The detailed sampling procedure is provided in the next section.

% make the conditional predictive distribution of  $y
% (\mathbb{s}) \mid \mathbb{s}, \tmmathbf{x}(\mathbb{s}),\tmmathbf{\theta}, \pi^\ast (\mathcal{D})$ for a new location $\mathbb{s}$ and covariate
% $\tmmathbf{x} (\mathbb{s})$. Given $\{ \tmmathbf{\theta}, \pi ^\ast(\mathcal{D})
% \}$, write $\tmmathbf{\theta}
% (\mathbb{s}) =
% \sum_{j = 1}^k \tmmathbf{\theta}_j \mathbb{I} (\mathbb{s} \in
% \mathcal{D}^{\ast}_j)$ as the predicted regression coefficient at
% $\mathbb{s}$. We make the conditional predictive distribution by $\mathbb{P}_{\tmmathbf{\theta}
% (\mathbb{s} )} \{\cdot \mid \tmmathbf{x}(\mathbb{s})\}$. 

% Following Section \ref{SEC:computation}, we can draw posterior samples of $\{ \tmmathbf{\theta}, \pi ^\ast(\mathcal{D})
% \}$, and the corresponding predicted regression coefficient, say $\{\tmmathbf{\theta}^{(m)}
% (\mathbb{s})\}_{m=1}^{M}$.
% The posterior predictive distribution of $y(\mathbb{s}) \mid \mathbb{s}, \tmmathbf{x}(\mathbb{s})$ can be then approximated by the collection of conditional predictions $\{y^{(m)}(\mathbb{s})\}_{m=1}^{M}$, where $y^{(m)}(\mathbb{s})\sim \mathbb{P}_{\tmmathbf{\theta}^{(m)}
% (\mathbb{s} )} \{\cdot \mid \tmmathbf{x}(\mathbb{s})\}$ are samples drawn from the conditional predictive distribution.

\subsection{Bayesian Computations}\label{SEC:computation}

We use Markov chain Monte Carlo (MCMC) algorithm to draw samples from
the posterior distribution $\mathbb{P} \{\pi^\ast (\mathcal{D}), k, \mathcal{T},
\tmmathbf{\theta} \mid \tmmathbf{s, x, y} \}$.
Given $\pi ^\ast(\mathcal{D})$, we draw sample of
$\tmmathbf{\theta}$ from distribution $\mathbb{P}\{\tmmathbf{\theta} \mid \pi^\ast
(\mathcal{D}), k, \mathcal{T}, \tmmathbf{s, x, y}\}$. This
step is standard if we take a conjugate prior of $\tmmathbf{\theta}$ in
(\ref{EQ:priortheta0}), so we omit the details. 
To sample $\pi ^\ast(\mathcal{D})$, we analytically integrate $\tmmathbf{\theta}$ out and sample from the collapsed conditional distribution of $\pi ^\ast(\mathcal{D}), k,
\mathcal{T} \mid \tmmathbf{s, x, y}$.
Borrowing ideas from {\cite{luo2021bayesian}}, at each MCMC iteration, we propose one
of four moves from the current state: birth, death, change, and hyper, with
probabilities $r_b (k), r_d (k), r_c (k) \infixand r_h (k)$, respectively. 

For the birth move, we split one cluster in the current $\pi ^\ast(\mathcal{D})$ into two
clusters by randomly removing an edge connecting two blocks in the same cluster in
$\mathcal{T}$. For a death move, an edge in $\mathcal{T}$ that connects two distinct clusters in
$\pi^\ast (\mathcal{D})$ is selected uniformly, and the two clusters are
then combined into a single cluster. For a change move, we first perform a death move, then a birth move. The purpose of the change move is to encourage a better mixing of the sampler. The cluster
number is unchanged after the change move. Finally, for a hyper move, we update the spanning tree $\mathcal{T}$. For every edge $e \in \mathcal{E}$, we sample a weight $w_e \sim \tmop{Unif} (0, 1
/ 2)$, if $e$ connects blocks within the same cluster, and $w_e \sim
\tmop{Unif} (1 / 2, 1)$ otherwise. Based on $\{w_e \}_{e \in \mathcal{E}}$,
Prim's MST algorithm {\citep{prim1957shortest}} is used to construct a new
spanning tree $\mathcal{T}_{\tmop{new}}$ with a computational complexity of $O(K^2\log K)$. It has been shown that
$\mathcal{T}_{\tmop{new}}$ is guaranteed to induce current partition $\pi^\ast
(\mathcal{D})$ {\citep{teixeira2015generative}}. The computational details of the Metropolis-Hastings (M-H) acceptance ratios for the four moves are deferred to Section \ref{SEC:MHcomputation} in the Supplementary Material.

% The M-H acceptance
% rate is 1. Note that when applying RST prior (\ref{DEF:spanningtree2}), this
% is an exact sampler {\citep{luo2021bayesian}}. When using UST prior
% (\ref{DEF:spanningtree1}), this is an approximated sampler used in
% {\cite{teixeira2019bayesian}}.

\subsection{Selection of hyperparameters $(K,\lambda)$}
\label{SEC:hyperparameterselection}
We propose to consider several candidate pairs of $(K,\lambda)$, and obtain the corresponding posterior clustering samples. 
Specifically, in the linear regression setting, where the consistency theory is established in Section \ref{SEC:mainresult}, the candidate pairs $(K,\lambda)$ can be chosen based on their order conditions for achieving consistency (e.g., see Corollary \ref{CL:FDGcontraction}). To determine the optimal pair, we propose to use Calinski–Harabasz (CH) index \citep{calinski1974dendrite}, an internal measure of clustering quality, and Watanabe-Akaike information criterion (WAIC, {\cite{watanabe2010asymptotic}}). We demonstrate the effectiveness of using the CH index and WAIC through simulation in Section \ref{SEC:CHWAIC}.
%The optimal pair is selected as the one that maximizes the CH index.

%The effectiveness of this approach using the CH index will be demonstrated in Section \ref{SEC:sensitivity}.

\subsection{Application to areal data}
\label{SEC:apptoareal}

We demonstrate how to apply the proposed Spat-RPM to areal data, where observations are areal units rather than individual locations. Let $\{\mathbb{D}_i,\tmmathbf{x}(\mathbb{D}_i),y(\mathbb{D}_i)\}_{i=1}^{n}$ denote an areal dataset, where $\mathbb{D}_i$ is the $i$-th areal unit, and $\tmmathbf{x}(\mathbb{D}_i)$ and $y(\mathbb{D}_i)$ are the corresponding covariate and response variable, respectively. To apply the proposed Spat-RPM, we represent each areal unit $\mathbb{D}_i$ by a point (e.g., the centroid of $\mathbb{D}_i$), say $\tmmathbf{s}_i$. The areal dataset can then be reformulated as $\{\tmmathbf{s}_i,\tmmathbf{x}(\tmmathbf{s}_i), y (\tmmathbf{s}_i)\}_{i = 1}^n$, which is the data structure Spat-RPM can be applied to directly. Following Sections \ref{SEC:Model1} – \ref{SEC:computation}, we can obtain posterior samples of the partition of $\mathcal{S}=\{\tmmathbf{s}_i\}_{i=1}^{n}$, say $\pi(\mathcal{S})$. Since each $\tmmathbf{s}_i$ represents an areal unit $\mathbb{D}_i$, the partition of $\{\mathbb{D}_i\}_{i=1}^{n}$ (and hence the partition of the domain) is induced directly from $\pi(\mathcal{S})$. 

In Section \ref{SEC:areal data}, we conduct a simulation on an areal dataset to demonstrate the effectiveness of the proposed approach.

\section{Main result}\label{SEC:mainresult}

In this section, we introduce the theoretical result of the proposed Spat-RPM. Note that priors specified in Sections \ref{SEC:Model1} and
\ref{SEC:Model2} assume a general likelihood
$\mathbb{P}_{\tmmathbf{\theta} (\tmmathbf{s}_i)} \{y (\tmmathbf{s}_i) \mid
\tmmathbf{x}(\tmmathbf{s}_i)\}$. For example, $\mathbb{P}_{\tmmathbf{\theta}
(\tmmathbf{s}_i)} \{y (\tmmathbf{s}_i) \mid \tmmathbf{x}(\tmmathbf{s}_i)\}$
can be a Poisson, Gaussian or multinomial distribution.
In this section, to simplify the theoretical analysis, we focus
on the Gaussian distribution case, where we write our model specifically as a
linear regression form, based on which the theoretical analysis is conducted.

\subsection{Linear regression setting}\label{SEC:lmmodel}

Under the linear regression setting, we write our model as
\begin{equation}
  y (\tmmathbf{s}_i) = \mu (\tmmathbf{s}_i) + \epsilon (\tmmathbf{s}_i),
  \label{EQ:linearmodel}
\end{equation}
where $\mu (\tmmathbf{s}_i) =\tmmathbf{x}^T (\tmmathbf{s}_i) \tmmathbf{\theta}
(\tmmathbf{s}_i)$ is the regression mean of $y (s_i)$, and $\{\epsilon
(\tmmathbf{s}_i)\}_{i = 1}^n$ are identically and independently distributed
(i.i.d.) mean zero Gaussian noises. Based on model (\ref{EQ:linearmodel}), we
assume an independent conjugate Zellner's g-prior on each $\tmmathbf{\theta}_j$. Let $\tmmathbf{\theta}= \{\tmmathbf{\theta}_j\}_{j =
1}^k$. We modify models (\ref{EQ:priortheta0}) and
(\ref{EQ:priory0}) as
\begin{gather}
  \tmmathbf{\theta} \mid \pi^\ast (\mathcal{D}), k, \mathcal{T},
  \tmmathbf{s}, \tmmathbf{x} \sim \prod_{j = 1}^k \mathbb{P}_{\tmop{Gaussian}}
  \bigg[ \tmmathbf{\theta}_j ; \tmmathbf{0}, \gamma n \sigma^2 \bigg\{
  \sum_{\tmmathbf{s}_i \in \mathcal{D}_j^\ast}
  \tmmathbf{x}(\tmmathbf{s}_i)\tmmathbf{x}^T (\tmmathbf{s}_i)
  \bigg\}^{\dagger} \bigg],\text{ and} \label{EQ:priortheta}\\
  \tmmathbf{y} \mid \pi^\ast (\mathcal{D}), k, \mathcal{T}, \tmmathbf{s},
  \tmmathbf{x}, \tmmathbf{\theta} \sim \prod_{j = 1}^k \prod_{\tmmathbf{s}_i
  \in \mathcal{D}_j^\ast} \mathbb{P}_{\tmop{Gaussian}} \{y (\tmmathbf{s}_i) ;
  \tmmathbf{x}^T (\tmmathbf{s}_i)\tmmathbf{\theta}_j, \sigma^2 \},
  \label{EQ:priory}
\end{gather}
where $\mathbb{P}_{\tmop{Gaussian}} (\cdummy ; \tmmathbf{a},
\tmmathbf{\Sigma})$ denotes the probability density function of the Gaussian
distribution with mean $\tmmathbf{a}$ and covariance $\tmmathbf{\Sigma}$, $\gamma>0$ is a hyperparameter controlling the prior variance, and
$\cdot^{\dagger}$ denotes the pseudoinverse. $\sigma^2$ is the variance of $\epsilon (\tmmathbf{s}_i)$, which is often assumed unknown and assigned an inverse Gamma prior in practice. For simplicity, we assume $\sigma^2$ is a fixed value and drop the notation $\sigma^2$ in the left-hand side of (\ref{EQ:priortheta}) and (\ref{EQ:priory}). However, we shall show that our theoretical results hold even when $\sigma^2$ is not fixed at its true value. Furthermore, in the simulation study (Section \ref{SEC:sensitivity}), the empirical results show that $\sigma^2$ has little impact on the fitting performance, provided that $(K,\lambda)$ are selected appropriately.

Under the linear regression setting, we predict the regression mean for a new
observed location $\mathbb{s}$ and covariate $\tmmathbf{x}
(\mathbb{s})$. Following notations in Section \ref{SEC:prediction}, we predict the regression mean by $\mu \{\mathbb{s},\tmmathbf{x}(\mathbb{s})\} =\tmmathbf{x}^T
(\mathbb{s}) \tmmathbf{\theta} (\mathbb{s} )$, for a given $\{ \tmmathbf{\theta}, \pi^\ast (\mathcal{D})
\}$.

\subsection{Notations}
\label{SEC:notationmain}
We introduce some notations in this section. Denote $\| \cdummy \|_2$ and $\|
\cdummy \|_{\infty}$ as $L_2$ norm and $L_{\infty}$ norm, respectively.
For two locations $\tmmathbf{s}_1, \tmmathbf{s}_2$, let $d
(\tmmathbf{s}_1, \tmmathbf{s}_2) = \|\tmmathbf{s}_1 -\tmmathbf{s}_2 \|_2$ be
the Euclidean distance. For two \ spatial domains $\mathbb{D}_1$ and
$\mathbb{D}_2$, we write $d (\mathbb{D}_1, \mathbb{D}_2) =
\inf_{\tmmathbf{s}_1 \in \mathbb{D}_1, \tmmathbf{s}_2 \in \mathbb{D}_2} d
(\tmmathbf{s}_1, \tmmathbf{s}_2)$. %For a constant $\delta > 0$ and a spatial domain $\mathbb{D}$, we define the $\delta$-neighborhood of $\mathbb{D}$ as $ \mathcal{N} (\mathbb{D}, \delta) = \{\tmmathbf{s} \in \mathbb{R}^2 : d
%(\tmmathbf{s}, \mathbb{D}) \leq \delta\}$.
The notation $|\mathbb{D}|$ is used as two ways: if $\mathbb{D}$ is a spatial
domain, $|\mathbb{D}|$ represents its area; if $\mathbb{D}$ is a set,
$|\mathbb{D}|$ denotes its cardinality. Furthermore, for a spatial domain
$\mathbb{D}$, we denote $\|\mathbb{D}\|$ as the number of observed locations within it.

We use $c$ and $C$ to denote some constants independent of partitions $\pi^{\ast}
(\mathcal{D})$ (hence $\pi (\mathcal{S})$). The
values of $c$ and $C$ may change from line to line.

For two positive series $a_n$ and $b_n$, we say $a_n \gg b_n$ if $a_n / b_n
\rightarrow \infty$, and $a_n \ll b_n$ if $a_n / b_n \rightarrow 0$. We write
$a_n \sim b_n$ if there exist constants $c, C > 0$, such that $c < a_n / b_n <
C, \text{ for all }n \geqslant 1$. We write $a_n = O (b_n)$, if there exits a
constant $c > 0$, such that $a_n / b_n \leqslant c$ for all $n \geqslant 1$.
We write $a_n = o (b_n)$ if $a_n \ll b_n$.

For a positive integer $a$, we use $\tmmathbf{I}_a$ to denote an $a \times a$
identity matrix. For a matrix $\tmmathbf{A}$ and a constant $c$, we say
$\tmmathbf{A}> c$, if $\lambda_{\min} (\tmmathbf{A}) > c$, and $\tmmathbf{A}<
c$ if $\lambda_{\max} (\tmmathbf{A}) < c$, where $\lambda_{\min}
(\tmmathbf{A})$ and $\lambda_{\max} (\tmmathbf{A})$ are the minimum and
maximum eigenvalues of $\tmmathbf{A}$, respectively.

\subsection{Assumptions and main theorems}

Recall that we assume $\mathcal{D}$ can be partitioned into $k_0$
heterogeneous sub-domains $\{\mathcal{D}_{l, 0} \}_{l = 1}^{k_0}$. For the sub-domain $\mathcal{D}_{l, 0}$, let
$\tmmathbf{\theta}_{l, 0}$ be the corresponding true regression coefficient. We consider two asymptotic regimes in the theoretical analysis.
\begin{itemize}
    \item \textbf{Fixed Data-Generating (FDG) regime}: The domain partition $\{\mathcal{D}_{l, 0} \}_{l = 1}^{k_0}$ and the coefficients $\{\tmmathbf{\theta}_{l,0}\}_{l=1}^{k_0}$ are fixed, and we study the asymptotic behavior as $n \to \infty$.
    \item \textbf{Varying Data-Generating (VDG) regime}: The number of sub-domains $k_0$ may grow with $n$, and both $\{\mathcal{D}_{l, 0} \}_{l = 1}^{k_0}$ and $\{\tmmathbf{\theta}_{l,0}\}_{l=1}^{k_0}$ may vary as $n \rightarrow \infty$.
\end{itemize}

The FDG regime captures many practical scenarios where the number of clusters is small to moderate, and the gaps between $\{\tmmathbf{\theta}_{l, 0}\}_{l=1}^{k_0}$ are fixed (i.e., there are abrupt changes between the regression coefficients). For example, in the human dorsolateral prefrontal cortex dataset \citep{maynard2021transcriptome}, a widely studied dataset in the spatial transcriptomics field, the underlying number of layers (clusters) is limited to $6$. In some neuroscience datasets, where the goal is to identify clusters of human brain subregions responding to specific stimuli, abrupt changes between clusters are frequently observed, and the number of true clusters is limited due to the structural properties of the human brain. Besides, in geosciences, geophysical properties or fluids (e.g., air and seawater) can undergo rapid changes at the boundaries between adjacent subsurface zones or fluid masses \citep{talley2011descriptive}. Additional examples of abrupt changes in environmental and economic sciences can be found in \cite{li2019spatial}. Thus, analysis under this regime provides interpretable insights that are directly applicable in practice.

The VDG regime, in contrast, offers greater flexibility in understanding how both $\{\mathcal{D}_{l, 0} \}_{l = 1}^{k_0}$ and $\{\tmmathbf{\theta}_{l, 0}\}_{l=1}^{k_0}$ affect clustering consistency. In particular, allowing $k_0 \to \infty$ enables the model to capture a fractal structure with an infinite number of clusters. Under the VDG regime, we allow the gaps between $\{\tmmathbf{\theta}_{l, 0}\}_{l=1}^{k_0}$ to go to $0$, enabling our model to capture smooth transitions between clusters in an asymptotic sense. Note that the FDG regime can be considered as a special case of the VDG regime, where the truth remains fixed as $n \to \infty$. Hence, in what follows, all assumptions and theorems are stated under the VDG regime, from which the results for the FDG regime follow as a special case.

Under the VDG regime, we assume that $\{ \tmmathbf{\theta}_{l, 0} \}_{l =
  1}^{k_0}$ are distributed within a ball centered at $\tmmathbf{o}_n$ with radius $r_n$, i.e., $\sup_{1
  \leqslant l \leqslant k_0} \| \tmmathbf{\theta}_{l, 0}-\tmmathbf{o}_n \|_2 \leqslant r_n$. For simplicity, we consider $\tmmathbf{o}_n=\tmmathbf{0}$, and it is easy to extend the result to the case $\tmmathbf{o}_n\neq \tmmathbf{0}$. Note that $r_n \rightarrow 0$ indicates that the gaps between $\{\tmmathbf{\theta}_{l, 0} \in \mathbb{R}^d\}_{l=1}^{k_0}$ go to $0$. We consider the scenario where
% Let $\tmmathbf{o}_n,r_n$ be the center and radius of the ball within which $\{ \tmmathbf{\theta}_{l, 0} \}_{l =
%   1}^{k_0}$ are distributed, i.e., $\sup_{1
%   \leqslant l \leqslant k_0} \| \tmmathbf{\theta}_{l, 0}-\tmmathbf{o}_n \|_2 \leqslant r_n$. For simplicity, we assume $\tmmathbf{o}_n=0$, and it is easy to extend the result to the case when $\tmmathbf{o}_n\neq 0$. Under the VDG regime, we consider the scenario where
  \begin{equation}
  \label{EQ:rnk0}
      k_0^{2+3/d}r_n^{-1}=O(n^{\tau}) \text{ for some }\tau<1/2.
  \end{equation}
  Equation (\ref{EQ:rnk0}) implies that $k_0$ cannot be too large and $r_n$ cannot be too small, and it ensures the existence of hyperparameters to achieve partition consistency. Further details are provided later in Assumption \ref{AS:hyperpara}. 
  
   We next introduce the required assumptions. We first make an assumption on the distribution
of the observed locations $\{\tmmathbf{s}_i \}_{i = 1}^n$.

\begin{assumption}
  \label{AS:distributionloc}The observed locations $\{\tmmathbf{s}_i \}_{i =
  1}^n$ are i.i.d. on spatial domain $\mathcal{D}$ with a probability density
  function $\mathbb{P}_{\mathcal{D}} (\cdummy)$ satisfying $0 <
  \inf_{\tmmathbf{s} \in \mathcal{D}} \mathbb{P}_{\mathcal{D}} (\tmmathbf{s})
  \leqslant \sup_{\tmmathbf{s} \in \mathcal{D}} \mathbb{P}_{\mathcal{D}}
  (\tmmathbf{s}) < \infty$.
\end{assumption}

Assumption \ref{AS:distributionloc} is standard in spatial literature
{\citep{yu2020estimation,yu2024distributed}}. Under Assumption
\ref{AS:distributionloc}, the observed locations are randomly scattered over
spatial domain $\mathcal{D}$.

Write $\partial \mathcal{D}_{l, 0}$ as the boundary of
sub-domain $\mathcal{D}_{l, 0}$. Let $N (\cdot, \delta, \| \cdot \|_2)$ be the
$\delta$-covering number and $| \partial \mathcal{D}_{l, 0} | = \lim_{\delta
\rightarrow 0} \delta N (\partial \mathcal{D}_{l, 0}, \delta, \| \cdot \|_2)$
be the length of $\partial \mathcal{D}_{l, 0}$. We make the following
assumption on $\{\mathcal{D}_{l, 0} \}_{l = 1}^{k_0}$.

\begin{assumption}
  \label{AS:lengthofboundary}There exist some constants $c, C > 0$, such that
  \begin{equation}
    c k_0^{- 1} \leqslant \inf_{1 \leqslant l \leqslant k_0} | \mathcal{D}_{l,
    0} | \leqslant \sup_{1 \leqslant l \leqslant k_0} | \mathcal{D}_{l, 0} |
    \leqslant C k_0^{- 1},\text{ }     c \leqslant \inf_{1 \leqslant l \leqslant k_0} \frac{| \partial
    \mathcal{D}_{l, 0} |}{| \mathcal{D}_{l, 0} |^{1 / 2}} \leqslant \sup_{1
    \leqslant l \leqslant k_0} \frac{| \partial \mathcal{D}_{l, 0} |}{|
    \mathcal{D}_{l, 0} |^{1 / 2}} \leqslant C, \label{EQ:boundary1}
  \end{equation}
  \begin{equation}
    \text{and }    N (\partial \mathcal{D}_{l, 0}, \delta,
    \| \cdot \|_2) \leqslant C |\partial \mathcal{D}_{l, 0} | \delta^{-1}, \text{ } \forall 1 \leqslant l \leqslant k_0, 0 < \delta < | \partial
    \mathcal{D}_{l, 0} |.
    \label{EQ:boundary3}
  \end{equation}
  Furthermore, for any two locations $\tmmathbf{s}, \tmmathbf{s}' \in
  \mathcal{D}_{l, 0}$, there exists a path connecting $\tmmathbf{s}$ and
  $\tmmathbf{s}'$, say $\mathcal{P} (\tmmathbf{s}, \tmmathbf{s}')$, such that
  $\mathcal{P} (\tmmathbf{s}, \tmmathbf{s}')$ is contained in $\mathcal{D}_{l,
  0}$, and
  \begin{equation}
    \label{EQ:pathdis} d \{\mathcal{P}(\tmmathbf{s}, \tmmathbf{s}'), \partial
    \mathcal{D}_{l, 0} \} = \min \{d (\tmmathbf{s}, \partial
    \mathcal{D}_{l, 0}), d (\tmmathbf{s}', \partial \mathcal{D}_{l, 0})\} \text{ or } d \{\mathcal{P}(\tmmathbf{s}, \tmmathbf{s}'), \partial
    \mathcal{D}_{l, 0} \} \geqslant C.
  \end{equation}
\end{assumption}

The first part of Equation (\ref{EQ:boundary1}) assumes that the areas of sub-domains $\{\mathcal{D}_{l,0}\}_{l=1}^{k_0}$ are of the same order as $k_0 \rightarrow \infty$. The second part of (\ref{EQ:boundary1}) assumes a relationship between the area and boundary length of $\mathcal{D}_{l,0}$, which is a standard regularization condition when the domain shape changes with $n$ (see e.g., \cite{Huang17042025}). Equation (\ref{EQ:boundary3}) is on the covering number of $\{\partial \mathcal{D}_{l, 0}\}_{l=1}^{k_0}$, which is the same as that in {\cite{willett2005faster,luo2021bast}} under the FDG regime. Equation (\ref{EQ:pathdis}) entails that the minimum distance from the path $\mathcal{P} (\tmmathbf{s}, \tmmathbf{s}')$ to the boundary is either at the endpoints, or is larger than some constant. This is a mild condition that is satisfied by many common shapes (e.g., finite union of regular polygons and circles), while ruling out sub-domains with arbitrarily thin bottlenecks (e.g., dumbbell shapes). In Section \ref{SEC:justificationofpath}, we justify Equation (\ref{EQ:pathdis}) by providing an example of $\mathcal{D}_{l,0}$ that satisfies it.

Under our model, the domain partition space consists of all partitions induced from the UST model in Section~\ref{SEC:Model1}. %which may not include the true partition. 
There may exist some blocks intersecting with boundaries $\{\partial \mathcal{D}_{l, 0}\}_{l=1}^{k_0}$ and containing misclustered locations, as our model assigns the same cluster membership for locations within the same block.  The following proposition is established under Assumption
\ref{AS:lengthofboundary}.

\begin{proposition}
  \label{PP:bestapproximation} Under Assumption \ref{AS:lengthofboundary},
  there exists a contiguous domain partition $\pi^\ast_0 (\mathcal{D}) =
  \{\mathcal{D}_{1, 0}^{\ast}, \ldots \mathcal{D}_{k_0, 0}^{\ast} \}$ in our partition model space, such
  that
  \begin{equation}
    |\mathcal{W}_{\pi_0^{\ast}} | = \sum_{j = 1}^{k_0} |\{B_m : B_m \subseteq
    \mathcal{D}_{j, 0}^{\ast}, B_m \subsetneq \mathcal{D}_{j, 0} \}| \leqslant
    c k_0^{1/2}K \label{EQ:bestpai}
  \end{equation}
  for some constant $c$, where
  $ \mathcal{W}_{\pi_0^{\ast}} = \cup_{j = 1}^{k_0} \{B_m : B_m \subseteq
     \mathcal{D}_{j, 0}^{\ast}, B_m \subsetneq \mathcal{D}_{j, 0} \}$ is the set of the blocks inside of $\mathcal{D}_{j,0}^{\ast}$, but not fully included in $\mathcal{D}_{j,0}$.     
    
\end{proposition}

We defer the proof of Proposition \ref{PP:bestapproximation} to Section \ref{SEC:proofofpp1}. %Under our model, there exist some blocks intersecting with $\mathcal{B}$. Those blocks always contain ``wrongly" clustered locations, since our model assigns the same cluster membership for locations within the same block. As a result, the true partition is not in our model's partition space. 
%We consider $\pi^\ast_0 (\mathcal{D})$ as the partition that is ``closest" to the true partition in our model's partition space. 
%Each cluster $\mathcal{D}_{j, 0}^{\ast}\in \pi^\ast_0 (\mathcal{D})$ is designed to approximate $\mathcal{D}_{j, 0}$, hence $\mathcal{W}_{\pi_0^{\ast}}$ is the set of blocks containing misclustered locations. 
In the construction of $\pi^\ast_0 (\mathcal{D})$, each cluster $\mathcal{D}_{j,0}^{\ast}\in \pi^\ast_0 (\mathcal{D})$ is designed to approximate $\mathcal{D}_{j,0}$. Thus, $\mathcal{W}_{\pi_0^{\ast}}$ is the set of blocks containing misclustered locations. Proposition
\ref{PP:bestapproximation} states that there exists a $\pi^\ast_0
(\mathcal{D})$ in our model's partition space, such that $|\mathcal{W}_{\pi_0^{\ast}}|$, the number of blocks containing misclustered locations, is upper bounded by $ck_0^{1/2}K$. In the following context, we refer to ``approximation error'' as the area of blocks in $\mathcal{W}_{\pi_0^{\ast}}$. Note that each block's area is $K^{-2}$. By Proposition \ref{PP:bestapproximation}, the approximation error is bounded by $ K^{- 2}
\times c k_0^{1/2} K = O(k_0^{1/2} K^{- 1})$. Hence, the larger $K$, the smaller the approximation error.

We make
following assumptions on $\{\tmmathbf{\theta}_{l, 0} \}_{l = 1}^{k_0}$ and
$\{\tmmathbf{x}(\tmmathbf{s}_i),y(\tmmathbf{s}_i)\}_{i = 1}^n$, respectively. Recall that we write $r_n$ as the radius of the ball within which $\{ \tmmathbf{\theta}_{l, 0} \}_{l =
  1}^{k_0}$ are distributed.

\begin{assumption}
  \label{AS:thetagap}We assume that $\{ \tmmathbf{\theta}_{l, 0} \}_{l =
  1}^{k_0}$ are well separated, i.e., $c k_0^{- 1 / d} r_n<\min_{l \neq l'} \|\tmmathbf{\theta}_{l, 0} -\tmmathbf{\theta}_{l', 0}
  \|_2 <C $ for some constants $c,C>0$.
\end{assumption}

\begin{assumption}
  \label{AS:covariates}Conditional on $\{ \tmmathbf{s}_i \}_{i = 1}^n$, we
  assume $\{ \tmmathbf{x}(\tmmathbf{s}_i) \}_{i = 1}^n$ are independent
  d-dimensional bounded random variables. Furthermore, we assume
  that there exist constants $c$ and $C$, such that $0 < c
  <\mathbb{E} \{\tmmathbf{x}(\tmmathbf{s}_i)\tmmathbf{x}^T (\tmmathbf{s}_i)
  \mid \tmmathbf{s}_i \} < C$ holds for $1 \leqslant i \leqslant n$. Conditional on $\{ \tmmathbf{s}_i \}_{i = 1}^n$ and $\{ \tmmathbf{x}(\tmmathbf{s}_i) \}_{i = 1}^n$, we assume $\{ y(\tmmathbf{s}_i) \}_{i = 1}^n$ are generated from Equation (\ref{EQ:linearmodel}).
\end{assumption}

Assumption \ref{AS:thetagap} ensures sufficient separation between clusters for the
identification of sub-domains $\{\mathcal{D}_{l, 0} \}_{l = 1}^{k_0}$. One typical example satisfying it is the quasi-uniform distribution \citep{wendland2004scattered} within the ball. Assumption \ref{AS:thetagap} further implies that $r_n < c k_0^{1/d}$, ensuring that the radius of the space of $\{ \tmmathbf{\theta}_{l, 0} \}_{l =
  1}^{k_0}$ is not excessively large. Although the case $r_n \geq c k_0^{1/d}$ is easier to achieve consistency and can be addressed following a similar proof strategy, we do not pursue it here, as our primary interest lies in how the shrinkage radius influences partition consistency.

We consider a random covariate design, and Assumption \ref{AS:covariates} is to avoid the collinearity of covariates, and
is similar to Assumption (A2) in {\cite{yu2024distributed}}. The boundness
conditions of the covariate $\tmmathbf{x} (\tmmathbf{s}_i)$ and $\mathbb{E}
\{\tmmathbf{x}(\tmmathbf{s}_i)\tmmathbf{x}^T (\tmmathbf{s}_i)\mid \tmmathbf{s}_i\}$ are standard
in linear regression setting, see also Assumption (C1) in
{\cite{luo2021bayesian}} and Assumption (A2) in {\cite{mu2020spatial}}.
%{Note that we do not assume $\tmmathbf{x} (\tmmathbf{s}_1)$ has the same conditional distribution as $\tmmathbf{x} (\tmmathbf{s}_2)$ for $\tmmathbf{s}_1 \neq \tmmathbf{s}_2$}. 

The next assumption is on the orders of 
%hyperparameters, which provide practical guidance for selecting 
the number of blocks hyperparameter $K$ and the Poisson
  hyperparameter $\lambda$ in (\ref{DEF:k}) in our prior model. 

\begin{assumption}
  \label{AS:hyperpara}We assume that there exist two positive values $\alpha_1,\alpha_2>0$, such that $K$ and $\lambda$ satisfy
  \begin{enumerate}[label=\theassumption.\arabic*,ref=\theassumption.\arabic*]
    \item \label{AS:lorder}
      $  k_0^{-1/2-1/d}r_n n^{1/2}\log^{-(1 + \alpha_1)/2} (n)\gg K \gg k_0^{3 / 2 + 2 / d} \log^{\alpha_2} (n) $
    \item \label{AS:lambda}
    $\lambda = o (1), \text{and }$
      $ k_0^{- 1 - 2 / d} r_n^2 n \gg \log
     (\lambda^{- 1}) \gg   n k_0^{1 / 2} K^{- 1} \log^{\alpha_2} (n) r_n^2 + K^2 \log (n)  $
  \end{enumerate}
\end{assumption}

  Under Equation (\ref{EQ:rnk0}), it can be verified algebraically that there exists a $K$ satisfying Assumption \ref{AS:lorder} (i.e., the upper bound in Assumption \ref{AS:lorder} exceeds the lower bound). Moreover, Assumption \ref{AS:lorder} guarantees the existence of a $\log(\lambda^{-1})$ such that Assumption \ref{AS:lambda} is satisfied (i.e., the upper bound in Assumption \ref{AS:lambda} exceeds the lower bound). In Remark \ref{RM:optimal rate}, we provide explicit orders of $(K,\lambda)$ that satisfy Assumption \ref{AS:hyperpara}, ensuring partition consistency with a relatively fast contraction rate.

Recall that after Proposition \ref{PP:bestapproximation}, we illustrate that a
larger $K$ indicates a smaller approximation error of $\pi^\ast_0
(\mathcal{D})$. However, since the number
of all the possible block partitions grows exponentially with $K^2$, a
larger $K$ also increases the difficulty in obtaining the correct
partition due to the curse of dimensionality. Thus, a trade-off analysis is necessary when choosing the value of $K$. Assumption \ref{AS:lorder} provides an order condition for $K$, ensuring a proper balance between the complexity of the partition space and the approximation error. 

% an upper bound condition for
% $K$, under which the partition consistency can be achieved while keeping the
% approximation error relatively small. Note that $K$ should satisfy $K \leq
% n^{1 / 2}$, otherwise, some blocks do not contain any observed locations. The
% order of $K$ by Assumption \ref{AS:lorder} is marginally smaller than $n^{1 / 2}$.

A similar trade-off exists for the order of $\lambda$. Since the model with a larger number of clusters provides enhanced flexibility for data fitting, the data likelihood typically prefers a larger number of
clusters. To avoid such ``overfitting", the Poisson hyperparameter $\lambda$
introduces a penalty for the number of clusters. A smaller $\lambda$ indicates that the prior favors fewer clusters, corresponding to a larger penalty on the cluster number. Since the likelihood contribution to the posterior grows with $n$, we propose to increase the penalty accordingly (i.e., $\lambda=o(1)$), such that the prior favours a smaller number of clusters to balance this effect. On the other hand, the rate of $\lambda$ going to zero must not be too rapid, otherwise, the
number of clusters can be underestimated. Assumption \ref{AS:lambda} provides the order condition for $\lambda$ to obtain partition consistency. Note that the condition $\lambda = o (1)$ implies that $\log(\lambda^{-1})$ is a positive value when $n$ is large, ensuring the validity of the order condition imposed on $\log(\lambda^{-1})$ in Assumption \ref{AS:lambda}.

A seemingly natural alternative choice for $\lambda$ is to place a fixed (i.e., independent of $n$) and nondegenerate prior on it. A similar idea is adopted to prove the consistency results for the number of clusters in Dirichlet process mixtures. Specifically, \cite{ascolani2023clustering} shows that a fixed, nondegenerate prior on the concentration parameter induces an $n$-dependent posterior term favoring fewer clusters, which counterbalances the likelihood’s preference for a larger number of clusters and yields consistency for the number of clusters. In our model, however, a fixed, nondegenerate prior on $\lambda$ is insufficient to achieve clustering consistency. We demonstrate more details in Section \ref{SEFC:lambdafurter} of Supplementary Material.

{Note that the order conditions of $(K,\lambda)$ in Assumption \ref{AS:hyperpara} depend on unknown quantities $(k_0,r_n)$. Under the FDG regime, $(k_0,r_n)$ are constants, so the order conditions in Assumption \ref{AS:hyperpara} can be applied directly to select $(K,\lambda)$. Under the VDG regime, where $(k_0,r_n)$ vary with $n$, we first estimate $(k_0,r_n)$ and then plug these estimates in to select $(K,\lambda)$. We describe $(k_0,r_n)$ estimation approach in Section \ref{SEC:k0rnestimation} of the Supplementary Material.}

Recall that we use $\Delta$ to denote the space of spanning trees induced from
$\mathcal{G}$. The next assumption is on the priors of our model.

\begin{assumption}
  \label{AS:prior}For $\pi^\ast_0 (\mathcal{D})$ in
    Proposition \ref{PP:bestapproximation}, we assume
    \[ \sup_{\mathcal{T}_1 \in \Delta, \mathcal{T}_2 \in \{\mathcal{T} \in
       \Delta : \pi^\ast_0 (\mathcal{D}) \text{ can be induced from }
       \mathcal{T}\}} 
       \frac{\mathbb{P}(\mathcal{T}_1)}{\mathbb{P}(\mathcal{T}_2)}  =
       O [\exp \{c k_0^{1/2}K \log (K)\}] \]
    for some constant $c$.
\end{assumption}

 Assumption \ref{AS:prior} assumes that the spanning tree's probability of inducing $\pi^\ast_0 (\mathcal{D})$ is not excessively small.
%the probability of the
%spanning tree capable of inducing $\pi^\ast_0 (\mathcal{D})$ not to be excessively small. 
For the UST prior specified in (\ref{DEF:spanningtree1}),
Assumption \ref{AS:prior} is satisfied immediately since the prior
ratio of any two spanning trees is $1$.

We next define a distance measure of two spatial domain partitions, say $\pi_1
(\mathcal{D}) = \{\mathcal{D}_{1 1}, \ldots, \mathcal{D}_{1 k_1} \}$ and
$\pi_2 (\mathcal{D}) = \{\mathcal{D}_{21}, \ldots, \mathcal{D}_{2 k_2} \}$,
where $k_1$ and $k_2$ are their respective number of clusters. We define the
``distance" between $\pi_1 (\mathcal{D})$ and $\pi_2 (\mathcal{D})$ as
\begin{equation}
  \epsilon \{ \pi_1 (\mathcal{D}), \pi_2 (\mathcal{D}) \} = 2 - | \mathcal{D}
  |^{- 1} \biggl[ \sum_{j = 1}^{k_1} \max_{l \in \{1, \ldots, k_2 \}} |
  \mathcal{D}_{1 j} \cap \mathcal{D}_{2 l} | + \sum_{l = 1}^{k_2} \max_{j \in
  \{1, \ldots, k_1 \}} | \mathcal{D}_{1 j} \cap \mathcal{D}_{2 l} | \biggr] .
  \label{DEF:epsilonspatial}
\end{equation}
Similarly, for two partitions of $\mathcal{S}$, say
$\pi_1 (\mathcal{S}) = \{\mathcal{S}_{1 1}, \ldots, \mathcal{S}_{1 k_1} \}$
and $\pi_2 (\mathcal{S}) = \{\mathcal{S}_{21}, \ldots, \mathcal{S}_{2 k_2} \}$,
where $k_1$ and $k_2$ are the number of clusters in each partition, respectively, we define the
``distance" between $\pi_1 (\mathcal{S})$ and $\pi_2 (\mathcal{S})$ as
\begin{equation}
  \epsilon_n \{ \pi_1 (\mathcal{S}), \pi_2 (\mathcal{S}) \} = 2 - n^{- 1}
  \biggl[ \sum_{j = 1}^{k_1} \max_{l \in \{1, \ldots, k_2 \}} | \mathcal{S}_{1
  j} \cap \mathcal{S}_{2 l} | + \sum_{l = 1}^{k_2} \max_{j \in \{1, \ldots,
  k_1 \}} | \mathcal{S}_{1 j} \cap \mathcal{S}_{2 l} | \biggr] .
  \label{DEF:epsilonnlocation}
\end{equation}

We can consider $ \epsilon_n(\cdot,\cdot)$ as a discrete version of $ \epsilon(\cdot,\cdot)$. \cite{van2000performance} first introduces the same distance measure as (\ref{DEF:epsilonnlocation}) (differing by a normalization term) for comparing two discrete set partitions. %``Markov cluster clusterings and clusterings of randomly generated test graphs". 
We extend $ \epsilon_n(\cdot,\cdot)$ to $ \epsilon(\cdot,\cdot)$ in this paper %where the former is on the discrete set partitions and the latter is 
for comparing two spatial domain partitions. The idea of $ \epsilon(\cdot,\cdot)$ (and $ \epsilon_n(\cdot,\cdot)$) is based on set matching. For each sub-domain $\mathcal{D}_{1 j}$ in $\pi_1(\mathcal{D})$, we find a ``best matched" sub-domain in $\pi_2(\mathcal{D})$, defined as the one sharing the largest intersection area with $\mathcal{D}_{1 j}$. The corresponding intersection area is then computed by $\max_{l \in \{1, \ldots, k_2 \}} |
  \mathcal{D}_{1 j} \cap \mathcal{D}_{2 l}|$.  If two partitions are close, the summation of the ``best matched" areas (i.e., $\sum_{j=1}^{k_1}\max_{l \in \{1, \ldots, k_2 \}} |
  \mathcal{D}_{1 j} \cap \mathcal{D}_{2 l}|$) is expected to be close to $|\mathcal{D}|$, leading to a small $\epsilon(\cdot,\cdot)$. The same rationale applies to $\sum_{l=1}^{k_2}\max_{j \in \{1, \ldots, k_1 \}} |
  \mathcal{D}_{1 j} \cap \mathcal{D}_{2 l}|$. Through the definition, we can see $ \epsilon(\cdot,\cdot)$ takes values in $[0,2)$ and equals $0$ when $\pi_1 (\mathcal{D})=\pi_2 (\mathcal{D})$. The same property holds for $ \epsilon_n(\cdot,\cdot)$. Roughly speaking, we can consider $\epsilon\{\pi_1 (\mathcal{D}),\pi_2 (\mathcal{D})\}/2$ ($\epsilon_{n}\{\pi_1 (\mathcal{S}),\pi_2 (\mathcal{S})\}/2$) as the ``mis-matched" percentage for $\pi_1 (\mathcal{D})$ and $ \pi_2 (\mathcal{D})$ ($\pi_1 (\mathcal{S})$ and $\pi_2 (\mathcal{S})$).  Furthermore, we have the following result for $ \epsilon(\cdot,\cdot)$ and $ \epsilon_n(\cdot,\cdot)$. 

\begin{proposition}
  \label{PP:distance} $\epsilon (\cdot, \cdot)$ and $\epsilon_n
  (\cdot, \cdot)$ defined in (\ref{DEF:epsilonspatial}) and
  (\ref{DEF:epsilonnlocation}) are distances, in the sense that they satisfy the axioms for a distance, i.e.,
  non-negativity, the identity of indiscernibles, symmetry, and triangle
  inequality.
\end{proposition}

We defer the proof of Proposition \ref{PP:distance} to Section
\ref{SEC:distanceproof}. Denote $\{\mathcal{S}_{l, 0} =
\{ \tmmathbf{s}_i \in \mathcal{D}_{l, 0} : 1 \leqslant i \leqslant n \}\}_{l=1}^{k_0}$ as the true partition of $\mathcal{S}$. Recall that we write
$\mathfrak{D}= [\{\tmmathbf{s}_i, \tmmathbf{x}(\tmmathbf{s}_i), y
(\tmmathbf{s}_i)\}_{i = 1}^n$ as the observed data. The following Theorem
\ref{TH:clustererror} establishes the partition consistency of the posterior domain partition
$\pi^\ast (\mathcal{D})$ and the observed location partition $\pi (\mathcal{S})$, with respect to distance $\epsilon
(\cdot, \cdot)$ and $\epsilon_n (\cdot, \cdot)$, respectively. 

\begin{theorem}
  \label{TH:clustererror}Let $0<\alpha_0<\alpha_2$ be a positive value. Under Assumptions
  \ref{AS:distributionloc}, \ref{AS:lengthofboundary}, \ref{AS:thetagap},
  \ref{AS:covariates}, \ref{AS:hyperpara} and \ref{AS:prior}, there exist some
  positive constants $c_1, c_2$ and $c_3$ (which are dependent on $\alpha_0$),
  such that with probability tending to $1$, we have
\begin{equation}
    \mathbb{P} (| \pi^{\ast} (\mathcal{D}) | = k_0 \mid \mathfrak{D})
    \geqslant 1 - c_1 \exp \{- c_2 n k_0^{1 / 2} K^{- 1} \log^{\alpha_0} (n)
    r_n^2 \}, \label{EQ:kconsistency}
  \end{equation}
  \begin{equation}
    \mathbb{P} (\epsilon [\pi^{\ast} (\mathcal{D}), \{\mathcal{D}_{l, 0} \}_{l
    = 1}^{k_0}] \leqslant c_3 k_0^{1 / 2 + 2 / d} K^{- 1} \log^{\alpha_0} (n)
    \mid \mathfrak{D}) \geqslant 1 - c_1 \exp \{- c_2 n k_0^{1 / 2} K^{- 1}
    \log^{\alpha_0} (n) r_n^2 \} \label{EQ:errorate0}
  \end{equation}
  and
  \begin{equation}
    \mathbb{P} (\epsilon_n [\pi (\mathcal{S}), \{\mathcal{S}_{l, 0} \}_{l =
    1}^{k_0}] \leqslant c_3 k_0^{1 / 2 + 2 / d} K^{- 1} \log^{\alpha_0} (n)
    \mid \mathfrak{D}) \geqslant 1 - c_1 \exp \{- c_2 n k_0^{1 / 2} K^{- 1}
    \log^{\alpha_0} (n) r_n^2 \} . \label{EQ:errorrate}
  \end{equation}
\end{theorem}

Under Assumption \ref{AS:hyperpara} and Equation (\ref{EQ:rnk0}), it can be verified by algebra that the right-hand sides of (\ref{EQ:kconsistency}) - (\ref{EQ:errorrate}) converge to $1$, and $k_0^{1 / 2 + 2 / d} K^{- 1} \log^{\alpha_0} (n) \rightarrow 0$. Thus, Theorem \ref{TH:clustererror} states the partition consistency of our model: with probability tending to $1$, the
posterior partition achieves the correct cluster number, and the posterior distributions of $\pi^\ast
(\mathcal{D})$ and $\pi
(\mathcal{D})$ achieve a contraction rate of $k_0^{1 / 2 + 2 / d} K^{- 1} \log^{\alpha_0} (n)$ around the truth.

Recall that $\pi^{\ast}_0(\mathcal{D})=\{ \mathcal{D}^{\ast}_{1, 0},\ldots,\mathcal{D}^{\ast}_{k_0, 0} \}$ in Proposition \ref{PP:bestapproximation}. For a given $\pi^{\ast}(\mathcal{D}) = \{\mathcal{D}_1^{\ast},
\ldots, \mathcal{D}^{\ast}_k \}$, we write $\mathcal{M} (\mathcal{D}^{\ast}_j) =
\tmop{argmax}_{l \in \{1, \ldots, k_0 \}} |\mathcal{D}^{\ast}_j \cap
\mathcal{D}^{\ast}_{l, 0} |$ as the index of the sub-domain in $\{
\mathcal{D}^{\ast}_{l, 0} \}_{l = 1}^{k_0}$ with the largest intersection area with
$\mathcal{D}^{\ast}_j$. $\mathcal{D}^{\ast}_{\mathcal{M}
(\mathcal{D}^{\ast}_j), 0}$ is considered as the ``best matched" sub-domain in $\{
\mathcal{D}^{\ast}_{l, 0} \}_{l = 1}^{k_0}$ for  $\mathcal{D}^{\ast}_j$. Thus, roughly speaking, we consider $\tmmathbf{\theta}_{\mathcal{M}(\mathcal{D}^{\ast}_j), 0}$ as the ``true" regression coefficient in $\mathcal{D}^{\ast}_j$. For a
new observed location $\mathbb{s}$ and covariate $\tmmathbf{x}
(\mathbb{s})$, write $\mu_0 \{\mathbb{s},\tmmathbf{x}(\mathbb{s})\}
=\tmmathbf{x}^T (\mathbb{s}) \tmmathbf{\theta}_0
(\mathbb{s})$ as the true regression mean, where
$\tmmathbf{\theta}_0 (\mathbb{s}) = \sum_{l = 1}^{k_0} \tmmathbf{\theta}_{l,0}
\mathbb{I} (\mathbb{s} \in \mathcal{D}_{l, 0})$ is the true regression coefficient. Recall the definition of $\mu \{\mathbb{s},\tmmathbf{x}(\mathbb{s}) \}$ in Section \ref{SEC:lmmodel}. The following Theorem \ref{TH:thetaconverge} states the posterior contraction rate of $\tmmathbf{\theta}$ and the prediction error of $\mu \{\mathbb{s},\tmmathbf{x}(\mathbb{s}) \}$.
\begin{theorem}
  \label{TH:thetaconverge}Under Assumptions \ref{AS:distributionloc},
  \ref{AS:lengthofboundary}, \ref{AS:thetagap}, \ref{AS:covariates},
  \ref{AS:hyperpara} and \ref{AS:prior} and for the same $\alpha_0$ in Theorem
  \ref{TH:clustererror}, with probability tending to $1$, we have
  \begin{gather}
    \mathbb{P} [\{\tmmathbf{\theta}_{\mathcal{M}(\mathcal{D}^{\ast}_j), 0}
    \}_{j = 1}^k =\{\tmmathbf{\theta}_{l, 0} \}_{l = 1}^{k_0} \mid
    \mathfrak{D}] \rightarrow 1,  \label{EQ:TH21}\\
    \mathbb{P} \{r_n^{- 1} k^{- 1} \sum_{j = 1}^k \|\tmmathbf{\theta}_j
    -\tmmathbf{\theta}_{\mathcal{M}(\mathcal{D}^{\ast}_j), 0} \|_2 > M_n
    k_0^{1 / 2 + 2 / d} K^{- 1} \log^{\alpha_0} (n) \mid \mathfrak{D}\}
    \rightarrow 0,  \label{EQ:TH22}\\
    \mathbb{P} \left\{ r_n^{- 2} \int_{\mathcal{D}}
    \|\tmmathbf{\theta}(\mathbb{s}) -\tmmathbf{\theta}_0 (\mathbb{s})\|_2^2
    \mathbb{P}(\mathbb{s}) d\mathbb{s}> M_n' k_0^{1 / 2 + 2 / d} K^{- 1}
    \log^{\alpha_0} (n) \mid \mathfrak{D} \right\} \rightarrow 0, \text{and} 
    \label{EQ:TH24}\\
    \mathbb{P} \left( r_n^{- 2} \iint [ \mu \{\mathbb{s},\tmmathbf{x}(\mathbb{s}) \} - \mu_0 \{\mathbb{s},\tmmathbf{x}(\mathbb{s}) \} ]^2 \mathbb{P} (\mathbb{s},
    \tmmathbf{x}) d\mathbb{s}
    d\tmmathbf{x} > M_n'' k_0^{1 /
    2 + 2 / d} K^{- 1} \log^{\alpha_0} (n) \mid \mathfrak{D} \right)
    \rightarrow 0  \label{EQ:TH23}
  \end{gather}
  for any sequences $M_n, M_n', M_n'' \rightarrow \infty$.
\end{theorem}

Equation (\ref{EQ:TH21}) states that with probability tending to 1, the set $\{\tmmathbf{\theta}_{\mathcal{M}(\mathcal{D}_j^\ast), 0} \}_{j = 1}^k$ is the same as the set $\{\tmmathbf{\theta}_{l, 0} \}_{l = 1}^{k_0}$.  Through the interpretation of $\tmmathbf{\theta}_{\mathcal{M}(\mathcal{D}^{\ast}_j), 0}
    $, $\|\tmmathbf{\theta}_j -\tmmathbf{\theta}_{\mathcal{M}(\mathcal{D}^{\ast}_{j}), 0}
\|_2$ specified in (\ref{EQ:TH22}) is considered as the distance between
$\tmmathbf{\theta}_j$ and its ``true" value. Thus, Equation (\ref{EQ:TH22}) states that averagely speaking, the rescaled regression coefficient achieves a contraction rate of $k_0^{1 / 2 + 2 / d} K^{- 1} \log^{\alpha_0} (n)$ around the ``truth". Equations (\ref{EQ:TH24}) - (\ref{EQ:TH23}) provide the {{posterior predictive contraction rates}} of the rescaled $\tmmathbf{\theta}(\mathbb{s})$ and $\mu \{\mathbb{s},\tmmathbf{x}(\mathbb{s}) \}$, respectively.

\begin{remark}
\label{RM:optimal rate}
Note that the contraction rates provided in Theorem \ref{TH:thetaconverge} coincide with that in Theorem \ref{TH:clustererror}, i.e., $k_0^{1 / 2 + 2 / d} K^{- 1} \log^{\alpha_0} (n)$. This suggests that the partition estimation error dominates the subsequent parameter estimation errors. On the other hand, note that this rate decreases as $K$ increases. Under Assumption \ref{AS:lorder}, the maximal admissible order of $K$  is $K \sim  k_0^{-1/2-1/d}r_n n^{1/2}\log^{-(1 + \alpha_b)/2} (n)$ for some $\alpha_b>0$. Correspondingly, we can set $\log(\lambda^{-1})\sim k_0^{-1-2/d}r_n^2 n \log^{-\alpha_p}(n)$ for some $0<\alpha_p<\alpha_b$. It can then be verified that Assumption \ref{AS:hyperpara} is satisfied. %with a resulting contraction rate $k_0^{1+3/d}r_n^{-1}n^{-1/2}\log^{\alpha_0+(1+\alpha_b)/2}(n)$ for our model.  
\end{remark}

Following the argument in Remark \ref{RM:optimal rate}, the next corollary states the optimal contraction rate of our model under the VDG regime, with the orders of $(K,\lambda)$ specified in Remark \ref{RM:optimal rate}.

\begin{corollary}
\label{CL:VDGcontraction}
    Under the specified orders of $(K,\lambda)$ in Remark \ref{RM:optimal rate}, and Assumptions \ref{AS:distributionloc} - \ref{AS:covariates} and
  \ref{AS:prior}, the contraction rates in Theorems \ref{TH:clustererror} - \ref{TH:thetaconverge} are $k_0^{1+3/d}r_n^{-1}n^{-1/2}\log^{\alpha_0+(1+\alpha_b)/2}(n)$ under the VDG regime.
\end{corollary}

The specified order of $K$ in Remark \ref{RM:optimal rate} implies that, when $k_0$ is large and $r_n$ is small, a smaller $K$ is required to achieve partition consistency. The rationale is that a large $k_0$ yields fewer locations per cluster, increasing the variance of the corresponding estimated regression coefficients, and thereby reducing distinguishability between clusters. Moreover, a small $r_n$ further exacerbates this effect. Hence, fewer blocks are needed such that the partition space is smaller, allowing the true signal to be extracted and partition consistency attained. On the other hand, the contraction rate in Corollary \ref{CL:VDGcontraction} is slower with a larger $k_0$ and a smaller $r_n$, which aligns with intuition. Note that the dimension $d$ also plays a role in the contraction rate. When $d$ is large, $\{ \tmmathbf{\theta}_{l, 0} \}_{l =1}^{k_0}$ are more sparsely distributed in the space, making them easier to distinguish.

Recall that under the FDG regime, both $r_n$ and $k_0$ are constant. Following the argument in Remark \ref{RM:optimal rate}, we obtain the following corollary regarding the optimal contraction rate in the FDG regime.

\begin{corollary}
\label{CL:FDGcontraction}
     Suppose $K \sim  n^{1/2}\log^{-(1 + \alpha_b)/2} (n)$ for some $\alpha_b>0$, and $\log(\lambda^{-1})\sim  n \log^{-\alpha_p}(n)$ for some $0<\alpha_p<\alpha_b$. Under Assumptions \ref{AS:distributionloc} - \ref{AS:covariates} and
  \ref{AS:prior}, the contraction rates in Theorems \ref{TH:clustererror} - \ref{TH:thetaconverge} are $n^{-1/2}\log^{\alpha_0+(1+\alpha_b)/2}(n)$ under the FDG regime.
\end{corollary}

Under the FDG regime, we compare the result of Corollary \ref{CL:FDGcontraction} with a change point detection result in a one-dimensional space. According to
Theorem 7 in {\cite{frick2014multiscale}}, the $\epsilon_{n}(\cdot,\cdot)$ of one-dimensional change point detection problem is of order $n^{-1}\log(n)$, which is smaller than our result. This difference arises because
partitioning in a spatial domain involves a substantially larger partition space
than the one-dimensional case, increasing the complexity of achieving
partition consistency. See also \cite{henglian} for a similar one-dimensional change point detection result under the Bayesian context.

As for $\tmmathbf{\theta}_j$, the posterior contraction rate $n^{- 1 / 2} \log^{\alpha_0 + (1 + \alpha_b)
/ 2} (n)$ is slightly slower than the classic parametric contraction
rate $n^{- 1 / 2}$. This is non-trivial, as it indicates that the unknown spatial partition caused by
spatial heterogeneity impacts the posterior contraction rate of $\tmmathbf{\theta}_{j} $ solely at the order of some power of $\log (n)$. Also note that the rate $n^{- 1 / 2} \log^{\alpha_0 + (1 + \alpha_b)
/ 2} (n)$ is the same as that in Corollary 6 of
{\cite{luo2021bayesian}} (if ignoring logarithmic terms), which considers the {posterior contraction rate} for the regression mean at $\{\tmmathbf{s}_i\}_{i=1}^{n}$.

\section{Simulation studies}\label{SEC:simu}

The simulation consists of three parts. In Section \ref{SEC:asymptotic}, we investigate the asymptotic properties of the proposed Spat-RPM model under the VDG regime on the domain $[0,1]^2$. In Section \ref{SEC:Ushape}, we consider a U-shape domain under the FDG regime. We examine the influence of hyperparameters $K,\lambda$, and $\sigma^2$, compare performance with the Bayesian spatially clustered varying coefficient (BSCC) model of \cite{luo2021bayesian}, and examine the effectiveness of using the CH index/WAIC to select $(K,\lambda)$. In Section \ref{SEC:areal data}, we apply Spat-RPM to generated areal data on a domain with a hole to demonstrate the model’s effectiveness for areal datasets. 

\subsection{Asymptotic analysis}
\label{SEC:asymptotic}
\begin{figure}[htbp]
		\centering
		\begin{tabular}{ccc}
             {\includegraphics[width=0.27\linewidth,height=0.15\textheight]{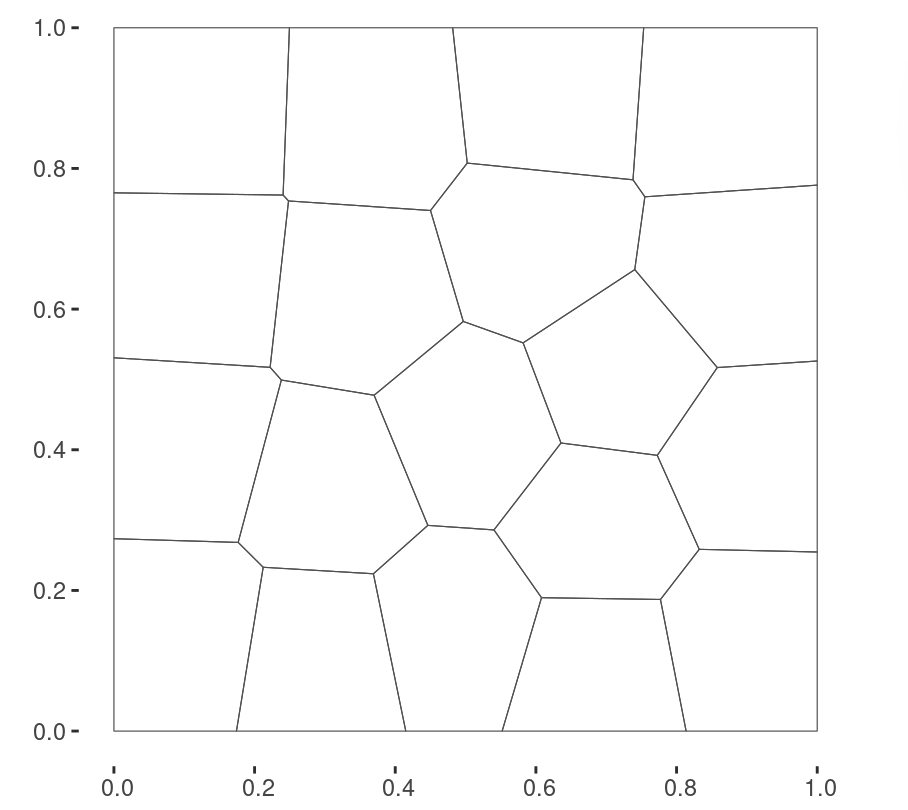}}&
			{\includegraphics[width=0.27\linewidth,height=0.15\textheight]{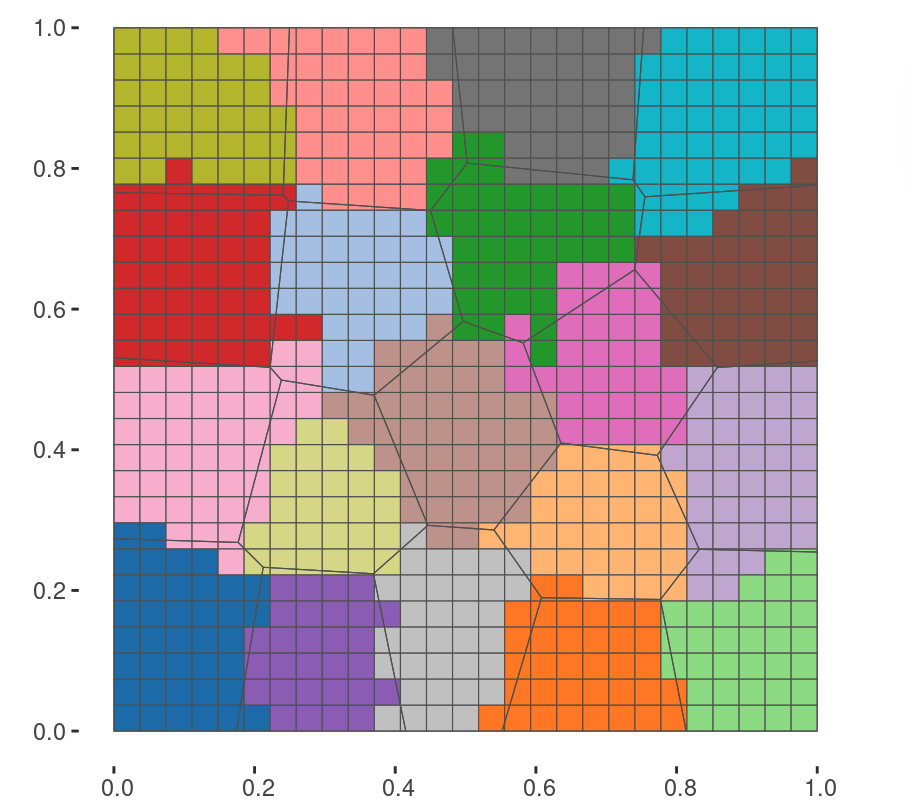}}&
			{\includegraphics[width=0.3\linewidth,height=0.15\textheight]{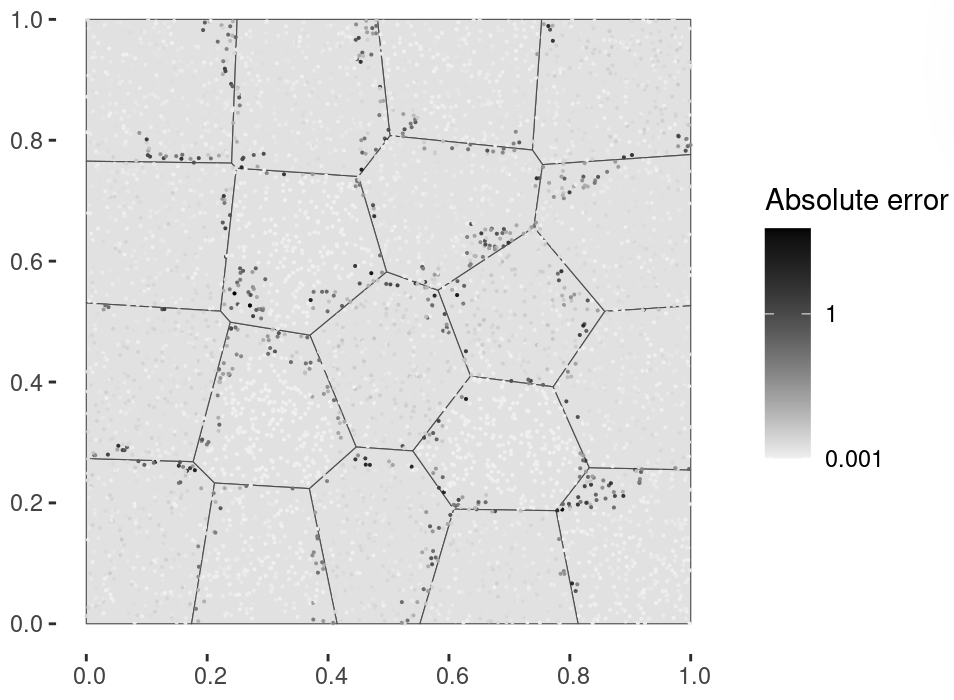}}\\
			{\small (a)} & {\small (b)} & {\small (c)}
		\end{tabular}
		\caption{Fitting results with $n=50000$ and $k_0=19$. (a) The generated true domain partition $\{\mathcal{D}_{l,0}\}_{l=1}^{k_0}$. (b) One randomly chosen posterior partition sample with different colors representing different clusters. (c) The spatial distribution of absolute prediction errors. The darker the color, the larger the absolute prediction error.}
		\label{FIG:n50000result}
        \vspace{-10pt}
	\end{figure}

Let $n$ be the sample size, the data is generated by
\begin{equation}
\label{EQ:simudata}
    y (\tmmathbf{s}_i) = \tmmathbf{x}^T (\tmmathbf{s}_i) \tmmathbf{\theta}
	(\tmmathbf{s}_i) + \epsilon (\tmmathbf{s}_i), 1 \leqslant i \leqslant n,
\end{equation}
where $\tmmathbf{x} (\tmmathbf{s}_i) = \{ x_1 (\tmmathbf{s}_i), x_2
	(\tmmathbf{s}_i) \}^T$ is a two dimensional vector, with $\{x_1 (\tmmathbf{s}_i)\}_{i=1}^{n}$ and $\{x_2 (\tmmathbf{s}_i)\}_{i=1}^{n}$ being $i.i.d.$ $\tmop{Unif}(-1,1)$. We focus on the domain $[0,1]^2$, where the true domain partition $\{\mathcal{D}_{l,0}\}_{l=1}^{k_0}$ is generated by the centroidal Voronoi tessellation (CVT) using Lloyd's algorithm \citep{lloyd1982least}. See Figure \ref{FIG:n50000result}(a) for the generated $\{\mathcal{D}_{l,0}\}_{l=1}^{k_0}$ with $k_0=19$. Given the domain partition, let $\tmmathbf{c}_{l,0}$ be the centroid of $\mathcal{D}_{l,0}$. We set $\tmmathbf{\theta}(\tmmathbf{s})=r_n\tmmathbf{c}_{l,0}$ if $\tmmathbf{s} \in \mathcal{D}_{l,0}$, and $r_n=5$. The noise $\{\epsilon (\tmmathbf{s}_i)\}_{i=1}^{n}$  are $i.i.d.$ Gaussian random variables with mean
	$0$ and variance $0.01$. The sampling locations $\{ \tmmathbf{s}_i\}_{i=1}^{n}$ are uniformly
	distributed within $[0,1]^2$.

	To investigate the asymptotic properties of our model, we compare the posterior samples
	with $n = (500,1000,5000,10000,30000,50000)$. Note that the covariate dimension $d=2$ under our simulation setting. Let $\lfloor a \rceil$ denote the nearest integer to $a$. We set $k_0=\lfloor5n^{1/8}\rceil=(11, 12, 14, 16, 18, 19)$, which satisfies Equation (\ref{EQ:rnk0}). We set $\gamma = 1$ throughout the simulation. We select the number of blocks hyperparameter $K$ and Poisson hyperparameter $\lambda$ with the orders in Remark \ref{RM:optimal rate}, i.e., $K = \lfloor c_b k_0^{-1}r_n n^{1 / 2}
	\log^{- (1 + \alpha_b) / 2} (n)\rceil$ and $\log (\lambda^{- 1}) = c_p k_0^{-2} r_n^2 n \log^{-
		\alpha_p} (n)$, where $c_b = 5, \alpha_b = 1$, $c_p = 0.1$ and $\alpha_p = 0.5$. We set the noise variance $\sigma^2 = 1$, which is different from the true variance $0.01$. We will see from the following result that this noise variance misspecification doesn't affect the partition consistency of our model, which aligns with our theoretical result. 
        
        We repeat the simulation $100$ times. For each repeat, we obtain posterior samples following strategies in Section \ref{SEC:computation}. We run $20000$ MCMC iterations with a burn-in period $15000$ and a thinning parameter $5$. Finally, for the sample size $n$ and the $r$-th repeat, we obtain posterior samples $\{ k_{n,r,s}, \pi^\ast_{n,r,s},
	\{ \tmmathbf{\theta}_{n,r,s, j} \}_{j = 1}^{k_{n,r,s}} \}_{s = 1}^M$,
	where $M=1000$, and $k_{n,r,s}, \pi^\ast_{n,r,s}$, $\{ \tmmathbf{\theta}_{n,r,s, j} \}_{j = 1}^{k_{n,r,s}}$ are the corresponding posterior cluster number, domain partition and regression coefficients, respectively, at the $s$-th MCMC sample.
	
\begin{figure}[htbp]
    \vspace{-5mm}
		\centering
		\begin{tabular}{cccc}
			{\includegraphics[width=0.23\linewidth,height=0.22\textheight]{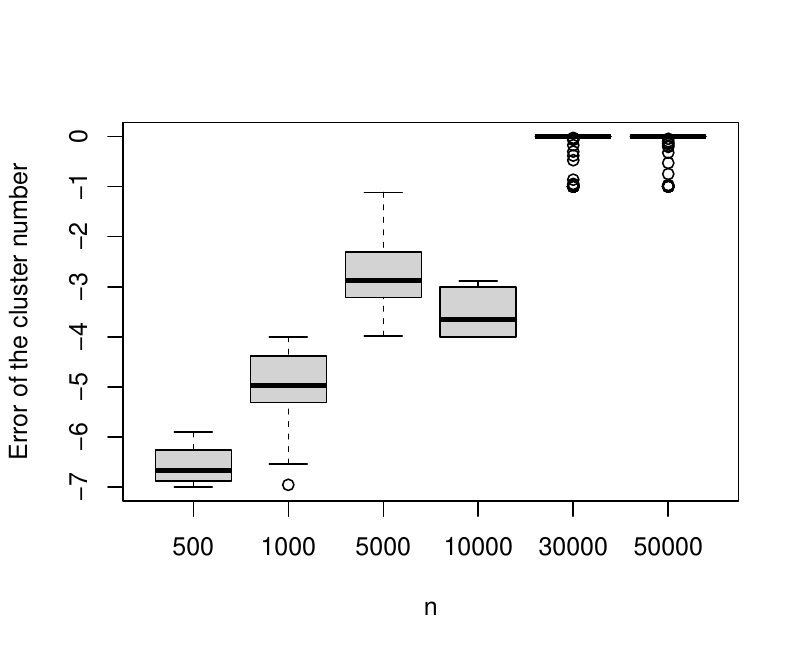}}&
			{\includegraphics[width=0.23\linewidth,height=0.22\textheight]{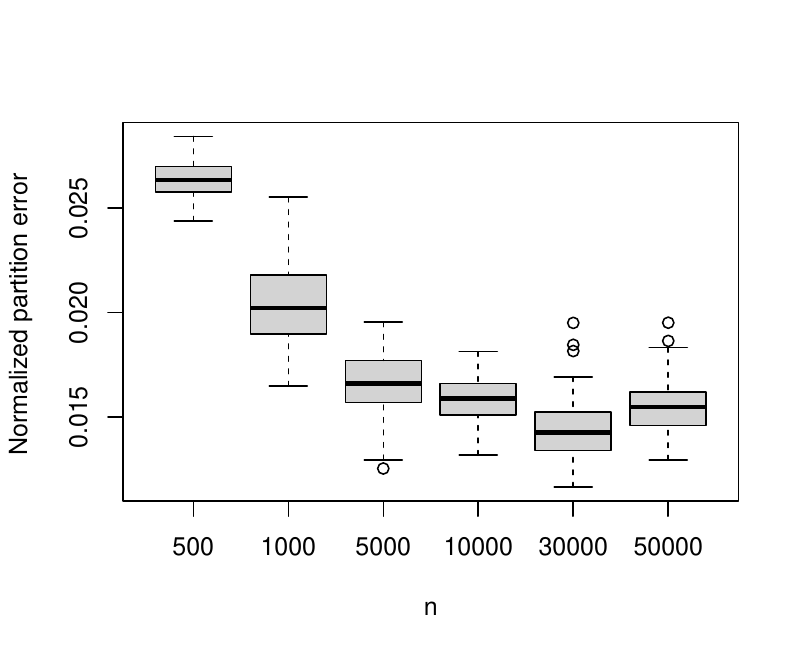}}&
			{\includegraphics[width=0.23\linewidth,height=0.22\textheight]{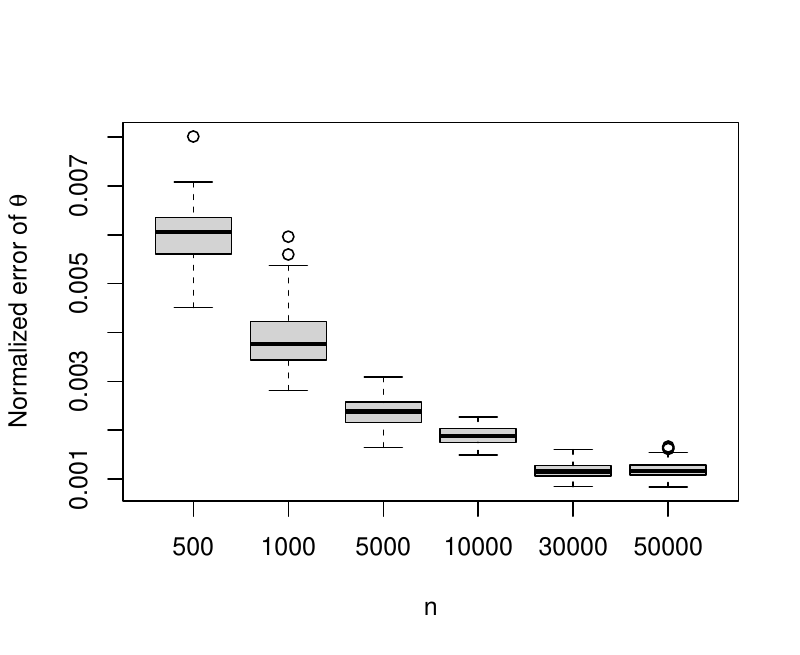}}&
			{\includegraphics[width=0.23\linewidth,height=0.22\textheight]{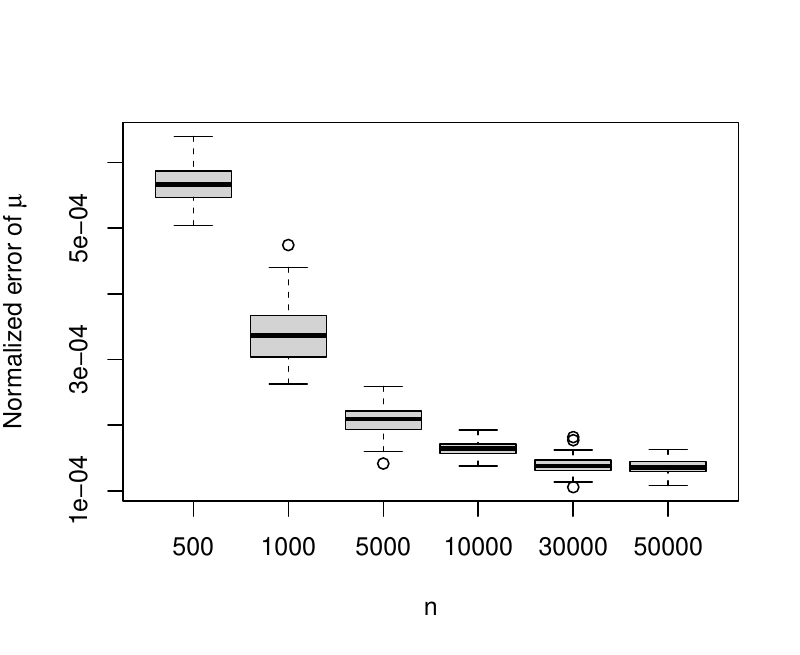}}\\
            {\small (a)} & {\small (b)} & {\small (c)} & {\small (d)}
		\end{tabular}
		\caption{Fitting results under different sample sizes $n$. (a) Boxplots of the mean error of cluster number. (b) - (d) Boxplots of ${\{ e_{n,r, 1} \}_{r = 1}^{100}}$, ${\{ e_{n, r, 2} \}_{r = 1}^{100}}$ and ${\{ e_{n,r,  3} \}_{r = 1}^{100}}$, respectively. }
		\label{FIG:symptotic_analysis}
        \vspace{-5mm}
	\end{figure}

	Figure \ref{FIG:symptotic_analysis}(a) shows boxplots of the mean error in the posterior number of clusters, computed as $M^{-1}\sum_{s=1}^{M}(k_{n,r,s}-k_0)$,
	across different sample sizes $n$ and repeated simulations. We can see that when $n$ is smaller than $10000$, the number
	of clusters is underestimated. However, when $n$ is larger than $30000$, the
	posterior number of clusters equals its true value in most of the repeats. This aligns with Equation (\ref{EQ:kconsistency}) in Theorem \ref{TH:clustererror}.

	Next, we evaluate the asymptotic properties of other metrics. For a given $(n,r)$
	and the corresponding posterior samples, we
	compute the following normalized errors according to Theorems \ref{TH:clustererror} - \ref{TH:thetaconverge} and Corollary \ref{CL:VDGcontraction}:
	\[ e_{n, r,1} = \sum_{s=1}^{M}\frac{ r_n n^{1 / 2}\epsilon [\pi^\ast_{n,r,s},\{\mathcal{D}_{l,0}\}_{l=1}^{k_0}]}{M k_0^{2.5}\log^{\alpha_0 + (1 +
			\alpha_b) / 2} (n)},\text{ and } e_{n, r,2} = \sum_{s=1}^{M}\frac{ n^{1 / 2} \sum_{j=1}^{k_{n,r,s}} \|
		\tmmathbf{\theta}_{n,r,s, j} - \tmmathbf{\theta}_{\mathcal{M}_{n,r,s, j}, 0}
		\|_2}{Mk_{n,r,s}k_0^{2.5}\log^{\alpha_0 + (1 + \alpha_b) / 2} (n)}, \]
where $\alpha_0 = 0.1$ and $\mathcal{M}_{n,r,s, j}$ is the index of the sub-domain in $\{\mathcal{D}_{l,0}\}_{l=1}^{k_0}$ with the largest intersection area with the $j$-th cluster, at the $s$-th MCMC sample. To study the prediction error of our model, we randomly generate $5000$ locations uniformly distributed in $[0,1]^2$, and the corresponding covariates. Following the procedure described in Section \ref{SEC:lmmodel}, we write $\tmmathbf{\mu}_{n,r,s}$ as the prediction vector of the regression mean at the sampled $5000$ locations, given the corresponding posterior sample. Let  $\tmmathbf{\mu}_0$ be the true
	value of $\tmmathbf{\mu}_{n,r,s}$. We define the normalized prediction error as 	\[ e_{n,r,3} = \sum_{s=1}^{M}\frac{ n^{1 / 2} \times 5000^{- 1} \| \tmmathbf{\mu}_{n,r,s} -
		\tmmathbf{\mu}_0 \|_2^2}{M r_n k_0^{2.5}\log^{\alpha_0 + (1 + \alpha_b) / 2} (n)}. \]
    
    According to Theorems
	\ref{TH:clustererror} - \ref{TH:thetaconverge} and Corollary \ref{CL:VDGcontraction}, the above three normalized
	errors are bounded by constants with a high probability when $n$ is large. Figures \ref{FIG:symptotic_analysis}(b) - (d) show boxplots of ${\{ e_{n, r, 1} \}_{r = 1}^{100}}$, ${\{ e_{n, r, 2} \}_{r = 1}^{100}}$, and ${\{ e_{n, r, 3} \}_{r = 1}^{100}}$ under different $n$, respectively. We can see that the three normalized errors first decrease as $n$ increases, then stay relatively
	stable when $n$ is larger than $30000$, which is
	consistent with our theory. %\kun{When $n$ is small, large normalized errors may result from the misspecification of $\sigma^2$ and having few locations within each block.} 
    Note that under our choice of the growth rate of $k_0$ and the hyperparameters, the contraction rate computed from Corollary \ref{CL:VDGcontraction} is $n^{-3/16}\log^{1.1}(n)$, which is relatively slow. Consequently, a sample size greater than $30000$ is required in our simulation setting to achieve consistency in estimating the number of clusters, and to ensure convergence to the theoretical asymptotic behavior.
	
	Specifically, we focus on a posterior sample when $n = 50000$ to check the fitting of our model. %For one randomly chosen MCMC sample, 
    Figures
	\ref{FIG:n50000result}(b) - (c) show a randomly chosen posterior sample of the partition, and the spatial distribution of the absolute prediction errors, computed as the absolute value of entries in $\tmmathbf{\mu}_{n,r,s}-\tmmathbf{\mu}_0$. We can see
	that the posterior partition result recovers the true partition well, except for
	some locations near boundaries as expected. We can also see from Figure \ref{FIG:n50000result}(c) that the absolute prediction errors are relatively small, except for those ``wrongly" clustered locations near boundaries.

	\subsection{U-shape domain}
    \label{SEC:Ushape}
\begin{figure}[htbp]
    \vspace{-7mm}
		\centering
		\begin{tabular}{ccc}
             {\includegraphics[width=0.28\linewidth,height=0.13\textheight]{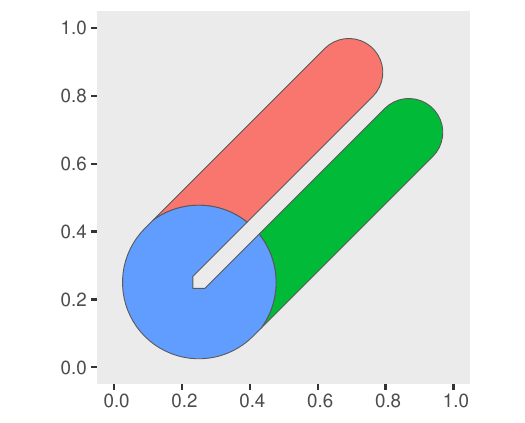}}&
			{\includegraphics[width=0.28\linewidth,height=0.13\textheight]{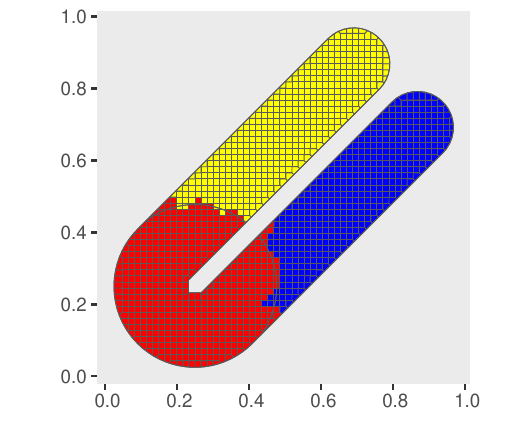}}&
            {\includegraphics[width=0.25\linewidth,height=0.16\textheight]{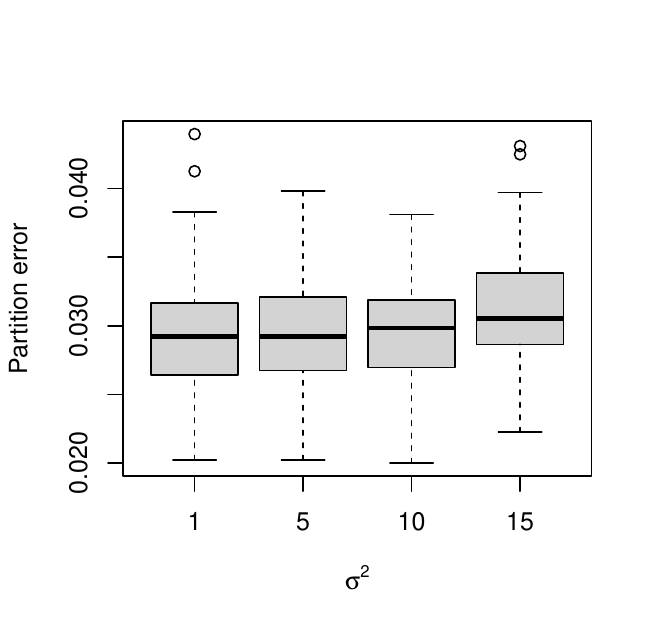}}
			%{\includegraphics[width=0.2\linewidth,height=0.16\textheight]{fig/CHvstrue.pdf}}
            \\
			{\small (a)} & {\small (b)} & {\small (c)}
		\end{tabular}
		\caption{(a) U-shape domain with $\{\mathcal{D}_{l,0}\}_{l=1}^{3}$ represented by different colors. (b) One randomly chosen posterior partition sample with different colors representing different clusters. (c) Boxplots of $\{e_{\cdot,\cdot,\sigma^2,r,1}\}_{r=1}^{100}$ under different $\sigma^2$.} %(d) The scatterplot of CH index and true partition error}
		\label{FIG:n4000result}
        \vspace{-10pt}
	\end{figure}

    This section conducts a simulation in the U-shape domain as shown in Figure \ref{FIG:n4000result}(a). U-shape is an irregular domain in the sense that Euclidean distance does not adequately capture spatial separation: locations on different arms of the “U” can appear close in terms of Euclidean distance, while being far apart in the actual geometry of the U-shape. The U-shape domain is partitioned into three
	sub-domains, $\{ \mathcal{D}_{l,0} \}_{l = 1}^3$, as indicated by different colors
	in Figure \ref{FIG:n4000result}(a): $\mathcal{D}_{1,0}$ is the upper arm, $\mathcal{D}_{2,0}$
	is the lower arm, and $\mathcal{D}_{3,0}$ is the middle circle. 

    We set $n=4000$ and generate data by (\ref{EQ:simudata}), with $\{\tmmathbf{x}^T (\tmmathbf{s}_i)\}_{i=1}^{n}$ following the same distribution as that in Section \ref{SEC:asymptotic}. We set $\tmmathbf{\theta} (\tmmathbf{s})=\tmmathbf{\theta}_{1, 0},\tmmathbf{\theta}_{2, 0} $ and $\tmmathbf{\theta}_{3, 0} $, for $\tmmathbf{s} \in
		\mathcal{D}_{1,0},\mathcal{D}_{2,0}$ and $\mathcal{D}_{3,0}$, respectively, where $\tmmathbf{\theta}_{1, 0} = (0, 0)^T,\tmmathbf{\theta}_{2, 0} = (-3, 4)^T$, and $\tmmathbf{\theta}_{3, 0} = (4, 3)^T$. We set the variance of $\epsilon (\tmmathbf{s}_i)$ to be $10$. To mimic the real data situations where the observed locations are not uniformly distributed, we sample locations $\{ \tmmathbf{s}_i\}_{i=1}^{n}$ with the density function $\mathbb{P}_{\mathcal{D}} (\tmmathbf{s})\propto \{1+0.2\cos(2 \pi s_1)\cos(2 \pi s_2)\}$, where $\tmmathbf{s}=(s_1,s_2)$.

We select $K$ and $\log(\lambda^{-1})$ following the orders in Corollary \ref{CL:FDGcontraction} under the FDG regime, i.e., $K = \lfloor c_b n^{1 / 2}
	\log^{- (1 + \alpha_b) / 2} (n)\rceil$ and $\log (\lambda^{- 1}) = c_p n \log^{-
		\alpha_p} (n)$, where $c_b, \alpha_b$, $c_p$ and $\alpha_p$ are the same as those in Section \ref{SEC:asymptotic}. Given an integer  $K$, to construct graph $\mathcal{G}$, we
	start from a mesh grid graph of $K^2$ square blocks with side length $K^{-1}$ in ${[0,
		1]^2}$. Then we remove blocks from the graph without observed locations. We also set the same $\sigma^2$ value as that in Section \ref{SEC:asymptotic}. Figure \ref{FIG:n4000result}(b) shows a randomly selected posterior partition result, which recovers the true partition well. The posterior values of $\{ \tmmathbf{\theta}_{j} \}_{j = 1}^3$
	under this posterior sample are $\tmmathbf{\theta}_{ 1} = (-0.02, -0.24)^T,
	\tmmathbf{\theta}_{2} = (-3.05, 3.96)^T$ and $\tmmathbf{\theta}_{ 3} =
	(3.87, 3.18)^T$, close to the true values. 

Next, under the current simulation setting, we conduct a sensitivity analysis to study the influence of hyperparameters $K,\lambda$, and $\sigma^2$. 

    \subsubsection{Sensitivity analysis}
    \label{SEC:sensitivity}
    We compute and compare the model fitting errors under different values of $K,\lambda$, and $\sigma^2$. The values of $(K,\lambda)$ are adjusted through $(c_b,c_p)$ described in the preceding paragraph. Specifically, we consider values of $(c_b,c_p,\sigma^2)\in \tmmathbf{C}_b\times\tmmathbf{C}_p\times \tmmathbf{C}_{\sigma^2}$, where $\tmmathbf{C}_b=\{1,3,5,7,9\},\tmmathbf{C}_p=\{0.01,0.05,0.1,1,2\}$ and $\tmmathbf{C}_{\sigma^2}=\{1,5,10,15\}$. We repeat the simulation for $100$ times. For the $r$-th repeat, and a given $(c_b,c_p,\sigma^2)$, we fit our model to the simulated data, and use the ``salso'' package in R~\cite{dahl2022search} to select the ``best" posterior sample. Based on this sample, we compute the domain partition error, the number of clusters, the estimation error of the regression coefficient, and the prediction error of the regression mean, denoted by $e_{K,\lambda,\sigma^2,r,1}$, $e_{K,\lambda,\sigma^2,r,2}$, $e_{K,\lambda,\sigma^2,r,3}$ and $e_{K,\lambda,\sigma^2,r,4}$, respectively.

    We first investigate the effect of $\sigma^2$. For the $r$-th repeat, write $e_{\cdot,\cdot,\sigma^2,r,1}=\min_{(K,\lambda)}\{e_{K,\lambda,\sigma^2,r,1}\}$ as the minimum partition error obtained under the given $\sigma^2$. Figure \ref{FIG:n4000result}(c) shows boxplots of $\{e_{\cdot,\cdot,\sigma^2,r,1}\}_{r=1}^{100}$ under different $\sigma^2$. We can see that the four boxplots are largely similar, with the case $\sigma^2=15$ showing a slightly higher mean. The corresponding mean values are computed as $0.0294, 0.0294, 0.0296$ and $0.0312$, respectively, suggesting that the differences are nearly negligible. This result indicates that the value of $\sigma^2$ has little impact on the fitting results, provided that $(K,\lambda)$ are chosen appropriately. Hence, in practice, we can fix $\sigma^2$ at a simple value (e.g., $1$ or the empirical variance of $\{y(\tmmathbf{s}_i)\}_{i=1}^n$), and focus on the selection of $(K,\lambda)$.
     %In practice, one simple way to set $\sigma^2$ is the empirical variance of $\{y(\tmmathbf{s}_i)\}_{i=1}^n$, which is approximately $15$ in this simulation setting. 

\begin{figure}[htbp]
    %\vspace{-5mm} 
		\centering
		\begin{tabular}{cccc}
			{\includegraphics[width=0.23\linewidth,height=0.22\textheight]{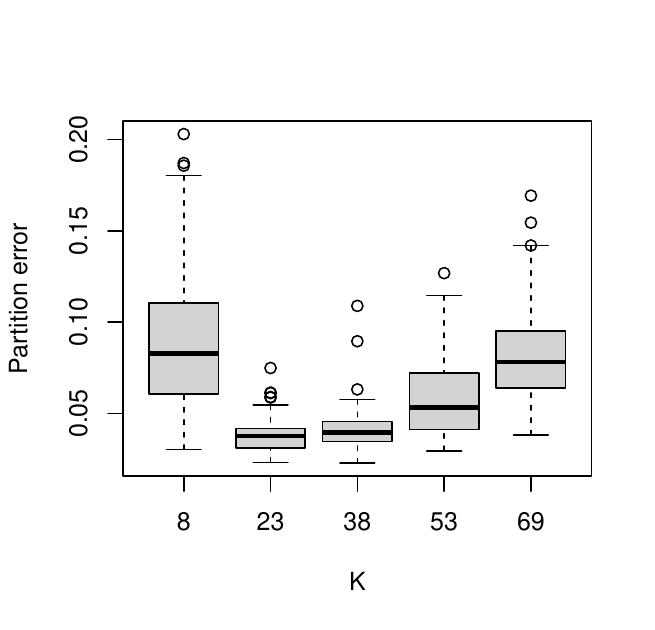}}&
			{\includegraphics[width=0.23\linewidth,height=0.22\textheight]{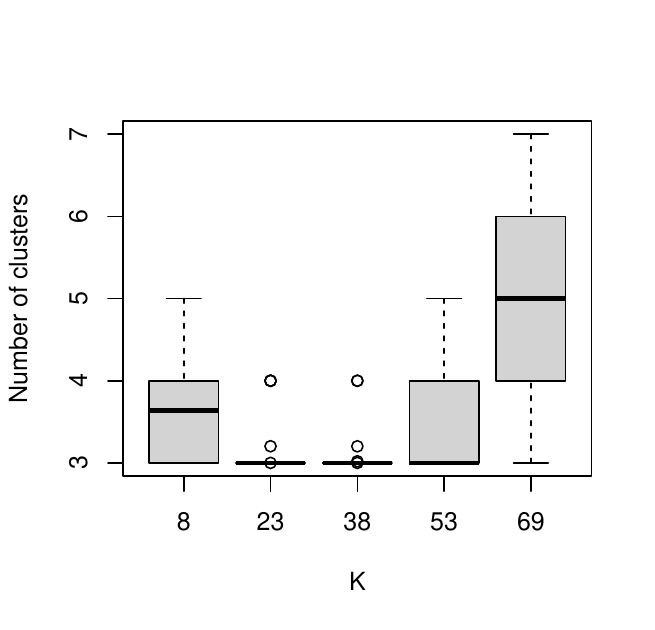}}&
			{\includegraphics[width=0.23\linewidth,height=0.22\textheight]{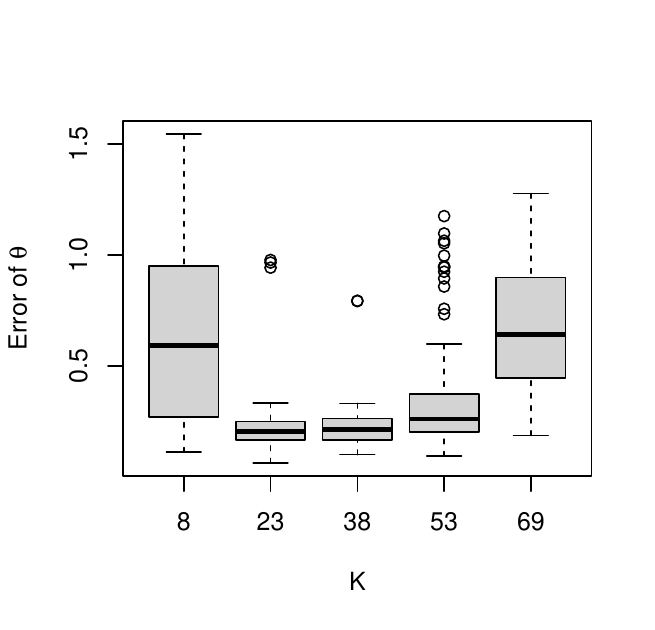}}&
			{\includegraphics[width=0.23\linewidth,height=0.22\textheight]{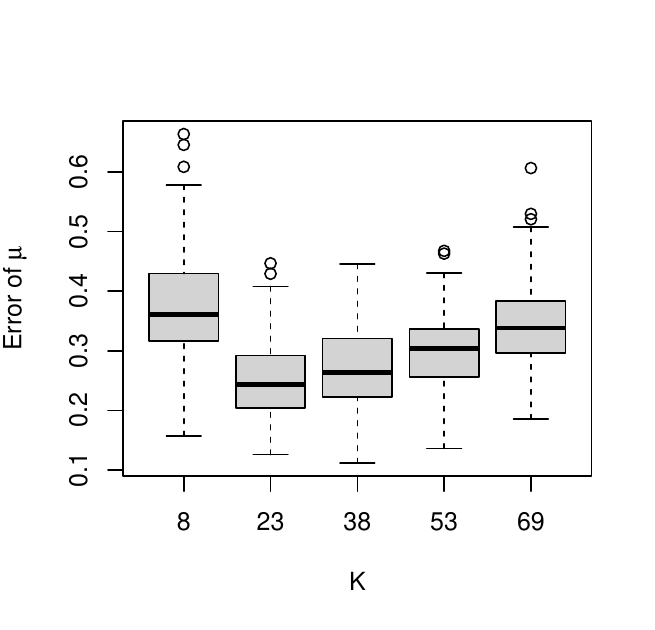}}
             \\
            {\small (a)} & {\small (b)} & {\small (c)} & {\small (d)}  \\
            			{\includegraphics[width=0.23\linewidth,height=0.22\textheight]{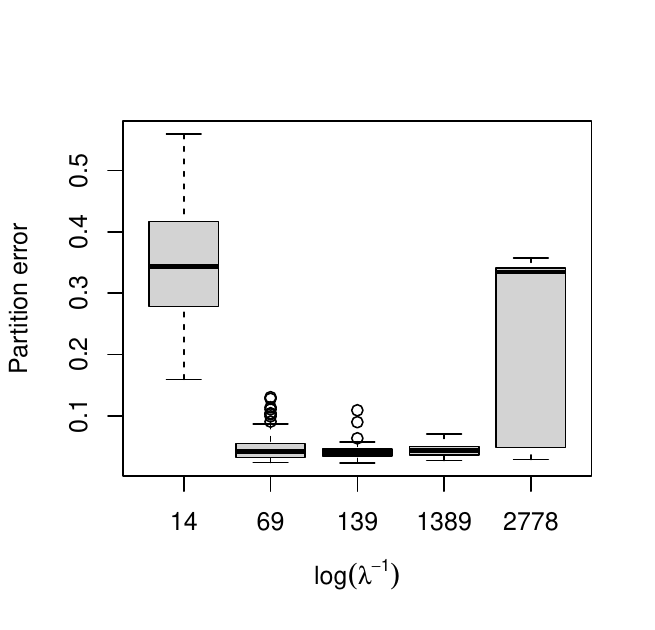}}&
			{\includegraphics[width=0.23\linewidth,height=0.22\textheight]{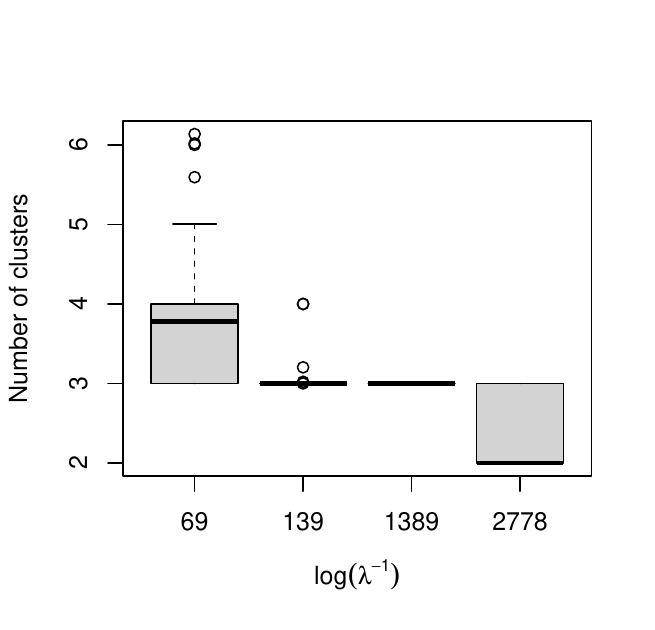}}&
			{\includegraphics[width=0.23\linewidth,height=0.22\textheight]{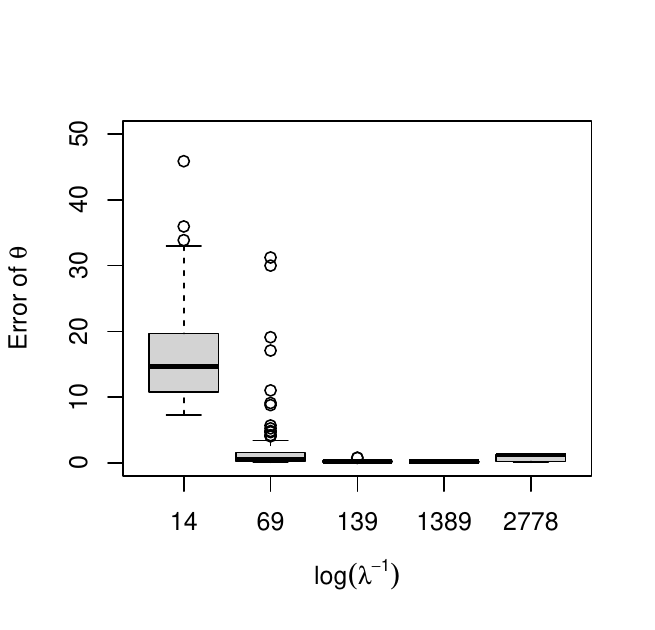}}&
			{\includegraphics[width=0.23\linewidth,height=0.22\textheight]{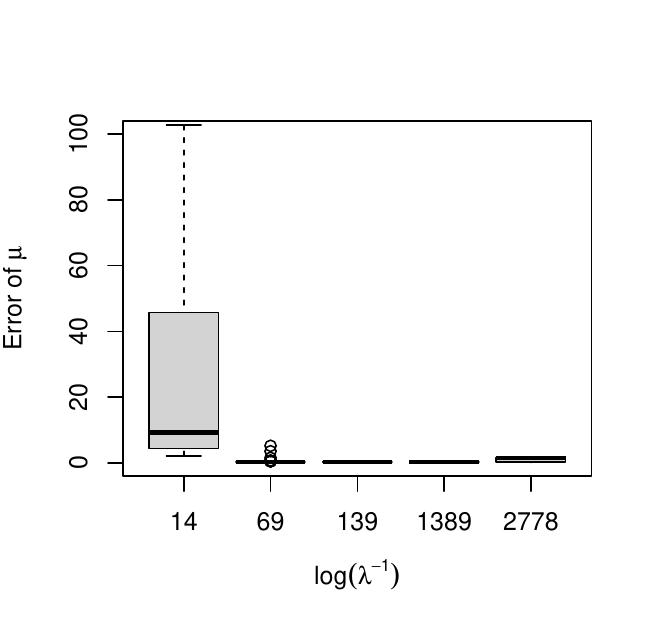}}\\
            {\small (e)} & {\small (f)} & {\small (g)} & {\small (h)}
		\end{tabular}
		\caption{Sensitivity analysis for $K$ and $\lambda$. (a) - (d) show boxplots of $e_{K,\lambda,\sigma^2,r,1}$, $e_{K,\lambda,\sigma^2,r,2}$, $e_{K,\lambda,\sigma^2,r,3}$ and $e_{K,\lambda,\sigma^2,r,4}$, respectively, with $c_p=0.1$ and $\sigma^2=1$ fixed. (e) - (h) show boxplots of $e_{K,\lambda,\sigma^2,r,1}$, $e_{K,\lambda,\sigma^2,r,2}$, $e_{K,\lambda,\sigma^2,r,3}$ and $e_{K,\lambda,\sigma^2,r,4}$, respectively, with $c_b=5$ and $\sigma^2=1$ fixed. The values of $K$ and $\lambda$ correspond to the constants in $\tmmathbf{C}_b$ and $\tmmathbf{C}_p$, respectively. In Figure (f), we omit the boxplot corresponding to $c_p=0.01$, where all repeats have more than $40$ clusters.}
		\label{FIG:sensitivity}
        %\vspace{-5mm} 
	\end{figure}

We next investigate the impact of $(K,\lambda)$.  Figures \ref{FIG:sensitivity}(a) - (d) show boxplots of $\{e_{K,\lambda,\sigma^2,r,q}\}_{r=1,q=1}^{100,4}$ under different values of $K$, with $c_p=0.1$ and $\sigma^2=1$ fixed. The results reveal a clear ``U-shaped" pattern: when $K$ is either too small or too large, the errors increase and the number of clusters is incorrectly estimated. This observation is consistent with the argument following Proposition \ref{PP:bestapproximation} and Assumption \ref{AS:hyperpara}: a small $K$ induces large approximation error, whereas a large $K$ complicates recovery of the correct partition due to the increased partition space.

 % Based on Figure \ref{FIG:n4000result}(c), we propose a hyperparameter selection strategy for our model. We first fix $\sigma^2$ using an initial guess, and then select $(K,\lambda)$ within a specified range of $(c_b,c_p)$. Figure \ref{FIG:n4000result}(c) indicates that the initial choice of $\sigma^2$ has little impact on the fitting results, provided that $(K,\lambda)$ are subsequently chosen appropriately. For example, a simple initial choice for $\sigma^2$ can be the empirical variance of $\{y(\tmmathbf{s}_i)\}_{i=1}^n$, which is approximately $15$ in our simulation setting.

 Figures \ref{FIG:sensitivity}(e) - (h) show boxplots of $\{e_{K,\lambda,\sigma^2,r,q}\}_{r=1,q=1}^{100,4}$ under different values of $\lambda$, with $c_b=5$ and $\sigma^2=1$ fixed. The results exhibit a similar ``U-shaped" pattern as Figures \ref{FIG:sensitivity}(a) – (d). A small $\log(\lambda^{-1})$ causes large errors from overestimating the number of clusters, whereas a large $\log(\lambda^{-1})$ causes large errors from underestimation.

\subsubsection{Comparison with BSCC}
\label{SEC:compareBSCC}
One key difference between these two models is
	that BSCC doesn't use the blocking technique and %constructs a graph (and thus
	%obtains partitions) 
    only provides clustering of the observed locations. 
	%In the proof in Section \ref{SEC:proof}, we find that blocking helps reduce the number of cluster spaces and control the noise effect, which is essential in obtaining cluster consistency. 
	We conduct the comparison for $n = 4000$ and $n=100$.
    % , with the results for the former presented in Section \ref{SEC:BSCCcompn4000} and those for the latter presented in Section \ref{SEC:BSCCcompn100}. 
    We repeat the simulation for $100$ times. For each repeat, we fit BSCC and our model to the
	same simulated data, respectively. %We use the same $(c_b,\alpha_b,c_p,\alpha_p,\sigma^2)$ as those specified in the setup at the beginning of Section \ref{SEC:Ushape}, and the default
	%hyperparameter setting for the BSCC model as specified in {\cite{luo2021bayesian}}. 
    We compare the posterior number of clusters, median absolute error (MedAE), and
	continuous ranked probability scores (CRPS, {\cite{gneiting2007strictly}}) for the regression mean prediction, and
	computing time between two models.

\begin{figure}[htbp]
		\centering
		\begin{tabular}{cccc}
			{\includegraphics[width=0.22\linewidth,height=0.22\textheight]{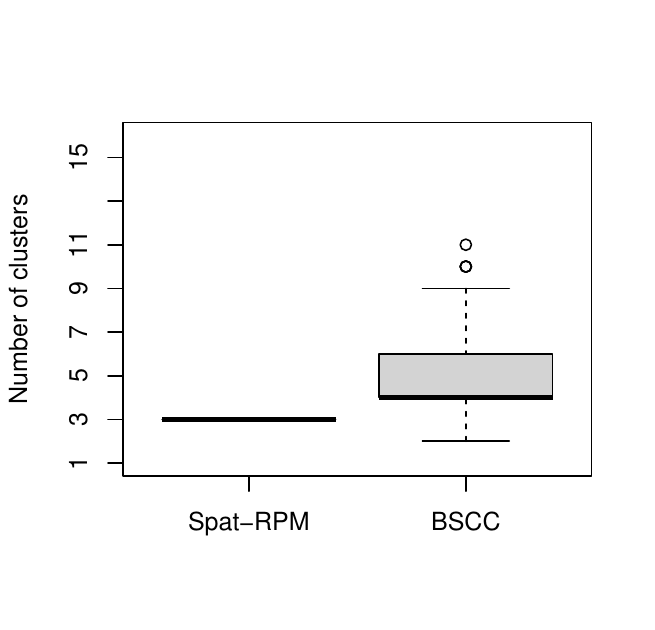}}&
			{\includegraphics[width=0.22\linewidth,height=0.22\textheight]{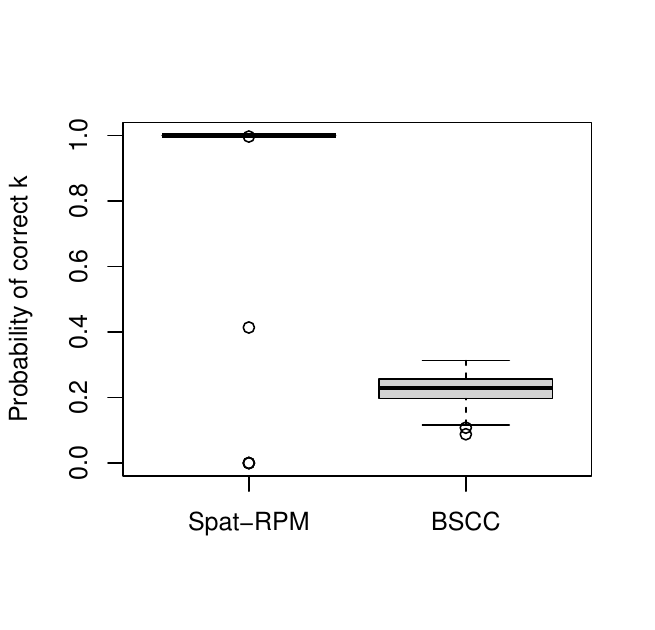}}
            &{\includegraphics[width=0.22\linewidth,height=0.22\textheight]{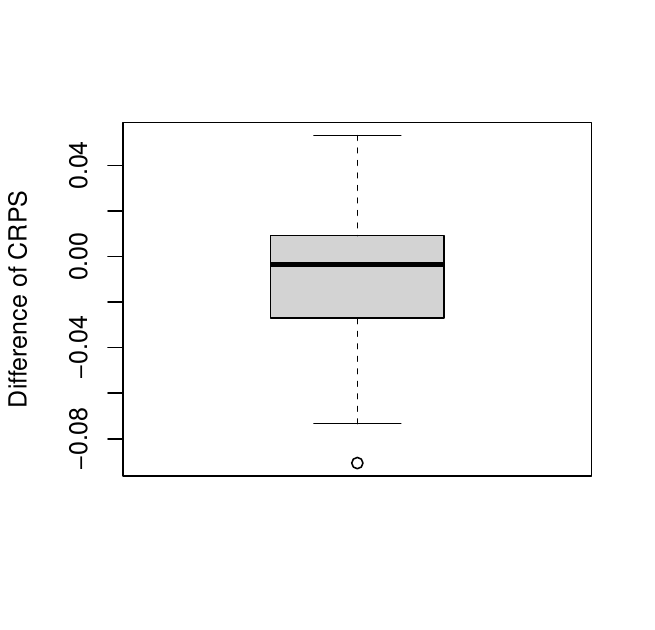}}&
            {\includegraphics[width=0.22\linewidth,height=0.22\textheight]{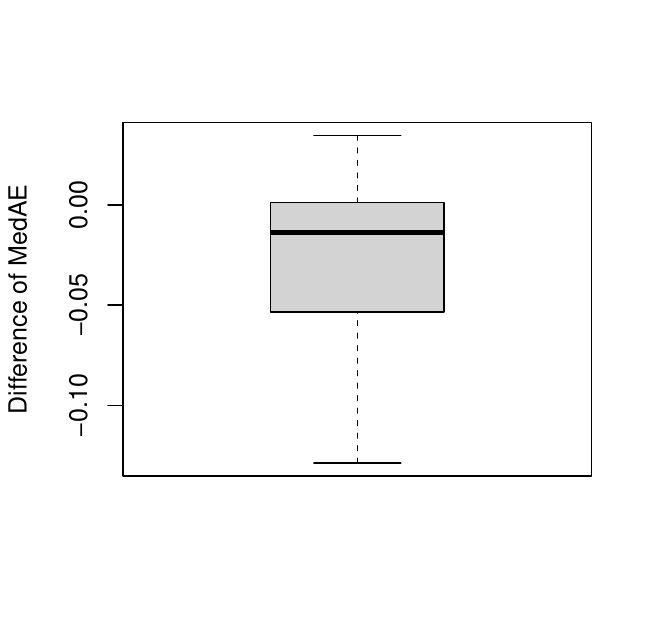}}
            \\
            {\small (a) } & {\small (b) } & {\small (c)} & {\small (d) }  
		\end{tabular}
		\caption{Comparison with BSCC under $n=4000$. (a) Boxplots of the posterior
			number of clusters for different MCMC samples, from one randomly selected
			repeat. (b) Boxplots of the posterior probability
			of the correct cluster number. 100 repeats are taken. (c) Boxplot of $\{ \tmop{CRPS}_{r, 1} - \tmop{CRPS}_{r, 2} \}_{r = 1}^{100}$.
			(d) Boxplot of $\{ \tmop{MedAE}_{r, 1} - \tmop{MedAE}_{r, 2} \}_{r = 1}^{100}$.}
		\label{FIG:modelcompare}
        %\vspace{-25pt}
	\end{figure}

\noindent \textbf{Comparison under $n=4000$.}
    % \subsubsection{Comparison under $n=4000$}
    % \label{SEC:BSCCcompn4000}
	% Note that BSCC model makes clusters on each dimension of the regression coefficient, we
	% only focus on its cluster result for the second dimension (i.e., the regression
	% coefficient for $x_2 (\tmmathbf{s}_i)$) for simplicity. 
     We use the same $(c_b,\alpha_b,c_p,\alpha_p,\sigma^2)$ as those specified in the setup at the beginning of Section \ref{SEC:Ushape}. Figures \ref{FIG:modelcompare}(a) - (b) show the posterior cluster number comparison between the two
	models. We can see from Figures \ref{FIG:modelcompare}(a) - (b) that the number
	of clusters in our model concentrates at the true value for all MCMC
	samples and repeats. On the contrary, BSCC tends to overestimate the
	number of clusters according to Figure \ref{FIG:modelcompare}(a). Figure \ref{FIG:modelcompare}(b) shows that for BSCC,
	the posterior probability of the correct cluster number is smaller than
	$0.3$ in most of repeats. The result indicates that our model shows significant improvement in terms of estimating the number of clusters, attributable to the blocking technique and hyperparameter selection guidelines in Assumption \ref{AS:hyperpara}.
    
	Since the prediction method of a new location is not provided in \cite{luo2021bayesian},  we compare prediction errors of the observed locations. For the $r$-th repeat, let $\tmop{CRPS}_{r,1}$ and $\tmop{CRPS}_{r,2}$ denote the CRPS values for our model and BSCC, respectively, and let $\tmop{MedAE}_{r,1}$ and $\tmop{MedAE}_{r,2}$ denote the corresponding MedAE values. We compute the difference of CRPS (MedAE) between two models as $\tmop{CRPS}_{r, 1} - \tmop{CRPS}_{r, 2}$ ($\tmop{MedAE}_{r,
		1} - \tmop{MedAE}_{r, 2}$) and the results are shown in Figures
	\ref{FIG:modelcompare}(c) - (d).
    
 %    Let $\tmmathbf{\mu}$ be the vector of the regression mean of the observed locations. Denote $\tmmathbf{\mu}_{r, s, 1}$ and $\tmmathbf{\mu}_{r, s, 2}$ as the predictions of $\tmmathbf{\mu}$ based on the $s$-th posterior sample from the $r$-th simulation repeat for our model and BSCC, respectively. We compute CRPS values for two models at
	% $r$-th repeat, say $\tmop{CRPS}_{r, 1}$ and $\tmop{CRPS}_{r, 2}$, based on
	% $\{ \tmmathbf{\mu}_{r, s, 1} \}_{s = 1}^M$ and $\{
	% \tmmathbf{\mu}_{r, s, 2} \}_{s = 1}^M$, respectively. We also compute
	% two models' MedAE as
	% \[ \tmop{MAE}_{r, 1} = n^{- 1} \Big\| M^{- 1} \sum_{s = 1}^M
	% (\tmmathbf{\mu}_{r, s, 1}  - \tmmathbf{\mu}_{r,0}) \Big\|_1, \infixand
	% \tmop{MAE}_{r, 2} = n^{- 1} \Big\| M^{- 1} \sum_{s = 1}^M
	% (\tmmathbf{\mu}_{r, s, 2} - \tmmathbf{\mu}_{r,0}) \Big\|_1, \]
	% where $\| \cdot \|_1$ is $L_1$ norm and $\tmmathbf{\mu}_{r,0}$ is the true value. 

	By definition, a negative difference indicates that our model performs better. From Figures \ref{FIG:modelcompare}(c) and (d), we observe that more than half of the simulation repeats show a negative CRPS difference, while nearly all repeats show a negative MedAE difference. To further evaluate this, we perform a one-sample t-test to determine whether the mean difference is significantly negative. For CRPS (MedAE), the t-test yields a mean difference of $-0.0085$ ($-0.025$) with a p-value $0.0074$ ($<0.001$). These results demonstrate that our model achieves higher prediction accuracy than BSCC with a relatively large sample size, likely due to the more accurate partition estimation.
    
    % Additionally, after comparing the theoretical results, we find that the posterior convergence rates of $\tmmathbf{\mu}$ are the same (ignoring logarithmic terms) for both models. This suggests that our model is more efficient, likely due to its more accurate cluster estimation.

Finally, we compare the computing times for MCMC sampling of the posterior distribution between the two models. Using the ``microbenchmark" package in R, we obtain the average computing time over 10 runs of the MCMC procedure. The average computing times for our and BSCC models are $37$ seconds and $71$ seconds, respectively. This demonstrates a noticeable reduction in computing time, which is attributed to the reduced partition space achieved through the blocking technique.

\begin{figure}[htbp]
		\centering
		\begin{tabular}{ccc}
			{\includegraphics[width=0.22\linewidth,height=0.22\textheight]{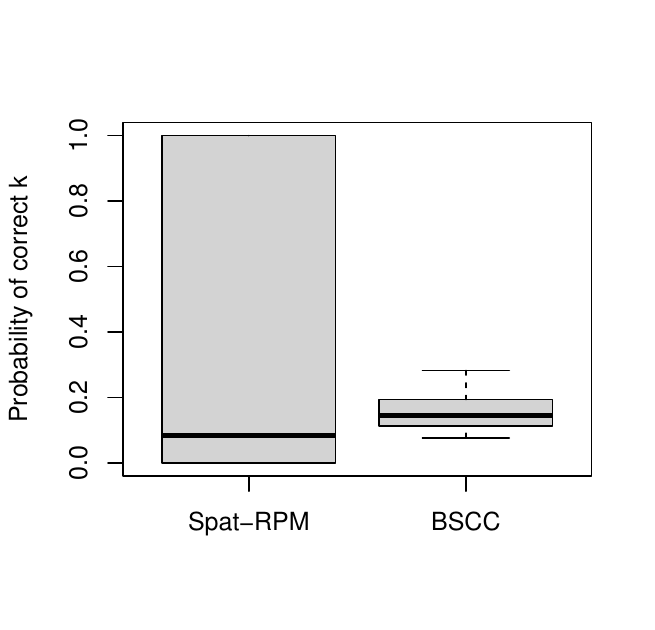}}&
            {\includegraphics[width=0.22\linewidth,height=0.22\textheight]{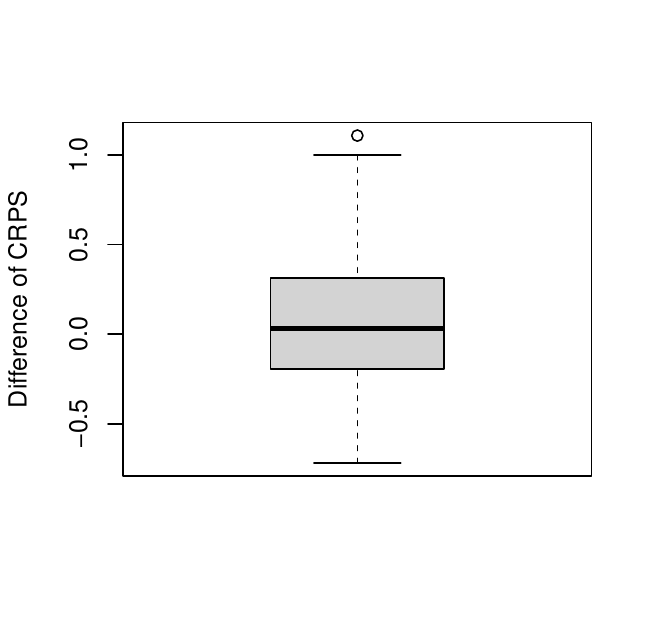}}
            &{\includegraphics[width=0.22\linewidth,height=0.22\textheight]{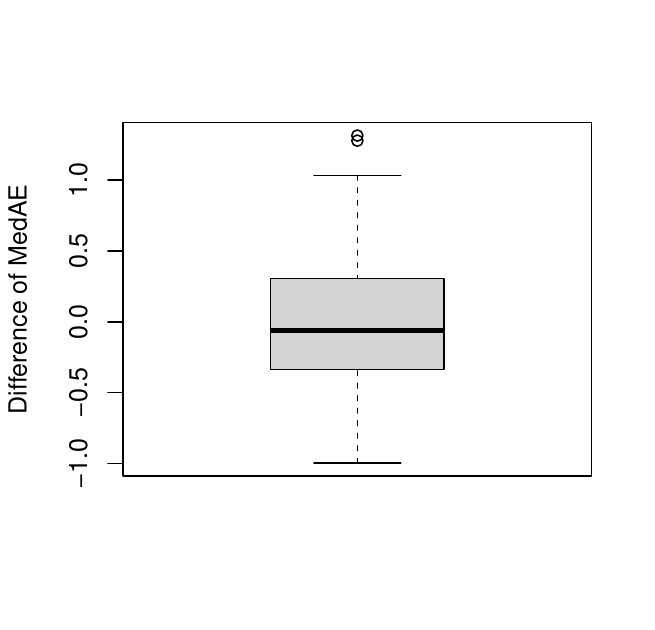}}\\
			{\small (a) } & {\small (b) } & {\small (c)}  
			
		\end{tabular}
		\caption{Comparison with BSCC under $n=100$. (a) Boxplots of the posterior probability
			of the correct cluster number. 100 repeats are taken. (b) Boxplot of $\{ \tmop{CRPS}_{r, 1} - \tmop{CRPS}_{r, 2} \}_{r = 1}^{100}$.
			(c) Boxplot of $\{ \tmop{MedAE}_{r, 1} - \tmop{MedAE}_{r, 2} \}_{r = 1}^{100}$.}
		\label{FIG:modelcomparen100}
        %\vspace{-25pt}
	\end{figure}

\noindent \textbf{Comparison under $n=100$.}
% \subsubsection{Comparison under $n=100$}
% \label{SEC:BSCCcompn100}
Following the setup at the beginning of Section \ref{SEC:Ushape}, we set $\sigma^2=1$, and select $(c_b,c_p)$ with the hyperparameter selection approach described in Section \ref{SEC:hyperparameterselection}. Figure \ref{FIG:modelcomparen100}(a) compares the posterior probability of the correct cluster number between the two models. When $n$ is small, only a few repeats of Spat-RPM achieve the correct cluster number, while all repeats of BSCC fail. Figures \ref{FIG:modelcomparen100}(b) – (c) show the differences in CRPS and MedAE, respectively. The medians of the two boxplots are close to $0$, suggesting that the performances of the two models are comparable. For CRPS (MedAE), the t-test yields a mean difference of $0.08$ ($-0.031$) with a p-value of $0.026$ ($0.95$). When the sample size is small, Spat-RPM does not exhibit its asymptotic properties and therefore shows no significant improvement over BSCC.

For the comparison of computing times, using ``microbenchmark" package in R, The average times over $10$ runs for our and BSCC models are $28$ seconds and $30.6$ seconds, respectively. Thus, when the sample size is small, Spat-RPM shows no significant improvement in computation.

% \subsubsection{Comparison with the BSCC model}\label{SEC:BSCCcompare}
% The result is deferred to Section \ref{SEC:compareBSCC} in the Supplementary Material.

% \subsubsection{Effectiveness of using the CH index/WAIC for hyperparameter selection}\label{SEC:effectiveCHWAIC}
%  The result is deferred to Section \ref{SEC:CHWAIC} in the Supplementary Material.

\subsubsection{Effectiveness of hyperparameter selection using CH index and WAIC}
\label{SEC:CHWAIC}
 We use the same setting as that in Section \ref{SEC:sensitivity}. 
We compute the CH index, WAIC and the true partition error under different combinations of $(c_b,c_p,\sigma^2)\in \tmmathbf{C}_b\times \tmmathbf{C}_p \times\tmmathbf{C}_{\sigma^2}$. For a fixed $\sigma^2\in \tmmathbf{C}_{\sigma^2}$, we compute the correlation coefficient between the negative CH index/WAIC and the true partition error, and the result is shown in Table \ref{TAB:CHWAIC} (we consider a negative CH index because a larger CH index indicates a better partition). 
\begin{table}[htbp]
\centering
\caption{Correlation coefficient between the negative CH index/WAIC and true partition error}
\label{TAB:CHWAIC}
\begin{tabular}{lllll}
\hline
Correlation coefficient  & $\sigma^2=1$ & $\sigma^2=5$   & $\sigma^2=10$    & $\sigma^2=15$ \\
\hline 
Negative CH index    & 0.8346                  & 0.5330 & 0.2981 & 0.1588                   \\ %this is the coeficient while dropping the case with only 1 cluster
%Negative CH index    & 0.8346                  & NA & NA & NA                   \\
WAIC        & -0.03365                & 0.9987 & 0.9997 & 0.9997           \\
\hline
\end{tabular}
\end{table}

We can see from Table \ref{TAB:CHWAIC} that when $\sigma^2=1$, the CH index exhibits a high correlation with the true partition error, whereas WAIC serves as a poor criterion for selecting $(K,\lambda)$, showing an almost zero correlation. This is because, for $\sigma^2=1$, some $(K,\lambda)$ pairs produce a large number (more than $70$) of small clusters—an extreme overfitting case—and WAIC fails to provide sufficient penalty in such situations. In contrast, the CH index performs well in these cases, as it inherently disfavors small clusters by definition. However, for $\sigma^2 \in (5,10,15)$, the CH index shows relatively low correlation, while WAIC exhibits a strong correlation with the true partition error. for $\sigma^2 \in (5,10,15)$, most partition results contain no more than $5$ clusters, making overfitting less of an issue for selecting $(K,\lambda)$. In such cases, the CH index, which evaluates partition quality with an emphasis on well-separated and compact clusters, may fail to select optimal hyperparameters, whereas WAIC performs well due to its emphasis on goodness of fit to the data. Hence, in practice, we recommend selecting $(K,\lambda)$ by considering both criteria.

\subsection{Areal data}
\label{SEC:areal data}
\begin{figure}[htbp]
		\centering
		\begin{tabular}{cc}
			{\includegraphics[width=0.28\linewidth,height=0.15\textheight]{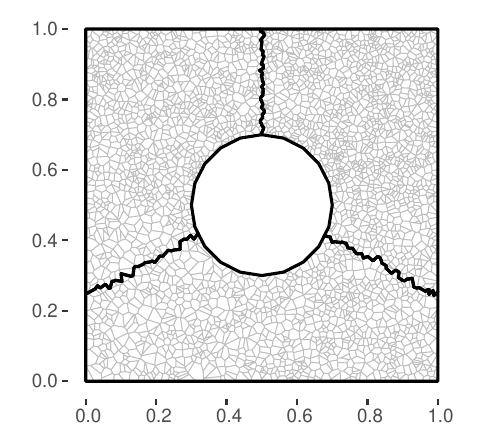}}&
			{\includegraphics[width=0.28\linewidth,height=0.15\textheight]{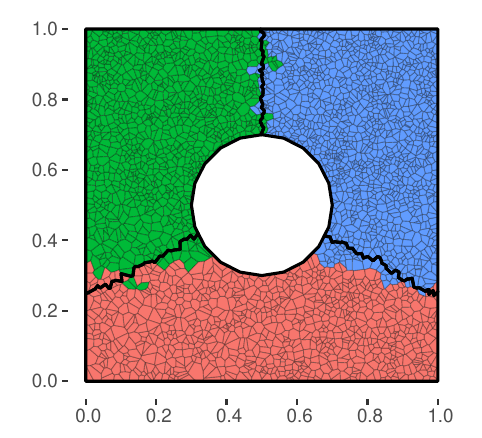}}\\
            {\small (a) } & {\small (b)}  
            \\
		\end{tabular}
		\caption{(a) Areal data within the domain $[0,1]^2$ containing a hole. (a) The distribution of areal units $\{\mathbb{D}_i\}_{i=1}^n$. The true boundary of the three clusters is denoted by black solid lines. (b) One randomly chosen posterior partition sample with different colors representing different clusters.}
		\label{FIG:holedata}
        %\vspace{-25pt}
        \vspace{-5mm}
	\end{figure}

In this section, we conduct a simulation study on an areal data generated within the domain $[0,1]^2$ containing a hole, following the fitting procedure described in Section \ref{SEC:apptoareal}. We sample locations from a Poisson point process with intensity function $4800\mathbb{I}(s_1\geq0.5,s_2\geq 0.5)+4000\mathbb{I}(s_1<0.5,s_2\geq 0.5)+3200\mathbb{I}(s_1\geq0.5,s_2< 0.5)+2400\mathbb{I}(s_1<0.5,s_2< 0.5)$, and remove those within the hole, yielding $n=3208$ locations. A CVT is then constructed based on these locations, with each CVT cell treated as an areal unit $\mathbb{D}_i$. Figure \ref{FIG:holedata}(a) displays the generated areal data, where each unit $\mathbb{D}_i$ is represented by a polygon. We partition $\{\mathbb{D}_i\}_{i=1}^n$ into three clusters, and denote $\{\mathcal{D}_{l,0}\}_{l=1}^3$ as the sub-domain corresponding to each cluster: $\mathcal{D}_{1,0}$ represents the upper-left region, $\mathcal{D}_{2,0}$ the upper-right region, and $\mathcal{D}_{3,0}$ the lower-middle region. The cluster boundaries are shown as black solid lines in Figure \ref{FIG:holedata}(a).

Each areal unit $\mathbb{D}_i$ is represented by its centroid $\tmmathbf{s}_i$. The covariate and response for the $i$-th areal unit, $\tmmathbf{x}(\mathbb{D}_i)$ and $y(\mathbb{D}_i)$, are set as $\tmmathbf{x}(\mathbb{D}_i)=\tmmathbf{x}(\tmmathbf{s}_i)$ and $y(\mathbb{D}_i)=y(\tmmathbf{s}_i)$, where $\tmmathbf{x}(\tmmathbf{s}_i)$ and $y(\tmmathbf{s}_i)$ are generated in the same manner as in the setup described at the beginning of Section \ref{SEC:Ushape}. We then apply Spat-RPM to $\{\mathbb{D}_i,\tmmathbf{x}(\mathbb{D}_i),y(\mathbb{D}_i)\}_{i=1}^{n}$ following Section \ref{SEC:apptoareal}, and posterior domain partition samples are obtained. Figure \ref{FIG:holedata}(b) shows one randomly chosen posterior partition sample. We can see that the posterior partition recovers $\{\mathcal{D}_{l,0}\}_{l=1}^3$ well, except for some areal units near boundaries.

% We then apply Spat-RPM to the dataset $[\{\tmmathbf{s}_i,\tmmathbf{x}(\tmmathbf{s}_i), y (\tmmathbf{s}_i)\}_{i = 1}^n$, using the same hyperparameters as before. Finally, posterior samples of the partition of $\mathcal{S}=\{\tmmathbf{s}_i\}_{i=1}^n$, say $\pi(\mathcal{S})$, are obtained.

% Since each areal unit is represented by its centroid $\tmmathbf{s}_i$, the partition $\pi(\mathcal{S})$ naturally induces a partition of the domain. Figure \ref{FIG:holedata}(b) shows the domain partition induced by a randomly selected posterior sample of $\pi(\mathcal{S})$. We can see that the posterior partition recovers $\{\mathcal{D}_{l,0}\}_{l=1}^3$ well, except for some areal units near boundaries.

	\section{Real data analysis}\label{SEC:realdata}
	
\begin{figure}[htbp]
		\centering
		\begin{tabular}{ccc}
			{\includegraphics[width=0.27\linewidth,height=0.18\textheight]{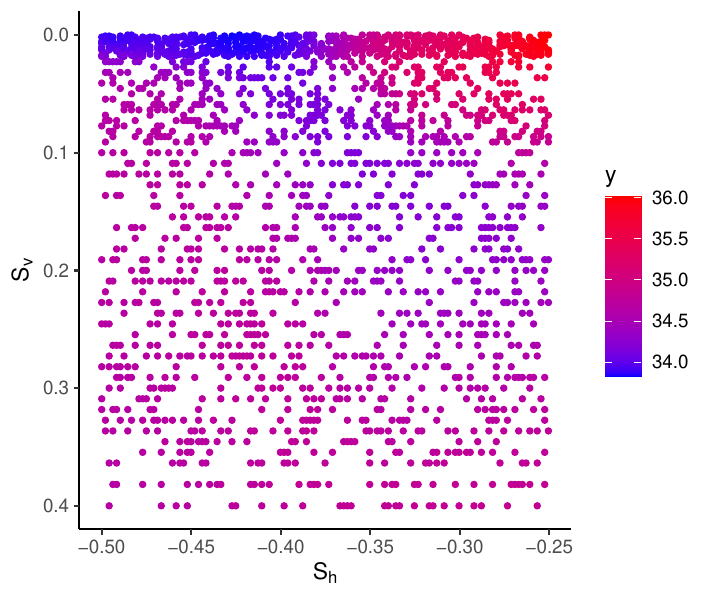}}&
            {\includegraphics[width=0.42\linewidth,height=0.18\textheight]{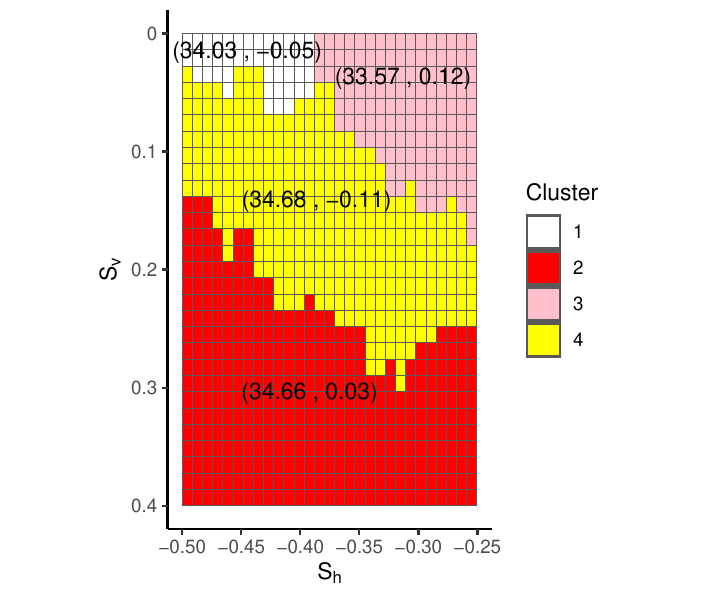}}&
			{\includegraphics[width=0.23\linewidth,height=0.18\textheight]{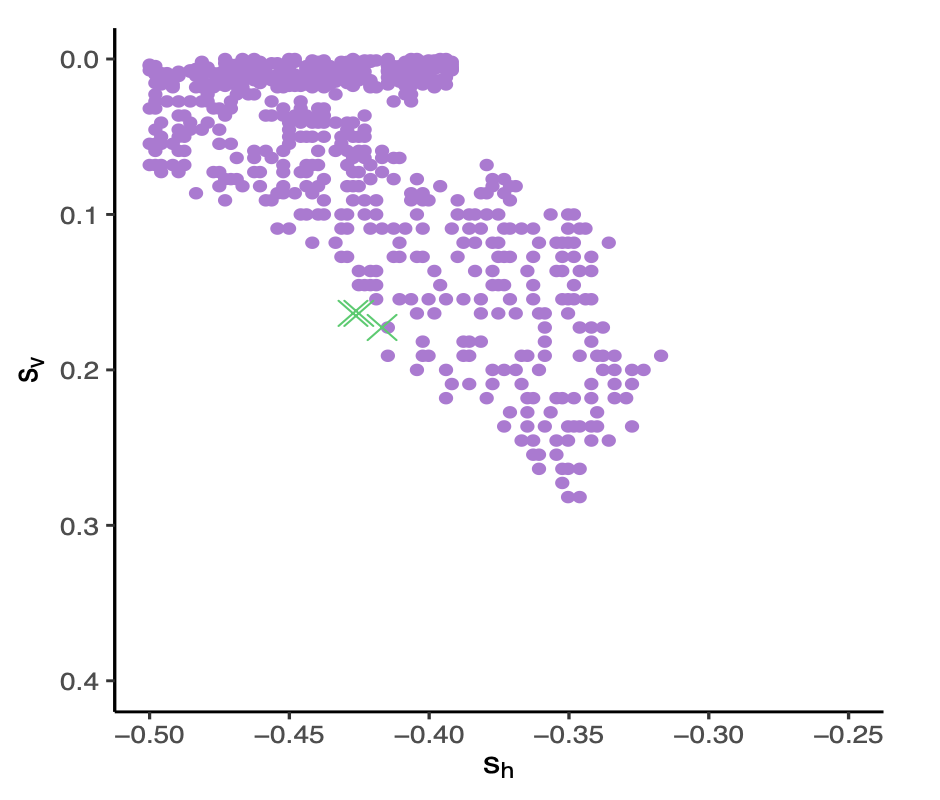}}
			\\
			{\small (a)} & {\small (b)} & {\small (c)} 
		\end{tabular}
		\caption{%Salinity data and partition results. 
        (a) The observed locations in salinity data. %Each point represents a
			%location, with the color indicating the value of salinity. 
            (b) Partition result from our model.
			Four clusters are obtained and represented by different colors. The number annotations within clusters are the corresponding posterior sample of $\{ \tmmathbf{\theta}(\tmmathbf{s}_i)
			 \}_{i=1}^{n}$. AAIW is identified as the area of clusters 1 and 4. (c) %Partition result from BSCC.
             The point identified by BSCC with negative $\theta_2 (\tmmathbf{s}_i)$ (hence as AAIW). }
		\label{FIG:salinitydata}
        \vspace{-5mm}
	\end{figure}

	In this section, we apply Spat-RPM to study the temperature-salinity (T-S) relationship of seawater in the Atlantic Ocean. The study aims to identify the Antarctic Intermediate Water (AAIW), characterized by a negative T-S relationship \citep{talley2011descriptive}. %Identifying AAIW provides crucial insights into ocean circulation and heat distribution, which are essential for understanding Earth's climate system. The T-S relationship is known to be homogeneous within certain
	% water masses, but can shift abruptly across boundaries. Consequently, the T-S
	% relationship is often modeled as a spatially piecewise constant function in
	% oceanography.
	The data on temperature and salinity is obtained from National Oceanographic  Data Center (https://www.nodc.noaa.gov/OC5/woa13/), and the detailed data description can be found in \cite{luo2021bayesian}. %We take the segment of the
	% Atlantic basin along $25^{\circ}$W between $60^{\circ}$S and the equator. We
	% apply the same rescaling procedure as in {\cite{luo2021bayesian}} to
	% eliminate the strong anisotropic spatial patterns of the original dataset.
	% Write $s_h$ and $s_v$ as the rescaled latitude and depth of the ocean, respectively. According
	% to the pilot study in {\cite{luo2021bayesian}}, AAIW is within $(s_h, s_v) \in
	% [- 0.5, 0.25] \times [0, 0.4]$. We thus only focus on the data within this
	% region. Finally, 
  %  we get $n = 1936$ observed locations and    
	% corresponding temperature and salinity. 
    The $n = 1936$ observed locations are shown in
	Figure \ref{FIG:salinitydata}(a).
	
	Let $\tmmathbf{s}_i = (s_{h, i}, s_{v, i})$ be the location of the $i$-th
	observation. Write $y (\tmmathbf{s}_i)$ as the salinity at $\tmmathbf{s}_i$
	and $\tmmathbf{x} (\tmmathbf{s}_i) = (1, \tmop{Temp} (\tmmathbf{s}_i))^T$,
	where $\tmop{Temp} (\tmmathbf{s}_i)$ is the temperature at $\tmmathbf{s}_i$.
	We model the T-S relationship as
	$ y (\tmmathbf{s}_i) = \tmmathbf{x}^T (\tmmathbf{s}_i) \tmmathbf{\theta}
	(\tmmathbf{s}_i) + \epsilon (\tmmathbf{s}_i)$,
	where $\tmmathbf{\theta} (\tmmathbf{s}_i) = (\theta_1 (\tmmathbf{s}_i),
	\theta_2 (\tmmathbf{s}_i))^T$ is the unknown piecewise constant regression coefficient and
	$\epsilon (\tmmathbf{s}_i)$ is a Gaussian noise.
    
	 We apply our model described in Section \ref{SEC:MODEL}. %and the detailed model and hyperparameter settings are described in Section~\ref{Ssec:realdata}.
      We use the
	sample variance of $\{ y (\tmmathbf{s}_i) \}_{i=1}^{n}$ as the value of $\sigma^2$ in
	 our model. We use the same values of $\gamma$ and $k_{\tmop{max}}$ as in simulation studies, as well as the same formulas 
	 to select hyperparameters $K$ and $\lambda$, i.e., $K = c_b n^{1
	 	/ 2} \log^{- (1 + \alpha_b) / 2} (n)$ and $\log (\lambda^{- 1}) = c_p n
	 \log^{- \alpha_p} (n)$ with $\alpha_b = 1$ and $\alpha_p = 0.5$. To decide the values of
	 $c_b$ and $c_p$, we adopt WAIC and select $(c_b, c_p) \in (1, 3, 5,
	 7) \times (0.01, 0.1, 0.5)$ as the values that minimize WAIC. We use MCMC to sample from the posterior distribution of
	partitions and $\{ \tmmathbf{\theta} (\tmmathbf{s}_i) \}_{i=1}^{n}$. The number of MCMC
	iterations, the burn-in period, and thinning parameter are the same as in the simulation
	setting. We use the ``salso'' package in R~\cite{dahl2022search} to obtain a point estimation of the partition from posterior
	samples, which is shown in Figure \ref{FIG:salinitydata}(b). We display the partition result obtained from the BSCC model by \cite{luo2021bayesian} in Figure \ref{FIG:salinitydata}(c) for comparison.  
    
    As shown in Figures \ref{FIG:salinitydata}(b) - (c), among the four clusters identified by our model, two clusters (clusters 1 and 4) have negative ${\theta_2 (\tmmathbf{s}_i)}$. Thus, the AAIW identified by our model corresponds to the areas of clusters 1 and 4. Compared to Figure \ref{FIG:salinitydata}(c), we can see that the majority of the AAIW region is the same for both models, while our model includes some areas with $S_h > -0.3$ as part of the AAIW. Specifically, we observe that clusters 1 and 4 correspond to shallow water and deep water areas, respectively. It is reasonable to observe some differences in the T-S relationship between these two areas.
	
	Although the AAIW identified by Spat-RPM and BSCC is consistent, these two models have fundamental differences. The BSCC model conducts partitioning exclusively at the observed locations, restricting its ability to identify the cluster memberships of unobserved locations. Moreover, due to the lack of partition consistency, AAIW in the BSCC model is estimated by checking signs of posterior samples $\{ \theta_2 (\tmmathbf{s}_i) \}_{i=1}^{n}$, which may be inconsistent with the partition result. In contrast, our model identifies the AAIW directly from the posterior domain partition, offering a simpler and more natural interpretation.

\bigskip    
\textbf{Acknowledgments.} The authors would like to thank the anonymous referees and the
Associate Editor for their helpful and detailed comments. The research of Kun Huang and Huiyan Sang was partially supported by NSF grant no. NSF DMS-2220231 and NSF SES-2521573.

\newpage

\def\theequation{S.\arabic{equation}}
\def\thefigure{S.\arabic{figure}}
\def\thetable{S.\arabic{table}}
\def\thesection{S.\arabic{section}}
\def\thepage{S\arabic{page}}
\def\thelemma{S.\arabic{lemma}}
\def\thecorollary{S.\arabic{corollary}}
\def\theproposition{S.\arabic{proposition}}

\setcounter{section}{0} \setcounter{equation}{0} \setcounter{figure}{0} %
\setcounter{page}{1}\setcounter{proposition}{0}\setcounter{lemma}{0}

\begin{center}
    \large\bf Supplementary Material for ``Consistent Bayesian Spatial Domain Partitioning Using Predictive Spanning Tree Methods''
\end{center}

\noindent This supplement is organized as follows. In Section \ref{SEC:MHcomputation}, we provide the computational details of M-H acceptance ratios in Section \ref{SEC:computation} of the main paper. In Section \ref{SEC:justificationofpath}, we justify Equation (\ref{EQ:pathdis}) in Assumption \ref{AS:lengthofboundary} using an example of a sub-domain shape. Section \ref{SEC:proof} contains all the proofs for theorems in the main paper. Section \ref{SEFC:lambdafurter} provides more details on why a fixed, nondegenerate prior on $\lambda$ is insufficient to achieve clustering consistency in our model. Section \ref{SEC:k0rnestimation} details the estimation procedure for $(k_0, r_n)$, which is used to select orders of $(K,\lambda)$ under the VDG regime.

\section{Computational details of M-H acceptance ratios}
\label{SEC:MHcomputation}
In this section, we compute M-H acceptance ratios for the four moves in Section \ref{SEC:computation} in the main paper. Let $\pi^\ast (\mathcal{D})$ be the current partition with $k$ clusters, and $\pi^\ast_{\tmop{new}} (\mathcal{D})$ be the new partition after a move.  For the birth move, the prior ratio is computed as
\begin{equation}
\label{EQ:priorratioforbirth}
    \frac{\mathbb{P}(k+1)\mathbb{P}(\mathcal{T})\mathbb{P}\{\pi_{\tmop{new}}^\ast(\mathcal{D})\mid k+1,\mathcal{T}\}}{\mathbb{P}(k)\mathbb{P}(\mathcal{T})\mathbb{P}\{\pi^\ast(\mathcal{D})\mid k,\mathcal{T}\}}=\frac{\lambda}{k+1}\times \frac{\binom{K^2-1}{k-1}}{\binom{K^2-1}{k}}=\frac{\lambda}{k+1}\times \frac{k}{K^2-k}.
\end{equation}
The proposal ratio is computed as
\begin{equation}
\label{EQ:proposalrationforbirth}
    \frac{r_d(k+1)}{r_b(k)}\times \frac{\binom{K^2-k}{1}}{\binom{k}{1}}=\frac{r_d(k+1)}{r_b(k)}\times \frac{K^2-k}{k}.
\end{equation}
Putting Equations (\ref{EQ:priorratioforbirth}) - (\ref{EQ:proposalrationforbirth}) together, the M-H acceptance ratio for the birth move is 
\[ \min \left\{ 1, \frac{\lambda}{k + 1} \times \frac{r_d (k + 1)}{r_b (k)}
   \times \frac{\mathbb{P}\{\tmmathbf{y} \mid \pi^\ast_{\tmop{new}}
   (\mathcal{D}), k + 1, \mathcal{T}, \tmmathbf{s},
   \tmmathbf{x}\}}{\mathbb{P}\{\tmmathbf{y} \mid \pi ^\ast(\mathcal{D}), k,
   \mathcal{T}, \tmmathbf{s}, \tmmathbf{x}\}} \right\}, \]
where $\mathbb{P} \{\tmmathbf{y} \mid \pi^\ast
(\mathcal{D}), k, \mathcal{T}, \tmmathbf{s}, \tmmathbf{x}\}$ is the integrated likelihood with $\tmmathbf{\theta}$ marginalized out. See the
closed form of $\mathbb{P} \{\tmmathbf{y} \mid \pi^\ast (\mathcal{D}), k,
\mathcal{T}, \tmmathbf{s}, \tmmathbf{x}\}$ in Equation (\ref{EQ:formlikelihood}) under a linear regression setting.

For the death move, the prior ratio is computed as
\begin{equation}
\label{EQ:priorratiofordeath}
    \frac{\mathbb{P}(k-1)\mathbb{P}(\mathcal{T})\mathbb{P}\{\pi_{\tmop{new}}^\ast(\mathcal{D})\mid k-1,\mathcal{T}\}}{\mathbb{P}(k)\mathbb{P}(\mathcal{T})\mathbb{P}\{\pi^\ast(\mathcal{D})\mid k,\mathcal{T}\}}=\frac{k}{\lambda}\times \frac{\binom{K^2-1}{k-1}}{\binom{K^2-1}{k-2}}=\frac{k}{\lambda}\times \frac{K^2-k+1}{k-1}.
\end{equation}
The proposal ratio is computed as
\begin{equation}
\label{EQ:proposalrationfordeath}
    \frac{r_b(k-1)}{r_d(k)}\times \frac{\binom{k-1}{1}}{\binom{K^2-k+1}{1}}=\frac{r_b(k-1)}{r_d(k)}\times \frac{k-1}{K^2-k+1}.
\end{equation}
Putting Equations (\ref{EQ:priorratiofordeath}) - (\ref{EQ:proposalrationfordeath}) together, the M-H acceptance ratio for the death move is 
\[ \min \left\{ 1, \frac{k}{\lambda} \times \frac{r_b (k - 1)}{r_d (k)} \times
   \frac{\mathbb{P}\{\tmmathbf{y} \mid \pi^\ast_{\tmop{new}} (\mathcal{D}),
   k - 1, \mathcal{T}, \tmmathbf{s}, \tmmathbf{x}\}}{\mathbb{P}\{\tmmathbf{y}
   \mid \pi^\ast (\mathcal{D}), k, \mathcal{T}, \tmmathbf{s},
   \tmmathbf{x}\}} \right\} . \]

For the change move, it is easy to see that both the prior ratio and the proposal ratio are $1$. Thus, the corresponding M-H acceptance rate is
\[ \min \left\{ 1, 
   \frac{\mathbb{P}\{\tmmathbf{y} \mid \pi^\ast_{\tmop{new}} (\mathcal{D}),
   k , \mathcal{T}, \tmmathbf{s}, \tmmathbf{x}\}}{\mathbb{P}\{\tmmathbf{y}
   \mid \pi^\ast (\mathcal{D}), k, \mathcal{T}, \tmmathbf{s},
   \tmmathbf{x}\}} \right\} . \]

Finally, for the hyper move, we first note that the current partition $\pi^\ast(\mathcal{D})$ remains unchanged after the hyper move, which makes the likelihood ratio equal to $1$. Under the UST prior, the prior ratio of two spanning trees is always $1$. However, the proposal ratio is difficult to compute because it depends on the mapping between $\{w_e \}_{e \in \mathcal{E}}$ and $\mathcal{T}$. In practice, we set the proposal ratio to $1$, and it has been discussed in \cite{teixeira2019bayesian} that this provides a good approximation for drawing posterior samples of $\mathcal{T}$.

% We next derive the prior ratio and proposal ratio under RST (\ref{DEF:spanningtree2}) and UST (\ref{DEF:spanningtree1}), respectively. Under the RST prior, the spanning tree $\mathcal{T}$ is determined by weights $\{w_e \}_{e \in \mathcal{E}}$, so updating $\mathcal{T}$ is equivalent to updating $\{w_e \}_{e \in \mathcal{E}}$. According to Equation (\ref{DEF:spanningtree2}), the prior ratio of two weights is always $1$. For the proposal ratio, since $\pi^\ast(\mathcal{D})$  is unchanged, the sets of cross-cluster edges and within-cluster edges are preserved. By the design of the proposal distribution, the proposal ratio is therefore also 1. See also \citep{luo2021bayesian} for the same sampling strategy applied to RST.
% \kun{I feel RST prior has some problems}

\section{Justification of Equation (\ref{EQ:pathdis})}
\label{SEC:justificationofpath}
\begin{figure}[htbp]
		\centering
		\begin{tabular}{cc}
             {\includegraphics[width=0.4\linewidth,height=0.15\textheight]{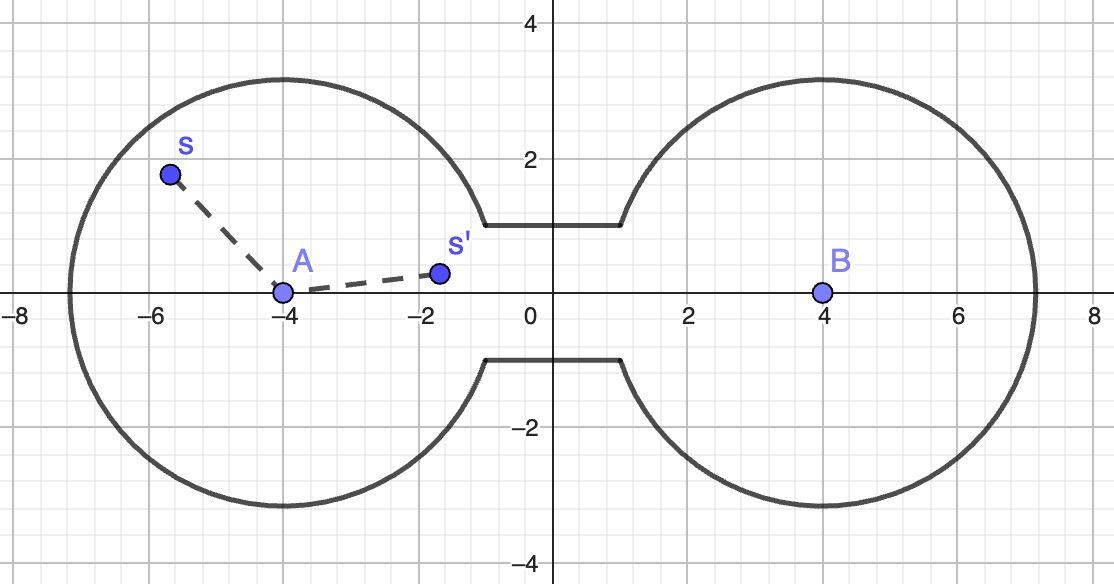}}&
			{\includegraphics[width=0.4\linewidth,height=0.15\textheight]{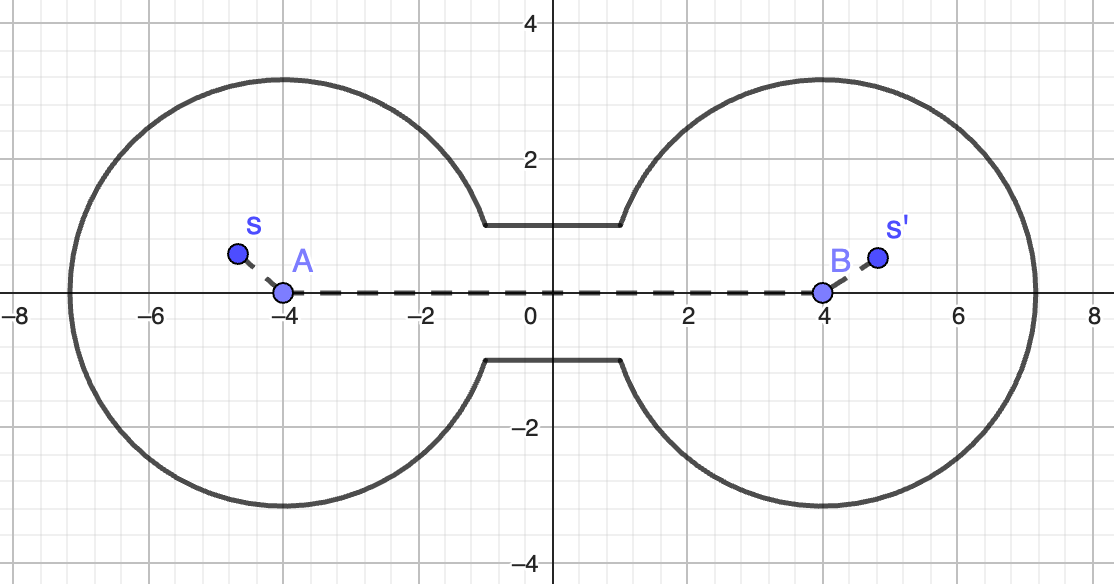}}
            \\
            {\small (a) $\tmmathbf{s},\tmmathbf{s}' \in \mathbb{D}_1$} & {\small (b) $\tmmathbf{s} \in \mathbb{D}_1$ and $\tmmathbf{s}'\in \mathbb{D}_3$}\\
            
			{\includegraphics[width=0.4\linewidth,height=0.15\textheight]{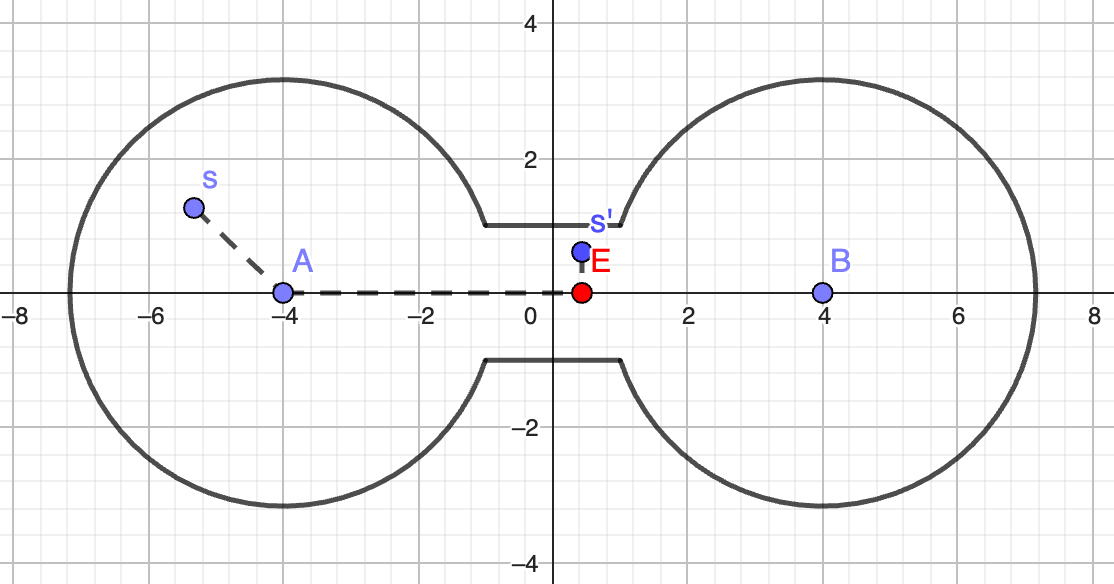}}&
            {\includegraphics[width=0.4\linewidth,height=0.15\textheight]{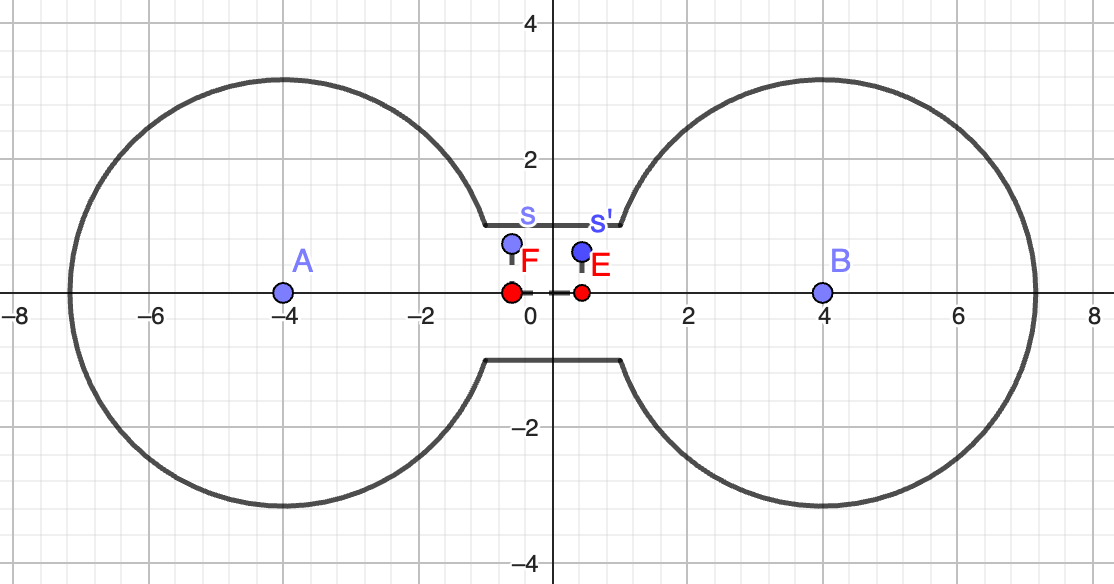}}
            \\
			 {\small (c) $\tmmathbf{s} \in \mathbb{D}_1$ and $\tmmathbf{s}'\in \mathbb{D}_2$} & {\small (d) $\tmmathbf{s},\tmmathbf{s}' \in \mathbb{D}_2$}
		\end{tabular}
		\caption{An example of $\mathcal{D}_{l,0}$ and four possible situations of $(\tmmathbf{s},\tmmathbf{s}')$. The constructed paths $\mathcal{P}(\tmmathbf{s},\tmmathbf{s}')$ are denoted by dashed lines.}
		\label{FIG:Boundary example}
        \vspace{-10pt}
	\end{figure}

In this section, we justify Equation (\ref{EQ:pathdis}) in Assumption \ref{AS:lengthofboundary} using an example of a sub-domain shape. Figure \ref{FIG:Boundary example} illustrates an example of $\mathcal{D}_{l,0}$, where the boundary is represented by the black solid line. Let $(x,y)$ denote the coordinates of a point. We can see that $\mathcal{D}_{l,0}$ consists of two incomplete circles centered at points $A$ and $B$, respectively, and connected by a narrow band in the region $-1<x<1$. With this example, we will illustrate the construction of $\mathcal{P}(\tmmathbf{s},\tmmathbf{s}')$ such that Equation (\ref{EQ:pathdis}) is satisfied. We will see from the construction that the constant $C$ in Equation (\ref{EQ:pathdis}) plays a role in ensuring that this shape satisfies it. Using the same construction technique, it is easy to verify that Equation (\ref{EQ:pathdis}) also holds for simpler shapes (e.g., a single circle or a regular polygon).

For the simplicity of notations, we refer to $\mathcal{D}_{l,0}$ as $\mathbb{D}$ in the following context, and we divide $\mathbb{D}$ into three components:
\[
\mathbb{D}_1 = \{(x,y)\in \mathbb{D}:x\leq -1\}
\]
\[
\mathbb{D}_2 = \{(x,y)\in \mathbb{D}:-1<x<1\},
\]
and
\[
\mathbb{D}_3 = \{(x,y)\in \mathbb{D}:x\geq 1\}.
\]

For two locations $\tmmathbf{s},\tmmathbf{s}'\in \mathbb{D}$, we will consider four situations, depicted as four sub-figures in Figure \ref{FIG:Boundary example}, and construct a path $\mathcal{P}(\tmmathbf{s},\tmmathbf{s}')$ (the dashed lines in Figure \ref{FIG:Boundary example}) that satisfies Equation (\ref{EQ:pathdis}), with the constant value $C$ set to $1$. The proofs for the other situations follow similarly due to the symmetry of $\mathbb{D}$, and are therefore omitted.

For the situation (a) where $\tmmathbf{s},\tmmathbf{s}' \in \mathbb{D}_1$, we construct path $\mathcal{P}(\tmmathbf{s},\tmmathbf{s}')$ as $\tmmathbf{s}-A-\tmmathbf{s}'$. It is then easy to verify that $d(\tmmathbf{s}-A,\partial\mathbb{D})=d(\tmmathbf{s},\partial\mathbb{D})$, and $d(A-\tmmathbf{s}',\partial\mathbb{D})=d(\tmmathbf{s}',\partial\mathbb{D})$. Hence, $d(\tmmathbf{s}-A-\tmmathbf{s}',\partial\mathbb{D})=\min\{d(\tmmathbf{s},\partial\mathbb{D}),d(\tmmathbf{s}',\partial\mathbb{D})\}$ and Equation (\ref{EQ:pathdis}) is satisfied.

For the situation (b) where $\tmmathbf{s} \in \mathbb{D}_1$ and $\tmmathbf{s}'\in \mathbb{D}_3$, we construct path $\mathcal{P}(\tmmathbf{s},\tmmathbf{s}')$ as $\tmmathbf{s}-A-B-\tmmathbf{s}'$. Similar to the situation (a), it can be verified that $d(\tmmathbf{s}-A,\partial\mathbb{D})=d(\tmmathbf{s},\partial\mathbb{D})$, and $d(B-\tmmathbf{s}',\partial\mathbb{D})=d(\tmmathbf{s}',\partial\mathbb{D})$. Besides, we can see $d(A-B,\partial\mathbb{D})=1$. Thus, $d(\tmmathbf{s}-A-B-\tmmathbf{s}',\partial\mathbb{D})=\min\{d(\tmmathbf{s},\partial\mathbb{D}),d(\tmmathbf{s}',\partial\mathbb{D}),1\}$, from which we conclude $d(\tmmathbf{s}-A-B-\tmmathbf{s}',\partial\mathbb{D})=\min\{d(\tmmathbf{s},\partial\mathbb{D}),d(\tmmathbf{s}',\partial\mathbb{D})\}$ or $d(\tmmathbf{s}-A-B-\tmmathbf{s}',\partial\mathbb{D})=1$, and Equation (\ref{EQ:pathdis}) is hence satisfied. 

For the situation (c) where $\tmmathbf{s} \in \mathbb{D}_1$ and $\tmmathbf{s}'\in \mathbb{D}_2$, we construct path $\mathcal{P}(\tmmathbf{s},\tmmathbf{s}')$ as $\tmmathbf{s}-A-E-\tmmathbf{s}'$, where $E$ (denoted by the red color point) is the foot of the perpendicular from point $\tmmathbf{s}'$ to the x-axis. It can be verified that $d(\tmmathbf{s}-A,\partial\mathbb{D})=d(\tmmathbf{s},\partial\mathbb{D})$, $d(A-E,\partial\mathbb{D})=1$, and $d(E-\tmmathbf{s}',\partial\mathbb{D})=d(\tmmathbf{s}',\partial\mathbb{D})$. Hence, $d(\tmmathbf{s}-A-E-\tmmathbf{s}',\partial\mathbb{D})=\min\{d(\tmmathbf{s},\partial\mathbb{D}),d(\tmmathbf{s}',\partial\mathbb{D}),1\}$ and Equation (\ref{EQ:pathdis}) is satisfied.

Finally, for the situation (d) where $\tmmathbf{s},\tmmathbf{s}' \in \mathbb{D}_2$, we construct path $\mathcal{P}(\tmmathbf{s},\tmmathbf{s}')$ as $\tmmathbf{s}-F-E-\tmmathbf{s}'$, where $F$ and $E$ (denoted by the red color points) are the feet of the perpendiculars from points $\tmmathbf{s}'$ and $\tmmathbf{s}$ to the x-axis, respectively. It can be verified that $d(\tmmathbf{s}-F,\partial\mathbb{D})=d(\tmmathbf{s},\partial\mathbb{D})$, $d(F-E,\partial\mathbb{D})=1$, and $d(E-\tmmathbf{s}',\partial\mathbb{D})=d(\tmmathbf{s}',\partial\mathbb{D})$. Hence, $d(\tmmathbf{s}-F-E-\tmmathbf{s}',\partial\mathbb{D})=\min\{d(\tmmathbf{s},\partial\mathbb{D}),d(\tmmathbf{s}',\partial\mathbb{D}),1\}$ and Equation (\ref{EQ:pathdis}) is satisfied.

\section{Proofs of main results in the main paper}\label{SEC:proof}

\subsection{Additional notations and definitions}\label{SEC:proofnotations}

Throughout the proof, in general, subscripts $i, (j, l), m$ denote the
index for locations, clusters in a partition, and blocks, respectively. $k$ is used as the number of clusters, and $k_0$ is the true number of
clusters. 
%We use $\cdot^{\ast}$ as to denote terms related to blocks (e.g., $\pi^{\ast}(\mathcal{D})$, and $ \{ \mathcal{D}^{\ast}_{l, 0} \}_{l = 1}^{k_0}$).

To simplify notations, we use $\pi^{\ast}$ ($\pi$) as the shorthand of $\pi^\ast
(\mathcal{D})$ $(\pi (\mathcal{S}))$. For two block partitions
$\pi^{\ast}_1$ and $\pi^{\ast}_2$, we write $\pi^{\ast}_1 \cap \pi^{\ast}_2$
as the intersection partition, such that two locations are within the same
cluster if and only if they are within the same cluster by both
$\pi^{\ast}_1$ and $\pi^{\ast}_2$. %We say a block $B_m$ is a ``boundary block'' if $B_m$ intersects with the boundary set $\mathcal{B}$. 
Let
$ \Xi^{\ast} = \{\pi^{\ast} : \mathbb{P}(\pi^{\ast}) > 0\} $
be the space of partitions under priors in Section \ref{SEC:Model1}, and we write
$ \tilde{\Xi}^{\ast} = \{\pi^{\ast} : \pi^{\ast} = \pi^{\ast}_1 \cap
   \pi^{\ast}_2  \text{ for some } \pi^{\ast}_1, \pi^{\ast}_2 \in \Xi^{\ast} \}
   $. 
It is easy to see that $\Xi^{\ast} \subseteq \tilde{\Xi}^{\ast}$. Recall the contiguous partition $\pi_0^{\ast}$ in Propsition \ref{PP:bestapproximation}. Under Assumption \ref{AS:prior}, we have $\pi_0^{\ast} \in
\Xi^{\ast}$.

For two domain partitions $\pi_1 (\mathcal{D}) = \{\mathcal{D}_{1 1}, \ldots,
\mathcal{D}_{1 k_1} \}$ and $\pi_2 (\mathcal{D}) = \{\mathcal{D}_{21}, \ldots,
\mathcal{D}_{2 k_2} \}$ with $k_1$ and $k_2$ denoting the number of clusters, we decompose $\epsilon \{ \pi_1 (\mathcal{D}), \pi_2
(\mathcal{D}) \} = \epsilon_1 \{ \pi_1 (\mathcal{D}), \pi_2 (\mathcal{D}) \} +
\epsilon_2 \{ \pi_1 (\mathcal{D}), \pi_2 (\mathcal{D}) \}$, where
\begin{equation}
  \epsilon_1 \{ \pi_1 (\mathcal{D}), \pi_2 (\mathcal{D}) \} = 1 - |
  \mathcal{D} |^{- 1} \biggl[ \sum_{j = 1}^{k_1} \max_{l \in \{1, \ldots, k_2
  \}} | \mathcal{D}_{1 j} \cap \mathcal{D}_{2 l} | \biggl],\text{ and}
  \label{EQ:epsilon1dis}
\end{equation}
\begin{equation}
  \epsilon_2 \{ \pi_1 (\mathcal{D}), \pi_2 (\mathcal{D}) \} = 1 - |
  \mathcal{D} |^{- 1} \biggl[ \sum_{l = 1}^{k_2} \max_{j \in \{1, \ldots, k_1
  \}} | \mathcal{D}_{1 j} \cap \mathcal{D}_{2 l} | \biggl] .
  \label{EQ:epsilon2dis}
\end{equation}
% With a slight abuse of notations, we write $\epsilon (\pi^{\ast}, \cdot) =
% \epsilon \{ \pi (\mathcal{D}), \cdot \}$, where $\pi (\mathcal{D})$ is the
% domain partition induced from $\pi^{\ast}$. Similarly, we write $\epsilon_1
% (\pi^{\ast}, \cdot)$ and $\epsilon_2 (\pi^{\ast}, \cdot)$ with the same
% meaning. 

Table \ref{TAB:notations} summarizes main notations used throughout the proof.

\begin{table}[h]
\caption{Commonly used notations throughout the proof}
  \begin{tabular}{ll}
  \hline
    Notation & Meaning\\
    \hline
    $\pi^{\ast} = \{ \mathcal{D}^{\ast}_1, \ldots, \mathcal{D}^{\ast}_k \}$ &
    Shorthand of $\pi^{\ast} (\mathcal{D})$, which is a domain partition under our model\\
    $\pi = \{ \mathcal{S}_1, \ldots, \mathcal{S}_k \}$ & Shorthand of $\pi
    (\mathcal{S})$, which is a partition of $\{\tmmathbf{s}_i\}_{i=1}^{n}$ and can be induced from
    $\pi^{\ast}$\\
    $\{ \mathcal{D}_{l, 0} \}_{l = 1}^{k_0}$ & True domain partition\\
$ \{ \mathcal{S}_{l, 0} \}_{l=1}^{k_0}$ & True partition of $\{\tmmathbf{s}_i\}_{i=1}^{n}$
    \\
    $\pi^{\ast}_0 = \{ \mathcal{D}^{\ast}_{1, 0}, \ldots,
    \mathcal{D}^{\ast}_{k_0, 0} \}$ & Domain partition defined in Proposition
    
    \ref{PP:bestapproximation}\\

    $\mathcal{M} (\mathcal{D}^{\ast}_j)$ & $\tmop{argmax}_{l \in \{1, \ldots,
    k_0 \}} |\mathcal{D}^{\ast}_j \cap \mathcal{D}^{\ast}_{l, 0} |$\\
    \hline
  \end{tabular}
  \label{TAB:notations}
  \vspace{-5mm}
\end{table}

\subsection{Proof framework}\label{SEC:proofframe}

For a given $\pi^{\ast}$, we decompose $\epsilon [\pi^{\ast}, \{
\mathcal{D}_{l, 0} \}_{l = 1}^{k_0}]$ as
\begin{equation}
  \epsilon [\pi^{\ast}, \{ \mathcal{D}_{l, 0} \}_{l = 1}^{k_0}]
  \overset{\text{Triangle inequality of $\epsilon (\cdot, \cdot)$}}{\leqslant}
  \epsilon (\pi^{\ast}, \pi_0^{\ast}) + \epsilon [\pi_0^{\ast}, \{
  \mathcal{D}_{l, 0} \}_{l = 1}^{k_0}] . \label{EQ:decomp}
\end{equation}
According to Proposition \ref{PP:bestapproximation}, it is easy to see that
$\epsilon [\pi_0^{\ast}, \{ \mathcal{D}_{l, 0} \}_{l = 1}^{k_0}] \leqslant c k_0^{1/2} K
\times K^{- 2} = c k_0^{1/2} K^{- 1} $. To
study $\epsilon [\pi^{\ast}, \{ \mathcal{D}_{l, 0} \}_{l = 1}^{k_0}]$, it then
remains to study $\epsilon (\pi^{\ast}, \pi_0^{\ast}) = \epsilon_1
(\pi^{\ast}, \pi_0^{\ast}) + \epsilon_2 (\pi^{\ast}, \pi_0^{\ast})$. We will
focus on the bound of $\epsilon_1 (\pi^{\ast}, \pi_0^{\ast})$, from which the
bound of $\epsilon_2 (\pi^{\ast}, \pi_0^{\ast})$ can be derived similarly. Note that by definition, $\epsilon_1 (\pi^{\ast},
\pi_0^{\ast}) K^2$ takes only integer values, and $0 \leqslant \epsilon_1
(\pi^{\ast}, \pi_0^{\ast}) K^2 \leqslant K^2$.

Recall the definition of $\mathcal{M} (\mathcal{D}^{\ast}_j)$ after Theorem
\ref{TH:clustererror}. Based on $\{\mathcal{M}(\mathcal{D}^{\ast}_j)\}_{j =
1}^{| \pi^{\ast} |}$ and the number of clusters, we divide the prior block partition space, $\Xi^{\ast}$, into
five categories:
\[ \Pi_1^{\ast} = \{\pi^{\ast} \in \Xi^{\ast} : | \pi^{\ast} | < k_0 \}, \text{ }
 \Pi_2^{\ast} = \{\pi^{\ast} \in \Xi^{\ast} : | \pi^{\ast} | > k_0 \}, \]
\[ \Pi_3^{\ast} = \left\{ \pi^{\ast} \in \Xi^{\ast} : | \pi^{\ast} | = k_0 
   \text{ and } \{\mathcal{M}(\mathcal{D}^{\ast}_j)\}_{j = 1}^{| \pi^{\ast} |}
   \subsetneq \{1, \ldots, k_0 \} \right\}, \]
   \[ \Pi_4^{\ast} = \cup_{q = \lfloor k_0^{1 / 2 + 2 / d} K \log^{\alpha_0} (n)
   \rfloor}^{K^2} \Pi^{\ast}_{\epsilon, q}, \text{and } \Pi_5^{\ast} = \cup_{q
   = 0}^{\lfloor k_0^{1 / 2 + 2 / d} K \log^{\alpha_0} (n) \rfloor - 1}
   \Pi^{\ast}_{\epsilon, q}, \text{where} \]
   \[\Pi^\ast_{\epsilon,q}=\left\{\pi^{\ast} \in \Xi^{\ast} : | \pi^{\ast} | = k_0, 
    \{\mathcal{M}(\mathcal{D}^{\ast}_j)\}_{j = 1}^{| \pi^{\ast} |}
   =\{1, \ldots, k_0 \} \text{ and }    \epsilon_1 (\pi^{\ast}, \pi_0^{\ast}) K^2=q\right\},\] and $\lfloor a \rfloor$ denotes the largest integer less than or equal to $a$. Both $\Pi_1^{\ast}$ and
$\Pi_3^{\ast}$ are sets of ``underfitted" partitions, in the sense that
$\lvert \{\mathcal{M}(\mathcal{D}^{\ast}_j)\}_{j = 1}^{| \pi^{\ast} |} \rvert
< k_0$ for $\pi^{\ast} \in \Pi_1^{\ast} \cup \Pi_3^{\ast}$. $\Pi_2^{\ast}$ is the set of ``overfitted" partitions, in the sense that
$|\pi^{\ast}|>k_0$ for $\pi^{\ast} \in \Pi_2^{\ast}$. $\Pi_4^\ast$ is the set of partitions with relatively large $\epsilon_1 (\pi^{\ast}, \pi_0^{\ast})$, although having the correct number of clusters. $\Pi_5^\ast$ is the set of ``good" partitions, in the sense that the number of clusters is correct and $\epsilon_1 (\pi^{\ast}, \pi_0^{\ast})$ is small. 

$\epsilon_1 (\pi^{\ast}, \pi_0^{\ast})$ takes different values when $\pi^{\ast}$ belongs to different categories.
We will study the posterior probability of $\pi^{\ast}$ in $\Pi^{\ast}_1,
\Pi_2^{\ast}, \Pi_3^{\ast}$, $\Pi_4^{\ast}$ and $\Pi_5^{\ast}$ respectively, based on which
we derive the posterior distribution of $\epsilon_1 (\pi^{\ast},
\pi_0^{\ast})$. To study the posterior distribution of $\pi^{\ast}$, we first
write out $\frac{\mathbb{P} (\pi^{\ast} \mid
\mathfrak{D})}{\mathbb{P} (\pi_0^{\ast} \mid \mathfrak{D})}$. Given a
partition $\pi^{\ast}$, by Bayesian rule, we have
\begin{eqnarray*}
  \mathbb{P} (\pi^{\ast} \mid \mathfrak{D}) & \propto & \sum_{\mathcal{T}}
  \mathbb{P} (\pi^{\ast}, \tmmathbf{x}, \tmmathbf{s}, \tmmathbf{y},
  \mathcal{T}) \propto \sum_{\mathcal{T}} \mathbb{P} (\tmmathbf{y} \mid
  \pi^{\ast}, \tmmathbf{x}, \tmmathbf{s}) \mathbb{P} (\mathcal{T}) \mathbb{P}
  (\pi^{\ast} \mid \mathcal{T}) .
\end{eqnarray*}
We then write the posterior ratio $\frac{\mathbb{P} (\pi^{\ast} \mid \mathfrak{D})}{\mathbb{P}
  (\pi_0^{\ast} \mid \mathfrak{D})}$ as
\begin{align}
   &\frac{\sum_{\mathcal{T}} \mathbb{P} (\tmmathbf{y} \mid \pi^{\ast},
  \tmmathbf{x}, \tmmathbf{s}) \mathbb{P} (\mathcal{T}) \mathbb{P} (\pi^{\ast}
  \mid \mathcal{T})}{\sum_{\mathcal{T}} \mathbb{P} (\tmmathbf{y} \mid
  \pi_0^{\ast}, \tmmathbf{x}, \tmmathbf{s}) \mathbb{P} (\mathcal{T})
  \mathbb{P} (\pi_0^{\ast} \mid \mathcal{T})} \overset{(i)}{\leqslant} c \exp \{c k_0^{1/2}K \log (K)\} \frac{\mathbb{P}
  (\tmmathbf{y} \mid \pi^{\ast}, \tmmathbf{x}, \tmmathbf{s}) 
  \sum_{\mathcal{T}} \mathbb{P} (\pi^{\ast} \mid \mathcal{T})}{\mathbb{P}
  (\tmmathbf{y} \mid \pi_0^{\ast}, \tmmathbf{x}, \tmmathbf{s}) 
  \sum_{\mathcal{T}} \mathbb{P} (\pi_0^{\ast} \mid \mathcal{T})} \nonumber\\
  = & c \exp \{c k_0^{1/2}K \log (K)\} \frac{\mathbb{P} (\tmmathbf{y} \mid
  \pi^{\ast}, \tmmathbf{x}, \tmmathbf{s})}{\mathbb{P} (\tmmathbf{y} \mid
  \pi_0^{\ast}, \tmmathbf{x}, \tmmathbf{s})} \times \frac{\sum_{\mathcal{T}}
  \mathbb{P} (\pi^{\ast} \mid \mathcal{T}) \mathbb{I} (\mathcal{T}
  \text{ can induce } \pi^{\ast})}{\sum_{\mathcal{T}} \mathbb{P} (\pi_0^{\ast}
  \mid \mathcal{T}) \mathbb{I} (\mathcal{T} \text{ can induce } \pi_0^{\ast})}
  \nonumber\\
  % & \overset{(i)}{=} & c \exp \{cK \log (K)\} \frac{\mathbb{P} (\tmmathbf{y}
  % \mid \pi^{\ast}, \tmmathbf{x}, \tmmathbf{s})}{\mathbb{P} (\tmmathbf{y} \mid
  % \pi_0^{\ast}, \tmmathbf{x}, \tmmathbf{s})} \times \frac{\sum_{\mathcal{T}}
  % \frac{\lambda^{| \pi^{\ast} |}}{| \pi^{\ast} | !} \left( \begin{array}{c}
  %   K^2 - 1\\
  %   | \pi^{\ast} | - 1
  % \end{array} \right)^{- 1} \mathbb{I} (\mathcal{T} \text{ can induce }
  % \pi^{\ast})}{\sum_{\mathcal{T}} \frac{\lambda^{| \pi_0^{\ast} |}}{|
  % \pi_0^{\ast} | !} \left( \begin{array}{c}
  %   K^2 - 1\\
  %   | \pi_0^{\ast} | - 1
  % \end{array} \right)^{- 1} \mathbb{I} (\mathcal{T} \text{ can induce }
  % \pi_0^{\ast})} \nonumber\\
  \overset{(ii)}{=}& c \exp \{c k_0^{1/2}K \log (K)\} \frac{\mathbb{P} (\tmmathbf{y} \mid
  \pi^{\ast}, \tmmathbf{x}, \tmmathbf{s}) \lambda^{| \pi^{\ast} |}}{\mathbb{P}
  (\tmmathbf{y} \mid \pi_0^{\ast}, \tmmathbf{x}, \tmmathbf{s}) \lambda^{|
  \pi_0^{\ast} |}} \times \frac{| \pi_0^{\ast} | !}{| \pi^{\ast} | !} 
  \frac{\left( \begin{array}{c}
    K^2 - 1\\
    | \pi_0^{\ast} | - 1
  \end{array} \right)}{\left( \begin{array}{c}
    K^2 - 1\\
    | \pi^{\ast} | - 1
  \end{array} \right)}  \frac{|\{\mathcal{T}: \mathcal{T} \text{ can induce }
  \pi^{\ast} \}|}{|\{\mathcal{T}: \mathcal{T} \text{ can induce } \pi_0^{\ast}
  \}|},  \label{EQ:gammaratio2}
\end{align}
where $(i)$ uses \text{Assumption} \ref{AS:prior}, and $(ii)$ uses Equations (\ref{DEF:k}) and (\ref{DEF:pi}). From Equation
(\ref{EQ:gammaratio2}), we can see that $\frac{\mathbb{P} (\tmmathbf{y} \mid
\pi^{\ast}, \tmmathbf{x}, \tmmathbf{s}) \lambda^{| \pi^{\ast} |}}{\mathbb{P}
(\tmmathbf{y} \mid \pi_0^{\ast}, \tmmathbf{x}, \tmmathbf{s}) \lambda^{|
\pi_0^{\ast} |}}$ and $\frac{|\{\mathcal{T}: \mathcal{T} \text{ can induce }
\pi^{\ast} \}|}{|\{\mathcal{T}: \mathcal{T} \text{ can induce } \pi_0^{\ast}
\}|}$ play essential roles in $\frac{\mathbb{P} (\pi^{\ast} \mid
\mathfrak{D})}{\mathbb{P} (\pi_0^{\ast} \mid \mathfrak{D})}$. 

To study $\frac{\mathbb{P} (\tmmathbf{y} \mid \pi^{\ast}, \tmmathbf{x},
\tmmathbf{s}) \lambda^{| \pi^{\ast} |}}{\mathbb{P} (\tmmathbf{y} \mid
\pi_0^{\ast}, \tmmathbf{x}, \tmmathbf{s}) \lambda^{| \pi_0^{\ast} |}}$, we
first define two events, say $\mathcal{A}_n$ and $\mathcal{E}_n$, and show
that $\mathbb{P} (\mathcal{A}_n \cap \mathcal{E}_n) \rightarrow 1$. The
definition of $\mathcal{A}_n$ and $\mathcal{E}_n$ are deferred to Section
\ref{SEC:AE}. Conditional on events $\mathcal{A}_n \cap \mathcal{E}_n$, we
give the bound of $\frac{\mathbb{P} (\tmmathbf{y} \mid \pi^{\ast},
\tmmathbf{x}, \tmmathbf{s}) \lambda^{| \pi^{\ast} |}}{\mathbb{P} (\tmmathbf{y}
\mid \pi_0^{\ast}, \tmmathbf{x}, \tmmathbf{s}) \lambda^{| \pi_0^{\ast} |}}$ by
the following proposition. 

\begin{proposition}
  \label{PP:likelihoodratio}Under events $\mathcal{A}_n \cap \mathcal{E}_n$
  and Assumptions \ref{AS:distributionloc}, \ref{AS:lengthofboundary},
  \ref{AS:thetagap}, \ref{AS:covariates}, \ref{AS:hyperpara} and
  \ref{AS:prior}, there exists a constant $c$, such that
  \[ \frac{\mathbb{P} (\tmmathbf{y} \mid \pi^{\ast}, \tmmathbf{x},
     \tmmathbf{s}) \lambda^{| \pi^{\ast} |}}{\mathbb{P} (\tmmathbf{y} \mid
     \pi^{\ast}_0, \tmmathbf{x}, \tmmathbf{s}) \lambda^{| \pi_0^{\ast} |}}
     \leqslant \left\{ \begin{array}{l}
       \exp (- c k_0^{- 1 - 2 / d} r_n^2 n), \text{if } \pi^{\ast} \in
       \Pi_1^{\ast}\\
       \exp \{- c \log (\lambda^{- 1}) \}, \text{if } \pi^{\ast} \in
       \Pi_2^{\ast}\\
       \exp (- ck_0^{- 1 - 2 / d} r_n^2 n), \text{if } \pi^{\ast} \in
       \Pi_3^{\ast}\\
       \exp \{- c k_0^{- 2 / d} n \epsilon_1 (\pi^{\ast}, \pi_0^{\ast}) r_n^2
       \}, \text{if } \pi^{\ast} \in \Pi_4^{\ast},
     \end{array} \right. . \]
\end{proposition}

We defer the proof of Proposition \ref{PP:likelihoodratio} to Section
\ref{SEC:likelihood}. We can see that the ratio
bounds for partitions in $\Pi_1^{\ast}$ and $\Pi_3^{\ast}$ share the same rate, since they are both sets of ``underfitted" partitions. As discussed after Assumption
\ref{AS:hyperpara}, we control the overfitted probability by Poisson
hyperparameter $\lambda$. Consequently, $\lambda$ appears in the ratio bound for the ``overfitted" $\pi^{\ast} \in \Pi_2^{\ast}$. For $\pi^{\ast} \in \Pi_4^\ast$, we can see that the larger $\epsilon_{1} (\pi^{\ast},\pi_{0}^{\ast})$, the
smaller $\frac{\mathbb{P} (\tmmathbf{y} \mid \pi^{\ast}, \tmmathbf{x},
\tmmathbf{s}) \lambda^{| \pi^{\ast} |}}{\mathbb{P} (\tmmathbf{y} \mid
\pi_0^{\ast}, \tmmathbf{x}, \tmmathbf{s}) \lambda^{| \pi_0^{\ast} |}}$.
{However, for $\pi^{\ast} \in \cup_{q = 0}^{k_0^{1 / 2 + 2 / d} K} \Pi_{\epsilon, q}^{\ast}$, a subset of
$\Pi_5^\ast$, it is not necessary that $\frac{\mathbb{P} (\tmmathbf{y} \mid \pi^{\ast},
\tmmathbf{x}, \tmmathbf{s}) \lambda^{| \pi^{\ast} |}}{\mathbb{P} (\tmmathbf{y}
\mid \pi_0^{\ast}, \tmmathbf{x}, \tmmathbf{s}) \lambda^{| \pi_0^{\ast} |}}$
converges to $0$. The reason is caused by the ``approximation error" between
$\pi_0^{\ast}$ and the true partition $\{\mathcal{D}_{l, 0} \}_{l = 1}^{k_0}$.
More details are in the proof in Section \ref{SEC:likelihood}.} The next proposition gives a bound of
$\frac{|\{\mathcal{T}: \mathcal{T} \text{ can induce } \pi^{\ast}
\}|}{|\{\mathcal{T}: \mathcal{T} \text{ can induce } \pi^{\ast}_0 \}|}$.

\begin{comment}
The blocking technique is essential in establishing Proposition
\ref{PP:likelihoodratio}. To bound $\frac{\mathbb{P} (\tmmathbf{y} \mid
\pi^{\ast}, \tmmathbf{x}, \tmmathbf{s}) \lambda^{| \pi^{\ast} |}}{\mathbb{P}
(\tmmathbf{y} \mid \pi_0^{\ast}, \tmmathbf{x}, \tmmathbf{s}) \lambda^{|
\pi_0^{\ast} |}}$ when $\pi^{\ast} \in \Pi_2^{\ast}$, it is necessary for
$\lambda$ to go to $0$. In that case, blocking reduces the effect of
$\{\epsilon (\tmmathbf{s}_i)\}_{i = 1}^n$, which assures the ``underfitted"
impact to dominate $\frac{\mathbb{P} (\tmmathbf{y} \mid \pi^{\ast},
\tmmathbf{x}, \tmmathbf{s}) \lambda^{| \pi^{\ast} |}}{\mathbb{P} (\tmmathbf{y}
\mid \pi_0^{\ast}, \tmmathbf{x}, \tmmathbf{s}) \lambda^{| \pi_0^{\ast} |}}$
when $\pi^{\ast} \in \Pi_1^{\ast} \cup \Pi_3^{\ast}$. We provide a more
detailed illustration in Remark \ref{RM:1} at the end of Section
\ref{SEC:likelihood}. 
\end{comment}

\begin{proposition}
  \label{PP:spanningtreeratio}Under our model and Assumption
  \ref{AS:lengthofboundary}, there exists a constant $C > 0$, such that
  \[ \frac{|\{\mathcal{T}: \mathcal{T} \text{can induce } \pi^{\ast}
     \}|}{|\{\mathcal{T}: \mathcal{T} \text{can induce } \pi^{\ast}_0 \}|}
     \leqslant \exp \{C k_0^{1 / 2} K \log (K)\} \times \left(
     \begin{array}{l}
       2 (K - 1)^2\\
       k - 1
     \end{array} \right) \]
  holds for all $\pi^{\ast} \in \Xi^{\ast}$, where $k$ is the number of clusters in $\pi^\ast$.
\end{proposition}

We defer the proof of Proposition \ref{PP:spanningtreeratio} to Section
\ref{SEC:treenumber}. Proposition \ref{PP:spanningtreeratio} provides an upper bound for the ratio of the numbers of spanning trees
inducing two partitions, for a
$K \times K$ mesh grid graph. Although this is a result of the mesh grid graph, it is worth
mentioning that it has the potential to be extended to a more general graph,
with some higher-level assumptions on graphs. This upper bound provides a reference potentially useful for studying other spanning-tree-based approaches.

The remaining proof consists of several parts. In Section \ref{SEC:AE}, we
detail the definitions of events $\mathcal{A}_n$ and $\mathcal{E}_n$. In
Section \ref{SEC:finalconsis}, making use of
Propositions \ref{PP:bestapproximation} - \ref{PP:distance} in the main paper, and Propositions \ref{PP:likelihoodratio} - \ref{PP:spanningtreeratio}, we prove
Theorem \ref{TH:clustererror}. In Section \ref{SEC:TH2proof}, we prove Theorem \ref{TH:thetaconverge} based on the proof of Theorem \ref{TH:clustererror}. In Section \ref{SEC:pre}, we introduce some basic lemmas regarding distribution and probability inequalities, which are useful for the later proof. Sections
\ref{SEC:proofofAn} and \ref{SEC:proofofEn} prove $\mathbb{P} (\mathcal{A}_n)
\rightarrow 1$ and $\mathbb{P} (\mathcal{E}_n) \rightarrow 1$, respectively. In Sections \ref{SEC:proofofpp1} -
\ref{SEC:treenumber}, we give the proof of Propositions \ref{PP:bestapproximation} - \ref{PP:distance}, and 
\ref{PP:likelihoodratio} - \ref{PP:spanningtreeratio}, respectively. 

\subsection{Definition of $\mathcal{A}_n$ and $\mathcal{E}_n$}\label{SEC:AE}

Recall that we use $\{B_m \}_{m = 1}^{K^2}$ to denote blocks and $\|B_m \|$ to
denote the number of locations in the $m$-th block. We define $\mathcal{A}_n
=\mathcal{A}_{1 n} \cap \mathcal{A}_{2 n}$ as events with respect to
$\tmmathbf{s}$ and $\tmmathbf{x}$, where
\begin{equation}
  \mathcal{A}_{1 n} = \left\{ c < \frac{\min_{1 \leqslant m \leqslant K^2}
  \|B_m \|}{n K^{- 2}} \leqslant \frac{\max_{1 \leqslant m \leqslant K^2}
  \|B_m \|}{n K^{- 2}} < C \right\}, \text{and} \label{EQ:A1n}
\end{equation}
\begin{equation}
  \mathcal{A}_{2 n} = \Big\{ 0 < c <\|B_m \|^{- 1}  \sum_{\tmmathbf{s}_i \in
  B_m} \tmmathbf{x}(\tmmathbf{s}_i)\tmmathbf{x}^T (\tmmathbf{s}_i) < C <
  \infty, m = 1, 2, \ldots, K^2 \Big\} \label{EQ:A2n}
\end{equation}
for some constants $c, C > 0$ to be determined later. We have the following
result.

\begin{lemma}
  \label{LM:An}Under Assumptions \ref{AS:distributionloc}, \ref{AS:covariates}
  and \ref{AS:hyperpara}, there exist constants $c, C > 0$, such that
  $\mathbb{P} (\mathcal{A}_n) \rightarrow 1$.
  
  \begin{proof}
    The proof is deferred to Section \ref{SEC:proofofAn}.
  \end{proof}
\end{lemma}

In what follows, we assume the constants $c$ and $C$ in (\ref{EQ:A1n}) -
(\ref{EQ:A2n}) are chosen such that $\mathbb{P} (\mathcal{A}_n) \rightarrow
1$. Under $\mathcal{A}_n$, we can replace the pseudoinverse notation in
(\ref{EQ:priortheta}) with inverse. For the simplicity of proof, we take the value of $\gamma$ in (\ref{EQ:priortheta}) to be $1$, and it is trivial to extend the proof to the case for any fixed $\gamma>0$. For a given $\pi^{\ast}$ (and $\pi =
\{\mathcal{S}_1, \ldots, \mathcal{S}_k \}$ induced from $\pi^{\ast}$), following
priors given by (\ref{EQ:priortheta}) and (\ref{EQ:priory}), and after
replacing the pseudoinverse notation with inverse, we have
\begin{equation}
  \tmmathbf{\theta} \mid \pi^{\ast}, k, \mathcal{T}, \tmmathbf{s},
  \tmmathbf{x} \sim \prod_{j = 1}^k \mathbb{P}_{\tmop{Gaussian}}
  \{\tmmathbf{\theta}_j ; \tmmathbf{0}, n \sigma^2 (\tmmathbf{x}_j^T
  \tmmathbf{x}_j)^{- 1} \}, \text{ and} \label{EQ:thetajdis}
\end{equation}
\begin{equation}
  \tmmathbf{y} \mid \pi^{\ast}, k, \mathcal{T}, \tmmathbf{s}, \tmmathbf{x},
  \tmmathbf{\theta} \sim \prod_{j = 1}^k \mathbb{P}_{\tmop{Gaussian}}
  (\tmmathbf{y}_j ; \tmmathbf{x}_j \tmmathbf{\theta}_j, \sigma^2
  \tmmathbf{I}_{|\mathcal{S}_j |}), \label{EQ:yjmodel}
\end{equation}
where $\tmmathbf{x}_j$ is a $|\mathcal{S}_j | \times d$ design matrix of
covariates in $\mathcal{S}_j$, and $\tmmathbf{y}_j = \{y (\tmmathbf{s}_i) :
\tmmathbf{s}_i \in \mathcal{S}_j \}$. Following Equations (\ref{EQ:thetajdis})
and (\ref{EQ:yjmodel}), $\tmmathbf{y}_j \mid \pi^{\ast}, k, \mathcal{T},
\tmmathbf{s}, \tmmathbf{x} \sim \tmop{Gaussian} [\tmmathbf{0}, \sigma^2
\{n\tmmathbf{x}_j (\tmmathbf{x}_j^T \tmmathbf{x}_j)^{- 1} \tmmathbf{x}_j^T
+\tmmathbf{I}_{|\mathcal{S}_j |} \}]$. We can thus write
\begin{eqnarray}
  \mathbb{P} (\tmmathbf{y} \mid \pi^{\ast}, \tmmathbf{x}, \tmmathbf{s}) & = &
  \mathbb{P} (\tmmathbf{y} \mid \pi^{\ast}, k, \mathcal{T}, \tmmathbf{s},
  \tmmathbf{x}) = \prod_{j = 1}^k \mathbb{P}_{\tmop{Gaussian}} [\tmmathbf{y}_j ;
  \tmmathbf{0}, \sigma^2 \{n\tmmathbf{\phi}_j +\tmmathbf{I}_{|\mathcal{S}_j |}
  \}] \nonumber\\
  & = & (2 \pi \sigma^2)^{- n / 2} \exp \left( - \frac{\tmmathbf{y}^T
  \tmmathbf{y}}{2 \sigma^2} \right)  (n + 1)^{- kd / 2} \exp \left\{
  \frac{n}{2 \sigma^2 (n + 1)} \tmmathbf{y}^T \tmmathbf{\phi}_{\pi^{\ast}}
  \tmmathbf{y} \right\},  \label{EQ:formlikelihood}
\end{eqnarray}
where $\tmmathbf{\phi}_j =\tmmathbf{x}_j  (\tmmathbf{x}_j^T \tmmathbf{x}_j)^{-
1} \tmmathbf{x}_j^T$ is a projection matrix,
$\tmmathbf{\phi}_{\pi^{\ast}} =\tmmathbf{P}_{\pi^{\ast}}^T \tmop{diag}
(\tmmathbf{\phi}_1, \ldots \tmmathbf{\phi}_k) \tmmathbf{P}_{\pi^{\ast}}$, and
$\tmmathbf{P}_{\pi^{\ast}}$ ia a permutation matrix such that
$\tmmathbf{P}_{\pi^{\ast}} \tmmathbf{y}= (\tmmathbf{y}^T_1, \ldots,
\tmmathbf{y}^T_k)^T$.

Recall the definitions of $\tilde{\Xi}^{\ast}$ and $\cap$ in Section
\ref{SEC:proofnotations}. We define $\mathcal{E}_n =\mathcal{E}_{1 n}
\cap \mathcal{E}_{2 n} \cap \mathcal{E}_{3 n} \cap \mathcal{E}_{4 n}$ as
events with respect to $\tmmathbf{\epsilon}$, where
\begin{equation}
  \mathcal{E}_{1 n} = \{\sup_{\pi^{\ast} \in \tilde{\Xi}^{\ast}}
  \tmmathbf{\epsilon}^T \tmmathbf{\phi}_{\pi^{\ast}} \tmmathbf{\epsilon}
  \leqslant C K^2 \log (n)\}, \label{EQ:epsilon1n}
\end{equation}
\begin{equation}
  \mathcal{E}_{2 n} = [\tmmathbf{\epsilon}^T (\tmmathbf{\phi}_{\pi^{\ast} \cap
  \pi^{\ast}_0} +\tmmathbf{\phi}_{\pi^{\ast}})\tmmathbf{\epsilon} \leqslant C
  \epsilon_1 (\pi^{\ast}, \pi_0^{\ast}) K^2 \log (n), \forall \pi^{\ast} \in
  \Pi_4^{\ast}], \label{EQ:epsilon2n}
\end{equation}
\begin{equation}
  \mathcal{E}_{3 n} = \Bigl\{ \sup_{1 \leqslant m \leqslant K^2} \bigl\|
  \sum_{\tmmathbf{s}_i \in B_m} \tmmathbf{x}(\tmmathbf{s}_i) \epsilon
  (\tmmathbf{s}_i) \bigr \| _2 \leqslant C \sqrt{n K^{- 2} \log (K)} \Bigr\}
  \label{EQ:epsilon3n}, \text{and }
\end{equation}
\begin{equation}
  \mathcal{E}_{4 n} = \Bigl\{ \sup_{1 \leqslant l \leqslant k_0} \bigl\|
  \sum_{\tmmathbf{s}_i \in \mathcal{D}^{\ast}_{l, 0}}
  \tmmathbf{x}(\tmmathbf{s}_i) \epsilon (\tmmathbf{s}_i) \bigr\|_2 \leqslant n^{1 /
  2} \log^{\alpha_0} (n) \Bigr\} \label{EQ:epsilon4n}
\end{equation}
for $\alpha_0$ in Theorem \ref{TH:clustererror} and some constant $C > 0$ to
be determined later. We have the following result.

\begin{lemma}
  \label{LM:En}Under our model and Assumptions \ref{AS:covariates} and
  \ref{AS:hyperpara}, there exists a constant $C > 0$, such that $\mathbb{P}
  (\mathcal{E}_n) \rightarrow 1$.
  
  \begin{proof}
    The proof is deferred to Section \ref{SEC:proofofEn}
  \end{proof}
\end{lemma}

In what follows, we assume that the constant $C$ in (\ref{EQ:epsilon3n}) are
chosen such that $\mathbb{P} (\mathcal{E}_n) \rightarrow 1$. Combining Lemmas
\ref{LM:An} and \ref{LM:En}, we conclude that under Assumptions
\ref{AS:distributionloc}, \ref{AS:covariates} and \ref{AS:hyperpara},
$\mathbb{P} (\mathcal{A}_n \cap \mathcal{E}_n) \rightarrow 1$. In the later
context, probabilities are considered while conditional on the events
$\mathcal{A}_n \cap \mathcal{E}_n$ by default.

\subsection{Proof of Theroem \ref{TH:clustererror}}\label{SEC:finalconsis}

Recall that we partition $\Xi^{\ast} = \Pi_1^{\ast} \cup \Pi_2^{\ast} \cup
\Pi_3^{\ast} \cup \Pi_4^{\ast} \cup \Pi_5^{\ast}$. We first study the
probability of $\pi^{\ast}$ belonging to $\Pi_1^{\ast}, \Pi_2^{\ast},
\Pi_3^{\ast}$ and $\Pi_4^{\ast}$, which is given by the following lemma.

\begin{lemma}
  \label{LM:probofpai}Under events $\mathcal{A}_n \cap \mathcal{E}_n$ and
  Assumptions \ref{AS:distributionloc}, \ref{AS:lengthofboundary},
  \ref{AS:thetagap}, \ref{AS:covariates}, \ref{AS:hyperpara} and
  \ref{AS:prior}, there exists a constant $c$, such that
  \begin{equation}
    \mathbb{P} (\pi^{\ast} \in \Pi_1^{\ast} \cup \Pi_2^{\ast} \cup
    \Pi_3^{\ast} \mid \mathfrak{D}) \leqslant c \exp \{- c \log (\lambda^{-
    1}) \}, \text{and} \label{EQ:posteriorconsistency1}
  \end{equation}
  \begin{equation}
    \mathbb{P} (\pi^{\ast} \in \Pi_4^{\ast} \mid \mathfrak{D}) \leqslant c
    \exp \{- cn k_0^{1 / 2} K^{- 1} \log^{\alpha_0} (n) r_n^2 \} .
    \label{EQ:posteriorconsistency2}
  \end{equation}
  \begin{proof}
    Recall the expression of $\frac{\mathbb{P} (\pi^{\ast} \mid
    \mathfrak{D})}{\mathbb{P} (\pi_0^{\ast} \mid \mathfrak{D})}$ in
    (\ref{EQ:gammaratio2}). In Proposition \ref{PP:spanningtreeratio}, we give
    a bound of $\frac{|\{\mathcal{T}: \mathcal{T} \text{can induce }
    \pi^{\ast} \}|}{|\{\mathcal{T}: \mathcal{T} \text{can induce }
    \pi^{\ast}_0 \}|}$. Combining with the fact that $\frac{| \pi_0^{\ast} |
    !}{| \pi^{\ast} | !} \left( \begin{array}{c}
      K^2 - 1\\
      | \pi_0^{\ast} | - 1
    \end{array} \right) \left( \begin{array}{c}
      K^2 - 1\\
      | \pi^{\ast} | - 1
    \end{array} \right)^{- 1} \leqslant c k_0^{k_0} K^{2 k_0}
    \overset{\text{Assumption \ref{AS:hyperpara}}}{\leqslant} c \exp \{ck_0^{1
    / 2} K \log (K)\}$, we have
    \begin{equation}
      \frac{\mathbb{P} (\pi^{\ast} \mid \mathfrak{D})}{\mathbb{P}
      (\pi_0^{\ast} \mid \mathfrak{D})} \leqslant c \exp \{ck_0^{1 / 2} K \log
      (K)\}  \frac{\mathbb{P} (\tmmathbf{y} \mid \pi^{\ast}, \tmmathbf{x},
      \tmmathbf{s}) \lambda^{| \pi^{\ast} |}}{\mathbb{P} (\tmmathbf{y} \mid
      \pi_0^{\ast}, \tmmathbf{x}, \tmmathbf{s}) \lambda^{| \pi_0^{\ast} |}}
      \times \left( \begin{array}{l}
        2 (K - 1)^2\\
        k - 1
      \end{array} \right) . \label{EQ:posteriorratio}
    \end{equation}
    Next, under events $\mathcal{A}_n \cap \mathcal{E}_n$, we will make use of
    Proposition \ref{PP:likelihoodratio} to bound $\mathbb{P} (\pi^{\ast} \in
    \Pi_1^{\ast} \cup \Pi_2^{\ast} \cup \Pi_3^{\ast} \mid \mathfrak{D})$, and
    $\mathbb{P} (\pi^{\ast} \in \Pi_4^{\ast} \mid \mathfrak{D})$,
    respectively.
    
    \tmtextbf{Proof of (\ref{EQ:posteriorconsistency1})}:
    
    Under events $\mathcal{A}_n \cap \mathcal{E}_n$, according to Proposition
    \ref{PP:likelihoodratio} and Assumption \ref{AS:hyperpara}, for
    $\pi^{\ast} \in \Pi_1^{\ast} \cup \Pi_2^{\ast} \cup \Pi_3^{\ast}$, we have
    $\frac{\mathbb{P} (\tmmathbf{y} \mid \pi^{\ast}, \tmmathbf{x},
    \tmmathbf{s}) \lambda^{| \pi^{\ast} |}}{\mathbb{P} (\tmmathbf{y} \mid
    \pi_0^{\ast}, \tmmathbf{x}, \tmmathbf{s}) \lambda^{| \pi_0^{\ast} |}}
    \leqslant \exp \{- c \log (\lambda^{- 1}) \}$. Thus, we write $\mathbb{P}
    (\pi^{\ast} \in \Pi_1^{\ast} \cup \Pi_2^{\ast} \cup \Pi_3^{\ast} \mid
    \mathfrak{D})$ as
    \begin{eqnarray*}
    & &\sum_{\pi^{\ast} \in \Pi_1^{\ast} \cup \Pi_2^{\ast} \cup \Pi_3^{\ast}}
      \mathbb{P} (\pi^{\ast} \mid \mathfrak{D})\\
       & \overset{\text{Equation }
      \left( \ref{EQ:posteriorratio} \right)}{\leqslant} & \sum_{\pi^{\ast}
      \in \Pi_1^{\ast} \cup \Pi_2^{\ast} \cup \Pi_3^{\ast}} c \exp \{- c \log
      (\lambda^{- 1}) + ck_0^{1 / 2} K \log (K) + c K^2 \log (K) \} \\
      & \overset{(i)}{\leqslant} & c \exp \{- c \log (\lambda^{- 1}) +
      ck_0^{1 / 2} K \log (K) + c K^2 \log (K) \} \\
      &\overset{\text{Assumption }
      \ref{AS:hyperpara}}{\leqslant}& c \exp \{- c \log (\lambda^{- 1}) \},
    \end{eqnarray*}
    where $(i)$ uses the fact that the maximum number of partitions is no
    larger than $K^{2 K^2}$. The proof of (\ref{EQ:posteriorconsistency1}) is
    completed.
    
    \tmtextbf{Proof of (\ref{EQ:posteriorconsistency2})}:
    
    Under events $\mathcal{A}_n \cap \mathcal{E}_n$, according to Proposition
    \ref{PP:likelihoodratio}, for $\pi^{\ast} \in \Pi_4^{\ast}$, we have
    $\frac{\mathbb{P} (\tmmathbf{y} \mid \pi^{\ast}, \tmmathbf{x},
    \tmmathbf{s}) \lambda^{| \pi^{\ast} |}}{\mathbb{P} (\tmmathbf{y} \mid
    \pi_0^{\ast}, \tmmathbf{x}, \tmmathbf{s}) \lambda^{| \pi_0^{\ast} |}}
    \leqslant \exp \{- ck_0^{- 2 / d} n \epsilon_1 (\pi^{\ast}, \pi_0^{\ast})
    r_n^2 \}$. Thus, we write $\mathbb{P} (\pi^{\ast} \in \Pi_4^{\ast} \mid \mathfrak{D})$ as
    \begin{eqnarray*}
       & \sum_{q =
      \lfloor k_0^{1 / 2 + 1 / d} K \log^{\alpha_0} (n) \rfloor}^{K^2}
      \sum_{\pi^{\ast} \in \Pi_{\epsilon, q}^{\ast}} \mathbb{P} (\pi^{\ast}
      \mid \mathfrak{D}) & \\
      \overset{(i)}{\leqslant} & \sum_{q = \lfloor k_0^{1 / 2 + 2 / d} K
      \log^{\alpha_0} (n) \rfloor}^{K^2} \sum_{\pi^{\ast} \in \Pi_{\epsilon,
      q}^{\ast}} c \exp \{- ck_0^{- 2 / d} n q K^{- 2} r_n^2 + ck_0^{1 / 2} K
      \log (K) + c k_0 \log (K) \}  & \\
      \overset{(ii)}{\leqslant} & \sum_{q = \lfloor k_0^{1 / 2 + 2 / d} K
      \log^{\alpha_0} (n) \rfloor}^{K^2} \sum_{\pi^{\ast} \in \Pi_{\epsilon,
      q}^{\ast}} c \exp \{- ck_0^{- 2 / d} n q K^{- 2} r_n^2 \}  & \\
      \overset{(iii)}{\leqslant} & \sum_{q = \lfloor k_0^{1 / 2 + 2 / d} K
      \log^{\alpha_0} (n) \rfloor}^{K^2} c \exp \{- ck_0^{- 2 / d} n q K^{- 2}
      r_n^2 + Cq \log (K) + q \log (k_0) \}  & \\
      \overset{(iv)}{\leqslant} & \sum_{q =
      \lfloor k_0^{1 / 2 + 2 / d} K \log^{\alpha_0} (n) \rfloor}^{K^2} c \exp
      \{- ck_0^{- 2 / d} n q K^{- 2} r_n^2 \}  & \\
      \overset{(v)}{\leqslant} & c \exp \{- cn k_0^{1 / 2} K^{- 1}
      \log^{\alpha_0} (n) r_n^2 \}, & 
    \end{eqnarray*}
    where $(i)$ uses \text{Equation } (\ref{EQ:posteriorratio}), $(ii)$ uses the fact that $\frac{k_0^{1 / 2} K \log (K) + k_0 \log
    (K)}{k_0^{- 2 / d} n q K^{- 2} r_n^2} = o (1)$ for $q \geqslant \lfloor
    k_0^{1 / 2 + 2 / d} K \log^{\alpha_0} (n) \rfloor$ from Assumption \ref{AS:hyperpara}, $(iii)$ uses the fact
    that $| \Pi_{\epsilon, q}^{\ast} | \leqslant C K^{2 q} k_0^q$, $(iv)$ uses \text{Assumption} \ref{AS:hyperpara}, and $(v)$
    uses the property of summation of a geometric sequence. The proof of
    (\ref{EQ:posteriorconsistency2}) is completed.
  \end{proof}
\end{lemma}

Following Lemma \ref{LM:probofpai}, we conclude that under events
$\mathcal{A}_n \cap \mathcal{E}_n$,
\begin{equation}
  \mathbb{P} (\pi^{\ast} \in \Pi_5^{\ast}) \geqslant 1 - c \exp \{- cn k_0^{1
  / 2} K^{- 1} \log^{\alpha_0} (n) r_n^2 \} .
  \label{EQ:posteriorconsistencytotal}
\end{equation}
We next derive some property for $\pi^{\ast} \in \Pi_5^{\ast}$.

\begin{lemma}
  \label{LM:ppofgoodpai}For $\forall \pi^{\ast} = \{\mathcal{D}^{\ast}_j \}_{j
  = 1}^{k_0} \in \Pi_5^{\ast}$, and under Assumptions \ref{AS:lengthofboundary} and \ref{AS:hyperpara}, there exists a uniform constant $c$, such that
  \begin{equation}
    \min_{1 \leqslant j \leqslant k_0} |\mathcal{D}^{\ast}_j \cap
    \mathcal{D}^{\ast}_{\mathcal{M}(\mathcal{D}^{\ast}_j), 0} | \geqslant c
    k_0^{- 1} . \label{EQ:minarea}
  \end{equation}
  Furthermore, for $l = 1, \ldots, k_0$, let
  
  \begin{gather}
    j_1 (l) = \tmop{argument}_{j \in \{1, \ldots, k_0 \}}
    \{\mathcal{M}(\mathcal{D}^{\ast}_j) = l\}, \text{and}  \label{EQ:j1l}\\
    j_2 (l) = \tmop{argmax}_{j \in \{1, \ldots, k_0 \}} |\mathcal{D}_j^{\ast}
    \cap \mathcal{D}^{\ast}_{l, 0} | .  \label{EQ:j2l}
  \end{gather}
  
  We conclude $j_1 (l) \equiv j_2 (l)$.
  
  \begin{proof}
    Firstly, from the definition of $\Pi_{\epsilon, q}^{\ast}$, we have
    $\epsilon_1 (\pi^{\ast}, \pi_0^{\ast}) \leqslant k_0^{1 / 2 + 2 / d} K
    \log^{\alpha_0} (n) \times K^{- 2} = k_0^{1 / 2 + 2 / d} K^{- 1}
    \log^{\alpha_0} (n)$ for $\forall \pi^{\ast} \in \Pi_5^{\ast}$. After
    simple algebra, we can rewrite $\epsilon_1 (\pi^{\ast}, \pi_0^{\ast})$
    defined in (\ref{EQ:epsilon1dis}) as
    \begin{eqnarray}
      \epsilon_1 (\pi^{\ast}, \pi_0^{\ast}) & = & |\mathcal{D}|^{- 1} \left\{
      \sum_{l = 1}^{k_0} |\mathcal{D}^{\ast}_{l, 0} | - \sum_{j = 1}^{k_0}
      |\mathcal{D}_j^{\ast} \cap
      \mathcal{D}^{\ast}_{\mathcal{M}(\mathcal{D}^{\ast}_j), 0} | \right\}
      \nonumber\\
      & = & |\mathcal{D}|^{- 1}  \left[ \sum_{j = 1}^{k_0}
      \{|\mathcal{D}^{\ast}_{\mathcal{M}(\mathcal{D}^{\ast}_j), 0} | -
      |\mathcal{D}_j^{\ast} \cap
      \mathcal{D}^{\ast}_{\mathcal{M}(\mathcal{D}^{\ast}_j), 0} |\} \right]
      \leqslant k_0^{1 / 2 + 2 / d} K^{- 1} \log^{\alpha_0} (n), 
      \label{EQ:e1bound}
    \end{eqnarray}
    where we use the fact that $| \pi^{\ast} | = k_0$ and
    $\{\mathcal{M}(\mathcal{D}^{\ast}_j)\}_{j = 1}^{| \pi^{\ast} |} = \{1,
    \ldots, k_0 \}$ (since $\pi^{\ast} \in \Pi_5^{\ast}$). On the other hand,
    since $\min_{l \in \{1, \ldots, k_0 \}} |\mathcal{D}^{\ast}_{l, 0} |
    \geqslant c k_0^{- 1}$, we conclude that $\min_{1 \leqslant j \leqslant
    k_0} |\mathcal{D}^{\ast}_j \cap
    \mathcal{D}^{\ast}_{\mathcal{M}(\mathcal{D}^{\ast}_j), 0} | \geqslant c
    k_0^{- 1}$ for some constant $c$ following (\ref{EQ:e1bound}) and
    Assumption \ref{AS:hyperpara}.
    
    Next, note that for a given $l$, if $j_1 (l) \neq j_2 (l)$, we have
    \[ \bigl| \mathcal{D}_{j_2 (l)}^{\ast} \cap
       \mathcal{D}^{\ast}_{\mathcal{M}(\mathcal{D}^{\ast}_{j_2 (l)}), 0}
       \bigr| \geqslant |\mathcal{D}_{j_2 (l)}^{\ast} \cap
       \mathcal{D}^{\ast}_{l, 0} | \geqslant |\mathcal{D}_{j_1 (l)}^{\ast}
       \cap \mathcal{D}^{\ast}_{l, 0} | \geqslant c k_0^{- 1}, \]
    and $\mathcal{M} (\mathcal{D}^{\ast}_{j_2 (l)}) \neq l$ since $\mathcal{M}
    (\mathcal{D}^{\ast}_{j_1 (l)}) = l$ and $j_1 (l) \neq j_2 (l)$. Thus,
    \[ \epsilon_1 (\pi^{\ast}, \pi_0^{\ast}) \geqslant |\mathcal{D}|^{- 1}
       \min \left\{ \bigl| \mathcal{D}_{j_2 (l)}^{\ast} \cap
       \mathcal{D}^{\ast}_{\mathcal{M}(\mathcal{D}^{\ast}_{j_2 (l)}), 0}
       \bigr|, |\mathcal{D}_{j_2 (l)}^{\ast} \cap \mathcal{D}^{\ast}_{l, 0} |
       \right\} \geqslant c k_0^{- 1}, \]
    which is contradictory to (\ref{EQ:e1bound}). We thus conclude $j_1 (l)
    \equiv j_2 (l)$. The lemma is proved.
  \end{proof}
\end{lemma}

We next give the proof of Theorem \ref{TH:clustererror}.

\begin{proof}
  \tmtextbf{Proof of (\ref{EQ:kconsistency}):} Following the definition of
  $\Pi^{\ast}_5$ and Equation (\ref{EQ:posteriorconsistencytotal}), together
  with the fact that $\mathbb{P} (\mathcal{A}_n \cap \mathcal{E}_n)
  \rightarrow 1$, we obtain (\ref{EQ:kconsistency}) immediately.
  
  \tmtextbf{Proof of (\ref{EQ:errorate0})}: Recall the decomposition
  (\ref{EQ:decomp}) of $\epsilon [\pi^{\ast}, \{\mathcal{D}_{l, 0} \}_{l =
  1}^{k_0}]$. Proposition \ref{PP:bestapproximation} has shown that $\epsilon
  [\pi_0^{\ast}, \{\mathcal{D}_{l, 0} \}_{l = 1}^{k_0}] \leqslant ck_0^{1 / 2}
  K^{- 1}$. We next show that under events $\mathcal{A}_n \cap \mathcal{E}_n$,
  for $\pi^{\ast} \in \Pi^{\ast}_5$, we have $\epsilon (\pi^{\ast},
  \pi_0^{\ast}) \leqslant c k_0^{1 / 2 + 2 / d} K^{- 1} \log^{\alpha_0} (n)$
  for some constant $c$, which leads to (\ref{EQ:errorate0}) together with the
  result in (\ref{EQ:posteriorconsistencytotal}).
  
  Firstly, for $\pi^{\ast} \in \Pi^{\ast}_5$, it is easy to see $\epsilon_1
  (\pi^{\ast}, \pi_0^{\ast}) \leqslant k_0^{1 / 2 + 2 / d} K \log^{\alpha_0}
  (n) \times K^{- 2} =  k_0^{1 / 2 + 2 / d} K^{- 1} \log^{\alpha_0}
  (n)$. It thus suffices to bound $\epsilon_2 (\pi^{\ast}, \pi_0^{\ast})$. We
  write
  \begin{eqnarray}
    \epsilon_2 (\pi^{\ast}, \pi_0^{\ast}) & = & |\mathcal{D}|^{- 1} \left[
    \sum_{l = 1}^{k_0} |\mathcal{D}^{\ast}_{l, 0} | - \sum_{l = 1}^{k_0}
    |\mathcal{D}_{j_2 (l)}^{\ast} \cap \mathcal{D}^{\ast}_{l, 0} | \right]
    \nonumber\\
    & \overset{(i)}{=} & |\mathcal{D}|^{- 1}
    \left[ \sum_{l = 1}^{k_0} |\mathcal{D}^{\ast}_{l, 0} | - \sum_{l =
    1}^{k_0} |\mathcal{D}_{j_1 (l)}^{\ast} \cap \mathcal{D}^{\ast}_{l, 0} |
    \right] = \epsilon_1 (\pi^{\ast}, \pi_0^{\ast}) \leqslant c k_0^{1 / 2 + 2
    / d} K^{- 1} \log^{\alpha_0} (n),  \label{EQ:e2bound}
  \end{eqnarray}
  where $(i)$ uses \text{Lemma} \ref{LM:ppofgoodpai}. Combining the above results, we conclude $\epsilon
  (\pi^{\ast}, \pi_0^{\ast}) \leqslant c k_0^{1 / 2 + 2 / d} K^{- 1}
  \log^{\alpha_0} (n)$ for some constant $c$ and Equation (\ref{EQ:errorate0})
  is proved.
  
  \tmtextbf{Proof of (\ref{EQ:errorrate}):} Denote $\pi_0$ as the partition of
  $\{\tmmathbf{s}_i \}_{i = 1}^n$ induced by $\pi_0^{\ast}$. Recall that we
  write $\{\mathcal{S}_{l, 0} \}_{l = 1}^{k_0}$ as the true partition of
  $\{\tmmathbf{s}_i \}_{i = 1}^n$. We decompose $\epsilon_n [\pi,
  \{\mathcal{S}_{l, 0} \}_{l = 1}^{k_0}] \leqslant \epsilon_n (\pi, \pi_0) +
  \epsilon_n [\pi_0, \{\mathcal{S}_{l, 0} \}_{l = 1}^{k_0}]$. Under
  $\mathcal{A}_n \cap \mathcal{E}_n$, it is easy to see that
  \[ \epsilon_n [\pi_0, \{\mathcal{S}_{l, 0} \}_{l = 1}^{k_0}] \leqslant
     ck_0^{1 / 2} K \times n K^{- 2} \times n^{- 1} = ck_0^{1 / 2} K^{- 1},
     \text{and} \]
  \[ \epsilon_n (\pi, \pi_0) \leqslant c n \epsilon (\pi^{\ast}, \pi_0^{\ast})
     \times n^{- 1} \overset{\text{Equation } \left( \ref{EQ:e2bound}
     \right)}{\leqslant} c k_0^{1 / 2 + 2 / d} K^{- 1} \log^{\alpha_0} (n) \]
  for $\pi^{\ast} \in \Pi^{\ast}_5$. (\ref{EQ:errorrate}) is thus proved
  following (\ref{EQ:posteriorconsistencytotal}).
\end{proof}

\subsection{Proof of Theorem \ref{TH:thetaconverge}}\label{SEC:TH2proof}

Note that Equation (\ref{EQ:TH21}) is obtained immediately from Equation
(\ref{EQ:posteriorconsistencytotal}). We thus focus on the proof of Equations
(\ref{EQ:TH22})-(\ref{EQ:TH23}) in this section. We first prove
(\ref{EQ:TH22}), based on which (\ref{EQ:TH24}) and (\ref{EQ:TH23}) can be
obtained.

\subsubsection{Proof of Equation (\ref{EQ:TH22})}

To begin with, write
\[ \Omega_n = \{\tmmathbf{\theta}=\{\tmmathbf{\theta}_j \}_{j = 1}^k : r_n^{-
   1} k^{- 1} \sum_{j = 1}^k \|\tmmathbf{\theta}_j
   -\tmmathbf{\theta}_{\mathcal{M}(\mathcal{D}^{\ast}_j), 0} \|_2 \leqslant
   M_n k_0^{1 / 2 + 2 / d} K^{- 1} \log^{\alpha_0} (n) \}, \]
for $M_n$ in Theorem \ref{TH:thetaconverge}. To prove (\ref{EQ:TH22}), it
suffices to show that with probability tending to $1$, we have
\begin{equation}
  \frac{\mathbb{P} (\tmmathbf{\theta} \in \Omega^c_n \mid
  \mathfrak{D})}{\mathbb{P} (\tmmathbf{\theta} \in \Omega_n \mid
  \mathfrak{D})} \rightarrow 0. \label{EQ:thetaratio}
\end{equation}
We write out the ratio above as
\begin{eqnarray}
  \frac{\mathbb{P} (\tmmathbf{\theta} \in \Omega^c_n \mid
  \mathfrak{D})}{\mathbb{P} (\tmmathbf{\theta} \in \Omega_n \mid
  \mathfrak{D})} & = & \frac{\sum_{\pi^{\ast} \in \Pi^{\ast}_5} \mathbb{P}
  (\tmmathbf{\theta} \in \Omega^c_n, \pi^{\ast} \mid \mathfrak{D}) +
  \sum_{\pi^{\ast} \in \Pi^{\ast c}_5} \mathbb{P} (\tmmathbf{\theta} \in
  \Omega^c_n, \pi^{\ast} \mid \mathfrak{D})}{\sum_{\pi^{\ast} \in
  \Pi^{\ast}_5} \mathbb{P} (\tmmathbf{\theta} \in \Omega_n, \pi^{\ast} \mid
  \mathfrak{D}) + \sum_{\pi^{\ast} \in \Pi^{\ast c}_5} \mathbb{P}
  (\tmmathbf{\theta} \in \Omega_n, \pi^{\ast} \mid \mathfrak{D})} \nonumber\\
  & \leqslant & \frac{\sum_{\pi^{\ast} \in \Pi^{\ast}_5} \mathbb{P}
  (\tmmathbf{\theta} \in \Omega^c_n, \pi^{\ast} \mid \mathfrak{D}) +\mathbb{P}
  (\pi^{\ast} \in \Pi^{\ast c}_5 \mid \mathfrak{D})}{\sum_{\pi^{\ast} \in
  \Pi^{\ast}_5} \mathbb{P} (\tmmathbf{\theta} \in \Omega_n, \pi^{\ast} \mid
  \mathfrak{D})} .  \label{EQ:thetaratio12}
\end{eqnarray}
In the proof of Theorem \ref{TH:clustererror}, we have already shown that
under events $\mathcal{A}_n \cap \mathcal{E}_n$, $\mathbb{P} (\pi^{\ast} \in
\Pi^{\ast c}_5 \mid \mathfrak{D}) \rightarrow 0$. To prove
(\ref{EQ:thetaratio}), it thus suffices to show that under $\mathcal{A}_n \cap
\mathcal{E}_n$, we have
\begin{equation}
  \sup_{\pi^{\ast} \in \Pi^{\ast}_5}  \frac{\mathbb{P} (\tmmathbf{\theta} \in
  \Omega^c_n, \pi^{\ast} \mid \mathfrak{D})}{\mathbb{P} (\tmmathbf{\theta} \in
  \Omega_n, \pi^{\ast} \mid \mathfrak{D})} \rightarrow 0.
  \label{EQ:thetaratio2}
\end{equation}
Recall that we use $\mathbb{P}_{\tmop{Gaussian}} (\cdummy ; \tmmathbf{\mu},
\tmmathbf{\Sigma})$ to denote the density function of the Gaussian variable
with mean $\tmmathbf{\mu}$ and covariance $\tmmathbf{\Sigma}$. For a partition
$\pi^{\ast}$ (and $\pi$ induced by $\pi^{\ast}$) we write $\tmmathbf{y}_j =
\{y (\tmmathbf{s}_i) : \tmmathbf{s}_i \in \mathcal{S}_j \}$ and
$\tmmathbf{x}_j$ as the $|\mathcal{S}_j | \times d$ design matrix for
covariates in the $j$-th cluster $\mathcal{S}_j \in \pi$. We have following
lemmas.

\begin{lemma}
  \label{LM:posthetadis}Given $\pi^{\ast} \in \Pi^{\ast}_5$ and under events
  $\mathcal{A}_n \cap \mathcal{E}_n$, we have
  \[ \frac{\mathbb{P} (\tmmathbf{\theta} \in \Omega^c_n, \pi^{\ast} \mid
     \mathfrak{D})}{\mathbb{P} (\tmmathbf{\theta} \in \Omega_n, \pi^{\ast}
     \mid \mathfrak{D})} = \frac{\int_{\tmmathbf{\theta} \in \Omega^c_n}
     \left\{ \prod_{j = 1}^{k_0} \mathbb{P}_{\tmop{Gaussian}}
     (\tmmathbf{\theta}_j ; \tmmathbf{\bar{\theta}}_j, \tmmathbf{\Sigma}_j)
     \right\} d\tmmathbf{\theta}}{1 - \int_{\tmmathbf{\theta} \in \Omega^c_n}
     \left\{ \prod_{j = 1}^{k_0} \mathbb{P}_{\tmop{Gaussian}}
     (\tmmathbf{\theta}_j ; \tmmathbf{\bar{\theta}}_j, \tmmathbf{\Sigma}_j)
     \right\} d\tmmathbf{\theta}}, \]
  where $\tmmathbf{\bar{\theta}}_j = \frac{n}{n + 1}  (\tmmathbf{x}_j^T
  \tmmathbf{x}_j)^{- 1} \tmmathbf{x}_j^T \tmmathbf{y}_j$, and
  $\tmmathbf{\Sigma}_j = \frac{n \sigma^2}{n + 1}  (\tmmathbf{x}_j^T
  \tmmathbf{x}_j)^{- 1}$.
  
  \begin{proof}
    By the Bayesian rule, we have
    \begin{equation}
      \frac{\mathbb{P} (\tmmathbf{\theta} \in \Omega^c_n, \pi^{\ast} \mid
      \mathfrak{D})}{\mathbb{P} (\tmmathbf{\theta} \in \Omega_n, \pi^{\ast}
      \mid \mathfrak{D})} = \frac{\int_{\tmmathbf{\theta} \in \Omega^c_n}
      \mathbb{P} (\tmmathbf{\theta} \mid \pi^{\ast}, \tmmathbf{x},
      \tmmathbf{s}) \mathbb{P} (\tmmathbf{y} \mid \tmmathbf{\theta},
      \pi^{\ast}, \tmmathbf{x}, \tmmathbf{s})
      d\tmmathbf{\theta}}{\int_{\tmmathbf{\theta} \in \Omega_n} \mathbb{P}
      (\tmmathbf{\theta} \mid \pi^{\ast}, \tmmathbf{x}, \tmmathbf{s})
      \mathbb{P} (\tmmathbf{y} \mid \tmmathbf{\theta}, \pi^{\ast},
      \tmmathbf{x}, \tmmathbf{s}) d\tmmathbf{\theta}} . \label{EQ:thetaratio3}
    \end{equation}
    Since $\pi^{\ast} \in \Pi^{\ast}_5$, we have $| \pi^{\ast} | = k_0$.
    Following priors (\ref{EQ:thetajdis}) - (\ref{EQ:yjmodel}), we have
    \[ \mathbb{P} (\tmmathbf{\theta} \mid \pi^{\ast}, \tmmathbf{x},
       \tmmathbf{s}) \mathbb{P} (\tmmathbf{y} \mid \tmmathbf{\theta},
       \pi^{\ast}, \tmmathbf{x}, \tmmathbf{s}) \propto \prod_{j = 1}^{k_0}
       \mathbb{P}_{\tmop{Gaussian}} (\tmmathbf{\theta}_j ;
       \tmmathbf{\bar{\theta}}_j, \tmmathbf{\Sigma}_j), \]
    where $\propto$ means we do not care about those constants independent of
    $\tmmathbf{\theta}$, since they are contained in both numerator and
    denominator of (\ref{EQ:thetaratio3}) and can be canceled out. Note that
    $\prod_{j = 1}^{k_0} \mathbb{P}_{\tmop{Gaussian}} (\tmmathbf{\theta}_j ;
    \tmmathbf{\bar{\theta}}_j, \tmmathbf{\Sigma}_j)$ is a probability density
    function of a multivariate Gaussian variable, we can thus write
    (\ref{EQ:thetaratio3}) as
    \begin{eqnarray*}
      \frac{\mathbb{P} (\tmmathbf{\theta} \in \Omega^c_n, \pi^{\ast} \mid
      \mathfrak{D})}{\mathbb{P} (\tmmathbf{\theta} \in \Omega_n, \pi^{\ast}
      \mid \mathfrak{D})} & = & \frac{\int_{\tmmathbf{\theta} \in \Omega^c_n}
      \left\{ \prod_{j = 1}^{k_0} \mathbb{P}_{\tmop{Gaussian}}
      (\tmmathbf{\theta}_j ; \tmmathbf{\bar{\theta}}_j, \tmmathbf{\Sigma}_j)
      \right\} d\tmmathbf{\theta}}{\int_{\tmmathbf{\theta} \in \Omega_n}
      \left\{ \prod_{j = 1}^{k_0} \mathbb{P}_{\tmop{Gaussian}}
      (\tmmathbf{\theta}_j ; \tmmathbf{\bar{\theta}}_j, \tmmathbf{\Sigma}_j)
      \right\} d\tmmathbf{\theta}},
    \end{eqnarray*}
    which proves the lemma.
  \end{proof}
\end{lemma}

The next lemma studies the property of $\tmmathbf{\bar{\theta}}_j
-\tmmathbf{\theta}_{\mathcal{M} (\mathcal{D}_j^{\ast}), 0}$.

\begin{lemma}
  \label{LM:thetan}Under $\mathcal{A}_n \cap \mathcal{E}_n$ and Assumption
  \ref{AS:covariates}, we have
  \[ \sup_{\pi^{\ast} \in \Pi^{\ast}_5}  \sum_{j = 1}^{k_0} \|
     \tmmathbf{\bar{\theta}}_j
     -\tmmathbf{\theta}_{\mathcal{M}(\mathcal{D}^{\ast}_j), 0} \|_2 \leqslant
     c M_n^{1 / 2} k_0^{3 / 2 + 2 / d} K^{- 1} \log^{\alpha_0} (n) r_n \]
  for some constant $c$.
  
  \begin{proof}
    For a given $\pi^{\ast} = \{\mathcal{D}^{\ast}_j \}_{j = 1}^{k_0} \in
    \Pi^{\ast}_5$, let $\tmmathbf{\mu}_{j, 0} =\mathbb{E} (\tmmathbf{y}_j \mid
    \tmmathbf{s}_j, \tmmathbf{x}_j)$ and $\tmmathbf{\epsilon}_j = \{\epsilon
    (\tmmathbf{s}_i)\}_{\tmmathbf{s}_i \in \mathcal{D}^{\ast}_j}$. We write
    \begin{eqnarray*}
      \sum_{j = 1}^{k_0} \| \tmmathbf{\bar{\theta}}_j
      -\tmmathbf{\theta}_{\mathcal{M}(\mathcal{D}_j^{\ast}), 0} \|_2 & = &
      \sum_{j = 1}^{k_0}  \left\| \frac{n}{n + 1} (\tmmathbf{x}_j^T
      \tmmathbf{x}_j)^{- 1} \tmmathbf{x}_j^T (\tmmathbf{\mu}_{j, 0}
      +\tmmathbf{\epsilon}_j)
      -\tmmathbf{\theta}_{\mathcal{M}(\mathcal{D}_j^{\ast}), 0} \right\|_2
      \leqslant I_1 + I_2 + I_3,
    \end{eqnarray*}
    where 
    \[ I_1 = \sum_{j = 1}^{k_0} \| (\tmmathbf{x}_j^T \tmmathbf{x}_j)^{- 1}
       \tmmathbf{x}_j^T (\tmmathbf{\mu}_{j, 0} -\tmmathbf{x}_j
       \tmmathbf{\theta}_{\mathcal{M}(\mathcal{D}_j^{\ast}), 0}) \|_2, \]
    \[ I_2 = \frac{1}{n + 1} \sum_{j = 1}^{k_0} \|
       \tmmathbf{\theta}_{\mathcal{M}(\mathcal{D}_j^{\ast}), 0} \|_2,
       \text{and } I_3 = \sum_{j = 1}^{k_0} \| (\tmmathbf{x}_j^T
       \tmmathbf{x}_j)^{- 1} \tmmathbf{x}_j^T \tmmathbf{\epsilon}_j \|_2 . \]
    For $I_1$, note that for $\pi$ induced by $\pi^{\ast} \in \Pi^{\ast}_5$,
    we have shown in the proof of Theorem \ref{TH:clustererror} that
    $\epsilon_n [\pi, \{\mathcal{S}_{l, 0} \}_{l = 1}^{k_0}] \leqslant c_3 
    k_0^{1 / 2 + 2 / d} K^{- 1} \log^{\alpha_0} (n)$, hence, the number of
    nonzero entries in $\{ \tmmathbf{\mu}_{j, 0} -\tmmathbf{x}_j
    \tmmathbf{\theta}_{\mathcal{M}(\mathcal{D}_j^{\ast}), 0} \}_{j = 1}^{k_0}$
    is no more than $c k_0^{1 / 2 + 2 / d} n K^{- 1} \log^{\alpha_0} (n)$, and
    $\| \tmmathbf{\mu}_{j, 0} -\tmmathbf{x}_j
    \tmmathbf{\theta}_{\mathcal{M}(\mathcal{D}_j^{\ast}), 0} \|_{\infty}
    \leqslant c r_n$. On the other hand, by event $\mathcal{A}_n$ and Equation
    (\ref{EQ:minarea}), we have $(\tmmathbf{x}_j^T \tmmathbf{x}_j)^{- 1}
    \leqslant cn^{- 1} k_0$. Together with the boundness assumption on
    $\tmmathbf{x} (\tmmathbf{s}_i)$ in Assumption \ref{AS:covariates}, we have
    \[ \sup_{\pi^{\ast} \in \Pi^{\ast}_5} I_1 \leqslant cn^{- 1} k_0 \times
       r_n \times k_0^{1 / 2 + 2 / d} n K^{- 1} \log^{\alpha_0} (n) = c k_0^{3
       / 2 + 2 / d} K^{- 1} \log^{\alpha_0} (n) r_n . \]
    It is easy to see that $\sup_{\pi^{\ast} \in \Pi^{\ast}_5} I_2 \leqslant
    cn^{- 1} k_0 r_n$. It thus remains to bound $I_3$.
    
    For $I_3$, we further decompose
    \[ I_3 \leqslant  \sum_{j = 1}^{k_0} \| (\tmmathbf{x}_j^T
       \tmmathbf{x}_j)^{- 1} \|_2 \| \tmmathbf{x}_j^T \tmmathbf{\epsilon}_j
       \|_2 \leqslant cn^{- 1} k_0 \sum_{j = 1}^{k_0} \| \tmmathbf{x}_j^T
       \tmmathbf{\epsilon}_j \|_2 \leqslant cn^{- 1} k_0 (I_{31} + I_{32} +
       I_{33}), \]
    where
    \[ I_{31} = \sum_{j = 1}^{k_0} \left\| \sum_{\tmmathbf{s}_i \in
       \mathcal{D}^{\ast}_{\mathcal{M}(\mathcal{D}^{\ast}_j), 0}}
       \tmmathbf{x}(\tmmathbf{s}_i) \epsilon (\tmmathbf{s}_i) \right\|_2, \]
    \[ I_{32} = \sum_{j = 1}^{k_0} \left\| \sum_{\tmmathbf{s}_i \in
       \mathcal{D}^{\ast}_j \setminus
       \mathcal{D}^{\ast}_{\mathcal{M}(\mathcal{D}^{\ast}_j), 0}}
       \tmmathbf{x}(\tmmathbf{s}_i) \epsilon (\tmmathbf{s}_i) \right\|_2,
       \text{ and } I_{33} = \sum_{j = 1}^{k_0} \left\| \sum_{\tmmathbf{s}_i
       \in \mathcal{D}^{\ast}_{\mathcal{M}(\mathcal{D}^{\ast}_j), 0} \setminus
       \mathcal{D}^{\ast}_j} \tmmathbf{x}(\tmmathbf{s}_i) \epsilon
       (\tmmathbf{s}_i) \right\|_2 . \]
    From $\mathcal{E}_{4 n}$ in (\ref{EQ:epsilon4n}), we have $\sup_{\pi^{\ast} \in \Pi^{\ast}_5}I_{31}
    \leqslant k_0 n^{1 / 2} \log^{\alpha_0} (n)$. Since $\pi^{\ast} \in
    \Pi^{\ast}_5$, the number of blocks in $\cup_{j=1}^{k_0}\{\mathcal{D}_j^{\ast} \setminus
    \mathcal{D}^{\ast}_{\mathcal{M} (\mathcal{D}^{\ast}_j), 0}\}$ is smaller
    than $\lfloor k_0^{1 / 2 + 2 / d} K \log^{\alpha_0} (n) \rfloor$.
    Combining the result given by $\mathcal{A}_n$ and $\mathcal{E}_{3 n}$, we
    have
    \[\sup_{\pi^{\ast} \in \Pi^{\ast}_5} I_{32} \leqslant   n^{1 / 2} K^{- 1} \log^{1 / 2} (K)
       \times k_0^{1 / 2 + 2 / d} K \log^{\alpha_0} (n)  \leqslant k_0^{1 /
       2 + 2 / d} n^{1 / 2} \log^{1 / 2 + \alpha_0} (n) . \]
    Similarly, it can be shown that $\sup_{\pi^{\ast} \in \Pi^{\ast}_5} I_{33}
    \leqslant k_0^{1 / 2 + 2 / d} n^{1 / 2} \log^{1 / 2 + \alpha_0} (n)$.
    Putting the result together and from Assumption \ref{AS:hyperpara}, the
    lemma is proved.
  \end{proof}
\end{lemma}

We next prove (\ref{EQ:TH22}).

\begin{proof}
  By the result of Lemmas \ref{LM:posthetadis} and \ref{LM:thetan}, to prove
  inequality (\ref{EQ:thetaratio2}), it suffices to show that under events
  $\mathcal{A}_n \cap \mathcal{E}_n$, $\sup_{\pi^{\ast} \in \Pi^{\ast}_5} 
  \int_{\tmmathbf{\theta} \in \Omega^c_n} \left\{ \prod_{j = 1}^{k_0}
  \mathbb{P}_{\tmop{Gaussian}} (\tmmathbf{\theta}_j ;
  \tmmathbf{\bar{\theta}}_j, \tmmathbf{\Sigma}_j) \right\} d\tmmathbf{\theta}
  \rightarrow 0$. Let $\{\tmmathbf{Z}_j \}_{j = 1}^{k_0}$ be independent
  random variables with probability density function
  $\mathbb{P}_{\tmop{Gaussian}} (\cdummy ; \tmmathbf{\bar{\theta}}_j,
  \tmmathbf{\Sigma}_j)$, we compute $\sup_{\pi^{\ast} \in \Pi^{\ast}_5} 
  \int_{\tmmathbf{\theta} \in \Omega^c_n} \left\{ \prod_{j = 1}^{k_0}
  \mathbb{P}_{\tmop{Gaussian}} (\tmmathbf{\theta}_j ;
  \tmmathbf{\bar{\theta}}_j, \tmmathbf{\Sigma}_j) \right\} d\tmmathbf{\theta}$
  as
  
  \begin{align*}
    & \sup_{\pi^{\ast} \in \Pi^{\ast}_5} \mathbb{P} \{ \sum_{j = 1}^{k_0}
    \|\tmmathbf{Z}_j -\tmmathbf{\theta}_{\mathcal{M}(\mathcal{D}_j^{\ast}), 0}
    \|_2 > M_n k_0^{3 / 2 + 2 / d} K^{- 1} \log^{\alpha_0} (n) r_n \}\\
    \overset{\text{Lemma } \ref{LM:thetan}}{\leqslant} & \sup_{\pi^{\ast} \in
    \Pi^{\ast}_5} \left[ \mathbb{P} \left\{ \sum_{j = 1}^{k_0}
    \|\tmmathbf{Z}_j - \tmmathbf{\bar{\theta}}_j \|_2 > \frac{M_n k_0^{3 / 2 +
    2 / d} K^{- 1} \log^{\alpha_0} (n) r_n}{2} \right\} \right]\\
    \leqslant & \sup_{\pi^{\ast} \in \Pi^{\ast}_5} \left[ \sum_{j = 1}^{k_0}
    \mathbb{P} \left\{ k_0 \|\tmmathbf{Z}_j - \tmmathbf{\bar{\theta}}_j \|_2 >
    \frac{M_n k_0^{3 / 2 + 2 / d} K^{- 1} \log^{\alpha_0} (n) r_n}{2} \right\}
    \right] .
  \end{align*}
  
  On the other hand, under $\mathcal{A}_n$, we have $\lambda_{\max}
  (\tmmathbf{\Sigma}_j) = \frac{n \sigma^2}{n + 1} \lambda_{\max}
  \{(\tmmathbf{x}_j^T \tmmathbf{x}_j)^{- 1} \} \leqslant \frac{cn \sigma^2}{(n
  + 1) |\mathcal{S}_j |} \leqslant c n^{- 1} k_0$ for some universal constant
  $c$ for all $j$. Applying Lemma \ref{LM:Gaussiantail}, we have
  
  \begin{align*}
    & \sum_{j = 1}^{k_0} \mathbb{P} \left\{ k_0 \|\tmmathbf{Z}_j -
    \tmmathbf{\bar{\theta}}_j \|_2 > \frac{M_n k_0^{3 / 2 + 2 / d} K^{- 1}
    \log^{\alpha_0} (n) r_n}{2} \right\}\\
    \leqslant & \sum_{j = 1}^{k_0} \frac{c K n^{- 1 / 2}}{M_n   k_0^{2 / d}
    \log^{\alpha_0} (n) r_n} \exp \left\{ - \frac{c M_n ^2 k_0^{4 / d} \log^{2
    \alpha_0} (n) r^2_n}{n^{- 1} K^2} \right\} .
  \end{align*}
  
  From Assumption \ref{AS:hyperpara}, the right-hand side of the inequality is
  bounded by a $\pi^{\ast}$-independent series going to $0$. Putting results
  together, (\ref{EQ:TH22}) is proved.
\end{proof}

\subsubsection{Proof of Equation (\ref{EQ:TH24})}

\begin{proof}
  Take the $M_n$ in $\Omega_n$ to be $(M_n')^{1 / 3}$, we write
  \[ \Omega_n' = \{\tmmathbf{\theta}=\{\tmmathbf{\theta}_j \}_{j = 1}^k :
     r_n^{- 1} k^{- 1} \sum_{j = 1}^k \|\tmmathbf{\theta}_j
     -\tmmathbf{\theta}_{\mathcal{M}(\mathcal{D}^{\ast}_j), 0} \|_2 \leqslant
     (M_n')^{1 / 3} k_0^{1 / 2 + 2 / d} K^{- 1} \log^{\alpha_0} (n) \} . \]
  Note that we have already shown that $\mathbb{P} (\pi^{\ast} \in
  \Pi^{\ast}_5 \mid \mathfrak{D}) \rightarrow 1, \text{and } \mathbb{P}
  (\tmmathbf{\theta} \in \Omega_n' \mid \mathfrak{D}) \rightarrow 1$. Together
  with Assumption \ref{AS:distributionloc}, it thus suffices to show that
  \[ \mathbb{P} \left\{ \int_{\mathcal{D}} \|\tmmathbf{\theta}(\mathbb{s})
     -\tmmathbf{\theta}_0 (\mathbb{s})\|_2^2 d\mathbb{s}> M_n' k_0^{1 / 2 + 2
     / d} K^{- 1} \log^{\alpha_0} (n) r_n^2 \mid \mathfrak{D}, \pi^{\ast} \in
     \Pi^{\ast}_5, \tmmathbf{\theta} \in \Omega'_n \right\} \rightarrow 0. \]
  Given a $\pi^{\ast} \in \Pi^{\ast}_5  \text{ and } \tmmathbf{\theta} \in
  \Omega'_n$, we next write (recall $j_1 (l)$ and $j_2 (l)$ in (\ref{EQ:j1l})
  - (\ref{EQ:j2l}))
  \begin{eqnarray*}
    \tmmathbf{\theta} (\mathbb{s}) -\tmmathbf{\theta}_0 (\mathbb{s}) & = &
    \sum_{j = 1}^{k_0} \tmmathbf{\theta}_j \mathbb{I} (\mathbb{s} \in
    \mathcal{D}^{\ast}_j) - \sum_{l = 1}^{k_0} \tmmathbf{\theta}_0
    (\mathbb{s}) \mathbb{I} (\mathbb{s} \in \mathcal{D}_{l, 0})\\
    & = & \sum_{l = 1}^{k_0} \{\tmmathbf{\theta}_{j_1 (l)}
    \mathbb{I}(\mathbb{s} \in \mathcal{D}^{\ast}_{j_1 (l)})
    -\tmmathbf{\theta}_{l, 0} \mathbb{I}(\mathbb{s} \in \mathcal{D}_{l, 0})\}
    = I_1 (\mathbb{s}) + I_2 (\mathbb{s}), \text{where}
  \end{eqnarray*}
  \[ I_1 (\mathbb{s}) = \sum_{l = 1}^{k_0} \{(\tmmathbf{\theta}_{j_1 (l)}
     -\tmmathbf{\theta}_{l, 0})\mathbb{I}(\mathbb{s} \in
     \mathcal{D}^{\ast}_{j_1 (l)})\}, \text{and } I_2 (\mathbb{s}) = \sum_{l =
     1}^{k_0} \tmmathbf{\theta}_{l, 0}  \{\mathbb{I}(\mathbb{s} \in
     \mathcal{D}^{\ast}_{j_1 (l)}) -\mathbb{I}(\mathbb{s} \in \mathcal{D}_{l,
     0})\} . \]
  Since $\tmmathbf{\theta} \in \Omega'_n$, we have
  \begin{eqnarray}
    \int_{\mathcal{D}} \| I_1 (\mathbb{s}) \|_2^2 d\mathbb{s} & = & \sum_{l = 1}^{k_0} \|
    \tmmathbf{\theta}_{j_1 (l)} -\tmmathbf{\theta}_{l, 0} \|_2^2  \int_{\mathcal{D}}
    \mathbb{I} (\mathbb{s} \in \mathcal{D}^{\ast}_{j_1 (l)}) d\mathbb{s}
    \leqslant \sum_{j = 1}^{k_0} c k_0^{- 1} \|\tmmathbf{\theta}_j
    -\tmmathbf{\theta}_{\mathcal{M}(\mathcal{D}^{\ast}_j), 0} \|_2^2
    \nonumber\\
    & \leqslant & c k_0^{- 1}  (M_n')^{2 / 3} k_0^{3 + 4 / d} K^{- 2}
    \log^{2 \alpha_0} (n) r_n^2 = c (M_n')^{2 / 3} k_0^{2 + 4 / d} K^{- 2}
    \log^{2 \alpha_0} (n) r_n^2 \label{EQ:I1snew} . 
  \end{eqnarray}
  On the other hand,
  \begin{eqnarray}
    \int_{\mathcal{D}} \|I_2 (\mathbb{s})\|_2^2 d\mathbb{s} &
    \overset{}{\leqslant} & r_n^2 \int_{\mathcal{D}} \left\{ \sum_{l =
    1}^{k_0}  | \mathbb{I}(\mathbb{s} \in \mathcal{D}^{\ast}_{j_1 (l)})
    -\mathbb{I}(\mathbb{s} \in \mathcal{D}_{l, 0}) | \right\}^2 d\mathbb{s}
    \nonumber\\
    & \overset{(i)}{\leqslant} & r_n^2 \int_{\mathcal{D}} \left\{ \sum_{l =
    1}^{k_0} 2 | \mathbb{I}(\mathbb{s} \in \mathcal{D}^{\ast}_{j_1 (l)})
    -\mathbb{I}(\mathbb{s} \in \mathcal{D}_{l, 0}) |^2 \right\} d\mathbb{s}
    \nonumber\\
    & = & 2 r_n^2 \sum_{l = 1}^{k_0} \{|\mathcal{D}^{\ast}_{j_1 (l)}
    \setminus \mathcal{D}_{l, 0} | + |\mathcal{D}_{l, 0} \setminus
    \mathcal{D}^{\ast}_{j_1 (l)} |\} \nonumber\\
    & = & 2 r_n^2 \sum_{l = 1}^{k_0} \{|\mathcal{D}^{\ast}_{j_1 (l)}
    \setminus \mathcal{D}_{l, 0} | + |\mathcal{D}_{l, 0} \setminus
    \mathcal{D}^{\ast}_{j_2 (l)} |\} \leqslant c k_0^{1 / 2 + 2 / d} K^{- 1}
    \log^{\alpha_0} (n) r_n^2 \label{EQ:I2snew}, 
  \end{eqnarray}
  where $(i)$ uses the fact that the summation $\sum_{l = 1}^{k_0}  |
  \mathbb{I}(\mathbb{s} \in \mathcal{D}^{\ast}_{j_1 (l)})
  -\mathbb{I}(\mathbb{s} \in \mathcal{D}_{l, 0}) |$ has at most 2 nonzero
  summands, and the last inequality is from (\ref{EQ:e2bound}). Combining
  result of (\ref{EQ:I1snew}) - (\ref{EQ:I2snew}), we conclude
  \begin{eqnarray*}
    & \mathbb{P} \left\{ \int_{\mathcal{D}} \|\tmmathbf{\theta}(\mathbb{s})
    -\tmmathbf{\theta}_0 (\mathbb{s})\|_2^2 d\mathbb{s}> M_n' k_0^{1 / 2 + 2 /
    d} K^{- 1} \log^{\alpha_0} (n) r_n^2 \mid \mathfrak{D}, \pi^{\ast} \in
    \Pi^{\ast}_5, \tmmathbf{\theta} \in \Omega'_n \right\} & \\
    \leqslant & \mathbb{P} \left\{ 2 \int_{\mathcal{D}} \|I_1
    (\mathbb{s})\|_2^2 d\mathbb{s}+ 2 \int_{\mathcal{D}} \|I_2
    (\mathbb{s})\|_2^2 d\mathbb{s}> M_n' k_0^{1 / 2 + 2 / d} K^{- 1}
    \log^{\alpha_0} (n) r_n^2 \mid \mathfrak{D}, \pi^{\ast} \in \Pi^{\ast}_5,
    \tmmathbf{\theta} \in \Omega'_n \right\}, & 
  \end{eqnarray*}
  which converges to $0$ according to (\ref{EQ:I1snew}) - (\ref{EQ:I2snew}).
  Equation (\ref{EQ:TH24}) is proved.
\end{proof}

\subsubsection{Proof of Equation (\ref{EQ:TH23})}

\begin{proof}
  Under Assumption \ref{AS:covariates}, $\tmmathbf{x} (\mathbb{s})$ is bounded.
  Together with Equation (\ref{EQ:TH24}), Equation (\ref{EQ:TH23}) is
  established immediately.
\end{proof}

\subsection{Preliminary lemmas}\label{SEC:pre}
\begin{lemma}
  \label{LM:Berstein}(Bernstein's inequality) Let $X_1, \ldots, X_n$ be
  independent zero-mean real-valued random variables and let $S_n = \sum_{i =
  1}^n X_i$. If there exists a constant $c > 0$, such that Cramer's condition
  \begin{equation}
    \mathbb{E} |X_i |^k \leqslant c^{k - 2} k!\mathbb{E}X_i^2 < \infty, i = 1,
    2, \ldots, n ; k = 3, 4, \ldots . \label{EQ:Cramer}
  \end{equation}
  holds, then
  \[ \mathbb{P} (|S_n | \geqslant t) \leqslant 2 \exp \left( - \frac{t^2}{4
     \sum_{i = 1}^n \mathbb{E}X_i^2 + 2 ct} \right), t > 0. \]
\end{lemma}

\begin{lemma}
  \label{LM:chisquarebound}(Lemma 1 of \cite{laurent2000adaptive}) Let
  $\chi^2_r$ be a chi-square distribution with a degree of freedom $r$. The
  following concentration inequalities hold for any $x>0$:
  \[ \mathbb{P} \left( \chi^2_r > r + 2 x + 2 \sqrt{rx} \right) \leq \exp (-
     x), \]
  and
  \[ \mathbb{P} \left( \chi^2_r < r - 2 \sqrt{rx} \right) \leq \exp (- x) . \]
\end{lemma}
From Lemma \ref{LM:chisquarebound}, we can see that for $x > r$, we have
\begin{equation}
  \mathbb{P} (\chi^2_r > 5 x) \leqslant \mathbb{P} \left( \chi^2_r > r + 2 x +
  2 \sqrt{rx} \right) \leq \exp (- x) . \label{EQ:chisqbound}
\end{equation}
\begin{lemma}
  \label{LM:binomialcoe}The binomial coefficient satisfies
  \[ \left( \frac{n}{k} \right)^k \leqslant \left( \begin{array}{c}
       n\\
       k
     \end{array} \right) \leqslant \left( \frac{en}{k} \right)^k \]
  and
  \[ \sum_{i = 0}^k \left( \begin{array}{c}
       n\\
       i
     \end{array} \right) \leqslant (n + 1)^k \]
  for k=0,...,n.
  \begin{proof}
      The proof is trivial and omitted.
  \end{proof}
\end{lemma}

\begin{lemma}
  \label{LM:chisquaredis}For a real symmetric matrix $\tmmathbf{A}$ satisfying
  $\tmmathbf{A}^2 =\tmmathbf{A}$ and a standard Gaussian random vector
  $\tmmathbf{e}$, we have
  \[ \tmmathbf{e}^T  \tmmathbf{A} \tmmathbf{e}
     \sim \chi_r^2, \]
  where $r$ is the number of positive eigenvalues of $\tmmathbf{A}$.
  
  \begin{proof}
    The proof is trivial by performing eigen-decomposition of $\tmmathbf{A}$
    and noting that the eigenvalues of $\tmmathbf{A}$ equal either $0$ or $1$.
  \end{proof}
\end{lemma}

\begin{lemma}
  \label{LM:Gaussiantail}Denote $\tmmathbf{Z}$ as a $d$-dimensional Gaussian
  random variable with mean $\tmmathbf{0}$ and positive definite covariance
  $\tmmathbf{\Sigma}$. For any $t > 0$, we have
  \[ \mathbb{P} (\|\tmmathbf{Z}\|_2 > t) \leqslant \sqrt{\frac{2}{\pi}} d^{3 /
     2} \lambda_{\max}^{1 / 2} (\tmmathbf{\Sigma}) t^{- 1} \exp \left\{ -
     \frac{t^2}{2 d \lambda_{\max} (\tmmathbf{\Sigma})} \right\} . \]
  \begin{proof}
    First, for a $1$-dimensional Gaussian variable $Z \sim \tmop{Gaussian} (0,
    1)$, we have
    \begin{eqnarray}
      \mathbb{P} (|Z| > t) & = & \mathbb{P} (Z > t) +\mathbb{P} (Z < t)
      \nonumber\\
      & = & \frac{2}{\sqrt{2 \pi}}  \int_t^{+ \infty} \exp \left\{ -
      \frac{1}{2} z^2 \right\} dz \leqslant \sqrt{\frac{2}{\pi}}  \int_t^{+
      \infty} \frac{z}{t} \exp \left\{ - \frac{1}{2} z^2 \right\} dz
      \nonumber\\
      & = & t^{- 1}  \sqrt{\frac{2}{\pi}} \exp \left( - \frac{t^2}{2} \right)
      .  \label{EQ:unigaussiantail}
    \end{eqnarray}
    Next, for $d \geqslant 1$, we write $\tilde{\tmmathbf{Z}}
    =\tmmathbf{\Sigma}^{- 1 / 2} \tmmathbf{Z}= (\tilde{Z}_1, \ldots,
    \tilde{Z}_d)^T$. It is easy to see that $\tilde{\tmmathbf{Z}} \sim
    \tmop{Gaussian} (\tmmathbf{0}, \tmmathbf{I}_d)$. Note that
    $\|\tmmathbf{Z}\|_2 = \|\tmmathbf{\Sigma}^{1 / 2}  \tilde{\tmmathbf{Z}}
    \|_2 \leqslant \lambda_{\max}^{1 / 2} (\tmmathbf{\Sigma}) \|
    \tilde{\tmmathbf{Z}} \|_2$, we thus have
    \begin{eqnarray*}
      \mathbb{P} (\|\tmmathbf{Z}\|_2 > t) & \leqslant & \mathbb{P} \{\|
      \tilde{\tmmathbf{Z}} \|_2 > \lambda_{\max}^{- 1 / 2} (\tmmathbf{\Sigma})
      t\}\\
      & \leqslant & \sum_{i = 1}^d \mathbb{P} \{| \tilde{Z}_i | > d^{- 1 / 2}
      \lambda_{\max}^{- 1 / 2} (\tmmathbf{\Sigma}) t\}\\
      & \overset{\text{Equation ( \ref{EQ:unigaussiantail})} }{\leqslant} & \sqrt{\frac{2}{\pi}} d^{3 / 2} \lambda_{\max}^{1 /
      2} (\tmmathbf{\Sigma}) t^{- 1} \exp \left\{ - \frac{t^2}{2 d
      \lambda_{\max} (\tmmathbf{\Sigma})} \right\} .
    \end{eqnarray*}
  \end{proof}
\end{lemma}

\subsection{Proof of Lemma \ref{LM:An}}\label{SEC:proofofAn}

To prove Lemma \ref{LM:An}, we will show that under Assumptions
\ref{AS:distributionloc}, \ref{AS:covariates} and \ref{AS:hyperpara},
$\mathbb{P} (\mathcal{A}_{1 n}) \rightarrow 1$, and $\mathbb{P}
(\mathcal{A}_{2 n}) \rightarrow 1$ for some constants $c, C > 0$. The proofs are
given by Lemmas \ref{LM:numberofpoints} and \ref{LM:covempirical},
respectively.

\begin{lemma}
  \label{LM:numberofpoints}Under Assumptions \ref{AS:distributionloc} and
  \ref{AS:hyperpara}, there exit positive constants $c$ and $C$, such that for
  $\mathcal{A}_{1 n}$ defined in (\ref{EQ:A1n}), we have $\mathbb{P}
  (\mathcal{A}_{1 n}) \rightarrow 1$.
  
  \begin{proof}
    We prove the last inequality in (\ref{EQ:A1n}), and the first inequality
    is similar. Given a block $B_m$, denote $\delta_{m, i}$ as the binary
    variable indicating whether $\tmmathbf{s}_i$ is within $B_m$ or not. We
    can see that
    \[ \|B_m \| = \sum_{i = 1}^n \delta_{m, i} . \]
    Under Assumption \ref{AS:distributionloc}, $\{\delta_{m, i} \}_{i = 1}^n$
    is a series of i.i.d. binary random variable with success probability $p_m
    = \int_{B_m} \mathbb{P}_{\mathcal{D}} (\tmmathbf{s}) d\tmmathbf{s}$. It is
    easy to see that $c = 1$ satisfies condition (\ref{EQ:Cramer}) in Lemma
    \ref{LM:Berstein} for random variables $\{\delta_{m, i} - p_m \}_{i =
    1}^n$, and $\mathbb{E} (\delta_{m, i} - p_m)^2 = p_m  (1 - p_m)$. Thus, by
    taking $t = a_1  \sqrt{np_m  (1 - p_m) \log (n)}$ in Lemma
    \ref{LM:Berstein}, where $a_1$ is a constant to be determined later, we
    have
    \begin{eqnarray*}
      \left. \mathbb{P} \{ |\|B_m \|- np_m | \geqslant a_1  \sqrt{np_m (1 -
      p_m) \log (n)} \right\} & \leqslant & 2 \exp \left\{ - \frac{a_1^2 \log
      (n)}{4 + 2 a_1  \sqrt{\frac{\log (n)}{np_m (1 - p_m)}}} \right\} .
    \end{eqnarray*}
    Next, from Assumptions \ref{AS:distributionloc} and \ref{AS:hyperpara}, we
    can see
    \[ \inf_{1 \leqslant m \leqslant K^2} p_m \sim \sup_{1 \leqslant m
       \leqslant K^2} p_m \sim K^{- 2}, \]
    leading to
    \[ \sup_{1 \leqslant m \leqslant K^2}  \sqrt{np_m  (1 - p_m) \log (n)}
       \sim \sqrt{n K^{- 2} \log (n)} \overset{\text{Assumption
       \ref{AS:hyperpara}}}{\ll} n K^{- 2} . \]
    Thus, for a given $a_1$, we can then find an $a_2$, such that $a_2 n K^{-
    2} - np_m > a_1  \sqrt{np_m  (1 - p_m) \log (n)}$, $m = 1, 2, \ldots,
    K^2$. Therefore,
    \begin{eqnarray*}
      \mathbb{P} \left\{ \frac{\max_{1 \leqslant m \leqslant K^2} \|B_m \|}{n
      K^{- 2}} \geqslant a_2 \right\} & \leqslant & \sum_{m = 1}^{K^2}
      \mathbb{P} \left\{ \frac{\|B_m \|}{n K^{- 2}} \geqslant a_2 \right\}\\
      & \leqslant & \sum_{m = 1}^{K^2} \mathbb{P} \{|\|B_m \|- np_m |
      \geqslant a_2 n K^{- 2} - np_m \}\\
      & \leqslant & \left. \sum_{m = 1}^{K^2} \mathbb{P} \{ |\|B_m \|- np_m |
      \geqslant a_1  \sqrt{np_m (1 - p_m) \log (n)} \right\}\\
      & \leqslant & 2 K^2 \exp \left\{ - \frac{a_1^2 \log (n)}{4 + 2 a_1 
      \sqrt{\frac{\log (n)}{np_m (1 - p_m)}}} \right\} .
    \end{eqnarray*}
    On the other hand, using the fact that $\frac{\log (n)}{np_m  (1 - p_m)} =
    o (1)$ (Assumption \ref{AS:hyperpara}), we can find a $a_1$, such that $2
    K^2 \exp \left\{ - \frac{a_1^2 \log (n)}{4 + 2 a_1  \sqrt{\frac{\log
    (n)}{np_m (1 - p_m)}}} \right\} < n^{- 1}$ holds when $n$ is large. Thus,
    the last inequality of (\ref{EQ:A1n}) is proved. The proof of the first
    inequality of (\ref{EQ:A1n}) is similar and is thus omitted.
  \end{proof}
\end{lemma}

Conditional on $\mathcal{A}_{1 n}$, we have the following result.

\begin{lemma}
  \label{LM:covempirical}Under Assumptions \ref{AS:distributionloc},
  \ref{AS:covariates} and \ref{AS:hyperpara}, there exists a constant $C > 0$,
  such that
  \[ \mathbb{P} \left\{ \sup_{1 \leqslant m \leqslant K^2} \left\|\frac{\sum_{\tmmathbf{s}_i \in B_m} [\tmmathbf{x}(\tmmathbf{s}_i)\tmmathbf{x}^T
     (\tmmathbf{s}_i) -\mathbb{E}\{\tmmathbf{x}(\tmmathbf{s}_i)\tmmathbf{x}^T
     (\tmmathbf{s}_i) \mid \tmmathbf{s}_i \}]}{\|B_m \|}
     \right\|_{\infty} \leqslant C \log^{-
     \alpha_1 / 2} (n) \mid \mathcal{A}_{1 n} \right\} \rightarrow 1. \]
  \begin{proof}
    Let $x_p (\tmmathbf{s}_i)$ be the $p$-th entry of $\tmmathbf{x}
    (\tmmathbf{s}_i)$. Since the dimension of $\tmmathbf{x} (\tmmathbf{s}_i)$
    is finite, it suffices to show that there exists $C_{pp'} > 0$, such that
    \[ \mathbb{P} \left\{ \sup_{1 \leqslant m \leqslant K^2} \left|\frac{ \sum_{\tmmathbf{s}_i \in B_m} [x_p (\tmmathbf{s}_i) x_{p'}
       (\tmmathbf{s}_i) -\mathbb{E}\{x_p (\tmmathbf{s}_i) x_{p'}
       (\tmmathbf{s}_i) \mid \tmmathbf{s}_i \}]}{\|B_m \|}
 \right| \leqslant C_{pp'} \log^{-
       \alpha_1 / 2} (n) \mid \mathcal{A}_{1 n} \right\} \rightarrow 1 \]
    for $\forall p, p' = 1, 2, \ldots, d$. Assumption \ref{AS:covariates}
    entails that Cramer's condition (\ref{EQ:Cramer}) holds for $x_p
    (\tmmathbf{s}_1) x_{p'} (\tmmathbf{s}_1)$, we can thus apply Lemma
    \ref{LM:Berstein} and obtain
    \begin{eqnarray*}
      & \mathbb{P} \left\{ \sup_{1 \leqslant m \leqslant K^2} \left|
      \frac{\sum_{\tmmathbf{s}_i \in B_m} [x_p (\tmmathbf{s}_i) x_{p'}
      (\tmmathbf{s}_i) -\mathbb{E}\{x_p (\tmmathbf{s}_i) x_{p'}
      (\tmmathbf{s}_i) \mid \tmmathbf{s}_i \}]}{\|B_m \|}
      \right | > C_{pp'} \log^{- \alpha_1 /
      2} (n) \mid \mathcal{A}_{1 n} \right\} & \\
      \leqslant & \sum_{m = 1}^{K^2} \mathbb{P} \left\{ \left|
      \frac{\sum_{\tmmathbf{s}_i \in B_m} [x_p (\tmmathbf{s}_i) x_{p'}
      (\tmmathbf{s}_i) -\mathbb{E}\{x_p (\tmmathbf{s}_i) x_{p'}
      (\tmmathbf{s}_i) \mid \tmmathbf{s}_i \}]}{ \|B_m \|} 
       \right| > C_{pp'} \log^{- \alpha_1 /
      2} (n) \mid \mathcal{A}_{1 n} \right\} & \\
      \overset{(i)}{\leqslant} & 2 \sum_{m =
      1}^{K^2} \exp \left[ - \frac{\|B_m \|C_{pp'}^2 \log^{- \alpha_1}
      (n)}{4\|B_m \|^{- 1}  \sum_{\tmmathbf{s}_i \in B_m} \mathbb{E}\{x^2_p
      (\tmmathbf{s}_i) x^2_{p'} (\tmmathbf{s}_i) \mid \tmmathbf{s}_i \}+ 2
      cC_{pp'} \log^{- \alpha_1 / 2} (n)} \right] & \\
      \overset{ \mathcal{A}_{1 n}}{\leqslant} & 2 K^2 \exp \left[
      - \frac{c n K^{- 2} C_{pp'}^2 \log^{- \alpha_1} (n)}{4\|B_m \|^{- 1} 
      \sum_{\tmmathbf{s}_i \in B_m} \mathbb{E}\{x^2_p (\tmmathbf{s}_i)
      x^2_{p'} (\tmmathbf{s}_i) \mid \tmmathbf{s}_i \}+ 2 cC_{pp'} \log^{-
      \alpha_1 / 2} (n)} \right], & 
    \end{eqnarray*}
    where $(i)$ uses \text{Lemma} \ref{LM:Berstein}. Since $n K^{- 2} \gg \log^{1 + \alpha_1} (n)$ according to Assumption
    \ref{AS:hyperpara}, we can find a $C_{pp'} > 0$, such that the the above
    inequality is smaller than $n^{- 1} = o (1)$. Iterating over $p$ and $p'$,
    the lemma is proved.
  \end{proof}
\end{lemma}

Together with Assumption \ref{AS:covariates}, an immediate result from Lemma
\ref{LM:covempirical} is that there exist constants $c, C > 0$, such that
$\mathbb{P} (\mathcal{A}_{2 n}) \rightarrow 1$, for $\mathcal{A}_{2 n}$
defined in (\ref{EQ:A2n}). Combining the results, Lemma \ref{LM:An} is proved.

\subsection{Proof of Lemma \ref{LM:En}}\label{SEC:proofofEn}

To prove Lemma \ref{LM:En}, we will show that under our model and Assumption
\ref{AS:hyperpara}, $\mathbb{P} (\mathcal{E}_{1 n}) \rightarrow 1$,
$\mathbb{P} (\mathcal{E}_{2 n}) \rightarrow 1$, $\mathbb{P} (\mathcal{E}_{3
n}) \rightarrow 1$ and $\mathbb{P} (\mathcal{E}_{4 n}) \rightarrow 1$, for
some constants $c, C > 0$. The proofs are given by Lemmas
\ref{LM:epsilonloosebnd}, \ref{LM:epsilon2n}, \ref{LM:epsilon3n} and
\ref{LM:epsilon4n}, respectively.

\begin{lemma}
  \label{LM:epsilonloosebnd}Under our model, there exists a constant $C > 0$,
  such that $\mathbb{P} (\mathcal{E}_{1 n}) \rightarrow 1$ for $\mathcal{E}_{1
  n}$ defined in (\ref{EQ:epsilon1n}).
  
  \begin{proof}
    It is easy to see that $\tmmathbf{\phi}_{\pi^{\ast}}$ is a real symmetric
    matrix with $\tmmathbf{\phi}_{\pi^{\ast}}^2
    =\tmmathbf{\phi}_{\pi^{\ast}}$, and the number of positive eigenvalues of
    $\tmmathbf{\phi}_{\pi^{\ast}}$ is smaller than $K^2 d$. Thus, by applying
    Lemma \ref{LM:chisquaredis}, $\sigma_0^{- 2} \tmmathbf{\epsilon}^T
    \tmmathbf{\phi}_{\pi^{\ast}} \tmmathbf{\epsilon}$ follows a chi-square
    distribution with degree of freedom no larger than $K^2 d$, where $\sigma_0^2$ is the true variance of $\epsilon(\tmmathbf{s}_i)$. Let $C_1$ be a
    constant to be determined later. Applying Equation (\ref{EQ:chisqbound}),
    we have
    \begin{eqnarray*}
    &&\mathbb{P} \{\sup_{\pi^{\ast} \in \Xi^{\ast}} \tmmathbf{\epsilon}^T
      \tmmathbf{\phi}_{\pi^{\ast}} \tmmathbf{\epsilon}> C_1 K^2 \log (n)\} \\
      &
      \leqslant & \sum_{\pi^{\ast} \in \tilde{\Xi}^{\ast}} \mathbb{P}
      \{\tmmathbf{\epsilon}^T \tmmathbf{\phi}_{\pi^{\ast}}
      \tmmathbf{\epsilon}> C_1 K^2 \log (n)\}\\
      & \leqslant & \sum_{\pi^{\ast} \in \tilde{\Xi}^{\ast}} \sum_{r =
      1}^{K^2 d} \mathbb{P} \{\chi_r^2 > C_1 \sigma_0^{- 2} K^2 \log (n)\}\\
      & \overset{\text{Equation } \left( \ref{EQ:chisqbound}
      \right)}{\leqslant} & | \tilde{\Xi}^{\ast} | K^2 d \exp \{- 5^{- 1} C_1
      \sigma_0^{- 2} K^2 \log (n)\}\\
      & \overset{(i)}{\leqslant} & \exp \{2 K^2 \log (K) + \log (K^2 d) -
      5^{- 1} C_1 \sigma_0^{- 2} K^2 \log (n)\},
    \end{eqnarray*}
    where $(i)$ uses the fact that $| \tilde{\Xi}^{\ast} | \leqslant K^{2
    K^2}$. We can see that we can find a large $C_1$, such that the above
    inequality goes to $0$. Setting the constant $C$ in $\mathcal{E}_{1 n}$ to
    be $C_1$, the lemma is proved.
  \end{proof}
\end{lemma}

\begin{lemma}
  \label{LM:epsilon2n}Under our model and Assumption \ref{AS:hyperpara}, there exists a constant $C > 0$,
  such that $\mathbb{P} (\mathcal{E}_{2 n}) \rightarrow 1$, for $\mathcal{E}_{2 n}$
  defined in (\ref{EQ:epsilon2n}).
  
  \begin{proof}
    It suffices to show that
    \begin{equation}
      \mathbb{P} [\tmmathbf{\epsilon}^T \tmmathbf{\phi}_{\pi^{\ast} \cap
      \pi^{\ast}_0} \tmmathbf{\epsilon} \leqslant C \epsilon_1 (\pi^{\ast},
      \pi_0^{\ast}) K^2 \log (n), \forall \pi^{\ast} \in \Pi_4^{\ast}]
      \rightarrow 1, \label{EQ:epsilon21}
    \end{equation}
    and
    \begin{equation}
      \mathbb{P} [\tmmathbf{\epsilon}^T \tmmathbf{\phi}_{\pi^{\ast}}
      \tmmathbf{\epsilon} \leqslant C \epsilon_1 (\pi^{\ast}, \pi_0^{\ast})
      K^2 \log (n), \forall \pi^{\ast} \in \Pi_4^{\ast}] \rightarrow 1.
      \label{EQ:epsilon22}
    \end{equation}
    We give proof of Equation (\ref{EQ:epsilon21}), and the proof of Equation
    (\ref{EQ:epsilon22}) is similar. Following similar arguments in Lemma
    \ref{LM:epsilonloosebnd}, $\sigma_0^{- 2} \tmmathbf{\epsilon}^T
    \tmmathbf{\phi}_{\pi^{\ast} \cap \pi^{\ast}_0} \tmmathbf{\epsilon}$
    follows a chi-square distribution with degree of freedom no larger than
    $k_0^2 d$. Let $C_1$ be a constant to be determined later. We have
    \begin{eqnarray*}
      & \mathbb{P} [\tmmathbf{\epsilon}^T \tmmathbf{\phi}_{\pi^{\ast} \cap
      \pi^{\ast}_0} \tmmathbf{\epsilon}> C_1 \epsilon_1 (\pi^{\ast},
      \pi_0^{\ast}) K^2 \log (n), \exists \pi^{\ast} \in \Pi_4^{\ast}] & \\
      \leqslant & \sum_{\pi^{\ast} \in \Pi_4^{\ast}} \mathbb{P}
      [\tmmathbf{\epsilon}^T \tmmathbf{\phi}_{\pi^{\ast} \cap \pi^{\ast}_0}
      \tmmathbf{\epsilon}> C_1 \epsilon_1 (\pi^{\ast}, \pi_0^{\ast}) K^2 \log
      (n)] & \\
      \leqslant & \sum_{q = \lfloor k_0^{1 / 2 + 2 / d} K \log^{\alpha_0} (n)
      \rfloor}^{K^2} \sum_{\pi^{\ast} \in \Pi_{\epsilon, q}^{\ast}} \mathbb{P}
      [\tmmathbf{\epsilon}^T \tmmathbf{\phi}_{\pi^{\ast} \cap \pi^{\ast}_0}
      \tmmathbf{\epsilon}> C_1 q \log
      (n)] & \\
      \overset{\text{Equation } \left( \ref{EQ:chisqbound} \right)}{\leqslant}
      & \sum_{q = \lfloor k_0^{1 / 2 + 2 / d} K \log^{\alpha_0} (n)
      \rfloor}^{K^2} \sum_{\pi^{\ast} \in \Pi_{\epsilon, q}^{\ast}} \exp \{-
      5^{- 1} \sigma_0^{- 2} C_1 q \log (n)\} & \\
      \overset{(i)}{\leqslant} & \sum_{q = \lfloor k_0^{1 / 2 + 2 / d} K
      \log^{\alpha_0} (n) \rfloor}^{K^2} \exp \{1 + 2 q \log (K) + q \log
      (k_0) - 5^{- 1} \sigma_0^{- 2} C_1 q \log (n)\}, & 
    \end{eqnarray*}
    where (i) uses the fact that
    \[ | \Pi_{\epsilon, q}^{\ast} | \leqslant \left( \begin{array}{l}
         K^2\\
         q
       \end{array} \right) k_0^q \overset{\text{Lemma }
       \ref{LM:binomialcoe}}{\leqslant} eK^{2 q} k_0^q . \]
    Note that Assumption \ref{AS:hyperpara} entails that $\log (K) = O \{\log
    (n)\}$. Thus, we can find a $C_1 > 100 \sigma_0^2 \log (K) \log^{- 1}
    (n)$, such that
    \begin{eqnarray*}
      & \mathbb{P} [\tmmathbf{\epsilon}^T \tmmathbf{\phi}_{\pi^{\ast} \cap
      \pi^{\ast}_0} \tmmathbf{\epsilon}> C_1 \epsilon_1 (\pi^{\ast},
      \pi_0^{\ast}) K^2 \log (n), \exists \pi^{\ast} \in \Pi_4^{\ast}] & \\
      \leqslant & \sum_{q = \lfloor k_0^{1 / 2 + 2 / d} K \log^{\alpha_0} (n)
      \rfloor}^{K^2} c \exp \{- 10 q \log (K) \} \rightarrow 0. & 
    \end{eqnarray*}
    Setting $C=C_1$, Equation (\ref{EQ:epsilon21}) is thus proved. The proof of Equation
    (\ref{EQ:epsilon22}) is omitted.
  \end{proof}
\end{lemma}

\begin{lemma}
  \label{LM:epsilon3n}Under our model and Assumptions \ref{AS:covariates} and
  \ref{AS:hyperpara}, there exists a constant $C > 0$, such that for
  $\mathcal{E}_{3 n}$ defined in (\ref{EQ:epsilon3n}), we have $\mathbb{P}
  (\mathcal{E}_{3 n}) \rightarrow 1$.
  
  \begin{proof}
    Let $x_p (\tmmathbf{s}_i)$ be the $p$-th entry of $\tmmathbf{x}
    (\tmmathbf{s}_i)$. Since the dimension of $\tmmathbf{x} (\tmmathbf{s}_i)$
    is finite, it suffices to show that
    \[ \mathbb{P} \left\{ \sup_{1 \leqslant m \leqslant K^2} \left|
       \sum_{\tmmathbf{s}_i \in B_m} x_p (\tmmathbf{s}_i) \epsilon
       (\tmmathbf{s}_i) \right| \leqslant C \sqrt{n K^{- 2} \log (K)} \right\}
       \rightarrow 1 \]
    Since we have shown that $\mathbb{P} (\mathcal{A}_n) \rightarrow 1$ in
    Lemma \ref{LM:An}, it suffices to show that
    \[ \mathbb{P} \left\{ \sup_{1 \leqslant m \leqslant K^2} \left|
       \sum_{\tmmathbf{s}_i \in B_m} x_p (\tmmathbf{s}_i) \epsilon
       (\tmmathbf{s}_i) \right| \leqslant C \sqrt{n K^{- 2} \log (K)} \mid
       \mathcal{A}_n \right\} \rightarrow 1. \]
    We write
    \begin{eqnarray*}
      &  & \mathbb{P} \left\{ \sup_{1 \leqslant m \leqslant K^2} \left|
      \sum_{\tmmathbf{s}_i \in B_m} x_p (\tmmathbf{s}_i) \epsilon
      (\tmmathbf{s}_i) \right| > C \sqrt{n K^{- 2} \log (K)} \mid
      \mathcal{A}_n \right\}\\
      & \leqslant & \sum_{m = 1}^{K^2} \mathbb{P} \left\{ \left|
      \sum_{\tmmathbf{s}_i \in B_m} x_p (\tmmathbf{s}_i) \epsilon
      (\tmmathbf{s}_i) \right| > C \sqrt{n K^{- 2} \log (K)} \mid
      \mathcal{A}_n \right\}\\
      & \overset{(i)}{\leqslant} & K^2 \times 2 \exp \left\{ - \frac{C^2 n
      K^{- 2} \log (K)}{cn K^{- 2} + 2 cC \sqrt{n K^{- 2} \log (K)}} \right\},
    \end{eqnarray*}
    where we apply Lemma \ref{LM:Berstein} in (i). Taking $C = 4 \sqrt{c}$,
    together with $\text{Assumption } (\ref{AS:hyperpara})$, we can see the
    right-hand side of the inequality goes to $0$. The lemma is proved.
  \end{proof}
\end{lemma}

\begin{lemma}
  \label{LM:epsilon4n}Under our model and Assumptions \ref{AS:covariates} and
  \ref{AS:hyperpara}, we have $\mathbb{P} (\mathcal{E}_{4 n}) \rightarrow 1$,
  for $\mathcal{E}_{4 n}$ defined in (\ref{EQ:epsilon4n}).
  
  \begin{proof}
    It suffices to show that for $p = 1, \ldots, d$,
    \[ \mathbb{P} \left\{ \sup_{1 \leqslant l \leqslant k_0} \left|
       \sum_{\tmmathbf{s}_i \in \mathcal{D}^{\ast}_{l, 0}} x_p
       (\tmmathbf{s}_i) \epsilon (\tmmathbf{s}_i) \right| > n^{1 / 2}
       \log^{\alpha_0} (n) \mid \mathcal{A}_n \right\} \rightarrow 0. \]
    The probability is bounded by
    \begin{eqnarray*}
      &  & \mathbb{P} \left\{ \sup_{1 \leqslant l \leqslant k_0} \left|
      \sum_{\tmmathbf{s}_i \in \mathcal{D}^{\ast}_{l, 0}} x_p (\tmmathbf{s}_i)
      \epsilon (\tmmathbf{s}_i) \right| > n^{1 / 2} \log^{\alpha_0} (n) \mid
      \mathcal{A}_n \right\}\\
      & \overset{\text{Lemma \ref{LM:Berstein}}}{\leqslant} & \sum_{l =
      1}^{k_0} 2 \exp \left\{ - \frac{n \log^{2 \alpha_0} (n)}{c \|
      \mathcal{D}^{\ast}_{l, 0} \| + c n^{1 / 2} \log^{\alpha_0} (n)}
      \right\}\\
      & = & \sum_{l = 1}^{k_0} 2\mathbb{I} \left\{ \| \mathcal{D}^{\ast}_{l,
      0} \| \leqslant \frac{n}{\log (n)} \right\} \exp \left\{ - \frac{n
      \log^{2 \alpha_0} (n)}{c \| \mathcal{D}^{\ast}_{l, 0} \| + c n^{1 / 2}
      \log^{\alpha_0} (n)} \right\}\\
      &&+ \sum_{l = 1}^{k_0} 2\mathbb{I} \left\{
      \| \mathcal{D}^{\ast}_{l, 0} \| > \frac{n}{\log (n)} \right\} \exp
      \left\{ - \frac{n \log^{2 \alpha_0} (n)}{c \| \mathcal{D}^{\ast}_{l, 0}
      \| + c n^{1 / 2} \log^{\alpha_0} (n)} \right\}\\
      & \leqslant & \sum_{l = 1}^{k_0} 2\mathbb{I} \left\{ \|
      \mathcal{D}^{\ast}_{l, 0} \| \leqslant \frac{n}{\log (n)} \right\} \exp
      \{ - c \log^{1 + 2 \alpha_0} (n) \}\\
      && + \sum_{l = 1}^{k_0} 2\mathbb{I}
      \left\{ \| \mathcal{D}^{\ast}_{l, 0} \| > \frac{n}{\log (n)} \right\}
      \exp \{ - c \log^{2 \alpha_0} (n) \}\\
      & \overset{(i)}{\leqslant} & n \exp \{ - c \log^{1 + 2 \alpha_0} (n)
      \} + \log (n) \times \exp \{ - c \log^{2 \alpha_0} (n) \} \rightarrow 0,
    \end{eqnarray*}
    where $(i)$ uses the fact that $k_0 \leqslant n$ from Assumption
    \ref{AS:hyperpara}, and the fact that the number of clusters with
    locations larger than $\frac{n}{\log (n)}$ is upper bounded by $\log (n)$.
    The lemma is proved.
  \end{proof}
\end{lemma}

Combining the results of Lemmas \ref{LM:epsilonloosebnd} - \ref{LM:epsilon4n},
we finish the proof of Lemma \ref{LM:En}.

\subsection{Proof of Proposition
\ref{PP:bestapproximation}}\label{SEC:proofofpp1}

Note that the left-hand side of Equation (\ref{EQ:bestpai}) is upper bounded by $K^2$. Thus, when $K \ll k_0^{1/2}$ or $K \sim k_0^{1/2}$, the proposition is proved immediately. In what follows, we consider the case when $K \gg k_0^{1/2}$.

Let $\mathbb{B}_l = \{ B_m : B_m \cap \partial \mathcal{D}_{l, 0} \neq
\emptyset \}$ be the set of blocks intersecting with $\partial \mathcal{D}_{l,
0}$, and $\mathbb{B}= \cup_{l = 1}^{k_0} \mathbb{B}_l$. We first bound $| \mathbb{B}_l |$ by the following lemma.

\begin{lemma}
  \label{LM:numberofboundaryblocks}Under Assumption \ref{AS:lengthofboundary} and with $K \gg k_0^{1/2}$, we have
  \begin{equation}
    \sup_{1 \leqslant l \leqslant k_0} | \mathbb{B}_l | \leqslant c k_0^{- 1 /
    2} K, \label{EQ:partialDbound}
  \end{equation}
  for some constant $c$. Consequently, we have $|\mathbb{B}| \leqslant \sum_{l =
  1}^{k_0} | \mathbb{B}_l | \leqslant c k_0^{1 / 2} K$.
  
  \begin{proof}
    From
    Equation (\ref{EQ:boundary1}) in Assumption \ref{AS:lengthofboundary}, it
    is easy to see
    \[  c k^{- 1 / 2}_0\leq c \inf_{1 \leqslant l \leqslant k_0} | \mathcal{D}_{l, 0} |^{1 / 2}
       {\leqslant \inf_{1 \leqslant l \leqslant k_0}}  | \partial
       \mathcal{D}_{l, 0} | {\leqslant \sup_{1 \leqslant l \leqslant k_0}}  |
       \partial \mathcal{D}_{l, 0} | \leqslant C \sup_{1 \leqslant l \leqslant
       k_0} | \mathcal{D}_{l, 0} |^{1 / 2} \leq  C k^{- 1 / 2}_0 , \]
    leading to $K^{- 1} {\ll \inf_{1 \leqslant
    l \leqslant k_0}}  | \partial \mathcal{D}_{l, 0} | \sim k_0^{- 1 / 2}$.
    Together with Equation (\ref{EQ:boundary3}), we have
    \[ \sup_{1 \leqslant l \leqslant k_0} N (\partial \mathcal{D}_{l, 0},
       K^{- 1}, \| \cdot \|_2) {\leqslant C K \sup_{1 \leqslant l \leqslant
       k_0}}  | \partial \mathcal{D}_{l, 0} | \leqslant C k^{- 1 / 2}_0 K. \]
    Since each block $B_m$ has a side length $K^{- 1}$, Equation
    (\ref{EQ:partialDbound}) is established. The lemma is proved.
  \end{proof}
\end{lemma}

With Lemma \ref{LM:numberofboundaryblocks}, we next prove Proposition
\ref{PP:bestapproximation}. Recall that in our model, the domain partition
$\pi^{\ast}$ is induced from the partition of blocks $\mathcal{V}= \{B_m \}_{m
= 1}^{K^2}$. In Section \ref{SEC:Model1}, the partition of blocks is induced
from the mesh grid graph $\mathcal{G}= \{\mathcal{V}, \mathcal{E}\}$. Thus, to
prove Proposition \ref{PP:bestapproximation}, it suffices to show that there
exists a contiguous partition of $\mathcal{V}$, say $\pi_0 (\mathcal{V}) =
\{\mathcal{V}_{1, 0}, \ldots, \mathcal{V}_{k_0, 0} \}$, such that
\begin{equation}
  \sum_{j = 1}^{k_0} |\{B_m : B_m \in \mathcal{V}_{j, 0}, B_m \subsetneq
  \mathcal{D}_{j, 0} \}| \leqslant c k_0 ^{1 / 2} K. \label{EQ:bestappV}
\end{equation}
The proof is given as follows.

\begin{proof}
  We conduct the proof by constructing $\pi_0 (\mathcal{V})$. For a block $B_m
  \in \mathbb{B}$, we write Neighbour($B_m$) as the set of blocks surrounding
  $B_m$ (see Figure \ref{FIG:neighbour} for an illustration). Define set
  \[ \mathbb{b}= \{B_m : \text{there exists a } B_{m'} \in \mathbb{B},
     \text{such that } B_m \in \text{Neighbour} (B_{m'})\} \cup \mathbb{B} \]
  as the set of blocks that are ``near'' $\mathbb{B}$. Denote
  $\cdot^c$ as the complement. We will first construct a partition $\pi^1 =
  \{\mathcal{V}^1_1, \ldots \mathcal{V}^1_{k_0} \}$ for blocks in
  $\mathbb{B}^c$. Then we filter $\pi^1$ to obtain a partition of a subset of
  blocks in $\mathbb{B}^c$, say $\pi^2 = \{\mathcal{V}^2_1, \ldots
  \mathcal{V}^2_{k_0} \}$. After filtration, we will show that each
  $\mathcal{V}^2_j$ is connected under the mesh grid $\mathcal{G}$. We then
  construct $\pi_0 (\mathcal{V})$ by extending $\pi^2$ to a partition of all
  blocks. Finally, we verify that $\pi_0 (\mathcal{V})$ is a contiguous
  partition and (\ref{EQ:bestappV}) is satisfied under $\pi_0 (\mathcal{V})$.
  
  \begin{figure}[h]
    \begin{center}
      \begin{tabular}{lll}
        $\ast$ & $\ast$ & $\ast$\\
        $\ast$ & $B_m$ & $\ast$\\
        $\ast$ & $\ast$ & $\ast$
      \end{tabular}
    \end{center}
    
    \
    \caption{\label{FIG:neighbour}Illustration of \text{Neighbour}($B_m$).
    Blocks in \text{Neighbour}($B_m$) are denoted by $\ast$.}
  \end{figure}
  
  To begin with, note that for a given block $B_m \in \mathbb{B}^c$, it is
  fully containd in some sub-domain of $\{\mathcal{D}_{l, 0} \}_{l =
  1}^{k_0}$. We construct $\pi^1 = \{\mathcal{V}^1_1, \ldots
  \mathcal{V}^1_{k_0} \}$ as
  \[ \mathcal{V}^1_j = \{B_m \in \mathbb{B}^c : B_m \subseteq \mathcal{D}_{j,
     0} \}, 1 \leqslant j \leqslant k_0 . \]
  For a given $j$ and two blocks $B_m, B_{m'} \in \{\mathcal{V}^1_j \cap
  \mathbb{b}^c \} \subseteq \{\mathcal{V}^1_j \cap \mathbb{B}^c \}$, through
  the definition of $\mathbb{b}^c$, we can see the center points of $B_m 
  \text{ and } B_{m'}$, say $\tmmathbf{c}_m$ and $\tmmathbf{c}_{m'}$, satisfy
  \[ d (\tmmathbf{c}_m, \partial
       \mathcal{D}_{j, 0}) \geqslant 1.5 K^{- 1}, \text{and } d
     (\tmmathbf{c}_{m'}, \partial
       \mathcal{D}_{j, 0}) \geqslant 1.5 K^{- 1} . \]
  According to Equation (\ref{EQ:pathdis}) in Assumption \ref{AS:lengthofboundary}, there exists a path
  $\mathcal{P} (\tmmathbf{c}_m, \tmmathbf{c}_{m'})$, such that
  \[ d \{\mathcal{P}(\tmmathbf{c}_m, \tmmathbf{c}_{m'}), \partial
       \mathcal{D}_{j, 0}\}
     \geqslant 1.5 K^{- 1} . \]
  Since each block is a $K^{- 1} \times K^{- 1}$ rectangle, the distance of
  any two points within one block is no larger than $\sqrt{2} K^{- 1}$. We
  thus conclude that for any block $B_{m''}$ intersecting with $\mathcal{P}
  (\tmmathbf{c}_m, \tmmathbf{c}_{m'})$, we have
  \[ B_{m''} \in \mathcal{V}_j^1 \cap \mathbb{B}^c . \]
  Hence, there exists a path from $\tmmathbf{c}_m  \text{ to }
  \tmmathbf{c}_{m'}$ and intersects with only blocks in $\mathcal{V}_j^1 \cap
  \mathbb{B}^c$. Together with the fact that $\{\mathcal{V}_j^1 \cap
  \mathbb{b}^c \} \subseteq \{\mathcal{V}_j^1 \cap \mathbb{B}^c \}$, we
  conclude that for each $1 \leqslant j \leqslant k_0$, there exists a
  connected component of $\mathcal{V}_j^1 \cap \mathbb{B}^c$, say
  $\mathcal{V}^2_j$, such that $\{\mathcal{V}_j^1 \cap \mathbb{b}^c \}
  \subseteq \mathcal{V}^2_j$. We then obtain our second partition $\pi^2 =
  \{\mathcal{V}^2_1, \ldots, \mathcal{V}^2_{k_0} \}$.
  
  Now that $\pi^2$ is a partition of subset of $\{B_m \}_{m = 1}^{K^2}$. We
  next expand $\mathcal{V}^2_j$ to $\mathcal{V}_{j, 0}$, such that
  $\mathcal{V}_{j, 0} \supseteq \mathcal{V}_j^2$, and $\{\mathcal{V}_{1, 0},
  \ldots, \mathcal{V}_{k_0, 0} \}$ is a partition of all the blocks $\{B_m
  \}_{m = 1}^{k_0}$. Since the mesh grid $\mathcal{G}$ and each
  $\mathcal{V}^2_j$ are connected, it is easy to see that there exists an
  expansion of $\{\mathcal{V}^2_1, \ldots, \mathcal{V}^2_{k_0} \}$, say
  $\{\mathcal{V}_{1, 0}, \ldots, \mathcal{V}_{k_0, 0} \}$, such that
  $\{\mathcal{V}_{1, 0}, \ldots, \mathcal{V}_{k_0, 0} \}$ is a contiguous
  partition. On the other hand, we have
  \begin{eqnarray*}
    \sum_{j = 1}^{k_0} |\{B_m : B_m \in \mathcal{V}_{j, 0}, B_m \subsetneq
    \mathcal{D}_{j, 0} \}| & \leqslant & | \cup_{j = 1}^{k_0}
    \{\mathcal{V}_{j, 0} \setminus \mathcal{V}_j^2 \} | \leqslant | \{ \cup_{j
    = 1}^{k_0} \mathcal{V}_j^2 \}^c |\\
    & \overset{}{\overset{(i)}{\leqslant}} & |\mathbb{b}| \overset{(i
    i)}{\leqslant} 9 |\mathbb{B}| \overset{(i i i)}{\leqslant} c k_0^{1 / 2}
    K,
  \end{eqnarray*}
  where $(i)$ uses the fact that $\{\mathcal{V}_j^1 \cap \mathbb{b}^c \}
  \subseteq \mathcal{V}^2_j$, $(i i)$ uses $| \text{Neighbour} (B_m) \cup B_m
  | \leqslant 9$, and ($i i i$) uses Lemma \ref{LM:numberofboundaryblocks}.
  Equation (\ref{EQ:bestappV}) is thus established and the proposition is
  proved.
\end{proof}

\subsection{Proof of Proposition \ref{PP:distance}}\label{SEC:distanceproof}

We give the proof for $\epsilon (\cdot, \cdot)$ in this section. The proof of
$\epsilon_n (\cdot, \cdot)$ is similar and thus omitted. Note that the proof
of non-negativity, identity of indiscernibles and symmetry axioms are trivial,
we thus only show the triangle inequality.

Recall that for two domain partition $\pi_1 (\mathcal{D})$ and $\pi_2
(\mathcal{D})$, we decompose $\epsilon \{ \pi_1 (\mathcal{D}), \pi_2
(\mathcal{D}) \} = \epsilon_1 \{ \pi_1 (\mathcal{D}), \pi_2 (\mathcal{D}) \} +
\epsilon_2 \{ \pi_1 (\mathcal{D}), \pi_2 (\mathcal{D}) \}$, where $\epsilon_1
\{ \pi_1 (\mathcal{D}), \pi_2 (\mathcal{D}) \}$ and $\epsilon_2 \{ \pi_1
(\mathcal{D}), \pi_2 (\mathcal{D}) \}$ are defined in (\ref{EQ:epsilon1dis})
and (\ref{EQ:epsilon2dis}), respectively. It is easy to see that $\epsilon_1
\{ \pi_1 (\mathcal{D}), \pi_2 (\mathcal{D}) \} = \epsilon_2 \{ \pi_2
(\mathcal{D}), \pi_1 (\mathcal{D}) \}$. Thus, to obtain the triangle
inequality for $\epsilon (\cdot, \cdot)$, it suffices to show the triangle
inequality for $\epsilon_1 (\cdot, \cdot)$. Before that, we first study some
properties of $\epsilon_1 (\cdot, \cdot)$.

\begin{lemma}
  \label{LM:triangleine0}For any domain partitions $\pi_1 (\mathcal{D}) =
  \{\mathcal{D}_{1 1}, \ldots, \mathcal{D}_{1 k_1} \}, \pi_2 (\mathcal{D}) =
  \{\mathcal{D}_{21}, \ldots, \mathcal{D}_{2 k_2} \}$ and $\pi_3 (\mathcal{D})
  = \{\mathcal{D}_{31}, \ldots, \mathcal{D}_{3 k_3} \}$, where $k_1, k_2$ and
  $k_3$ are the number of clusters in $\pi_1 (\mathcal{D})$, $\pi_2
  (\mathcal{D})$ and $\pi_3 (\mathcal{D})$, respectively, we have
  \begin{equation}
    \epsilon_1 \{ \pi_1 (\mathcal{D}), \pi_2 (\mathcal{D}) \} - \epsilon_1 \{
    \widetilde{\pi } (\mathcal{D}), \pi_2 (\mathcal{D}) \} \geqslant 0,
    \label{EQ:nestedmonotone}
  \end{equation}
  where $\widetilde{\pi } (\mathcal{D}) = \pi_1 (\mathcal{D}) \cap \pi_3
  (\mathcal{D}) = \{ \tilde{\mathcal{D}}_1, \ldots
  \tilde{\mathcal{D}}_{\tilde{k}} \}$, with $\tilde{k}$ denoting the corresponding number of clusters. Furthermore, we have
  \begin{equation}
    \epsilon_1 \{ \pi_1 (\mathcal{D}), \pi_3 (\mathcal{D}) \} - \epsilon_1 \{
    \widetilde{\pi } (\mathcal{D}), \pi_3 (\mathcal{D}) \} \geqslant
    \epsilon_1 \{ \pi_1 (\mathcal{D}), \pi_2 (\mathcal{D}) \} - \epsilon_1 \{
    \widetilde{\pi } (\mathcal{D}), \pi_2 (\mathcal{D}) \} .
    \label{EQ:etutaande}
  \end{equation}
  \begin{proof}
    Following (\ref{EQ:epsilon1dis}), we have
    \[ \epsilon_1 \{ \pi_1 (\mathcal{D}), \pi_2 (\mathcal{D}) \} = 1 - |
       \mathcal{D} |^{- 1} \sum_{j = 1}^{k_1} \max_{l \in \{1, \ldots, k_2 \}}
       | \mathcal{D}_{1 j} \cap \mathcal{D}_{2 l} |, \]
    and
    \[ \epsilon_1 \{ \widetilde{\pi } (\mathcal{D}), \pi_2 (\mathcal{D}) \} =
       1 - | \mathcal{D} |^{- 1} \sum_{j = 1}^{\tilde{k}} \max_{l \in \{1,
       \ldots, k_2 \}} | \tilde{\mathcal{D}}_j \cap \mathcal{D}_{2 l} | . \]
    Since $\tilde{\pi} (\mathcal{D})$ is nested in $\pi_1 (\mathcal{D})$, for
    each $j = 1, \ldots, k_1$, there exists an index set $\mathcal{I} (j)$,
    such that $\cup_{j' \in \mathcal{I} (j)} \tilde{\mathcal{D}}_{j'}
    =\mathcal{D}_{1 j}$. Thus, we write
\begin{eqnarray}
    &&\epsilon_1 \{ \pi_1 (\mathcal{D}), \pi_2 (\mathcal{D}) \} - \epsilon_1
      \{ \widetilde{\pi } (\mathcal{D}), \pi_2 (\mathcal{D}) \} \nonumber\\
     &=& |
      \mathcal{D} |^{- 1} \sum_{j = 1}^{k_1} \left[ \left\{ \sum_{j' \in
      \mathcal{I} (j)} \max_{l \in \{1, \ldots, k_2 \}} |
      \tilde{\mathcal{D}}_{j'} \cap \mathcal{D}_{2 l} | \right\} - \max_{l \in
      \{1, \ldots, k_2 \}} | \mathcal{D}_{1 j} \cap \mathcal{D}_{2 l} |
      \right],\label{EQ:elager00}
\end{eqnarray}
    % \begin{equation}
    %   \epsilon_1 \{ \pi_1 (\mathcal{D}), \pi_2 (\mathcal{D}) \} - \epsilon_1
    %   \{ \widetilde{\pi } (\mathcal{D}), \pi_2 (\mathcal{D}) \} = |
    %   \mathcal{D} |^{- 1} \sum_{j = 1}^{k_1} \left[ \left\{ \sum_{j' \in
    %   \mathcal{I} (j)} \max_{l \in \{1, \ldots, k_2 \}} |
    %   \tilde{\mathcal{D}}_{j'} \cap \mathcal{D}_{2 l} | \right\} - \max_{l \in
    %   \{1, \ldots, k_2 \}} | \mathcal{D}_{1 j} \cap \mathcal{D}_{2 l} |
    %   \right], \label{EQ:elager00}
    % \end{equation}
    from which it is easy to see that $\epsilon_1 \{ \pi_1 (\mathcal{D}), \pi_2 (\mathcal{D}) \} - \epsilon_1
      \{ \widetilde{\pi } (\mathcal{D}), \pi_2 (\mathcal{D}) \}$
    \begin{equation}
       \geqslant |
      \mathcal{D} |^{- 1} \sum_{j = 1}^{k_1} \left[ \left\{ \max_{l \in \{1,
      \ldots, k_2 \}} \sum_{j' \in \mathcal{I} (j)} | \tilde{\mathcal{D}}_{j'}
      \cap \mathcal{D}_{2 l} | \right\} - \max_{l \in \{1, \ldots, k_2 \}} |
      \mathcal{D}_{1 j} \cap \mathcal{D}_{2 l} | \right] = 0.
      \label{EQ:elager0}
    \end{equation}
    Equation (\ref{EQ:nestedmonotone}) is thus proved.
    
    Next, substituting $\pi_2 (\mathcal{D})$ in (\ref{EQ:elager00}) with
    $\pi_3 (\mathcal{D})$, we have
\begin{eqnarray}
    &&\epsilon_1 \{ \pi_1 (\mathcal{D}), \pi_3 (\mathcal{D}) \} - \epsilon_1
      \{ \widetilde{\pi } (\mathcal{D}), \pi_3 (\mathcal{D}) \} \nonumber \\
      &=&|
      \mathcal{D} |^{- 1} \sum_{j = 1}^{k_1} \left[ \left\{ \sum_{j' \in
      \mathcal{I} (j)} \max_{l \in \{1, \ldots, k_3 \}} |
      \tilde{\mathcal{D}}_{j'} \cap \mathcal{D}_{3 l} | \right\} - \max_{l \in
      \{1, \ldots, k_3 \}} | \mathcal{D}_{1 j} \cap \mathcal{D}_{3 l} |
      \right] \label{EQ:elarger02}
\end{eqnarray}
    % \begin{equation}
    %   \epsilon_1 \{ \pi_1 (\mathcal{D}), \pi_3 (\mathcal{D}) \} - \epsilon_1
    %   \{ \widetilde{\pi } (\mathcal{D}), \pi_3 (\mathcal{D}) \} = |
    %   \mathcal{D} |^{- 1} \sum_{j = 1}^{k_1} \left[ \left\{ \sum_{j' \in
    %   \mathcal{I} (j)} \max_{l \in \{1, \ldots, k_3 \}} |
    %   \tilde{\mathcal{D}}_{j'} \cap \mathcal{D}_{3 l} | \right\} - \max_{l \in
    %   \{1, \ldots, k_3 \}} | \mathcal{D}_{1 j} \cap \mathcal{D}_{3 l} |
    %   \right] \label{EQ:elarger02}
    % \end{equation}
    for the same $\mathcal{I} (j)$ as define above. Since $\widetilde{\pi }
    (\mathcal{D}) = \pi_1 (\mathcal{D}) \cap \pi_3 (\mathcal{D})$, we can see
    that $\max_{l \in \{1, \ldots, k_3 \}} | \tilde{\mathcal{D}}_{j'} \cap
    \mathcal{D}_{3 l} | = | \tilde{\mathcal{D}}_{j'} |$. Thus, we rewrite
    (\ref{EQ:elarger02}) as
    \begin{equation}
      \epsilon_1 \{ \pi_1 (\mathcal{D}), \pi_3 (\mathcal{D}) \} - \epsilon_1
      \{ \widetilde{\pi } (\mathcal{D}), \pi_3 (\mathcal{D}) \} = |
      \mathcal{D} |^{- 1} \sum_{j = 1}^{k_1} \left[ \left\{ \sum_{j' \in
      \mathcal{I} (j)} | \tilde{\mathcal{D}}_{j'} | \right\} - \max_{l \in
      \{1, \ldots, k_3 \}} | \mathcal{D}_{1 j} \cap \mathcal{D}_{3 l} |
      \right] . \label{EQ:elarger03}
    \end{equation}
    On the other hand, let $\tilde{j}' (j) = \tmop{argmax}_{j' \in \mathcal{I}
    (j)} | \tilde{\mathcal{D}}_{j'} |$, we can see that $\max_{l \in \{1,
    \ldots, k_3 \}} | \mathcal{D}_{1 j} \cap \mathcal{D}_{3 l} | = |
    \tilde{\mathcal{D}}_{\tilde{j}' (j)} |$. Thus, we rewrite
    (\ref{EQ:elarger03}) and (\ref{EQ:elager00}) as
    \[ \epsilon_1 \{ \pi_1 (\mathcal{D}), \pi_3 (\mathcal{D}) \} - \epsilon_1
       \{ \widetilde{\pi } (\mathcal{D}), \pi_3 (\mathcal{D}) \} = |
       \mathcal{D} |^{- 1} \sum_{j = 1}^{k_1} \left[ \left\{ \sum_{j' \in
       \mathcal{I} (j) \backslash \tilde{j}' (j)} | \tilde{\mathcal{D}}_{j'} |
       \right\} \right], \]
    and
    \begin{eqnarray*}
    &&\epsilon_1 \{ \pi_1 (\mathcal{D}), \pi_2 (\mathcal{D}) \} - \epsilon_1
      \{ \widetilde{\pi } (\mathcal{D}), \pi_2 (\mathcal{D}) \}\\
       & = & |
      \mathcal{D} |^{- 1} \sum_{j = 1}^{k_1} \left[ \left\{ \sum_{j' \in
      \mathcal{I} (j) \backslash \tilde{j}' (j)} \max_{l \in \{1, \ldots, k_2
      \}} | \tilde{\mathcal{D}}_{j'} \cap \mathcal{D}_{2 l} | \right\}
      \right]\\
      &  & + | \mathcal{D} |^{- 1} \sum_{j = 1}^{k_1} \{ \max_{l \in \{1,
      \ldots, k_2 \}} | \tilde{\mathcal{D}}_{ \tilde{j}' (j) l} \cap
      \mathcal{D}_{2 l} | - \max_{l \in \{1, \ldots, k_2 \}} | \mathcal{D}_{1
      j} \cap \mathcal{D}_{2 l} | \}\\
      & \overset{(i)}{\leqslant} & | \mathcal{D} |^{- 1} \sum_{j = 1}^{k_1}
      \left[ \left\{ \sum_{j' \in \mathcal{I} (j) \backslash \tilde{j}' (j)}
      \max_{l \in \{1, \ldots, k_2 \}} | \tilde{\mathcal{D}}_{j'} \cap
      \mathcal{D}_{2 l} | \right\} \right],
    \end{eqnarray*}
    where $(i)$ is because $\tilde{\mathcal{D}}_{\tilde{j}' (j)} \subseteq
    \mathcal{D}_{1 j}$. Note that $| \tilde{\mathcal{D}}_{j'} | \geqslant
    \max_{l \in \{1, \ldots, k_2 \}} | \tilde{\mathcal{D}}_{j'} \cap
    \mathcal{D}_{2 l} |$, Equation (\ref{EQ:etutaande}) is proved.
  \end{proof}
\end{lemma}

Making use of Lemma \ref{LM:triangleine0}, we prove the triangle inequality
for $\epsilon_1 (\cdot, \cdot)$ as follows.

\begin{proof}
  For any domain partitions $\pi_1 (\mathcal{D})$, $\pi_2 (\mathcal{D})$ and
  $\pi_3 (\mathcal{D})$, let $\widetilde{\pi } (\mathcal{D}) = \pi_1
  (\mathcal{D}) \cap \pi_3 (\mathcal{D})$. We have
  \begin{eqnarray*}
    \epsilon_1 \{ \pi_1 (\mathcal{D}), \pi_3 (\mathcal{D}) \} & = & \epsilon_1
    \{ \pi_1 (\mathcal{D}), \pi_3 (\mathcal{D}) \} - \epsilon_1 \{
    \widetilde{\pi } (\mathcal{D}), \pi_3 (\mathcal{D}) \}\\
    & \overset{\text{Equation } \left( \ref{EQ:etutaande} \right)}{\geqslant}
    & \epsilon_1 \{ \pi_1 (\mathcal{D}), \pi_2 (\mathcal{D}) \} - \epsilon_1
    \{ \widetilde{\pi } (\mathcal{D}), \pi_2 (\mathcal{D}) \}\\
    & \overset{(i)}{\geqslant} & \epsilon_1 \{ \pi_1 (\mathcal{D}), \pi_2
    (\mathcal{D}) \} - \epsilon_1 \left\{ {\pi_3}  (\mathcal{D}), \pi_2
    (\mathcal{D}) \right\},
  \end{eqnarray*}
  where $(i)$ is due to $\epsilon_1 \{ \widetilde{\pi } (\mathcal{D}), \pi_2
  (\mathcal{D}) \} \leqslant \epsilon_1 \left\{ {\pi_3}  (\mathcal{D}), \pi_2
  (\mathcal{D}) \right\}$ from $\tmop{Equation} \left( \ref{EQ:nestedmonotone}
  \right)$. The triangle inequality for $\epsilon_1 (\cdot, \cdot)$ is thus
  proved.
\end{proof}

Combining the result above, the proposition is proved.

\subsection{Proof of Proposition \ref{PP:likelihoodratio}}\label{SEC:likelihood}

This section gives proof of Proposition \ref{PP:likelihoodratio}. We first
give some technical lemmas. Write $\tmmathbf{\mu}_0 =\mathbb{E} (\tmmathbf{y}
\mid \tmmathbf{s}, \tmmathbf{x})$ as the true regression mean.

\begin{lemma}
  \label{LM:bestapp}Under the event $\mathcal{A}_n$ and Assumptions
  \ref{AS:lengthofboundary}, \ref{AS:covariates} and \ref{AS:hyperpara}, there exists a constant
  $c$, such that
  \[ \|(\tmmathbf{\phi}_{\pi^{\ast}_0} -\tmmathbf{I}_n)\tmmathbf{\mu}_0 \|^2_2
     \leqslant c n k_0^{1 / 2} K^{- 1} r_n^2 . \]
  \begin{proof}
    Recall the definition of $\tmmathbf{P}_{\pi^{\ast}}$ under
    (\ref{EQ:formlikelihood}), let $\tmmathbf{X}_{\pi_0^{\ast}}
    =\tmmathbf{P}_{\pi^{\ast}_0}^T \tmop{diag} (\tmmathbf{x}_1, \ldots,
    \tmmathbf{x}_{k_0})$. We can see that
    \[ \tmmathbf{\phi}_{\pi^{\ast}_0} \tmmathbf{\mu}_0
       =\tmmathbf{X}_{\pi_0^{\ast}}  \hat{\tmmathbf{\theta}}, \]
    where
    \[ \hat{\tmmathbf{\theta}} = \tmop{argmin}_{\tmmathbf{\theta} \in
       \mathbb{R}^{k_0 d}} \|\tmmathbf{X}_{\pi_0^{\ast}}
       \tmmathbf{\theta}-\tmmathbf{\mu}_0 \|^2_2 . \]
    From (\ref{EQ:bestpai}), for $\tmmathbf{\theta}_0 =
    (\tmmathbf{\theta}^T_{1, 0}, \ldots, \tmmathbf{\theta}^T_{k_0, 0})^T$,
    where $\{\tmmathbf{\theta}_{l, 0} \}_{l = 1}^{k_0}$ are the true values of
    $\{\tmmathbf{\theta}_l \}_{l = 1}^{k_0}$, we can see
    \[ \|\tmmathbf{X}_{\pi_0^{\ast}} \tmmathbf{\theta}_0 -\tmmathbf{\mu}_0
       \|^2_2 \leqslant C n K^{- 2} \times k_0^{1 / 2} K \times r_n^2 = C n
       k_0^{1 / 2} K^{- 1} r_n^2 \]
    under $\mathcal{A}_n$. Therefore, we conclude
    \[ \|(\tmmathbf{\phi}_{\pi^{\ast}_0} -\tmmathbf{I}_n)\tmmathbf{\mu}_0
       \|^2_2 \leqslant \|\tmmathbf{X}_{\pi_0^{\ast}} \tmmathbf{\theta}_0
       -\tmmathbf{\mu}_0 \|^2_2 \leqslant  C n k_0^{1 / 2} K^{- 1} r_n^2, \]
    which proves the lemma.
  \end{proof}
\end{lemma}

Recall that we write $\pi_0^{\ast} = \{\mathcal{D}^{\ast}_{l, 0} \}_{l =
1}^{k_0}$ and $\mathcal{M} (\mathcal{D}^{\ast}_j) = \tmop{argmax}_{l \in \{1,
\ldots, k_0 \}} |\mathcal{D}^{\ast}_j \cap \mathcal{D}^{\ast}_{l, 0} |$ as the
index of the sub-domain in $\{\mathcal{D}^{\ast}_{l, 0} \}_{l = 1}^{k_0}$ with
the largest intersection area with $\mathcal{D}^{\ast}_j$. After simple
algebra, it can be shown that $\epsilon_1 (\pi^{\ast}, \pi_0^{\ast})$ defined
in (\ref{EQ:epsilon1dis}) can be re-written as
\begin{equation}
 \epsilon_1 (\pi^{\ast}, \pi_0^{\ast}) = \sum_{j = 1}^k \left\{ \sum_{l \in
   \{1, \ldots, k_0 \} \setminus \mathcal{M}(\mathcal{D}^{\ast}_j)}
   |\mathcal{D}^{\ast}_j \cap \mathcal{D}^{\ast}_{l, 0} | \right\} \label{EQ:anothere1}
\end{equation}
where we use the fact $|\mathcal{D}| = 1$ since $\mathcal{D}= [0, 1]^2$.
Recall $\alpha_0$ in Theorem \ref{TH:clustererror}. The next
lemma gives a lower bound of $\|(\tmmathbf{I}_n
-\tmmathbf{\phi}_{\pi^{\ast}})\tmmathbf{\mu}_0 \|^2_2$.

\begin{lemma}
  \label{LM:underfitted}Under the event $\mathcal{A}_n$ and Assumption
  \ref{AS:thetagap}, there exists a uniform constant $c > 0$, such that
  \[ \|(\tmmathbf{I}_n -\tmmathbf{\phi}_{\pi^{\ast}})\tmmathbf{\mu}_0 \|^2_2
     \geqslant c k_0^{- 2 / d} r_n^2 n \epsilon_1 (\pi^{\ast}, \pi_0^{\ast})
  \]
  holds for all $\pi^{\ast} \in \Xi^{\ast}$ with $\epsilon_1 (\pi^{\ast},
  \pi_0^{\ast}) K^2 \geqslant \lfloor k_0^{1 / 2} K \log^{\alpha_0} (n)
  \rfloor$.
  
  \begin{proof}
    Let $\bar{\mathcal{D}}^{\ast}_{l, 0} = \{B_m : B_m \in
    \mathcal{D}^{\ast}_{l, 0}, B_m \subseteq \mathcal{D}_{l, 0} \}$. Given a
    partition $\pi^{\ast} = \{ \mathcal{D}^{\ast}_1, \ldots,
    \mathcal{D}^{\ast}_k \}$, let $n_{j l}$ and $\bar{n}_{j l}$ be the number
    of locations within $\mathcal{D}^{\ast}_j \cap \mathcal{D}^{\ast}_{l, 0}$
    and $\mathcal{D}^{\ast}_j \cap \bar{\mathcal{D}}^{\ast}_{l, 0}$,
    respectively. Write $\overline{\tmmathbf{X}}_{j l}$ be the $\bar{n}_{j l}
    \times d$ design matrix generated by the locations within
    $\mathcal{D}^{\ast}_j \cap \bar{\mathcal{D}}^{\ast}_{l, 0}$. We can see
\begin{eqnarray*}
  \|(\tmmathbf{I}_n -\tmmathbf{\phi}_{\pi^{\ast}})\tmmathbf{\mu}_0 \|^2_2&\geqslant& 
  \sum_{j = 1}^k \sum_{l = 1}^{k_0} \|
       \overline{\tmmathbf{X}}_{j l} (\tmmathbf{\theta}_{l, 0} -
       \tmmathbf{\theta}_j) \|_2^2\\
       &\geqslant& \sum_{j = 1}^k \sum_{l = 1}^{k_0}
       (\tmmathbf{\theta}_{l, 0} - \tmmathbf{\theta}_j)^T
       \overline{\tmmathbf{X}}_{j l}^T \overline{\tmmathbf{X}}_{j l}
       (\tmmathbf{\theta}_{l, 0} - \tmmathbf{\theta}_j)
       \overset{\mathcal{A}_n}{\geqslant} \sum_{j = 1}^k \sum_{l = 1}^{k_0} c
       \bar{n}_{j l} \| \tmmathbf{\theta}_{l, 0} - \tmmathbf{\theta}_j \|_2^2,
\end{eqnarray*}   
    % \[ \|(\tmmathbf{I}_n -\tmmathbf{\phi}_{\pi^{\ast}})\tmmathbf{\mu}_0 \|^2_2
    %    \geqslant \sum_{j = 1}^k \sum_{l = 1}^{k_0} \|
    %    \overline{\tmmathbf{X}}_{j l} (\tmmathbf{\theta}_{l, 0} -
    %    \tmmathbf{\theta}_j) \|_2^2 \geqslant \sum_{j = 1}^k \sum_{l = 1}^{k_0}
    %    (\tmmathbf{\theta}_{l, 0} - \tmmathbf{\theta}_j)^T
    %    \overline{\tmmathbf{X}}_{j l}^T \overline{\tmmathbf{X}}_{j l}
    %    (\tmmathbf{\theta}_{l, 0} - \tmmathbf{\theta}_j)
    %    \overset{\mathcal{A}_n}{\geqslant} \sum_{j = 1}^k \sum_{l = 1}^{k_0} c
    %    \bar{n}_{j l} \| \tmmathbf{\theta}_{l, 0} - \tmmathbf{\theta}_j \|_2^2,
    % \]
    where $\tmmathbf{\theta}_j$ is the $j$-th regression coefficent computed
    from $\tmmathbf{\phi}_{\pi^{\ast}} \tmmathbf{\mu}_0$. For a given $j$ and
    $\tmmathbf{\theta}_j$, write $l_1 (j)$ and $l_2 (j)$ be the index of the
    minimum two values of $\{ \| \tmmathbf{\theta}_{l, 0} -
    \tmmathbf{\theta}_j \|_2 \}_{l = 1}^{k_0}$, and we assume $\bar{n}_{j l_1
    (j)} \leqslant \bar{n}_{j l_2 (j)}$ without loss of generality. For any $l
    \notin \{ l_1 (j), l_2 (j) \}$, we have
    \[ \| \tmmathbf{\theta}_{l, 0} - \tmmathbf{\theta}_j \|_2 \geqslant
       \frac{\| \tmmathbf{\theta}_{l_1 (j), 0} - \tmmathbf{\theta}_j \|_2 + \|
       \tmmathbf{\theta}_{l_2 (j), 0} - \tmmathbf{\theta}_j \|_2}{2} \geqslant
       \frac{\| \tmmathbf{\theta}_{l_1 (j), 0} - \tmmathbf{\theta}_{l_2 (j),
       0} \|_2}{2} . \]
    Hence, we lower bound $\sum_{l = 1}^{k_0}  \bar{n}_{j l} \| \tmmathbf{\theta}_{l, 0} -
      \tmmathbf{\theta}_j \|_2^2$ by
    \begin{eqnarray}
       & & \bar{n}_{j l_1 (j)} \|
      \tmmathbf{\theta}_{l_1 (j), 0} - \tmmathbf{\theta}_j \|_2^2 + \bar{n}_{j
      l_2 (j)} \| \tmmathbf{\theta}_{l_2 (j), 0} - \tmmathbf{\theta}_j \|_2^2
      + \sum_{l \notin \{ l_1 (j), l_2 (j) \}}  \frac{\bar{n}_{j l}}{4} \|
      \tmmathbf{\theta}_{l_1 (j), 0} - \tmmathbf{\theta}_{l_2 (j), 0} \|_2^2
      \nonumber\\
      & \geqslant & \bar{n}_{j l_1 (j)} (\| \tmmathbf{\theta}_{l_1 (j), 0} -
      \tmmathbf{\theta}_j \|_2^2 + \| \tmmathbf{\theta}_{l_2 (j), 0} -
      \tmmathbf{\theta}_j \|_2^2) + \| \tmmathbf{\theta}_{l_1 (j), 0} -
      \tmmathbf{\theta}_{l_2 (j), 0} \|_2^2 \sum_{l \notin \{ l_1 (j), l_2 (j)
      \}}  \frac{\bar{n}_{j l}}{4} \nonumber\\
      & \geqslant & \frac{\bar{n}_{j l_1 (j)}}{2} \| \tmmathbf{\theta}_{l_1
      (j), 0} - \tmmathbf{\theta}_{l_2 (j), 0} \|_2^2 + \|
      \tmmathbf{\theta}_{l_1 (j), 0} - \tmmathbf{\theta}_{l_2 (j), 0} \|_2^2
      \sum_{l \notin \{ l_1 (j), l_2 (j) \}}  \frac{\bar{n}_{j l}}{4}
      \nonumber\\
      & = & 4^{- 1} \| \tmmathbf{\theta}_{l_1 (j), 0} -
      \tmmathbf{\theta}_{l_2 (j), 0} \|_2^2 \left( \bar{n}_{j l_1 (j)} +
      \sum_{l \neq l_2 (j)} \bar{n}_{j l} \right) \overset{\text{Assumption
      \ref{AS:thetagap}}}{\geqslant} c k_0^{- 2 / d} r_n^2 \left( \bar{n}_{j
      l_1 (j)} + \sum_{l \neq l_2 (j)} \bar{n}_{j l} \right) \nonumber\\
      &  & \quad\quad\quad \quad \geqslant c k_0^{- 2 / d} r_n^2  \left( \sum_{l = 1}^{k_0}
      \bar{n}_{j l} - \max_{1 \leqslant l \leqslant k_0} \bar{n}_{j l} \right)
      . \label{EQ:underfitted1} 
    \end{eqnarray}
    Next, note that from Proposition \ref{PP:bestapproximation} and
    $\mathcal{A}_n$, we have \
    \begin{align}
      &\left| \left( \sum_{j = 1}^k \sum_{l = 1}^{k_0} n_{j l} - \sum_{j = 1}^k
      \max_{1 \leqslant l \leqslant k_0} n_{j l} \right) - \left( \sum_{j =
      1}^k \sum_{l = 1}^{k_0} \bar{n}_{j l} - \sum_{j = 1}^k \max_{1 \leqslant
      l \leqslant k_0} \bar{n}_{j l} \right) \right| \nonumber\\   
      \leqslant & c k_0^{1 / 2} K
      \times n K^{- 2} = c n k_0^{1 / 2} K^{- 1} . \label{EQ:underfitted2}
    \end{align}
    On the other hand, let $l^{\ast} (j) = \tmop{argmax}_{1 \leqslant l
    \leqslant k_0} n_{j l}$. We have
    \begin{eqnarray}
       \sum_{j = 1}^k \sum_{l = 1}^{k_0} n_{j l} - \sum_{j = 1}^k
      \max_{1 \leqslant l \leqslant k_0} n_{j l}  & = &  \sum_{j
      = 1}^k \sum_{l \in \{ 1, \ldots, k_0 \} \backslash l^{\ast} (j)} n_{j l}  \nonumber\\
      & \overset{\mathcal{A}_n}{\geqslant} & c n \times \left( \sum_{j = 1}^k
      \sum_{l \in \{ 1, \ldots, k_0 \} \backslash l^{\ast} (j)}
      |\mathcal{D}^{\ast}_j \cap \mathcal{D}^{\ast}_{l, 0} | \right) \nonumber\\
      &\overset{\text{Equation \ref{EQ:anothere1}}}{\geqslant}& c n \epsilon_1 (\pi^{\ast},
      \pi_0^{\ast})
       \label{EQ:underfitted3} 
    \end{eqnarray}
    Together (\ref{EQ:underfitted1}) - (\ref{EQ:underfitted3}) and the fact
    that $n \epsilon_1 (\pi^{\ast}, \pi_0^{\ast}) \geqslant c n k_0^{1 / 2}
    K^{- 1} \log^{\alpha_0} (n)$, the lemma is proved.
  \end{proof}
\end{lemma}
Next, under events $\mathcal{A}_n \cap \mathcal{E}_n$, we give the proof of
Proposition \ref{PP:likelihoodratio} as follows.

\begin{proof}
  We first consider cases when $\pi^{\ast} \in \Pi_1^{\ast} \cup \Pi_2^{\ast}
  \cup \Pi_3^{\ast}$. Following Equation (\ref{EQ:formlikelihood}), we have
  \begin{equation}
    \frac{\mathbb{P} (\tmmathbf{y} \mid \pi^{\ast}, \tmmathbf{x},
    \tmmathbf{s}) \lambda^{| \pi^{\ast} |}}{\mathbb{P} (\tmmathbf{y} \mid
    \pi^{\ast}_0, \tmmathbf{x}, \tmmathbf{s}) \lambda^{| \pi_0^{\ast} |}} =
    \exp \left[ (| \pi^{\ast}_0 | - | \pi^{\ast} |) \left\{ \log (\lambda^{-
    1}) + \frac{d \log (n + 1^{\nosymbol})}{2} \right\} +
    \frac{n\tmmathbf{y}^T (\tmmathbf{\phi}_{\pi^{\ast}}
    -\tmmathbf{\phi}_{\pi_0^{\ast}})\tmmathbf{y}}{2 \sigma^2 (n + 1)} 
    \right]. \label{EQ:Likelihoodr}
  \end{equation}
  Write $\tilde{\pi}^{\ast} = \pi^{\ast} \cap \pi^{\ast}_0$, we bound
  $\tmmathbf{y}^T  (\tmmathbf{\phi}_{\pi^{\ast}}
  -\tmmathbf{\phi}_{\pi_0^{\ast}}) \tmmathbf{y}$ by
  \begin{eqnarray}
    \tmmathbf{y}^T  (\tmmathbf{\phi}_{\pi^{\ast}}
    -\tmmathbf{\phi}_{\pi_0^{\ast}}) \tmmathbf{y} & = & \tmmathbf{y}^T 
    (\tmmathbf{\phi}_{\tilde{\pi}^{\ast}} -\tmmathbf{\phi}_{\pi^{\ast}_0})
    \tmmathbf{y}+\tmmathbf{y}^T  (\tmmathbf{\phi}_{\pi^{\ast}}
    -\tmmathbf{\phi}_{\tilde{\pi}^{\ast}}) \tmmathbf{y} \nonumber\\
    & \overset{(i)}{=} & \|(\tmmathbf{\phi}_{\tilde{\pi}^{\ast}}
    -\tmmathbf{\phi}_{\pi^{\ast}_0})\tmmathbf{y}\|^2_2 -
    \|(\tmmathbf{\phi}_{\tilde{\pi}^{\ast}}
    -\tmmathbf{\phi}_{\pi^{\ast}})\tmmathbf{y}\|^2_2 \nonumber\\
    & \leqslant & 2\| (\tmmathbf{\phi}_{\tilde{\pi}^{\ast}}
    -\tmmathbf{\phi}_{\pi^{\ast}_0}) \tmmathbf{\mu}_0  \|^2_2 + 2\| 
    (\tmmathbf{\phi}_{\tilde{\pi}^{\ast}} -\tmmathbf{\phi}_{\pi^{\ast}_0})
    \tmmathbf{\epsilon}\|^2_2 - \|(\tmmathbf{\phi}_{\tilde{\pi}^{\ast}}
    -\tmmathbf{\phi}_{\pi^{\ast}})\tmmathbf{y}\|^2_2 \nonumber\\
    & \overset{(ii)}{\leqslant} & 8\| (\tmmathbf{I}_n
    -\tmmathbf{\phi}_{\pi^{\ast}_0}) \tmmathbf{\mu}_0  \|^2_2 + 4\|
    \tmmathbf{\phi}_{\tilde{\pi}^{\ast}} \tmmathbf{\epsilon} \|^2 + 4\|
    \tmmathbf{\phi}_{\pi^{\ast}_0} \tmmathbf{\epsilon}\|^2_2 -
    \|(\tmmathbf{\phi}_{\tilde{\pi}^{\ast}}
    -\tmmathbf{\phi}_{\pi^{\ast}})\tmmathbf{y}\|^2_2 \nonumber\\
    & \overset{(iii)}{\leqslant} & c n k_0^{1 / 2} K^{- 1} r_n^2 + CK^2 \log
    (n) - \|(\tmmathbf{\phi}_{\tilde{\pi}^{\ast}}
    -\tmmathbf{\phi}_{\pi^{\ast}})\tmmathbf{y}\|^2_2, \label{EQ:Rsquaredif} 
  \end{eqnarray}
  where $(i)$ uses the fact that $(\tmmathbf{\phi}_{\tilde{\pi}^{\ast}}
  -\tmmathbf{\phi}_{\pi^{\ast}_0})^2 =\tmmathbf{\phi}_{\tilde{\pi}^{\ast}}
  -\tmmathbf{\phi}_{\pi_0^{\ast}}$ and $(\tmmathbf{\phi}_{\tilde{\pi}^{\ast}}
  -\tmmathbf{\phi}_{\pi^{\ast}})^2 =\tmmathbf{\phi}_{\tilde{\pi}^{\ast}}
  -\tmmathbf{\phi}_{\pi^{\ast}}$ since $\tilde{\pi}^{\ast}$ in nested in
  $\pi^{\ast}$ and $\pi^{\ast}_0$, $(ii)$ is because
  \[ \|(\tmmathbf{\phi}_{\tilde{\pi}^{\ast}}
     -\tmmathbf{\phi}_{\pi^{\ast}_0})\tmmathbf{\mu}_0 \|^2_2 \leqslant 2
     \|(\tmmathbf{\phi}_{\tilde{\pi}^{\ast}} -\tmmathbf{I}_n)\tmmathbf{\mu}_0
     \|^2_2 + 2\|(\tmmathbf{I}_n
     -\tmmathbf{\phi}_{\pi^{\ast}_0})\tmmathbf{\mu}_0 \|^2_2 \leqslant 4
     \|(\tmmathbf{I}_n -\tmmathbf{\phi}_{\pi^{\ast}_0})\tmmathbf{\mu}_0
     \|^2_2, \]
  and $(iii)$ uses result from Lemma \ref{LM:bestapp} and $\mathcal{E}_{1 n}$.
  We next discuss $\pi^{\ast} \in \Pi_1^{\ast}, \Pi_2^{\ast},  
  \Pi_3^{\ast}$ and $\Pi_4^{\ast}$, respectively.
  
  \noindent \tmtextbf{Cases when} $\pi^{\ast} \in \Pi_1^{\ast}$:
  
  Since each block has an area $K^{- 2}$, together with Assumption
  \ref{AS:lengthofboundary}, we can see for each $D_{l, 0}, l = 1, \ldots,
  k_0$, the number of blocks intersecting with it is of the order $k_0^{- 1}
  K^2$. Combining with the proof of Proposition \ref{PP:bestapproximation} and Assumption \ref{AS:hyperpara}, we
  further conclude that the number of blocks within each $D^{\ast}_{l, 0}, l =
  1, \ldots, k_0$ is larger than $c k_0^{- 1} K^2$. Similarly, it is easy to
  derive that the number of blocks within each $D^{\ast}_{l, 0}, l = 1,
  \ldots, k_0$ is smaller than $Ck_0^{- 1} K^2$ for another constant $C$.
  
  Based on the above result, we can see that there exists a constant $c$, such
  that $\epsilon_1 (\pi^{\ast}, \pi_0^{\ast}) K^2 > c (k_0 - k) k_0^{- 1} K^2$
  holds for $\pi^{\ast} \in \Pi_1^{\ast}$. Since Assumption \ref{AS:hyperpara}
  guarantees that $K \gg k_0^{3 / 2} \log^{\alpha_0} (n)$, we can apply Lemma
  \ref{LM:underfitted} and obtain
  \begin{equation}
    \|(\tmmathbf{I}_n -\tmmathbf{\phi}_{\pi^{\ast}})\tmmathbf{\mu}_0 \|^2_2
    \geqslant c n (k_0 - k) k_0^{- 1 - 2 / d} r_n^2, \forall \pi^{\ast} \in
    \Pi_1^{\ast} . \label{EQ:pi1underfitted}
  \end{equation}
  Next, applying Equation (\ref{EQ:Rsquaredif}), we have
  \begin{eqnarray*}
    \tmmathbf{y}^T  (\tmmathbf{\phi}_{\pi^{\ast}}
    -\tmmathbf{\phi}_{\pi_0^{\ast}}) \tmmathbf{y} & \leqslant & c n k_0^{1 /
    2} K^{- 1} r_n^2 + CK^2 \log (n) - \|(\tmmathbf{\phi}_{\tilde{\pi}^{\ast}}
    -\tmmathbf{\phi}_{\pi^{\ast}})\tmmathbf{y}\|^2_2\\
    & \leqslant & c n k_0^{1 / 2} K^{- 1} r_n^2 + CK^2 \log (n) -
    \{\|(\tmmathbf{\phi}_{\tilde{\pi}^{\ast}}
    -\tmmathbf{\phi}_{\pi^{\ast}})\tmmathbf{\mu}_0 \|_2
    -\|(\tmmathbf{\phi}_{\tilde{\pi}^{\ast}}
    -\tmmathbf{\phi}_{\pi^{\ast}})\tmmathbf{\epsilon}\|_2 \}^2 .
  \end{eqnarray*}
  Note that
  \begin{eqnarray}
    \|(\tmmathbf{\phi}_{\tilde{\pi}^{\ast}}
    -\tmmathbf{\phi}_{\pi^{\ast}})\tmmathbf{\mu}_0 \|_2 & \geqslant &
    \|(\tmmathbf{I}_n -\tmmathbf{\phi}_{\pi^{\ast}})\tmmathbf{\mu}_0 \|_2 -
    \|(\tmmathbf{\phi}_{\tilde{\pi}^{\ast}} -\tmmathbf{I}_n)\tmmathbf{\mu}_0
    \|_2 \nonumber\\
    & \overset{(i)}{\geqslant} & cn^{1 / 2} (k_0 - k)^{1 / 2} k_0^{- 1 / 2 -
    1 / d} r_n - Cn^{1 / 2} k_0^{1 / 4} K^{- 1 / 2} r_n \nonumber\\
    &\overset{\text{Assumption \ref{AS:hyperpara}}}{\geqslant}& cn^{1 / 2} (k_0
    - k)^{1 / 2} k_0^{- 1 / 2 - 1 / d} r_n,  \label{EQ:faimiu2lowerbound}
  \end{eqnarray}
  where $(i)$ uses Equation (\ref{EQ:pi1underfitted}) and Lemma
  \ref{LM:bestapp}. Besides, under $\mathcal{E}_{1 n}$, we have
  \begin{equation}
    \|(\tmmathbf{\phi}_{\tilde{\pi}^{\ast}}
    -\tmmathbf{\phi}_{\pi^{\ast}})\tmmathbf{\epsilon}\|_2 \leqslant C K
    \log^{1 / 2} (n) . \label{EQ:faiepsilonupper}
  \end{equation}
  Putting the result together, we conclude
  \begin{eqnarray*}
    \tmmathbf{y}^T  (\tmmathbf{\phi}_{\pi^{\ast}}
    -\tmmathbf{\phi}_{\pi_0^{\ast}}) \tmmathbf{y} & \leqslant & - c n (k_0 -
    k) k_0^{- 1 - 2 / d} r_n^2 .
  \end{eqnarray*}
  Plugging the result into Equation (\ref{EQ:Likelihoodr}), we have
  \begin{eqnarray*}
    \frac{\mathbb{P} (\tmmathbf{y} \mid \pi^{\ast}, \tmmathbf{x},
    \tmmathbf{s}) \lambda^{| \pi^{\ast} |}}{\mathbb{P} (\tmmathbf{y} \mid
    \pi^{\ast}_0, \tmmathbf{x}, \tmmathbf{s}) \lambda^{| \pi_0^{\ast} |}} &
    \leqslant & \exp \left[ (k_0 - k) \left\{ \log (\lambda^{- 1}) + \frac{d
    \log (n + 1)}{2} - c k_0^{- 1 - 2 / d} r_n^2 n \right\} \right]\\
    & \overset{\text{Assumption } \ref{AS:hyperpara}}{\leqslant} & \exp (- c
    k_0^{- 1 - 2 / d} r_n^2 n) .
  \end{eqnarray*}
  \tmtextbf{Cases when} $\pi^{\ast} \in \Pi_2^{\ast}$:
  
  From Equation (\ref{EQ:Rsquaredif}), we can see that
  \[ \tmmathbf{y}^T  (\tmmathbf{\phi}_{\pi^{\ast}}
     -\tmmathbf{\phi}_{\pi_0^{\ast}}) \tmmathbf{y} \leqslant c n k_0^{1 / 2}
     K^{- 1} r_n^2 + CK^2 \log (n) . \]
  Thus, together with Assumption \ref{AS:hyperpara}, we have
  \begin{eqnarray*}
    \frac{\mathbb{P} (\tmmathbf{y} \mid \pi^{\ast}, \tmmathbf{x},
    \tmmathbf{s}) \lambda^{| \pi^{\ast} |}}{\mathbb{P} (\tmmathbf{y} \mid
    \pi^{\ast}_0, \tmmathbf{x}, \tmmathbf{s}) \lambda^{| \pi_0^{\ast} |}} &
    \leqslant & \exp \left[ - \left\{ \log (\lambda^{- 1}) + \frac{d \log (n +
    1)}{2} \right\} + c n k_0^{1 / 2} K^{- 1} r_n^2 + CK^2 \log (n) \right]\\
    & \overset{}{\leqslant} & \exp [- c
    \log (\lambda^{- 1})] ,
  \end{eqnarray*}
  where the last inequality uses Assumption \ref{AS:hyperpara}.

  \noindent\tmtextbf{Cases when} $\pi^{\ast} \in \Pi_3^{\ast}$:
  
  Using the similar arguments as the cases of $\Pi_1^{\ast}$, it is easy to
  see that there exists a constant $c$, such that $\epsilon_1 (\pi^{\ast},
  \pi_0^{\ast}) K^2 > c k_0^{- 1} K^2$ holds for any $\pi^{\ast} \in
  \Pi_3^{\ast}$. Following the similar argument as in cases of $\Pi_1^{\ast}$,
  we obtain
  \[ \begin{array}{ll}
       \frac{\mathbb{P} (\tmmathbf{y} \mid \pi^{\ast}, \tmmathbf{x},
       \tmmathbf{s}) \lambda^{| \pi^{\ast} |}}{\mathbb{P} (\tmmathbf{y} \mid
       \pi^{\ast}_0, \tmmathbf{x}, \tmmathbf{s}) \lambda^{| \pi_0^{\ast} |}} &
       \leqslant
     \end{array} \exp (- c k_0^{- 1 - 2 / d} r_n^2 n) . \]
     
  \noindent\tmtextbf{Cases when} $\pi^{\ast} \in \Pi_4^{\ast}$:
  
  Write $\tilde{\pi}^{\ast} = \pi^{\ast} \cap \pi^{\ast}_0$, following the
  same arguments in (\ref{EQ:Rsquaredif}), we have
  \begin{eqnarray*}
  && \tmmathbf{y}^T  (\tmmathbf{\phi}_{\pi^{\ast}}
    -\tmmathbf{\phi}_{\pi_0^{\ast}}) \tmmathbf{y}\\
    & \leqslant & 8\|
    (\tmmathbf{I}_n -\tmmathbf{\phi}_{\pi^{\ast}_0}) \tmmathbf{\mu}_0  \|^2_2
    + 4\| \tmmathbf{\phi}_{\tilde{\pi}^{\ast}} \tmmathbf{\epsilon} \|^2 + 4\|
    \tmmathbf{\phi}_{\pi^{\ast}_0} \tmmathbf{\epsilon}\|^2_2 -
    \|(\tmmathbf{\phi}_{\tilde{\pi}^{\ast}}
    -\tmmathbf{\phi}_{\pi^{\ast}})\tmmathbf{y}\|^2_2\\
    & \overset{(i)}{\leqslant} & c n k_0^{1 / 2} K^{- 1} r_n^2 + C \epsilon_1
    (\pi^{\ast}, \pi_0^{\ast}) K^2 \log (n) -
    \|(\tmmathbf{\phi}_{\tilde{\pi}^{\ast}}
    -\tmmathbf{\phi}_{\pi^{\ast}})\tmmathbf{y}\|^2_2\\
    & \leqslant & c n k_0^{1 / 2} K^{- 1} r_n^2 + C \epsilon_1 (\pi^{\ast},
    \pi_0^{\ast}) K^2 \log (n) - \{\|(\tmmathbf{\phi}_{\tilde{\pi}^{\ast}}
    -\tmmathbf{\phi}_{\pi^{\ast}})\tmmathbf{\mu}_0 \|_2
    -\|(\tmmathbf{\phi}_{\tilde{\pi}^{\ast}}
    -\tmmathbf{\phi}_{\pi^{\ast}})\tmmathbf{\epsilon}\|\}^2,
  \end{eqnarray*}
  where $(i)$ uses Lemma \ref{LM:bestapp} and $\mathcal{E}_{2 n}$. On the
  other hand, from Lemmas \ref{LM:bestapp} and \ref{LM:underfitted}, we can
  see
  \begin{eqnarray*}
    \|(\tmmathbf{\phi}_{\tilde{\pi}^{\ast}}
    -\tmmathbf{\phi}_{\pi^{\ast}})\tmmathbf{\mu}_0 \|_2 & \geqslant &
    \|(\tmmathbf{I}_n -\tmmathbf{\phi}_{\pi^{\ast}})\tmmathbf{\mu}_0 \|_2 -
    \|(\tmmathbf{\phi}_{\tilde{\pi}^{\ast}} -\tmmathbf{I}_n)\tmmathbf{\mu}_0
    \|_2\\
    & \geqslant & c k^{- 1 / d}_0 n^{1 / 2} \epsilon_1^{1 / 2} (\pi^{\ast},
    \pi_0^{\ast}) r_n - c n^{1 / 2} k_0^{1 / 4} K^{- 1 / 2} r_n,
  \end{eqnarray*}
  and
  \[ \|(\tmmathbf{\phi}_{\tilde{\pi}^{\ast}}
     -\tmmathbf{\phi}_{\pi^{\ast}})\tmmathbf{\epsilon}\|_2 \leqslant
     \|\tmmathbf{\phi}_{\tilde{\pi}^{\ast}} \tmmathbf{\epsilon}\|_2 +
     \|\tmmathbf{\phi}_{\pi^{\ast}} \tmmathbf{\epsilon}\|_2 \leqslant C
     \epsilon^{1 / 2}_1 (\pi^{\ast}, \pi_0^{\ast}) K \log^{1 / 2} (n) \]
  under $\mathcal{E}_{2 n}$. Since $\pi^{\ast} \in \Pi_4^{\ast}$ and according
  to Assumption \ref{AS:hyperpara}, we have
  \[ c k^{- 1 / d}_0 n^{1 / 2} \epsilon_1^{1 / 2} (\pi^{\ast}, \pi_0^{\ast})
     r_n \gg c n^{1 / 2} k_0^{1 / 4} K^{- 1 / 2} r_n + C \epsilon^{1 / 2}_1
     (\pi^{\ast}, \pi_0^{\ast}) K \log^{1 / 2} (n), \]
  leading to
  \[ \{\|(\tmmathbf{\phi}_{\tilde{\pi}^{\ast}}
     -\tmmathbf{\phi}_{\pi^{\ast}})\tmmathbf{\mu}_0 \|_2
     -\|(\tmmathbf{\phi}_{\tilde{\pi}^{\ast}}
     -\tmmathbf{\phi}_{\pi^{\ast}})\tmmathbf{\epsilon}\|_2 \}^2 \geqslant c
     k_0^{- 2 / d} n \epsilon_1 (\pi^{\ast}, \pi_0^{\ast}) r_n^2 . \]
  Combining the result, we obtain
  \begin{eqnarray*}
    \tmmathbf{y}^T  (\tmmathbf{\phi}_{\pi^{\ast}}
    -\tmmathbf{\phi}_{\pi_0^{\ast}}) \tmmathbf{y} & \leqslant & c n k_0^{1 /
    2} K^{- 1} r_n^2 + C \epsilon_1 (\pi^{\ast}, \pi_0^{\ast}) K^2 \log (n) -
    c k_0^{- 2 / d} n \epsilon_1 (\pi^{\ast}, \pi_0^{\ast}) r_n^2\\
    & \overset{(i)}{\leqslant} & - c k_0^{- 2 / d} n \epsilon_1 (\pi^{\ast},
    \pi_0^{\ast}) r_n^2,
  \end{eqnarray*}
  where $(i)  \text{ uses the fact that } \epsilon_1 (\pi^{\ast}, \pi_0^{\ast})
  K^2 \geqslant \lfloor k_0^{1 / 2 + 2 / d} K \log^{\alpha_0} (n) \rfloor$.
  Plugging this result into Equation (\ref{EQ:Likelihoodr}), we have
  \begin{eqnarray*}
    \frac{\mathbb{P} (\tmmathbf{y} \mid \pi^{\ast}, \tmmathbf{x},
    \tmmathbf{s}) \lambda^{| \pi^{\ast} |}}{\mathbb{P} (\tmmathbf{y} \mid
    \pi^{\ast}_0, \tmmathbf{x}, \tmmathbf{s}) \lambda^{| \pi_0^{\ast} |}} & =
    & \exp \left\{ \frac{n\tmmathbf{y}^T (\tmmathbf{\phi}_{\pi^{\ast}}
    -\tmmathbf{\phi}_{\pi_0^{\ast}})\tmmathbf{y}}{2 \sigma^2 (n + 1)} \right\}
    \leqslant \exp \{ - c k_0^{- 2 / d} n \epsilon_1 (\pi^{\ast},
    \pi_0^{\ast}) r_n^2 \} .
  \end{eqnarray*}
  Combining the result, the proposition is proved.
\end{proof}

\subsection{Graph results and
theory}\label{SEC:treenumber}\label{SEC:parnuber}

This section derives some general graph results and gives proof of Proposition
\ref{PP:spanningtreeratio}. Section \ref{SEC:generalgraph} provides a general
result on the number of spanning trees, given a graph $\mathcal{G}$. Section
\ref{SEC:spanningtreeratio} applies result of Section \ref{SEC:generalgraph}
to our model setting to prove Proposition \ref{PP:spanningtreeratio}.

\subsubsection{General result of graph}\label{SEC:generalgraph}

The result in this section is a general graph result, which is independent of
our model. For generality, we take some new notations, which may be different
from the main paper. Let $\mathcal{G}= (\mathcal{V}_0, \mathcal{E})$ be a
spatial graph, where $\mathcal{V}_0$ is a vertex set, and the edge set
$\mathcal{E}$ is a subset of $\{(\tmmathbf{v}_i, \tmmathbf{v}_{i'}) :
\tmmathbf{v}_i, \tmmathbf{v}_{i'} \in \mathcal{V}, \tmmathbf{v}_i \neq
\tmmathbf{v}_{i'} \}$. We say a vertex set $\mathcal{V} \subseteq
\mathcal{V}_0$ is connected under $\mathcal{G}$, if there exists a path from
$\tmmathbf{v}_1$ to $\tmmathbf{v}_2$ with all the vertexes in path contained
in $\mathcal{V}$, for any two vertexes $\tmmathbf{v}_1, \tmmathbf{v}_2 \in
\mathcal{V}$. Based on graph $\mathcal{G}$, we define the distance of two
vertexes $\tmmathbf{v}_1, \tmmathbf{v}_2 \in \mathcal{V}_0$ as
\[ d_{\mathcal{G}}  (\tmmathbf{v}_1, \tmmathbf{v}_2) = \min_{\mathcal{P} \in
   \mathcal{P}_{\mathcal{G}} (\tmmathbf{v}_1, \tmmathbf{v}_2)} |\mathcal{P}|,
\]
where $\mathcal{P}_{\mathcal{G}} (\tmmathbf{v}_1, \tmmathbf{v}_2)$ is the set
containing all paths from $\tmmathbf{v}_1$ to $\tmmathbf{v}_2$ under graph
$\mathcal{G}$. For two vertex sets $\mathcal{V}_1$ and $\mathcal{V}_2$, we
define their distance as
\[ d_{\mathcal{G}}  (\mathcal{V}_1, \mathcal{V}_2) = \min_{\tmmathbf{v}_1 \in
   \mathcal{V}_1, \tmmathbf{v}_2 \in \mathcal{V}_2} d_{\mathcal{G}} 
   (\tmmathbf{v}_1, \tmmathbf{v}_2) . \]
Given a connected set $\mathcal{V}  \subseteq \mathcal{V}_0$, we write $S
(\mathcal{V})$ as the set of spanning trees of $\mathcal{V}$ under
$\mathcal{G}$. For a general vertex set $\mathcal{V}$ (which is not necessary
to be connected), it is easy to see that there exists a unique decomposition
$\mathcal{V}= \cup_j \mathcal{V}_j$, such that each $\mathcal{V}_j$ is
connected, and $d_{\mathcal{G}}  (\mathcal{V}_j, \mathcal{V}_{j'}) > 1$ for $j
\neq j'$. Based on the decomposition, we define an operator of $\mathcal{V}$,
say $\mathcal{H} (\mathcal{V})$, as $\mathcal{H} (\mathcal{V}) = \prod_j
\mathcal{H} (\mathcal{V}_j)$, where $\mathcal{H} (\mathcal{V}_j) = |S
(\mathcal{V}_j) |$. For a set of $m$ disjoint vertex sets $\{\mathcal{V}_j
\}_{j = 1}^m$, we write $b (\{\mathcal{V}_j \}_{j = 1}^m) = \sum_{j = 1}^m
\sum_{j' = 1}^m \mathbb{I} (j' > j) b (\{\mathcal{V}_j, \mathcal{V}_{j'} \})$,
where $b (\{\mathcal{V}_j, \mathcal{V}_{j'} \})$ is the number of edges (under
graph $\mathcal{G}$) connecting $\mathcal{V}_j$ and $\mathcal{V}_{j'}$.

\begin{figure}[htp]
		\centering
		\resizebox{150pt}{150pt}{\includegraphics{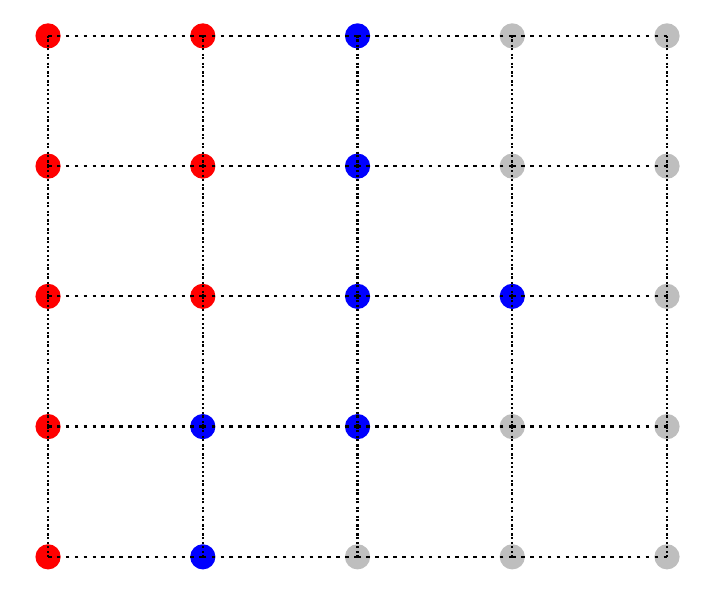}}
		\caption{One graph example. In this example, the graph $\mathcal{G}$ is a $5\times 5$ mesh grid. We denote $\mathcal{V}_1$, $\mathcal{V}_2$ and $\mathcal{V}_3$ by vertices with red, blue and gray colors, respectively. We can see $\{\mathcal{V}_j\}_{j=1}^{3}$ are three connected vertex sets and $d_{\mathcal{G}}(\mathcal{V}_1,\mathcal{V}_2)=1$, $d_{\mathcal{G}}(\mathcal{V}_1,\mathcal{V}_3)=2$ and $d_{\mathcal{G}}(\mathcal{V}_2,\mathcal{V}_3)=1$. According to the definition, $b (\{\mathcal{V}_j \}_{j=1}^{3}) = b (\{\mathcal{V}_1,\mathcal{V}_{2} \})+b (\{\mathcal{V}_1,\mathcal{V}_{3} \})+b (\{\mathcal{V}_2,\mathcal{V}_{3} \})$, where $b (\{\mathcal{V}_1,\mathcal{V}_{2} \})=6$, $b (\{\mathcal{V}_1,\mathcal{V}_{3} \})=0$ and $b (\{\mathcal{V}_2,\mathcal{V}_{3} \})=7$ since the number of edges connecting $\{\mathcal{V}_1,\mathcal{V}_{2} \}$, $\{\mathcal{V}_1,\mathcal{V}_{3} \}$ and $\{\mathcal{V}_2,\mathcal{V}_{3} \}$ are $6$, $0$ and $7$, respectively.}
		\label{FIG:meshgrid}
	\end{figure}

Note that $\mathcal{H} (\mathcal{V})$ defined above is the number of spanning
trees of $\mathcal{V}$, if $\mathcal{V}$ is connected. We will study the
property of $\mathcal{H} (\cdummy)$, because $\mathcal{H} (\cdummy)$ has a
close relationship with the number of spanning trees inducing a particular
partition. We have the following lemma.

\begin{lemma}
  \label{LM:connectednumber}Let $\mathcal{V}_1$ and $\mathcal{V}_2$ be two
  connected vertex sets, and $\mathcal{V}_1 \cap \mathcal{V}_2 = \emptyset$.
  Write $\mathcal{V}=\mathcal{V}_1 \cup \mathcal{V}_2$, we have
  \[ \mathcal{H} (\mathcal{V}_1) \mathcal{H} (\mathcal{V}_2) \leqslant
     \mathcal{H} (\mathcal{V}) \leqslant \mathcal{H} (\mathcal{V}_1)
     \mathcal{H} (\mathcal{V}_2) \times \exp [b (\{\mathcal{V}_1,
     \mathcal{V}_2 \})\{\log (|\mathcal{V}_1 |) + \log (|\mathcal{V}_2 |) +
     \log (2)\}] . \]
  \begin{proof}
    If $d_{\mathcal{G}}  (\mathcal{V}_1, \mathcal{V}_2) > 1$, it is easy to
    see that $b (\{\mathcal{V}_1, \mathcal{V}_2 \}) = 0$ and $\mathcal{H}
    (\mathcal{V}) =\mathcal{H} (\mathcal{V}_1) \mathcal{H} (\mathcal{V}_2)$ by the 
    definition. The inequality holds immediately. We next consider the case
    when $d_{\mathcal{G}}  (\mathcal{V}_1, \mathcal{V}_2) = 1$, hence
    $\mathcal{V}$ is also connected and $S (\mathcal{V})$ is well defined.
    
    If $d_{\mathcal{G}}  (\mathcal{V}_1, \mathcal{V}_2) = 1$, there exists a
    pair $(\tmmathbf{s}_1, \tmmathbf{s}_2)$, with $\tmmathbf{s}_1 \in
    \mathcal{V}_1$, $\tmmathbf{s}_2 \in \mathcal{V}_2$, such that
    $\tmmathbf{s}_1 \text{ is connected to } \tmmathbf{s}_2$ by an edge $e \in
    \mathcal{E}$. For any $\mathcal{T}_1 \in S (\mathcal{V}_1), \mathcal{T}_2
    \in S (\mathcal{V}_2)$, we can thus construct a spanning tree, say
    $\mathcal{T} (\mathcal{T}_1, \mathcal{T}_2)$ by connecting $\mathcal{T}_1$
    and $\mathcal{T}_2$ with $e$. It is easy to see that $\mathcal{T}
    (\mathcal{T}_1, \mathcal{T}_2) \in S (\mathcal{V})$ and $\mathcal{T}
    (\mathcal{T}_1, \mathcal{T}_2) \neq \mathcal{T} (\mathcal{T}_1',
    \mathcal{T}_2')$ if $(\mathcal{T}_1, \mathcal{T}_2) \neq (\mathcal{T}_1',
    \mathcal{T}_2')$. Thus, we have $\mathcal{H} (\mathcal{V}_1) \mathcal{H}
    (\mathcal{V}_2)=|S(\mathcal{V}_1)||S(\mathcal{V}_2)| \leqslant |S(\mathcal{V}) |=\mathcal{H} (\mathcal{V})$.
    
    We next prove the second inequality. For any spanning tree $\mathcal{T}
    \in S (\mathcal{V})$, we split it by removing all the edges connecting
    $(\tmmathbf{s}_1, \tmmathbf{s}_2)$, where $\tmmathbf{s}_1 \in
    \mathcal{V}_1$ and $\tmmathbf{s}_2 \in \mathcal{V}_2$. We can see that the
    number of removed edges is smaller than $b (\{\mathcal{V}_1, \mathcal{V}_2
    \})$. After the removal, we will obtain some ``sub'' spanning trees, the
    vertexes of which are either subset of $\mathcal{V}_1$ or subset of
    $\mathcal{V}_2$. Write $\tilde{S}_i (\mathcal{T})$ as the set of ``sub''
    spanning trees obtained from $\mathcal{T}$ with vertexes belonging to
    $\mathcal{V}_i$, $i = 1, 2$. Let $A = \{\{ \tilde{S}_1 (\mathcal{T}),
    \tilde{S}_2 (\mathcal{T})\}: \mathcal{T} \in S (\mathcal{V})\}$ be the
    space of $\{ \tilde{S}_1 (\mathcal{T}), \tilde{S}_2 (\mathcal{T})\}$. We
    can see that each $\mathcal{T} \in S (\mathcal{V})$ corresponds to an
    element $a \in A$. We will then bound $\mathcal{H} (\mathcal{V}) = |S
    (\mathcal{V}) |$ by
    \begin{eqnarray*}
       &  & |A| \times \max_{a \in A}
      [|\{\mathcal{T} \in S (\mathcal{V}) : \mathcal{T}
      \text{ can induce a}\}|]\\
      & \leqslant & |\{ \tilde{S}_1 (\mathcal{T}) : \mathcal{T} \in S
      (\mathcal{V})\}| \times |\{ \tilde{S}_2 (\mathcal{T}) : \mathcal{T} \in
      S (\mathcal{V})\}| \times \max_{a \in A} [|\{\mathcal{T} \in S
      (\mathcal{V}) : \mathcal{T} \text{ can induce a}\}|] .
    \end{eqnarray*}
    Since $\mathcal{V}_i$ is connected, we can always add some edges between
    those sub spanning trees in $\tilde{S}_i (\mathcal{T})$ to join them into
    a spanning tree consisting all the vertexes of $\mathcal{V}_i$. Thus, $\tilde{S}_i (\mathcal{T})$ is within the set of $\text{``sub" spanning trees obtained after cutting no more than } b\{\mathcal{V}_1,
       \mathcal{V}_2 \} \text{ edges from } \mathcal{T}_i \in S
       (\mathcal{V}_i)$.
    % \[ \tilde{S}_i (\mathcal{T}) \in
    %    \{\text{``sub" spanning trees obtained after cutting no more than } b\{\mathcal{V}_1,
    %    \mathcal{V}_2 \} \text{ edges from } \mathcal{T}_i \in S
    %    (\mathcal{V}_i)\} . \]
    Together with Lemma \ref{LM:binomialcoe}, we hence conclude that
    \begin{eqnarray*}
      |\{ \tilde{S}_i (\mathcal{T}) : \mathcal{T} \in S (\mathcal{V})\}| &
      \leqslant & \mathcal{H} (\mathcal{V}_i) \times \sum_{j = 0}^{b
      (\{\mathcal{V}_1, \mathcal{V}_2 \})} \left( \begin{array}{l}
        |\mathcal{V}_i | - 1\\
        j
      \end{array} \right)\\
      & \leqslant & \mathcal{H} (\mathcal{V}_i) \times \exp \{b
      (\{\mathcal{V}_1, \mathcal{V}_2 \}) \log (|\mathcal{V}_i |)\} .
    \end{eqnarray*}
    On the other hand, note that if $\mathcal{T} \text{ can induce a } \in A$,
    $\mathcal{T}$ can be converted back by adding back those removed edges
    from $a$. Since the number of removed edges is no more than $b
    (\mathcal{V}_1, \mathcal{V}_2)$, we conclude that
    \[ \max_{a \in A} |\{\mathcal{T} \in S (\mathcal{V}) : \mathcal{T}
       \text{ can induce a}\}| \leqslant 2^{b (\{\mathcal{V}_1, \mathcal{V}_2
       \})} = \exp \{\log (2) b (\{\mathcal{V}_1, \mathcal{V}_2 \})\} . \]
    Putting the result together, the lemma is proved.
  \end{proof}
\end{lemma}

The following lemma extends Lemma \ref{LM:connectednumber} to a more general
case.

\begin{lemma}
  \label{LM:Hvineq}Let $\{\mathcal{V}_i \}_{i = 1}^m$ be $m$ connected vertex
  sets, with $\mathcal{V}_m \cap \mathcal{V}_{m'} = \emptyset$ if $m \neq m'$.
  Write $\mathcal{V}= \cup_{i = 1}^m \mathcal{V}_i$, we have
  \begin{equation}
    \prod_{i = 1}^m \mathcal{H} (\mathcal{V}_i) \leqslant \mathcal{H}
    (\mathcal{V}) \leqslant \left\{ \prod_{i = 1}^m \mathcal{H}(\mathcal{V}_i)
    \right\} \times \exp \left[ b (\{\mathcal{V}_i \}_{i = 1}^m) \left\{ 2
    \log \left( \sum_{i = 1}^m |\mathcal{V}_i | \right) + \log (2) \right\}
    \right] . \label{EQ:Hv}
  \end{equation}
  \begin{proof}
    We will prove the lemma by induction. It is easy to see that inequality
    holds for $m = 1$. Besides, from Lemma \ref{LM:connectednumber}, we can
    see that the inequality holds for $m = 2$. Next, we prove that if
    (\ref{EQ:Hv}) holds for $m = m_0 \geqslant 2$, it also holds for $m = m_0
    + 1$. Let $\{\mathcal{V}_i \}_{i = 1}^{m_0 + 1}$ be $m_0 + 1$ connected
    vertex sets with $\mathcal{V}_m \cap \mathcal{V}_{m'} = \emptyset$ if $m
    \neq m'$. We discuss the following two cases of $\mathcal{V}_{m_0 + 1}$:
    
    \tmtextbf{If $\min_{1 \leqslant i \leqslant m_0} \{d_{\mathcal{G}}
    (\mathcal{V}_i, \mathcal{V}_{m_0 + 1})\} > 1$}:
    
    From the definition of $\mathcal{H} (\cdummy)$, we can see
    \[ \mathcal{H} (\mathcal{V}) =\mathcal{H} (\mathcal{V}_{m_0 + 1})
       \mathcal{H}(\cup_{i = 1}^{m_0} \mathcal{V}_i) . \]
    We can thus apply Equation (\ref{EQ:Hv}) to $\mathcal{H}(\cup_{i =
    1}^{m_0} \mathcal{V}_i$), obtaining
    \[ \mathcal{H} (\mathcal{V}) =\mathcal{H} (\mathcal{V}_{m_0 + 1})
       \mathcal{H}(\cup_{i = 1}^{m_0} \mathcal{V}_i) \geqslant \prod_{i =
       1}^{m_0 + 1} \mathcal{H} (\mathcal{V}_i), \]
    and
    \begin{eqnarray*}
      \mathcal{H} (\mathcal{V}) & = & \mathcal{H} (\mathcal{V}_{m_0 + 1})
      \mathcal{H}(\cup_{i = 1}^{m_0} \mathcal{V}_i)\\
      & \leqslant & \left\{ \prod_{i = 1}^{m_0 + 1}
      \mathcal{H}(\mathcal{V}_i) \right\} \times \exp \left[ b
      (\{\mathcal{V}_i \}_{i = 1}^{m_0}) \left\{ \log (2) + 2 \log \left(
      \sum_{i = 1}^{m_0} |\mathcal{V}_i | \right) \right\} \right]\\
      & \leqslant & \left\{ \prod_{i = 1}^{m_0 + 1}
      \mathcal{H}(\mathcal{V}_i) \right\} \times \exp \left[ b
      (\{\mathcal{V}_i \}_{i = 1}^{m_0 + 1}) \left\{ \log (2) + 2 \log \left(
      \sum_{i = 1}^{m_0 + 1} |\mathcal{V}_i | \right) \right\} \right] .
    \end{eqnarray*}
    The inequality is thus proved.
    
    \tmtextbf{If $\min_{1 \leqslant i \leqslant m_0} \{d_{\mathcal{G}}
    (\mathcal{V}_i, \mathcal{V}_{m_0 + 1})\} = 1$:}
    
    Without loss of generality, we assume $d_{\mathcal{G}} 
    (\mathcal{V}_{m_0}, \mathcal{V}_{m_0 + 1}) = 1$. Write
    $\tilde{\mathcal{V}} =\mathcal{V}_{m_0} \cup \mathcal{V}_{m_0 + 1}$, we
    can see that $\tilde{\mathcal{V}}$ is connected. So we can view
    $\mathcal{V}$ as a union of $m_0$ connected vertex sets, i.e.,
    $\mathcal{V}=\{\cup_{i = 1}^{m_0 - 1} \mathcal{V}_i \} \cup
    \tilde{\mathcal{V}}$. By applying (\ref{EQ:Hv}) to $\mathcal{H}
    (\mathcal{V}) =\mathcal{H} (\{ \cup_{i = 1}^{m_0 - 1} \mathcal{V}_i \}
    \cup \tilde{\mathcal{V}}$), we have
    \[ \mathcal{H} (\mathcal{V}) \geqslant \left\{ \prod_{i = 1}^{m_0 - 1}
       \mathcal{H}(\mathcal{V}_i) \right\} \mathcal{H} (\tilde{\mathcal{V}})
       \overset{(i)}{\geqslant} \left\{ \prod_{i = 1}^{m_0 + 1}
       \mathcal{H}(\mathcal{V}_i) \right\} \]
    where in $(i)$ we apply Equation (\ref{EQ:Hv}) to $\mathcal{H}
    (\tilde{\mathcal{V}}) =\mathcal{H} (\mathcal{V}_{m_0} \cup
    \mathcal{V}_{m_0 + 1})$. On the other hand, Equation (\ref{EQ:Hv}) also
    entails an upper bound of $\mathcal{H} (\mathcal{V})$ as
    \begin{eqnarray*}
       &  & \left\{ \prod_{i = 1}^{m_0 - 1}
      \mathcal{H}(\mathcal{V}_i) \right\} \mathcal{H} (\tilde{\mathcal{V}})
      \times \exp \left[ b (\{\mathcal{V}_1, \ldots, \mathcal{V}_{m_0 - 1},
      \tilde{\mathcal{V}} \}) \left\{ \log (2) + 2 \log \left( \sum_{i =
      1}^{m_0 - 1} |\mathcal{V}_i | + | \tilde{\mathcal{V}} | \right) \right\}
      \right]\\
      & \overset{(ii)}{\leqslant} & \left\{ \prod_{i = 1}^{m_0 + 1}
      \mathcal{H}(\mathcal{V}_i) \right\} \times \exp [b (\{\mathcal{V}_{m_0},
      \mathcal{V}_{m_0 + 1} \})\{\log (2) + 2 \log (| \tilde{\mathcal{V}}
      |)\}]\\
      &  & \times \exp \left[ b (\{\mathcal{V}_1, \ldots, \mathcal{V}_{m_0 -
      1}, \tilde{\mathcal{V}} \}) \left\{ \log (2) + 2 \log \left( \sum_{i =
      1}^{m_0 - 1} |\mathcal{V}_i | + | \tilde{\mathcal{V}} | \right) \right\}
      \right]\\
      & \leqslant & \left\{ \prod_{i = 1}^{m_0 + 1}
      \mathcal{H}(\mathcal{V}_i) \right\} \times \exp \left[ b
      (\{\mathcal{V}_i \}_{i = 1}^{m_0 + 1}) \left\{ \log (2) + 2 \log \left(
      \sum_{i = 1}^{m_0 + 1} |\mathcal{V}_i | \right) \right\} \right],
    \end{eqnarray*}
    where in $(ii)$ we apply Equation (\ref{EQ:Hv}) to $\mathcal{H}
    (\tilde{\mathcal{V}}) =\mathcal{H} (\mathcal{V}_{m_0} \cup
    \mathcal{V}_{m_0 + 1})$. Thus, by induction, the lemma is proved.
  \end{proof}
\end{lemma}

We furthermore extend Lemma \ref{LM:Hvineq} to the following general
result, where we do not require each $\mathcal{V}_i$ to be connected.

\begin{lemma}
  \label{LM:HVineq2}Let $\{\mathcal{V}_i \}_{i = 1}^m$ be $m$ vertex sets,
  with $\mathcal{V}_m \cap \mathcal{V}_{m'} = \emptyset$ if $m \neq m'$. Write
  $\mathcal{V}= \cup_{i = 1}^m \mathcal{V}_i$, we have
  \begin{equation}
    \prod_{i = 1}^m \mathcal{H} (\mathcal{V}_i) \leqslant \mathcal{H}
    (\mathcal{V}) \leqslant \left\{ \prod_{i = 1}^m \mathcal{H}(\mathcal{V}_i)
    \right\} \times \exp \left[ b (\{\mathcal{V}_i \}_{i = 1}^m) \left\{ 2
    \log \left( \sum_{i = 1}^m |\mathcal{V}_i | \right) + \log (2) \right\}
    \right] .
  \end{equation}
  \begin{proof}
    We first perform decomposition on $\mathcal{V}_i$ such that $\mathcal{V}_i
    = \cup_{j = 1}^{n_i} \mathcal{V}_{ij}$, where $d_{\mathcal{G}} 
    (\mathcal{V}_{ij}, \mathcal{V}_{ij'}) > 1 \text{ if }j \neq j'$ holds for
    all $i = 1, \ldots, m$, each $\mathcal{V}_{ij}$ is connected, and $n_i$ is
    the number of connected vertex sets for $\mathcal{V}_i$. By the definition
    of $\mathcal{H} (\cdummy)$, we have
    \begin{eqnarray*}
      \prod_{i = 1}^m \mathcal{H} (\mathcal{V}_i) & = & \prod_{i = 1}^m
      \prod_{j = 1}^{n_i} \mathcal{H} (\mathcal{V}_{ij}) .
    \end{eqnarray*}
    On the other hand, note that $\mathcal{V}= \cup_{i = 1}^m \cup_{j =
    1}^{n_i} \mathcal{V}_{ij}$ is a union of connected vertex sets
    $\{\mathcal{V}_{ij} \}$, after applying Lemma \ref{LM:Hvineq}, we have
    \begin{eqnarray*}
      \prod_{i = 1}^m \prod_{j = 1}^{n_i} \mathcal{H} (\mathcal{V}_{ij})
      \leqslant & \mathcal{H} (\mathcal{V}), & 
    \end{eqnarray*}
    and
    \begin{eqnarray*}
      \mathcal{H} (\mathcal{V}) & \leqslant & \prod_{i = 1}^m \prod_{j =
      1}^{n_i} \mathcal{H} (\mathcal{V}_{ij})
      \times \exp \left[ b (\{\mathcal{V}_{ij} \}_{i = 1, j = 1}^{m,
      n_i}) \left\{ \log (2) + 2 \log \left( \sum_{i = 1}^m |\mathcal{V}_i |
      \right) \right\} \right]\\
      & \overset{(i)}{=} & \prod_{i = 1}^m \mathcal{H} (\mathcal{V}_i) \times
      \exp \left[ b (\{\mathcal{V}_i \}_{i = 1}^m) \left\{ \log (2) + 2 \log
      \left( \sum_{i = 1}^m |\mathcal{V}_i | \right) \right\} \right],
    \end{eqnarray*}
    where the equality $(i)$ uses the fact that $d_{\mathcal{G}} 
    (\mathcal{V}_{ij}, \mathcal{V}_{ij'}) > 1 \text{ if }j \neq j'$. Combining
    the result above, the lemma is proved.
  \end{proof}
\end{lemma}

\subsubsection{Proof of Proposition
\ref{PP:spanningtreeratio}}\label{SEC:spanningtreeratio}

Recall that in our model, the domain partition $\pi ^{\ast}$ is induced from
the contiguous partition of blocks $\mathcal{V} = \{ B_m \}_{m = 1}^{K^2}$,
say $\pi (\mathcal{V}) = \{ \mathcal{V}_1, \ldots, \mathcal{V}_k \}$. In
Section \ref{SEC:proofofpp1}, we have shown that $\pi_0^{\ast}$ in Proposition
\ref{PP:bestapproximation} is induced from $\pi_0 (\mathcal{V}) = \{ \mathcal{V}_{1, 0},
\ldots, \mathcal{V}_{k_0, 0} \}$. We give the proof of Proposition
\ref{PP:spanningtreeratio} as follows.

\begin{proof}
  Based on results in Section \ref{SEC:generalgraph}, we have
  \begin{eqnarray*}
    &  & |\{\mathcal{T}: \mathcal{T} \text{can induce } \pi^{\ast} \}|\\
    & \overset{(i)}{\leqslant} & \left. \left\{ \prod_{j = 1}^k
    \mathcal{H}(\mathcal{V}_j) \right\} \times \left( \begin{array}{l}
      2 (K - 1)^2\\
      k - 1
    \end{array} \right) = \left[ \prod_{j = 1}^k \mathcal{H} \right. \{
    \cup_{l = 1}^{k_0} (\mathcal{V}_j \cap \mathcal{V}_{l, 0}) \} \right]
    \times \left( \begin{array}{l}
      2 (K - 1)^2\\
      k - 1
    \end{array} \right)\\
    & \overset{ \text{Lemma } \ref{LM:HVineq2}}{\leqslant} &
    \left[ \prod_{j = 1}^k \left\{ \left[ \prod_{l = 1}^{k_0}
    \mathcal{H}\{(\mathcal{V}_j \cap \mathcal{V}_{l, 0})\} \right] \times \exp
    (b [\{(\mathcal{V}_j \cap \mathcal{V}_{l, 0})\}_{l = 1}^{k_0}] \times \{4
    \log (K) + \log (2)\}) \right\} \right] \\
    &&\times \left( \begin{array}{l}
      2 (K - 1)^2\\
      k - 1
    \end{array} \right)\\
    & \overset{(ii)}{\leqslant} & \left( \prod_{j = 1}^k  \left[
    \prod_{l = 1}^{k_0} \mathcal{H}\{(\mathcal{V}_j \cap \mathcal{V}_{l, 0})\}
    \right]\right) \times \exp \{c k_0^{1/2}K \log (K)\}   \times \left(
    \begin{array}{l}
      2 (K - 1)^2\\
      k - 1
    \end{array} \right)\\
    & \overset{ \text{Lemma } \ref{LM:HVineq2}}{\leqslant} &
    \left. \left( \prod_{l = 1}^{k_0} [\mathcal{H} \{ \cup_{j = 1}^k
    (\mathcal{V}_j \cap \mathcal{V}_{l, 0}) \} \right.] \right) \times \exp
    \{c k_0^{1/2} K \log (K)\} \times \left( \begin{array}{l}
      2 (K - 1)^2\\
      k - 1
    \end{array} \right)\\
    & \overset{(iii)}{\leqslant} & |\{\mathcal{T}: \mathcal{T} \text{can
    induce } \pi^{\ast}_0 \}| \times \exp \{c k_0^{1/2}K \log (K)\} \times \left(
    \begin{array}{l}
      2 (K - 1)^2\\
      k - 1
    \end{array} \right) .
  \end{eqnarray*}
  For inequality $(i)$, we use the fact that each $\mathcal{V}_j$ is
  connected, thus $\mathcal{H} (\mathcal{V}_j)$ is the number of spanning
  trees of $\mathcal{V}_j$. $2 (K - 1)^2$ is the total number of graph edges,
  and $\left( \begin{array}{l}
    2 (K - 1)^2\\
    k - 1
  \end{array} \right)$ is the number of possible ways of cutting edges of
  $\mathcal{T} \in \{\mathcal{T}: \mathcal{T} \text{can induce } \pi^{\ast}
  \}$ to obtain $\pi^{\ast}$. For inequality $(i i)$, we use the fact that $\sum_{j=1}^{k}\sum_{l \neq l'}b
  [\{\mathcal{V}_j \cap \mathcal{V}_{l, 0}, \mathcal{V}_j \cap
  \mathcal{V}_{l', 0} \}] \leqslant c k_0^{1/2}K$ according to (\ref{EQ:bestappV}). For inequality $(iii)$, we use the fact that $\cup_{j = 1}^k
  (\mathcal{V}_j \cap \mathcal{V}_{l, 0})$ is connected since $\pi_0^{\ast}$
  is contiguous.
  
  Thus, for any partition $\pi^{\ast} \in \Xi^{\ast}$, we obtain
  \begin{equation}
    \frac{|\{\mathcal{T}: \mathcal{T} \text{can induce } \pi^{\ast}
    \}|}{|\{\mathcal{T}: \mathcal{T} \text{can induce } \pi^{\ast}_0 \}|}
    \leqslant \exp \{c k_0^{1 / 2} K \log (K)\} \times \left( \begin{array}{l}
      2 (K - 1)^2\\
      k - 1
    \end{array} \right) \tmmathbf{} \label{EQ:treeratio} .
  \end{equation}
\end{proof}

% \section{Additional details of real data analysis}\label{Ssec:realdata}

% Identifying AAIW provides crucial insights into ocean circulation and heat distribution, which are essential for understanding Earth's climate system. The T-S relationship is known to be homogeneous within certain
% water masses, but can shift abruptly across boundaries. Consequently, the T-S
% 	 relationship is often modeled as a spatially piecewise constant function in
% 	 oceanography. We take the segment of the
% 	 Atlantic basin along $25^{\circ}$W between $60^{\circ}$S and the equator. We
% 	 apply the same rescaling procedure as in {\cite{luo2021bayesian}} to
% 	 eliminate the strong anisotropic spatial patterns of the original dataset.
% 	 Write $s_h$ and $s_v$ as the rescaled latitude and depth of the ocean, respectively. According
% 	 to the pilot study in {\cite{luo2021bayesian}}, AAIW is within $(s_h, s_v) \in
% 	 [- 0.5, 0.25] \times [0, 0.4]$. We thus only focus on the data within this
% 	 region. Finally, 
%    we get $n = 1936$ observed locations and    
% 	 corresponding temperature and salinity. 

\section{More details on placing a prior on $\lambda$}
\label{SEFC:lambdafurter}
Recall that, in the main paper, we demonstrate that a fixed, nondegenerate prior on the concentration parameter of Dirichlet process mixtures induces an $n$-dependent posterior term favoring fewer clusters, and yields consistency for the number of clusters \citep{ascolani2023clustering}. However, this is not the case in our model, and we provide more details in this section. Suppose we place a prior on $\lambda$, denoted by $\mathbb{P}(\lambda)$. Then we compute the posterior ratio $\frac{\mathbb{P}(\pi^{\ast}\mid \mathfrak{D})}{\mathbb{P}(\pi^{\ast}_0\mid \mathfrak{D})}$ as
\begin{eqnarray}
   &  & \frac{\int \mathbb{P} (\pi^{\ast}, \lambda \mid
  \mathfrak{D}) d \lambda}{\int \mathbb{P} (\pi^{\ast}_0, \lambda \mid
  \mathfrak{D}) d \lambda} = \frac{\int \mathbb{P} (\pi^{\ast}, \lambda,
  \mathfrak{D}) d \lambda}{\int \mathbb{P} (\pi^{\ast}_0, \lambda,
  \mathfrak{D}) d \lambda} \nonumber\\
  & = & \frac{\int \mathbb{P} (\lambda) \mathbb{P} (\pi^{\ast} \mid \lambda)
  d \lambda}{\int \mathbb{P} (\lambda) \mathbb{P} (\pi^{\ast}_0 \mid \lambda)
  d \lambda} \times \frac{\mathbb{P} (\tmmathbf{y} \mid
  \pi^{\ast},\tmmathbf{x},\tmmathbf{s})}{\mathbb{P} (\tmmathbf{y} \mid \pi^{\ast}_0,\tmmathbf{x},\tmmathbf{s})}\nonumber\\
  & = & \frac{\int \mathbb{P} (\lambda) \frac{\exp (- \lambda) \lambda^k}{k!}
  \left| \left\{ \mathcal{T} : \mathcal{T} \text{can induce } \pi^{\ast}
  \right\} \right| \left( \begin{array}{l}
    K^2 - 1\\
    k - 1
  \end{array} \right)^{- 1} d \lambda}{\int \mathbb{P} (\lambda) \frac{\exp (-
  \lambda) \lambda^{k_0}}{k_0 !} \left| \left\{ \mathcal{T} : \mathcal{T}
  \text{can induce } \pi^{\ast}_0 \right\} \right| \left( \begin{array}{l}
    K^2 - 1\\
    k_0 - 1
  \end{array} \right)^{- 1} d \lambda} \times \frac{\mathbb{P} (\tmmathbf{y}
  \mid \pi^{\ast},\tmmathbf{x},\tmmathbf{s})}{\mathbb{P} (\tmmathbf{y} \mid \pi^{\ast}_0,\tmmathbf{x},\tmmathbf{s})}\nonumber\\
  & = & \underset{C_{\lambda} (k_0, k)}{\underbrace{\frac{\int \mathbb{P}
  (\lambda) \exp (- \lambda) \lambda^k d \lambda}{\int \mathbb{P} (\lambda)
  \exp (- \lambda) \lambda^{k_0} d \lambda}}} \times \frac{k_0 !}{k!}
  \frac{\left| \left\{ \mathcal{T} : \mathcal{T} \text{can induce } \pi^{\ast}
  \right\} \right|}{\left| \left\{ \mathcal{T} : \mathcal{T} \text{can induce
  } \pi^{\ast}_0 \right\} \right|} \frac{\left( \begin{array}{l}
    K^2 - 1\\
    k - 1
  \end{array} \right)^{- 1}}{\left( \begin{array}{l}
    K^2 - 1\\
    k_0 - 1
  \end{array} \right)^{- 1}} \frac{\mathbb{P} (\tmmathbf{y} \mid
  \pi^{\ast},\tmmathbf{x},\tmmathbf{s})}{\mathbb{P} (\tmmathbf{y} \mid \pi^{\ast}_0,\tmmathbf{x},\tmmathbf{s})} .\nonumber
  %\label{EQ:Clambda}
\end{eqnarray}

Following the derivation above, $C_\lambda(k_0,k)$ incorporates information about $\lambda$. However, different from the case in \cite{ascolani2023clustering}, $C_\lambda(k_0,k)$ is independent of $n$ under a fixed, nondegenerate prior $\mathbb{P}(\lambda)$. Consequently, $C_\lambda(k_0,k)$ does not favor fewer clusters when $n \rightarrow \infty$, and therefore cannot counterbalance the likelihood’s preference for a larger number of clusters.

Thus, to ensure that $C_\lambda(k_0,k)$ depends on $n$ and favors fewer clusters, we require $\lambda$ to depend on and decay with $n$ in Assumption \ref{AS:lambda}.

\section{Estimation of $(k_0,r_n)$ under the VDG regime}
\label{SEC:k0rnestimation}

Recall that in the main paper, we demonstrate that the order conditions on $(K,\lambda)$ in Assumption \ref{AS:hyperpara} depend on the unknown quantities $(k_0,r_n)$, which makes them impractical to apply under the VDG regime. Thus, under the VDG regime, we propose to first estimate $(k_0,r_n)$ and then plug these estimates into Assumption \ref{AS:hyperpara} to select $(K,\lambda)$. 

We detail the estimation procedure in this section. Section \ref{SEC:k0knownrn} first addresses the estimation of $k_0$ assuming that the order of $r_n$ is known. Section \ref{SEC:rn} then proposes an approach for estimating the order of $r_n$, and justifies it by Lemma \ref{PP:rnconsistency}. Section \ref{SEC:addsimu} conducts a simulation study to demonstrate the effectiveness of the proposed approach. Section \ref{SEC:addproof} provides the proof of Lemma \ref{PP:rnconsistency}. Recall that in the main paper, we assume that $(k_0, r_n)$ satisfy Equation (\ref{EQ:rnk0}).

\subsection{Estimation of $k_0$ with a known $r_n$ order}\label{SEC:k0knownrn}

% Under Equation (\ref{EQ:rnk0}), we have $k_0 = O \left( n^{\frac{d \tau}{2 d +
% 3}} r_n^{\frac{d}{2 d + 3}} \right)$ for some $\tau < 1 / 2$. 
Note that Assumption \ref{AS:lorder} imposes more stringent requirements on $K$ as $k_0$ increases: the upper bound on $K$ decreases with $k_0$, while the lower bound increases with $k_0$. On the other hand, let $k_{0,\tmop{max}}= n^{\frac{d}{4d+6}} r_n^{\frac{d}{2d+3}}$. According to Equation (\ref{EQ:rnk0}), $k_0$ is required to satisfy $k_0 \ll k_{0,\tmop{max}}$. 
% For $k_0 \sim n^{\frac{d}{4d+6}} r_n^{\frac{d}{2d+3}}$, the lower bound in Assumption \ref{AS:lorder} coincides with the upper bound, up to logarithmic factors. 
% On the other hand, Equation (\ref{EQ:rnk0}) entails that the maximal admissible order of $k_0$, say $k_{0,\tmop{max}}$, satisfies $k_{0,\tmop{max}} \ll n^{\frac{d}{4d+6}} r_n^{\frac{d}{2d+3}}$. 
Consequently, we take $k_{0,\tmop{max}}$ as a reference order for $k_0$ to derive a ``safe" choice of $(K,\lambda)$ that guarantees consistency. Specifically, we propose to set
\begin{equation}
  K \sim n^{\frac{3 d + 4}{8 d + 12}} r_n^{\frac{3 d + 4}{4 d + 6}}, \log
  (\lambda^{- 1}) \sim n^{\frac{3 d + 4}{4 d + 6}} r_n^{\frac{3 d + 4}{2 d +
  3}} \log^{1 + \alpha} (n) \label{EQ:firsthyper}
\end{equation}
for some $\alpha > 0$. With the orders in (\ref{EQ:firsthyper}) and together with Equation (\ref{EQ:rnk0}), it can be shown that Assumption \ref{AS:hyperpara} is satisfied.

% Since we don't know the true order of $k_0$, a safe way is to assume a
% largest $k_0$ order (under Equation (\ref{EQ:rnk0})), and derive the
% corresponding order of $(K, \lambda)$ to achieve consistency.

Following the above argument and Theorem \ref{TH:clustererror}, the hyperparameter orders specified in (\ref{EQ:firsthyper}) guarantee consistent estimation of $k_0$. In what follows, we refer to the corresponding estimated number of clusters as the first-step estimate. However, the orders in (\ref{EQ:firsthyper}) do not attain the optimal contraction rate stated in Corollary 1. After obtaining the first-step estimate, we refit the model using the orders in Remark 2, which yields the optimal rate.

Note that the order of $K$ specified in (\ref{EQ:firsthyper}) is smaller than that in Remark \ref{RM:optimal rate}. Hence, refitting the model using the order in Remark \ref{RM:optimal rate} can be viewed as a refinement step that incorporates the information of $k_0$. Theoretically, the posterior number of clusters in the refitted model concentrates at the same value as the first-step estimate, since both hyperparameter orders yield consistent estimation. In practice, we can fix the number of clusters in the refitted model at the first-step estimate, and run the MCMC algorithm with only change and hyper moves.

% Note that in the first step, we
% assume $k_0$ to be large, hence the selected $K$ is small. In the second step,
% the selected $k_0$ will be smaller than that in the first step. So the $K$
% used in the second step will be larger than that in the first step. We can
% consider the second step as a $k_0$-tailored step and it is a refinement of
% the result in the first step.

\subsection{Estimation of $r_n$ order}\label{SEC:rn}

We take a blocking strategy to estimate the order of $r_n$. Similar to Section \ref{SEC:Model1}, we first divide the entire domain $\mathcal{D}$ into $K_r^2$ blocks $\{ B_m
\}_{m = 1}^{K_r^2}$. Within each block $B_m$, write $\tmmathbf{x}_m$, 
$\tmmathbf{y}_m$ as the corresponding covariates and responses, respectively. We
then estimate a regression coefficient within each block by
\[ \hat{\tmmathbf{\theta}}_m = (\tmmathbf{x}_m^T \tmmathbf{x}_m)^{- 1}
   \tmmathbf{x}_m^T \tmmathbf{y}_m, \]
where the inverse can be replaced by pseudoinverse if $\tmmathbf{x}_m^T \tmmathbf{x}_m$ is not invertible. Based on $\{ \hat{\tmmathbf{\theta}}_m \}_{m =
1}^{K^2_r}$, we propose to estimate $r_n$ by
\begin{equation}
\hat{r}_n = \sqrt{n^{- 1} \sum_{m = 1}^{K_r^2} \| B_m \| \|
   \hat{\tmmathbf{\theta}}_m \|_2^2} . 
   \label{EQ:rnhat}
\end{equation}

The rationale of (\ref{EQ:rnhat}) is as follows. Recall that we assume$\{ \tmmathbf{\theta}_{l, 0} \}_{l = 1}^{k_0}$ are distributed within a ball centered at $\tmmathbf{0}$ (we can always center the data) with radius $r_n$. For a given $K_r$, we divide blocks into two categories: those intersecting the true domain partitioning boundaries and those that do not. On the one hand, it is desirable for the number of boundary-intersecting blocks to be small, since $\hat{\tmmathbf{\theta}}_m$ computed on such a block may not be close to any element of $\{ \tmmathbf{\theta}_{l, 0} \}_{l = 1}^{k_0}$. On the other hand, for a block $B_m$ that does not intersect the boundaries, $\hat{\tmmathbf{\theta}}_m$ is expected to be close to some $\tmmathbf{\theta}_{l,0}$. Combining these two observations, the quantity $n^{- 1} \sum_{m = 1}^{K_r^2} \| B_m \| \|
\hat{\tmmathbf{\theta}}_m \|_2^2$ is approximately the same order as $k_0^{-1}\sum_{l=1}^{k_0}\|\tmmathbf{\theta}_{l,0}\|_2^2$, which is also expected to be of the same order as $r_n^2$.

Following the above argument, $K_r$ plays an important role in computing $\hat{r}_n$. On the one hand, under Assumption \ref{AS:lengthofboundary}, it can be shown that the proportion of blocks intersecting the boundaries is of the order $k_0^{1/2}K_r^{-1}$. Thus, a small $K_r$ yields a large proportion of blocks intersecting the boundaries, which is undesirable. On the other hand, a large $K_r$ yields smaller blocks and thus fewer locations within each block, leading to a larger variance of $\hat{\tmmathbf{\theta}}_m$. In this paper, we propose to set
\begin{equation}
  K_r \sim n^{\frac{d}{8 d + 8}} , \label{EQ:Krorder}
\end{equation}
which balances these two effects when computing $\hat{r}_n$.

The rationale for the choice of $K_r$ in (\ref{EQ:Krorder}) is as follows. Since the proportion of blocks intersecting the boundaries is of the order $k_0^{1/2}K_r^{-1}$, ensuring that this proportion converges to $0$ requires $K_r \gg k_0^{1/2}$. On the other hand, Equation (\ref{EQ:rnk0}) and Assumption \ref{AS:thetagap} imply that $k_0^{1/2} = O\left(n^{\frac{\tau d}{4d+4}}\right)$ for some $\tau < 1/2$. Hence, (\ref{EQ:Krorder}) specifies a slowly increasing order for $K_r$, while guaranteeing the proportion of blocks intersecting the boundaries converges to $0$.

To establish that $\hat{r}_n$ consistently estimates the order of $r_n$, we impose an additional condition on $r_n$. We assume
\begin{equation}
  r_n \gg n^{- \frac{3 d + 4}{8 d + 8}} . \label{EQ:rnadditional}
\end{equation}
With Equation (\ref{EQ:rnadditional}), we have the following result.

\begin{lemma}
  \label{PP:rnconsistency}Under Assumptions \ref{AS:distributionloc} - \ref{AS:covariates}, and
  Equations (\ref{EQ:rnk0}) and (\ref{EQ:Krorder}) - (\ref{EQ:rnadditional}), there exist constants
  $c, C > 0$, such that
  \begin{equation}
    \mathbb{P} \left( c < \frac{\hat{r}_n}{r_n} < C \right) \rightarrow 1.
    \label{EQ:rnhatconsistent}
  \end{equation}
\end{lemma}

Lemma \ref{PP:rnconsistency} justifies the proposed estimator $\hat{r}_n$: with probability tending to $1$, $\hat{r}_n$ is of the same order as $r_n$. Compared with the main paper, the only additional condition required for establishing Lemma \ref{PP:rnconsistency} is Equation (\ref{EQ:rnadditional}). Note that Equation (\ref{EQ:rnk0}) in the main paper implies $r_n \gg n^{-1/2}$. Thus, Equation (\ref{EQ:rnadditional}) strengthens Equation (\ref{EQ:rnk0}) by a slight margin, particularly when $d$ is small. The proof of Lemma \ref{PP:rnconsistency} is deferred to Section \ref{SEC:addproof}.

\subsection{Simulation}
\label{SEC:addsimu}
%\subsection{Effectiveness of $(k_0,r_n)$ estimation approach}
Sections \ref{SEC:simuk0} and \ref{SEC:simurn} evaluate the approaches proposed in Sections \ref{SEC:k0knownrn} and \ref{SEC:rn}, respectively.

% We adopt the same data-generating setting as in Section \ref{SEC:asymptotic}, except that the orders of $(k_0,r_n)$ are modified to reflect different parameters of interest. 

\subsubsection{Effectiveness of $k_0$ estimation with a known $r_n$ order}
\label{SEC:simuk0}
\begin{figure}[htbp]
		\centering
		\begin{tabular}{cc}
			{\includegraphics[width=0.32\linewidth,height=0.22\textheight]{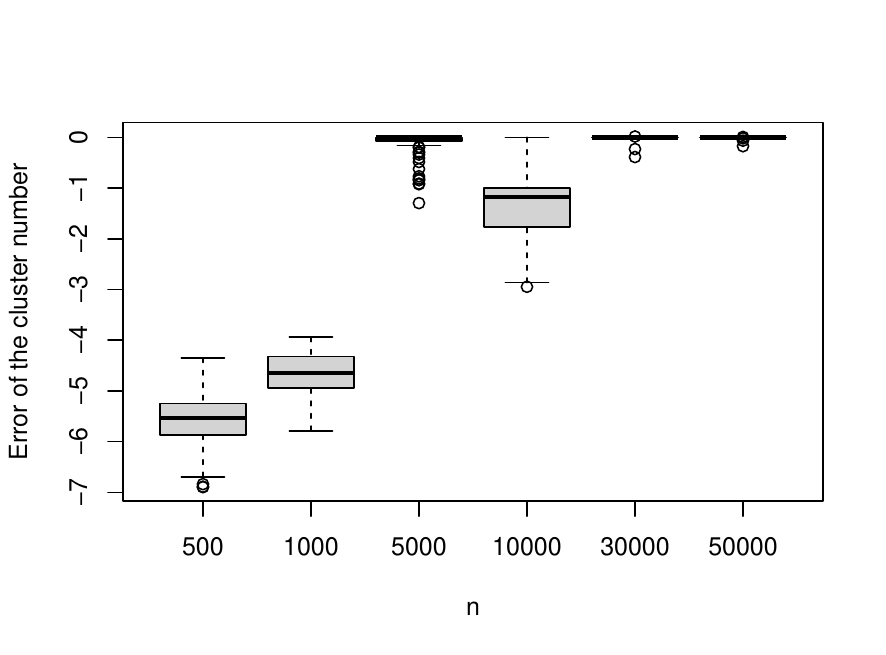}}&
			{\includegraphics[width=0.32\linewidth,height=0.22\textheight]{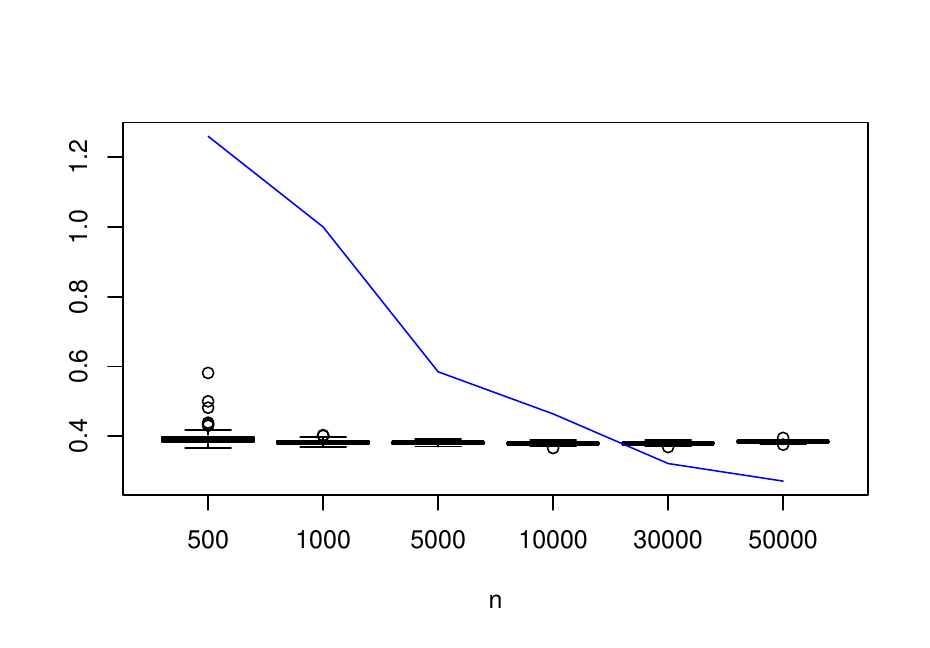}}
            \\
            {\small (a) } & {\small (b) }  
		\end{tabular}
		\caption{(a) Boxplots of the mean error in the posterior number of clusters across different $n$ and repeated simulations. (b) Boxplots of $\hat{r}_n / r_n$ across different $n$ and repeated simulations. The solid blue line represents values of $r_n$ under different $n$.}
		\label{FIG:k0rn}
	\end{figure}

We adopt the same data-generating setting as in Section \ref{SEC:asymptotic}, and examine posterior cluster number error under different sample sizes $n = (500,1000,5000,10000,30000,50000)$. Noting that $r_n$ is a constant in Section \ref{SEC:asymptotic}, and following Equation (\ref{EQ:firsthyper}), we set $K\sim n^{5/14}$ and $\log (\lambda^{- 1})\sim n^{5/7} \log^{1.5} (n)$. Figure \ref{FIG:k0rn}(a) shows boxplots of the mean error in the posterior number of clusters. We can see that after $n$ is larger than $30000$, the posterior number of clusters equals its true value in most repeats. This aligns with the arguments in Section \ref{SEC:k0knownrn}: we can use $k_0$-independent orders of $(K,\lambda)$ to obtain a consistent estimator of $k_0$ under the VDG regime.

%$\log (\lambda^{- 1}) = 0.001 n^{5/7} \log^{1.5} (n)$, $K=\lfloor 0.5n^{5/14} \rfloor$
	% \begin{figure}[htbp]
	% 	\centering
	% 		{\includegraphics[width=0.32\linewidth,height=0.22\textheight]{AOS/fig/k0asy.pdf}}
	% 	\caption{Boxplots of }
	% 	\label{FIG:k0asy}
 %        %\vspace{-25pt}
	% \end{figure}

\subsubsection{Effectiveness of $r_n$ order estimation}
\label{SEC:simurn}
% Same as Section \ref{SEC:asymptotic}, we consider $n = (500,1000,5000,10000,30000,50000)$. We set 

We adopt the same data-generating setting as in Section \ref{SEC:asymptotic}, except that we set $k_0 \equiv 10$ and $r_n = 10n^{-1/3}$ so that $r_n$ vanishes, making its estimation more challenging. Following Equation (\ref{EQ:Krorder}), we set $K_r=\lfloor 5n^{1/12} \rfloor$, compute $\hat{r}_n$ via Equation (\ref{EQ:rnhat}), and examine the ratio $\hat{r}_n / r_n$ under different sample sizes $n = (500,1000,5000,10000,30000,50000)$. We repeat the simulation $100$ times.

Figure \ref{FIG:k0rn}(b) shows boxplots of $\hat{r}_n / r_n$ and values of $r_n$. Although $r_n$ decreases substantially as $n$ increases, the ratio $\hat{r}_n/r_n$ remains approximately constant. This suggests that $\hat{r}_n$ provides an effective estimate of the order of $r_n$, consistent with Lemma \ref{PP:rnconsistency}. Note that the value of $\hat{r}_n/r_n$ is approximately $0.4$. We emphasize that we do not aim for $\hat{r}_n/r_n$ to converge to $1$, which is impractical since $r_n$ is defined only up to a multiplicative constant.

	% \begin{figure}[htbp]
	% 	\centering
	% 		{\includegraphics[width=0.32\linewidth,height=0.22\textheight]{AOS/fig/rn.pdf}}
	% 	\caption{Boxplots of $\hat{r}_n / r_n$ across different sample sizes $n$ and repeated simulations. The solid blue line represents values of $r_n$ under different $n$.}
	% 	\label{FIG:rn}
 %        %\vspace{-25pt}
	% \end{figure}

\subsection{Proof of Lemma \ref{PP:rnconsistency}}
\label{SEC:addproof}
Following Equations (\ref{EQ:A1n}) - (\ref{EQ:A2n}) in Section \ref{SEC:AE}, we define $\mathcal{A}_n$ based on blocks $\{B_m\}_{m=1}^{K_r^2}$. Following the similar argument in the proof of Lemma \ref{LM:An}, it can be shown that $\mathbb{P} (\mathcal{A}_n) \rightarrow 1$ with the $K_r$ order in
(\ref{EQ:Krorder}). Thus, it suffices to show that $\mathbb{P} \left( c <
\frac{\hat{r}_n}{r_n} < C \mid \mathcal{A}_n \right) \rightarrow 1$. To
simplify notations, in the following proof and by default, we will consider
the probability while conditional on $\{ \tmmathbf{s}_i, \tmmathbf{x}
(\tmmathbf{s}_i) \}_{i = 1}^n$ and $\mathcal{A}_n$, and all randomness
arises from $\{ \epsilon (\tmmathbf{s}_i) \}_{i = 1}^n$. We decompose
\[ \frac{\hat{r}^2_n}{r^2_n} = \frac{\hat{r}^2_{n, 1}}{r^2_n} +
   \frac{\hat{r}^2_{n, 2}}{r_n^2}, \]
where
\[ \hat{r}^2_{n, 1} = n^{- 1} \sum_{m = 1}^{K_r^2} \| B_m \| \mathbb{E} \|
   \hat{\tmmathbf{\theta}}_m \|_2^2, \]
and
\[ \hat{r}^2_{n, 2} = n^{- 1} \sum_{m = 1}^{K_r^2} \{ \| B_m \| (\|
   \hat{\tmmathbf{\theta}}_m \|_2^2 -\mathbb{E} \|
   \hat{\tmmathbf{\theta}}_m \|_2^2) \} . \]
Before studing $\hat{r}^2_{n, 1}$ and $\hat{r}^2_{n, 2}$, we first show a
result regarding $\{ \tmmathbf{\theta}_{l, 0} \}_{l = 1}^{k_0}$ in the
following lemma.

\begin{lemma}
  \label{LM:theta2norm}Under Assumption \ref{AS:thetagap}, we have
  \[ k_0^{- 1} \sum_{l = 1}^{k_0} \| \tmmathbf{\theta}_{l, 0} \|_2^2 \sim
     r_n^2 . \]
  \begin{proof}
    Since $\{ \tmmathbf{\theta}_{l, 0} \}_{l = 1}^{k_0}$ are distributed
    within a ball centered at $\tmmathbf{0}$ with radius $r_n$, it is
    obvious that $k_0^{- 1} \sum_{l = 1}^{k_0} \| \tmmathbf{\theta}_{l, 0}
    \|_2^2 \leqslant r_n^2$. We will next show that $k_0^{- 1} \sum_{l =
    1}^{k_0} \| \tmmathbf{\theta}_{l, 0} \|_2^2 > c r_n^2$ for some constant
    $c$.
    
    Let $c_1$ be the constant in Assumption \ref{AS:thetagap} such that
    $\min_{l \neq l'} \| \tmmathbf{\theta}_{l, 0} - \tmmathbf{\theta}_{l', 0}
    \|_2 > c_1 k_0^{- 1 / d} r_n$. Write
    \[ \tmop{Ball} (p_1, p_2) = \{ \tmmathbf{\theta} \in \mathbb{R}^d : p_1
       c_1 k_0^{- 1 / d} r_n \leqslant \| \tmmathbf{\theta} \|_2 < p_2 c_1
       k_0^{- 1 / d} r_n \} \]
    for intergers $0 \leqslant p_1 \leqslant p_2$. We can see that the space
    $\{ \tmmathbf{\theta} \in \mathbb{R}^d : \| \tmmathbf{\theta} \|_2
    \leqslant r_n \}$ is contained within
    \[ \cup_{p = 1}^{\lfloor c_1^{- 1} k_0^{1 / d} \rfloor + 1} \tmop{Ball} (p
       - 1, p) . \]
    Correspondingly, we divide $\{ \tmmathbf{\theta}_{l, 0} \}_{l = 1}^{k_0}$ into
    \[ \{ \tmmathbf{\theta}_{l, 0} \}_{l = 1}^{k_0} = \cup_{p = 1}^{\lfloor
       c_1^{- 1} k_0^{1 / d} \rfloor + 1} S_p, \]
    where $S_p$ is the set of $\tmmathbf{\theta}_{l, 0}$'s that are contained
    in $\tmop{Ball} (p - 1, p)$. We claim that there exists a constant $c$,
    such that $| S_p | \leqslant c (p + 1)^{d - 1}$, $\forall p \in \{ 1,
    \ldots, \lfloor c_1^{- 1} k_0^{1 / d} \rfloor + 1 \}$.
    
    We prove the claim as follows. For a given $p \geqslant 2$ and every
    $\tmmathbf{\theta} \in S_p$, we construct a ball centered at
    $\tmmathbf{\theta}$ with radius $2^{- 1} c_1 k_0^{- 1 / d} r_n$. From
    the construction and Assumption \ref{AS:thetagap}, those balls are
    disjoint, and are contained in $\tmop{Ball} (p - 2, p + 1)$. By comparing
    volumes between those balls and $\tmop{Ball} (p - 2, p + 1)$, we have
    \begin{eqnarray*}
      | S_p | \times c \times 2^{- d} c_1^d k_0^{- 1} r_n^d & \leqslant & c
      \times \{ (p + 1)^d - (p - 2)^d \} \times c_1^d k_0^{- 1} r_n^d\\
      & \leqslant & c \times 3 d (p + 1)^{d - 1} \times c_1^d k_0^{- 1}
      r_n^d,
    \end{eqnarray*}
    leading to
    \begin{equation}
      | S_p | \leqslant 3 \times 2^d \times d (p + 1)^{d - 1} = c (p + 1)^{d -
      1} . \label{EQ:Sp}
    \end{equation}
    Applying the similar argument, it is easy to prove the case for $p = 1$.
    The claim is thus proved.
    
    Next, let $c_2$ be the constant in (\ref{EQ:Sp}) such that $| S_p |
    \leqslant c_2 (p + 1)^{d - 1}, \forall p \in \{ 1, \ldots, \lfloor c_1^{-
    1} k_0^{1 / d} \rfloor + 1 \}$. We write $\tilde{p}$ as the integer such
    that
    \begin{equation}
      c_2 (\tilde{p} + 1)^d \leqslant k_0 < c_2 (\tilde{p} + 2)^d .
      \label{EQ:k0ptuta}
    \end{equation}
    On the other hand, note that
    \[ \sum_{p = 1}^{\tilde{p}} | S_p | \leqslant \sum_{p = 1}^{\tilde{p} +
       1} c_2 p^{d - 1} \leqslant c_2 (\tilde{p} + 1)^d \leqslant k_0, \]
    which means at most $c_2 (\tilde{p} + 1)^d$ elements of $\{
    \tmmathbf{\theta}_{l, 0} \}_{l = 1}^{k_0}$ are within $\tmop{Ball} (0,
    \tilde{p})$. Note further that this also corresponds to the case when $k_0^{- 1} \sum_{l =
    1}^{k_0} \| \tmmathbf{\theta}_{l, 0} \|_2^2$ achieves its minimum. We thus derive the lower bound as
    % placement of $\{
    % \tmmathbf{\theta}_{l, 0} \}_{l = 1}^{k_0}$ with minimum $k_0^{- 1} \sum_{l =
    % 1}^{k_0} \| \tmmathbf{\theta}_{l, 0} \|_2^2$. We can thus derive the lower bound of $k_0^{- 1} \sum_{l =
    % 1}^{k_0} \| \tmmathbf{\theta}_{l, 0} \|_2^2$ as
    \begin{eqnarray*}
      k_0^{- 1} \sum_{l = 1}^{k_0} \| \tmmathbf{\theta}_{l, 0} \|_2^2 &
      \geqslant & k_0^{- 1} \sum_{p = 1}^{\tilde{p}} c_2 (p + 1)^{d - 1}
      \times \{ (p - 1) c_1 k_0^{- 1 / d} r_n \}^2\\
      & \geqslant & c_1^2 k_0^{- 1} k_0^{- 2 / d} r_n^2 \sum_{p =
      1}^{\tilde{p}} c_2 (p - 1)^{d + 1}\\
      & \geqslant & c_1^2 k_0^{- 1} k_0^{- 2 / d} r_n^2 c_2 \left( 1 +
      \frac{(\tilde{p} - 1)^{d + 2}}{d + 2} \right) \overset{\text{Equation
      (\ref{EQ:k0ptuta})}}{\geqslant} c r_n^2 .
    \end{eqnarray*}
    The lemma is proved.
  \end{proof}
\end{lemma}

The next two lemmas study properties of $\frac{\hat{r}^2_{n, 1}}{r_n^2}$ and
$\frac{\hat{r}^2_{n, 2}}{r_n^2}$, respectively.

\begin{lemma}
  \label{LM:rn1}Under $\mathcal{A}_n$, Assumptions \ref{AS:distributionloc} - \ref{AS:covariates},
  and Equations (\ref{EQ:rnk0}) and (\ref{EQ:Krorder}) - (\ref{EQ:rnadditional}), we have
  \[ \frac{\hat{r}^2_{n, 1}}{r_n^2} \sim 1. \]
  \begin{proof}
    Under $\mathcal{A}_n$, there exist constants $c, C > 0$, such that
    \[ c \| \hat{\tmmathbf{\theta}}_m \|_2^2 \leqslant \| B_m \|^{- 1} \|
       \tmmathbf{x}_m \hat{\tmmathbf{\theta}}_m \|_2^2 \leqslant C \|
       \hat{\tmmathbf{\theta}}_m \|_2^2 \]
    for $\forall m \in \{ 1, \ldots, K_r^2 \}$. Thus, it suffices to show that
    \[ n^{- 1} \sum_{m = 1}^{K_r^2} \mathbb{E} \| \tmmathbf{x}_m
       \hat{\tmmathbf{\theta}}_m \|_2^2 \sim r_n^2 . \]
    Write $P_m = \tmmathbf{x}_m (\tmmathbf{x}_m^T \tmmathbf{x}_m)^{- 1}
    \tmmathbf{x}_m^T$ as the projection matrix constructed from
    $\tmmathbf{x}_m$, and write $\tmmathbf{\mu}_m$ and $\tmmathbf{\epsilon}_m$
    as the regression means and noises within block $B_m$. We compute
    \begin{eqnarray*}
      \mathbb{E} \| \tmmathbf{x}_m \hat{\tmmathbf{\theta}}_m \|_2^2 & = &
      \mathbb{E} \| P_m \tmmathbf{y}_m \|_2^2 = \| P_m \tmmathbf{\mu}_m \|_2^2
      +\mathbb{E} \| P_m \tmmathbf{\epsilon}_m \|_2^2
       \overset{(i)}{=}  \| P_m \tmmathbf{\mu}_m \|_2^2 + d \sigma_0^2,
    \end{eqnarray*}
    where $(i)$ is because $\sigma_0^{- 2} \tmmathbf{\epsilon}_m^T P_m
    \tmmathbf{\epsilon}_m$ follows a chi-square distribution with degree of
    freedom $d$ by Lemma \ref{LM:chisquaredis}. Hence,
    \begin{eqnarray*}
      n^{- 1} \sum_{m = 1}^{K_r^2} \mathbb{E} \| \tmmathbf{x}_m
      \hat{\tmmathbf{\theta}}_m \|_2^2 & = & n^{- 1} \sum_{m = 1}^{K_r^2}
      (\| P_m \tmmathbf{\mu}_m \|_2^2 + d \sigma_0^2)\\
      & \leqslant & c r_n^2 + n^{- 1} K_r^2 d \sigma_0^2 \leqslant c r_n^2,
    \end{eqnarray*}
    where the last inequality uses (\ref{EQ:Krorder}) and
    (\ref{EQ:rnadditional}). It next remains to show the lower bound of $n^{-
    1} \sum_{m = 1}^{K_r^2} \mathbb{E} \| \tmmathbf{x}_m
    \hat{\tmmathbf{\theta}}_m \|_2^2$.
    
    For each sub-domain $\mathcal{D}_{l, 0}$, we write $\mathcal{F}_l$ as the set of
    blocks fully contained in $\mathcal{D}_{l, 0}$. According to Assumption
    \ref{AS:lengthofboundary} and and the arguments after Equation
    (\ref{EQ:Krorder}) that $K_r \gg k_0^{1 / 2}$, $| \mathcal{F}_l |$ is lower bounded by $c k_0^{- 1} K_r^2$ for some $c>0$. Thus,
    \begin{eqnarray*}
      n^{- 1} \sum_{m = 1}^{K_r^2} \mathbb{E} \| \tmmathbf{x}_m
      \hat{\tmmathbf{\theta}}_m \|_2^2 & = & n^{- 1} \sum_{m = 1}^{K_r^2}
      (\| P_m \tmmathbf{\mu}_m \|_2^2 + d \sigma_0^2)\\
      & \geqslant & n^{- 1} \sum_{l = 1}^{k_0} \sum_{B_m \in \mathcal{F}_l}
      \| P_m \tmmathbf{\mu}_m \|_2^2 = n^{- 1} \sum_{l = 1}^{k_0} \sum_{B_m
      \in \mathcal{F}_l} \| \tmmathbf{x}_m \tmmathbf{\theta}_{l, 0} \|_2^2\\
      & \overset{\mathcal{A}_n}{\geqslant} & c n^{- 1} \sum_{l = 1}^{k_0}
      k_0^{- 1} K_r^2 \times \| B_m \| \times \| \tmmathbf{\theta}_{l, 0}
      \|_2^2\\
      & \overset{\mathcal{A}_n}{\geqslant} & c k_0^{- 1}  \sum_{l = 1}^{k_0}
      \| \tmmathbf{\theta}_{l, 0} \|_2^2 \overset{\text{Lemma
      \ref{LM:theta2norm}}}{\geqslant} c r_n^2 .
    \end{eqnarray*}
    The lemma is proved by combining the above results.
  \end{proof}
\end{lemma}

\begin{lemma}
  \label{LM:rn2}Under $\mathcal{A}_n$, Assumptions \ref{AS:distributionloc} - \ref{AS:covariates},
  and Equations (\ref{EQ:Krorder}) - (\ref{EQ:rnadditional}), we have
  \[ \frac{\hat{r}^2_{n, 2}}{r^2_n} = o_p (1) . \]
  \begin{proof}
    Note that for $m \neq m'$, $\hat{\tmmathbf{\theta}}_m$ is independent
    from $\hat{\tmmathbf{\theta}}_{m'}$. On the other hand, we have
    \begin{eqnarray*}
      \mathbb{E} \| \hat{\tmmathbf{\theta}}_m \|_2^4 &
      \overset{\mathcal{A}_n}{\leqslant} & c \| B_m \|^{- 2} \mathbb{E} \|
      \tmmathbf{x}_m \hat{\tmmathbf{\theta}}_m \|_2^4\\
      & \leqslant & c \| B_m \|^{- 2} \mathbb{E} \| P_m \tmmathbf{\mu}_m +
      P_m \tmmathbf{\epsilon}_m \|_2^4\\
      & \leqslant & c \| B_m \|^{- 2} \| P_m \tmmathbf{\mu}_m \|_2^4 + c \|
      B_m \|^{- 2} \mathbb{E} \| P_m \tmmathbf{\epsilon}_m \|_2^4\\
      & \leqslant & c r_n^4 + c \| B_m \|^{- 2},
    \end{eqnarray*}
    where the last inequality uses $\mathcal{A}_n$ and the fact that
    $\sigma_0^{- 2} \tmmathbf{\epsilon}_m^T P_m \tmmathbf{\epsilon}_m$ follows
    a chi-square distribution. Thus, we compute
    \begin{eqnarray*}
      \mathbb{E} \frac{\hat{r}^4_{n, 2}}{r^4_n} & = & n^{- 2} \sum_{m =
      1}^{K_r^2} \frac{\| B_m \|^2 \mathbb{E} (\|
      \hat{\tmmathbf{\theta}}_m \|_2^2 -\mathbb{E} \|
      \hat{\tmmathbf{\theta}}_m \|_2^2)^2}{r_n^4}\\
      & \leqslant & n^{- 2} \sum_{m = 1}^{K_r^2} (c \| B_m \|^2 + r_n^{- 4})
      \overset{\mathcal{A}_n}{\leqslant} c n^{- 2} \times K_r^2 \times (n^2
      K_r^{- 4} + r_n^{- 4})\\
      & = & c (K_r^{- 2} + n^{- 2} K_r^2 r_n^{- 4}),
    \end{eqnarray*}
    which converges to $0$ according to (\ref{EQ:Krorder}) and
    (\ref{EQ:rnadditional}). The lemma is proved.
  \end{proof}
\end{lemma}

Combining results of Lemmas \ref{LM:rn1} - \ref{LM:rn2}, Lemma
\ref{PP:rnconsistency} is proved.

% \section{Additional simulation results}
% \label{SEC:addsimu}
% This section provides additional simulation results. In Section \ref{SEC:compareBSCC}, we compare the performance of our model with BSCC in {\cite{luo2021bayesian}}. In Section \ref{SEC:CHWAIC}, we examine the effectiveness of using the CH index/WAIC to select $(K,\lambda)$, as proposed in Section \ref{SEC:hyperparameterselection}.

\newpage

\bibliographystyle{apalike} % Style BST file (imsart-number.bst or imsart-nameyear.bst)
\bibliography{clusterrefs}% Bibliography file (usually '*.bib')

\end{document}